\documentclass[longauth,traditabstract]{aa} 

\def\setsymbol#1#2{\expandafter\def\csname #1\endcsname{#2}}
\def\getsymbol#1{\csname #1\endcsname}

\def\Planck{{\it Planck\/}}

\def\allearlypapers{\nocite{planck2011-1.1, planck2011-1.3, planck2011-1.4, planck2011-1.5, planck2011-1.6, planck2011-1.7, planck2011-1.10, planck2011-1.10sup, planck2011-5.1a, planck2011-5.1b, planck2011-5.2a, planck2011-5.2b, planck2011-5.2c, planck2011-6.1, planck2011-6.2, planck2011-6.3a, planck2011-6.4a, planck2011-6.4b, planck2011-6.6, planck2011-7.0, planck2011-7.2, planck2011-7.3, planck2011-7.7a, planck2011-7.7b, planck2011-7.12, planck2011-7.13}}

\newbox\tablebox    \newdimen\tablewidth
\def\leaderfil{\leaders\hbox to 5pt{\hss.\hss}\hfil}
\def\tablenote#1 #2\par{\begingroup \parindent=0.8em
    \abovedisplayshortskip=0pt\belowdisplayshortskip=0pt
    \noindent
    $$\hss\vbox{\hsize\tablewidth \hangindent=\parindent \hangafter=1 \noindent
    \hbox to \parindent{\sup{\rm #1}\hss}\strut#2\strut\par}\hss$$
    \endgroup}

\def\L2{\ifmmode L_2\else $L_2$\fi}

\def\DeltaT{\ifmmode \Delta T\else $\Delta T$\fi}
\def\deltat{\ifmmode \Delta t\else $\Delta t$\fi}
\def\fknee{\ifmmode f_{\rm knee}\else $f_{\rm knee}$\fi}
\def\Fmax{\ifmmode F_{\rm max}\else $F_{\rm max}$\fi}
\def\solar{\ifmmode{\rm M}_{\mathord\odot}\else${\rm M}_{\mathord\odot}$\fi}

\def\inv{\ifmmode^{-1}\else$^{-1}$\fi}
\def\mo{\ifmmode^{-1}\else$^{-1}$\fi}
\def\sup#1{\ifmmode ^{\rm #1}\else $^{\rm #1}$\fi}
\def\expo#1{\ifmmode \times 10^{#1}\else $\times 10^{#1}$\fi}
\def\,{\thinspace}
\def\lsim{\mathrel{\raise .4ex\hbox{\rlap{$<$}\lower 1.2ex\hbox{$\sim$}}}}
\def\gsim{\mathrel{\raise .4ex\hbox{\rlap{$>$}\lower 1.2ex\hbox{$\sim$}}}}

\def\simprop{\mathrel{\raise .4ex\hbox{\rlap{$\propto$}\lower 1.2ex\hbox{$\sim$}}}}
\def\deg{\ifmmode^\circ\else$^\circ$\fi}
\def\pdeg{\ifmmode $\setbox0=\hbox{$^{\circ}$}\rlap{\hskip.11\wd0 .}$^{\circ}
          \else \setbox0=\hbox{$^{\circ}$}\rlap{\hskip.11\wd0 .}$^{\circ}$\fi}
\def\arcs{\ifmmode {^{\scriptstyle\prime\prime}}
          \else $^{\scriptstyle\prime\prime}$\fi}
\def\arcm{\ifmmode {^{\scriptstyle\prime}}
          \else $^{\scriptstyle\prime}$\fi}
\newdimen\sa  \newdimen\sb
\def\parcs{\sa=.07em \sb=.03em
     \ifmmode \hbox{\rlap{.}}^{\scriptstyle\prime\kern -\sb\prime}\hbox{\kern -\sa}
     \else \rlap{.}$^{\scriptstyle\prime\kern -\sb\prime}$\kern -\sa\fi}
\def\parcm{\sa=.08em \sb=.03em
     \ifmmode \hbox{\rlap{.}\kern\sa}^{\scriptstyle\prime}\hbox{\kern-\sb}
     \else \rlap{.}\kern\sa$^{\scriptstyle\prime}$\kern-\sb\fi}
\def\ra[#1 #2 #3.#4]{#1\sup{h}#2\sup{m}#3\sup{s}\llap.#4}
\def\dec[#1 #2 #3.#4]{#1\deg#2\arcm#3\arcs\llap.#4}
\def\deco[#1 #2 #3]{#1\deg#2\arcm#3\arcs}
\def\rra[#1 #2]{#1\sup{h}#2\sup{m}}

\def\dots{\relax\ifmmode \ldots\else $\ldots$\fi}
\def\WHzsr{\ifmmode $W\,Hz\mo\,sr\mo$\else W\,Hz\mo\,sr\mo\fi}
\def\mHz{\ifmmode $\,mHz$\else \,mHz\fi}
\def\GHz{\ifmmode $\,GHz$\else \,GHz\fi}
\def\mKs{\ifmmode $\,mK\,s$^{1/2}\else \,mK\,s$^{1/2}$\fi}
\def\muKs{\ifmmode \,\mu$K\,s$^{1/2}\else \,$\mu$K\,s$^{1/2}$\fi}
\def\muKRJs{\ifmmode \,\mu$K$_{\rm RJ}$\,s$^{1/2}\else \,$\mu$K$_{\rm RJ}$\,s$^{1/2}$\fi}
\def\muKHz{\ifmmode \,\mu$K\,Hz$^{-1/2}\else \,$\mu$K\,Hz$^{-1/2}$\fi}
\def\MJysr{\ifmmode \,$MJy\,sr\mo$\else \,MJy\,sr\mo\fi}
\def\MJysrmK{\ifmmode \,$MJy\,sr\mo$\,mK$_{\rm CMB}\mo\else \,MJy\,sr\mo\,mK$_{\rm CMB}\mo$\fi}
\def\microns{\ifmmode \,\mu$m$\else \,$\mu$m\fi}

\def\muK{\ifmmode \,\mu$K$\else \,$\mu$\hbox{K}\fi}
\def\microK{\ifmmode \,\mu$K$\else \,$\mu$\hbox{K}\fi}
\def\muW{\ifmmode \,\mu$W$\else \,$\mu$\hbox{W}\fi}
\def\kms{\ifmmode $\,km\,s$^{-1}\else \,km\,s$^{-1}$\fi}
\def\kmsMpc{\ifmmode $\,\kms\,Mpc\mo$\else \,\kms\,Mpc\mo\fi}
\setsymbol{LFI:center:frequency:70GHz:units}{70.3\,GHz}
\setsymbol{LFI:center:frequency:44GHz:units}{44.1\,GHz}
\setsymbol{LFI:center:frequency:30GHz:units}{28.5\,GHz}

\setsymbol{LFI:center:frequency:70GHz}{70.3}
\setsymbol{LFI:center:frequency:44GHz}{44.1}
\setsymbol{LFI:center:frequency:30GHz}{28.5}

\setsymbol{LFI:center:frequency:LFI18:Rad:M:units}{71.7\GHz}
\setsymbol{LFI:center:frequency:LFI19:Rad:M:units}{67.5\GHz}
\setsymbol{LFI:center:frequency:LFI20:Rad:M:units}{69.2\GHz}
\setsymbol{LFI:center:frequency:LFI21:Rad:M:units}{70.4\GHz}
\setsymbol{LFI:center:frequency:LFI22:Rad:M:units}{71.5\GHz}
\setsymbol{LFI:center:frequency:LFI23:Rad:M:units}{70.8\GHz}
\setsymbol{LFI:center:frequency:LFI24:Rad:M:units}{44.4\GHz}
\setsymbol{LFI:center:frequency:LFI25:Rad:M:units}{44.0\GHz}
\setsymbol{LFI:center:frequency:LFI26:Rad:M:units}{43.9\GHz}
\setsymbol{LFI:center:frequency:LFI27:Rad:M:units}{28.3\GHz}
\setsymbol{LFI:center:frequency:LFI28:Rad:M:units}{28.8\GHz}
\setsymbol{LFI:center:frequency:LFI18:Rad:S:units}{70.1\GHz}
\setsymbol{LFI:center:frequency:LFI19:Rad:S:units}{69.6\GHz}
\setsymbol{LFI:center:frequency:LFI20:Rad:S:units}{69.5\GHz}
\setsymbol{LFI:center:frequency:LFI21:Rad:S:units}{69.5\GHz}
\setsymbol{LFI:center:frequency:LFI22:Rad:S:units}{72.8\GHz}
\setsymbol{LFI:center:frequency:LFI23:Rad:S:units}{71.3\GHz}
\setsymbol{LFI:center:frequency:LFI24:Rad:S:units}{44.1\GHz}
\setsymbol{LFI:center:frequency:LFI25:Rad:S:units}{44.1\GHz}
\setsymbol{LFI:center:frequency:LFI26:Rad:S:units}{44.1\GHz}
\setsymbol{LFI:center:frequency:LFI27:Rad:S:units}{28.5\GHz}
\setsymbol{LFI:center:frequency:LFI28:Rad:S:units}{28.2\GHz}

\setsymbol{LFI:center:frequency:LFI18:Rad:M}{71.7}
\setsymbol{LFI:center:frequency:LFI19:Rad:M}{67.5}
\setsymbol{LFI:center:frequency:LFI20:Rad:M}{69.2}
\setsymbol{LFI:center:frequency:LFI21:Rad:M}{70.4}
\setsymbol{LFI:center:frequency:LFI22:Rad:M}{71.5}
\setsymbol{LFI:center:frequency:LFI23:Rad:M}{70.8}
\setsymbol{LFI:center:frequency:LFI24:Rad:M}{44.4}
\setsymbol{LFI:center:frequency:LFI25:Rad:M}{44.0}
\setsymbol{LFI:center:frequency:LFI26:Rad:M}{43.9}
\setsymbol{LFI:center:frequency:LFI27:Rad:M}{28.3}
\setsymbol{LFI:center:frequency:LFI28:Rad:M}{28.8}
\setsymbol{LFI:center:frequency:LFI18:Rad:S}{70.1}
\setsymbol{LFI:center:frequency:LFI19:Rad:S}{69.6}
\setsymbol{LFI:center:frequency:LFI20:Rad:S}{69.5}
\setsymbol{LFI:center:frequency:LFI21:Rad:S}{69.5}
\setsymbol{LFI:center:frequency:LFI22:Rad:S}{72.8}
\setsymbol{LFI:center:frequency:LFI23:Rad:S}{71.3}
\setsymbol{LFI:center:frequency:LFI24:Rad:S}{44.1}
\setsymbol{LFI:center:frequency:LFI25:Rad:S}{44.1}
\setsymbol{LFI:center:frequency:LFI26:Rad:S}{44.1}
\setsymbol{LFI:center:frequency:LFI27:Rad:S}{28.5}
\setsymbol{LFI:center:frequency:LFI28:Rad:S}{28.2}

\setsymbol{LFI:white:noise:sensitivity:70GHz:units}{152.6\muKs}
\setsymbol{LFI:white:noise:sensitivity:44GHz:units}{173.1\muKs}
\setsymbol{LFI:white:noise:sensitivity:30GHz:units}{146.8\muKs}

\setsymbol{LFI:white:noise:sensitivity:70GHz}{152.6}
\setsymbol{LFI:white:noise:sensitivity:44GHz}{173.1}
\setsymbol{LFI:white:noise:sensitivity:30GHz}{146.8}

\setsymbol{LFI:white:noise:sensitivity:LFI18:Rad:M:units}{512.0\muKs}
\setsymbol{LFI:white:noise:sensitivity:LFI19:Rad:M:units}{581.4\muKs}
\setsymbol{LFI:white:noise:sensitivity:LFI20:Rad:M:units}{590.8\muKs}
\setsymbol{LFI:white:noise:sensitivity:LFI21:Rad:M:units}{455.2\muKs}
\setsymbol{LFI:white:noise:sensitivity:LFI22:Rad:M:units}{492.0\muKs}
\setsymbol{LFI:white:noise:sensitivity:LFI23:Rad:M:units}{507.7\muKs}
\setsymbol{LFI:white:noise:sensitivity:LFI24:Rad:M:units}{462.2\muKs}
\setsymbol{LFI:white:noise:sensitivity:LFI25:Rad:M:units}{413.6\muKs}
\setsymbol{LFI:white:noise:sensitivity:LFI26:Rad:M:units}{478.6\muKs}
\setsymbol{LFI:white:noise:sensitivity:LFI27:Rad:M:units}{277.7\muKs}
\setsymbol{LFI:white:noise:sensitivity:LFI28:Rad:M:units}{312.3\muKs}
\setsymbol{LFI:white:noise:sensitivity:LFI18:Rad:S:units}{465.7\muKs}
\setsymbol{LFI:white:noise:sensitivity:LFI19:Rad:S:units}{555.6\muKs}
\setsymbol{LFI:white:noise:sensitivity:LFI20:Rad:S:units}{623.2\muKs}
\setsymbol{LFI:white:noise:sensitivity:LFI21:Rad:S:units}{564.1\muKs}
\setsymbol{LFI:white:noise:sensitivity:LFI22:Rad:S:units}{534.4\muKs}
\setsymbol{LFI:white:noise:sensitivity:LFI23:Rad:S:units}{542.4\muKs}
\setsymbol{LFI:white:noise:sensitivity:LFI24:Rad:S:units}{399.2\muKs}
\setsymbol{LFI:white:noise:sensitivity:LFI25:Rad:S:units}{392.6\muKs}
\setsymbol{LFI:white:noise:sensitivity:LFI26:Rad:S:units}{418.6\muKs}
\setsymbol{LFI:white:noise:sensitivity:LFI27:Rad:S:units}{302.9\muKs}
\setsymbol{LFI:white:noise:sensitivity:LFI28:Rad:S:units}{285.3\muKs}

\setsymbol{LFI:white:noise:sensitivity:LFI18:Rad:M}{512.0}
\setsymbol{LFI:white:noise:sensitivity:LFI19:Rad:M}{581.4}
\setsymbol{LFI:white:noise:sensitivity:LFI20:Rad:M}{590.8}
\setsymbol{LFI:white:noise:sensitivity:LFI21:Rad:M}{455.2}
\setsymbol{LFI:white:noise:sensitivity:LFI22:Rad:M}{492.0}
\setsymbol{LFI:white:noise:sensitivity:LFI23:Rad:M}{507.7}
\setsymbol{LFI:white:noise:sensitivity:LFI24:Rad:M}{462.2}
\setsymbol{LFI:white:noise:sensitivity:LFI25:Rad:M}{413.6}
\setsymbol{LFI:white:noise:sensitivity:LFI26:Rad:M}{478.6}
\setsymbol{LFI:white:noise:sensitivity:LFI27:Rad:M}{277.7}
\setsymbol{LFI:white:noise:sensitivity:LFI28:Rad:M}{312.3}
\setsymbol{LFI:white:noise:sensitivity:LFI18:Rad:S}{465.7}
\setsymbol{LFI:white:noise:sensitivity:LFI19:Rad:S}{555.6}
\setsymbol{LFI:white:noise:sensitivity:LFI20:Rad:S}{623.2}
\setsymbol{LFI:white:noise:sensitivity:LFI21:Rad:S}{564.1}
\setsymbol{LFI:white:noise:sensitivity:LFI22:Rad:S}{534.4}
\setsymbol{LFI:white:noise:sensitivity:LFI23:Rad:S}{542.4}
\setsymbol{LFI:white:noise:sensitivity:LFI24:Rad:S}{399.2}
\setsymbol{LFI:white:noise:sensitivity:LFI25:Rad:S}{392.6}
\setsymbol{LFI:white:noise:sensitivity:LFI26:Rad:S}{418.6}
\setsymbol{LFI:white:noise:sensitivity:LFI27:Rad:S}{302.9}
\setsymbol{LFI:white:noise:sensitivity:LFI28:Rad:S}{285.3}

\setsymbol{LFI:knee:frequency:70GHz:units}{29.5\mHz}
\setsymbol{LFI:knee:frequency:44GHz:units}{56.2\mHz}
\setsymbol{LFI:knee:frequency:30GHz:units}{113.7\mHz}

\setsymbol{LFI:knee:frequency:70GHz}{29.5}
\setsymbol{LFI:knee:frequency:44GHz}{56.2}
\setsymbol{LFI:knee:frequency:30GHz}{113.7}

\setsymbol{LFI:knee:frequency:LFI18:Rad:M:units}{16.3\mHz}
\setsymbol{LFI:knee:frequency:LFI19:Rad:M:units}{15.1\mHz}
\setsymbol{LFI:knee:frequency:LFI20:Rad:M:units}{18.7\mHz}
\setsymbol{LFI:knee:frequency:LFI21:Rad:M:units}{37.2\mHz}
\setsymbol{LFI:knee:frequency:LFI22:Rad:M:units}{12.7\mHz}
\setsymbol{LFI:knee:frequency:LFI23:Rad:M:units}{34.6\mHz}
\setsymbol{LFI:knee:frequency:LFI24:Rad:M:units}{46.2\mHz}
\setsymbol{LFI:knee:frequency:LFI25:Rad:M:units}{24.9\mHz}
\setsymbol{LFI:knee:frequency:LFI26:Rad:M:units}{67.6\mHz}
\setsymbol{LFI:knee:frequency:LFI27:Rad:M:units}{187.4\mHz}
\setsymbol{LFI:knee:frequency:LFI28:Rad:M:units}{122.2\mHz}
\setsymbol{LFI:knee:frequency:LFI18:Rad:S:units}{17.7\mHz}
\setsymbol{LFI:knee:frequency:LFI19:Rad:S:units}{22.0\mHz}
\setsymbol{LFI:knee:frequency:LFI20:Rad:S:units}{8.7\mHz}
\setsymbol{LFI:knee:frequency:LFI21:Rad:S:units}{25.9\mHz}
\setsymbol{LFI:knee:frequency:LFI22:Rad:S:units}{15.8\mHz}
\setsymbol{LFI:knee:frequency:LFI23:Rad:S:units}{129.8\mHz}
\setsymbol{LFI:knee:frequency:LFI24:Rad:S:units}{100.9\mHz}
\setsymbol{LFI:knee:frequency:LFI25:Rad:S:units}{38.9\mHz}
\setsymbol{LFI:knee:frequency:LFI26:Rad:S:units}{58.9\mHz}
\setsymbol{LFI:knee:frequency:LFI27:Rad:S:units}{104.4\mHz}
\setsymbol{LFI:knee:frequency:LFI28:Rad:S:units}{40.7\mHz}

\setsymbol{LFI:knee:frequency:LFI18:Rad:M}{16.3}
\setsymbol{LFI:knee:frequency:LFI19:Rad:M}{15.1}
\setsymbol{LFI:knee:frequency:LFI20:Rad:M}{18.7}
\setsymbol{LFI:knee:frequency:LFI21:Rad:M}{37.2}
\setsymbol{LFI:knee:frequency:LFI22:Rad:M}{12.7}
\setsymbol{LFI:knee:frequency:LFI23:Rad:M}{34.6}
\setsymbol{LFI:knee:frequency:LFI24:Rad:M}{46.2}
\setsymbol{LFI:knee:frequency:LFI25:Rad:M}{24.9}
\setsymbol{LFI:knee:frequency:LFI26:Rad:M}{67.6}
\setsymbol{LFI:knee:frequency:LFI27:Rad:M}{187.4}
\setsymbol{LFI:knee:frequency:LFI28:Rad:M}{122.2}
\setsymbol{LFI:knee:frequency:LFI18:Rad:S}{17.7}
\setsymbol{LFI:knee:frequency:LFI19:Rad:S}{22.0}
\setsymbol{LFI:knee:frequency:LFI20:Rad:S}{8.7}
\setsymbol{LFI:knee:frequency:LFI21:Rad:S}{25.9}
\setsymbol{LFI:knee:frequency:LFI22:Rad:S}{15.8}
\setsymbol{LFI:knee:frequency:LFI23:Rad:S}{129.8}
\setsymbol{LFI:knee:frequency:LFI24:Rad:S}{100.9}
\setsymbol{LFI:knee:frequency:LFI25:Rad:S}{38.9}
\setsymbol{LFI:knee:frequency:LFI26:Rad:S}{58.9}
\setsymbol{LFI:knee:frequency:LFI27:Rad:S}{104.4}
\setsymbol{LFI:knee:frequency:LFI28:Rad:S}{40.7}

\setsymbol{LFI:slope:70GHz:units}{$-1.03$\mHz}
\setsymbol{LFI:slope:44GHz:units}{$-0.89$\mHz}
\setsymbol{LFI:slope:30GHz:units}{$-0.87$\mHz}

\setsymbol{LFI:slope:70GHz}{$-1.03$}
\setsymbol{LFI:slope:44GHz}{$-0.89$}
\setsymbol{LFI:slope:30GHz}{$-0.87$}

\setsymbol{LFI:slope:LFI18:Rad:M:units}{$-1.04$\mHz}
\setsymbol{LFI:slope:LFI19:Rad:M:units}{$-1.09$\mHz}
\setsymbol{LFI:slope:LFI20:Rad:M:units}{$-0.69$\mHz}
\setsymbol{LFI:slope:LFI21:Rad:M:units}{$-1.56$\mHz}
\setsymbol{LFI:slope:LFI22:Rad:M:units}{$-1.01$\mHz}
\setsymbol{LFI:slope:LFI23:Rad:M:units}{$-0.96$\mHz}
\setsymbol{LFI:slope:LFI24:Rad:M:units}{$-0.83$\mHz}
\setsymbol{LFI:slope:LFI25:Rad:M:units}{$-0.91$\mHz}
\setsymbol{LFI:slope:LFI26:Rad:M:units}{$-0.95$\mHz}
\setsymbol{LFI:slope:LFI27:Rad:M:units}{$-0.87$\mHz}
\setsymbol{LFI:slope:LFI28:Rad:M:units}{$-0.88$\mHz}
\setsymbol{LFI:slope:LFI18:Rad:S:units}{$-1.15$\mHz}
\setsymbol{LFI:slope:LFI19:Rad:S:units}{$-1.00$\mHz}
\setsymbol{LFI:slope:LFI20:Rad:S:units}{$-0.95$\mHz}
\setsymbol{LFI:slope:LFI21:Rad:S:units}{$-0.92$\mHz}
\setsymbol{LFI:slope:LFI22:Rad:S:units}{$-1.01$\mHz}
\setsymbol{LFI:slope:LFI23:Rad:S:units}{$-0.95$\mHz}
\setsymbol{LFI:slope:LFI24:Rad:S:units}{$-0.73$\mHz}
\setsymbol{LFI:slope:LFI25:Rad:S:units}{$-1.16$\mHz}
\setsymbol{LFI:slope:LFI26:Rad:S:units}{$-0.79$\mHz}
\setsymbol{LFI:slope:LFI27:Rad:S:units}{$-0.82$\mHz}
\setsymbol{LFI:slope:LFI28:Rad:S:units}{$-0.91$\mHz}

\setsymbol{LFI:slope:LFI18:Rad:M}{$-1.04$}
\setsymbol{LFI:slope:LFI19:Rad:M}{$-1.09$}
\setsymbol{LFI:slope:LFI20:Rad:M}{$-0.69$}
\setsymbol{LFI:slope:LFI21:Rad:M}{$-1.56$}
\setsymbol{LFI:slope:LFI22:Rad:M}{$-1.01$}
\setsymbol{LFI:slope:LFI23:Rad:M}{$-0.96$}
\setsymbol{LFI:slope:LFI24:Rad:M}{$-0.83$}
\setsymbol{LFI:slope:LFI25:Rad:M}{$-0.91$}
\setsymbol{LFI:slope:LFI26:Rad:M}{$-0.95$}
\setsymbol{LFI:slope:LFI27:Rad:M}{$-0.87$}
\setsymbol{LFI:slope:LFI28:Rad:M}{$-0.88$}
\setsymbol{LFI:slope:LFI18:Rad:S}{$-1.15$}
\setsymbol{LFI:slope:LFI19:Rad:S}{$-1.00$}
\setsymbol{LFI:slope:LFI20:Rad:S}{$-0.95$}
\setsymbol{LFI:slope:LFI21:Rad:S}{$-0.92$}
\setsymbol{LFI:slope:LFI22:Rad:S}{$-1.01$}
\setsymbol{LFI:slope:LFI23:Rad:S}{$-0.95$}
\setsymbol{LFI:slope:LFI24:Rad:S}{$-0.73$}
\setsymbol{LFI:slope:LFI25:Rad:S}{$-1.16$}
\setsymbol{LFI:slope:LFI26:Rad:S}{$-0.79$}
\setsymbol{LFI:slope:LFI27:Rad:S}{$-0.82$}
\setsymbol{LFI:slope:LFI28:Rad:S}{$-0.91$}

\setsymbol{LFI:FWHM:70GHz:units}{13\parcm01}
\setsymbol{LFI:FWHM:44GHz:units}{27\parcm92}
\setsymbol{LFI:FWHM:30GHz:units}{32\parcm65}

\setsymbol{LFI:FWHM:70GHz}{13.01}
\setsymbol{LFI:FWHM:44GHz}{27.92}
\setsymbol{LFI:FWHM:30GHz}{32.65}

\setsymbol{LFI:FWHM:LFI18:units}{13\parcm39}
\setsymbol{LFI:FWHM:LFI19:units}{13\parcm01}
\setsymbol{LFI:FWHM:LFI20:units}{12\parcm75}
\setsymbol{LFI:FWHM:LFI21:units}{12\parcm74}
\setsymbol{LFI:FWHM:LFI22:units}{12\parcm87}
\setsymbol{LFI:FWHM:LFI23:units}{13\parcm27}
\setsymbol{LFI:FWHM:LFI24:units}{22\parcm98}
\setsymbol{LFI:FWHM:LFI25:units}{30\parcm46}
\setsymbol{LFI:FWHM:LFI26:units}{30\parcm31}
\setsymbol{LFI:FWHM:LFI27:units}{32\parcm65}
\setsymbol{LFI:FWHM:LFI28:units}{32\parcm66}

\setsymbol{LFI:FWHM:LFI18}{13.39}
\setsymbol{LFI:FWHM:LFI19}{13.01}
\setsymbol{LFI:FWHM:LFI20}{12.75}
\setsymbol{LFI:FWHM:LFI21}{12.74}
\setsymbol{LFI:FWHM:LFI22}{12.87}
\setsymbol{LFI:FWHM:LFI23}{13.27}
\setsymbol{LFI:FWHM:LFI24}{22.98}
\setsymbol{LFI:FWHM:LFI25}{30.46}
\setsymbol{LFI:FWHM:LFI26}{30.31}
\setsymbol{LFI:FWHM:LFI27}{32.65}
\setsymbol{LFI:FWHM:LFI28}{32.66}

\setsymbol{LFI:FWHM:uncertainty:LFI18:units}{0.170\arcm}
\setsymbol{LFI:FWHM:uncertainty:LFI19:units}{0.174\arcm}
\setsymbol{LFI:FWHM:uncertainty:LFI20:units}{0.170\arcm}
\setsymbol{LFI:FWHM:uncertainty:LFI21:units}{0.156\arcm}
\setsymbol{LFI:FWHM:uncertainty:LFI22:units}{0.164\arcm}
\setsymbol{LFI:FWHM:uncertainty:LFI23:units}{0.171\arcm}
\setsymbol{LFI:FWHM:uncertainty:LFI24:units}{0.652\arcm}
\setsymbol{LFI:FWHM:uncertainty:LFI25:units}{1.075\arcm}
\setsymbol{LFI:FWHM:uncertainty:LFI26:units}{1.131\arcm}
\setsymbol{LFI:FWHM:uncertainty:LFI27:units}{1.266\arcm}
\setsymbol{LFI:FWHM:uncertainty:LFI28:units}{1.287\arcm}

\setsymbol{LFI:FWHM:uncertainty:LFI18}{0.170}
\setsymbol{LFI:FWHM:uncertainty:LFI19}{0.174}
\setsymbol{LFI:FWHM:uncertainty:LFI20}{0.170}
\setsymbol{LFI:FWHM:uncertainty:LFI21}{0.156}
\setsymbol{LFI:FWHM:uncertainty:LFI22}{0.164}
\setsymbol{LFI:FWHM:uncertainty:LFI23}{0.171}
\setsymbol{LFI:FWHM:uncertainty:LFI24}{0.652}
\setsymbol{LFI:FWHM:uncertainty:LFI25}{1.075}
\setsymbol{LFI:FWHM:uncertainty:LFI26}{1.131}
\setsymbol{LFI:FWHM:uncertainty:LFI27}{1.266}
\setsymbol{LFI:FWHM:uncertainty:LFI28}{1.287}

\setsymbol{HFI:center:frequency:100GHz:units}{100\,GHz}
\setsymbol{HFI:center:frequency:143GHz:units}{143\,GHz}
\setsymbol{HFI:center:frequency:217GHz:units}{217\,GHz}
\setsymbol{HFI:center:frequency:353GHz:units}{353\,GHz}
\setsymbol{HFI:center:frequency:545GHz:units}{545\,GHz}
\setsymbol{HFI:center:frequency:857GHz:units}{857\,GHz}

\setsymbol{HFI:center:frequency:100GHz}{100}
\setsymbol{HFI:center:frequency:143GHz}{143}
\setsymbol{HFI:center:frequency:217GHz}{217}
\setsymbol{HFI:center:frequency:353GHz}{353}
\setsymbol{HFI:center:frequency:545GHz}{545}
\setsymbol{HFI:center:frequency:857GHz}{857}

\setsymbol{HFI:Ndetectors:100GHz}{8}
\setsymbol{HFI:Ndetectors:143GHz}{11}
\setsymbol{HFI:Ndetectors:217GHz}{12}
\setsymbol{HFI:Ndetectors:353GHz}{12}
\setsymbol{HFI:Ndetectors:545GHz}{3}
\setsymbol{HFI:Ndetectors:857GHz}{4}

\setsymbol{HFI:FWHM:Maps:100GHz:units}{9\parcm88}
\setsymbol{HFI:FWHM:Maps:143GHz:units}{7\parcm18}
\setsymbol{HFI:FWHM:Maps:217GHz:units}{4\parcm87}
\setsymbol{HFI:FWHM:Maps:353GHz:units}{4\parcm65}
\setsymbol{HFI:FWHM:Maps:545GHz:units}{4\parcm72}
\setsymbol{HFI:FWHM:Maps:857GHz:units}{4\parcm39}
\setsymbol{HFI:FWHM:Maps:100GHz}{9.88}
\setsymbol{HFI:FWHM:Maps:143GHz}{7.18}
\setsymbol{HFI:FWHM:Maps:217GHz}{4.87}
\setsymbol{HFI:FWHM:Maps:353GHz}{4.65}
\setsymbol{HFI:FWHM:Maps:545GHz}{4.72}
\setsymbol{HFI:FWHM:Maps:857GHz}{4.39}

\setsymbol{HFI:beam:ellipticity:Maps:100GHz}{1.15}
\setsymbol{HFI:beam:ellipticity:Maps:143GHz}{1.01}
\setsymbol{HFI:beam:ellipticity:Maps:217GHz}{1.06}
\setsymbol{HFI:beam:ellipticity:Maps:353GHz}{1.05}
\setsymbol{HFI:beam:ellipticity:Maps:545GHz}{1.14}
\setsymbol{HFI:beam:ellipticity:Maps:857GHz}{1.19}

\setsymbol{HFI:FWHM:Mars:100GHz:units}{9\parcm37}
\setsymbol{HFI:FWHM:Mars:143GHz:units}{7\parcm04}
\setsymbol{HFI:FWHM:Mars:217GHz:units}{4\parcm68}
\setsymbol{HFI:FWHM:Mars:353GHz:units}{4\parcm43}
\setsymbol{HFI:FWHM:Mars:545GHz:units}{3\parcm80}
\setsymbol{HFI:FWHM:Mars:857GHz:units}{3\parcm67}

\setsymbol{HFI:FWHM:Mars:100GHz}{9.37}
\setsymbol{HFI:FWHM:Mars:143GHz}{7.04}
\setsymbol{HFI:FWHM:Mars:217GHz}{4.68}
\setsymbol{HFI:FWHM:Mars:353GHz}{4.43}
\setsymbol{HFI:FWHM:Mars:545GHz}{3.80}
\setsymbol{HFI:FWHM:Mars:857GHz}{3.67}

\setsymbol{HFI:beam:ellipticity:Mars:100GHz}{1.18}
\setsymbol{HFI:beam:ellipticity:Mars:143GHz}{1.03}
\setsymbol{HFI:beam:ellipticity:Mars:217GHz}{1.14}
\setsymbol{HFI:beam:ellipticity:Mars:353GHz}{1.09}
\setsymbol{HFI:beam:ellipticity:Mars:545GHz}{1.25}
\setsymbol{HFI:beam:ellipticity:Mars:857GHz}{1.03}

\setsymbol{HFI:CMB:relative:calibration:100GHz}{$\lsim 1\%$}
\setsymbol{HFI:CMB:relative:calibration:143GHz}{$\lsim 1\%$}
\setsymbol{HFI:CMB:relative:calibration:217GHz}{$\lsim 1\%$}
\setsymbol{HFI:CMB:relative:calibration:353GHz}{$\lsim 1\%$}
\setsymbol{HFI:CMB:relative:calibration:545GHz}{}
\setsymbol{HFI:CMB:relative:calibration:857GHz}{}

\setsymbol{HFI:CMB:absolute:calibration:100GHz}{$\lsim 2\%$}
\setsymbol{HFI:CMB:absolute:calibration:143GHz}{$\lsim 2\%$}
\setsymbol{HFI:CMB:absolute:calibration:217GHz}{$\lsim 2\%$}
\setsymbol{HFI:CMB:absolute:calibration:353GHz}{$\lsim 2\%$}
\setsymbol{HFI:CMB:absolute:calibration:545GHz}{}
\setsymbol{HFI:CMB:absolute:calibration:857GHz}{}

\setsymbol{HFI:FIRAS:gain:calibration:accuracy:statistical:100GHz}{}
\setsymbol{HFI:FIRAS:gain:calibration:accuracy:statistical:143GHz}{}
\setsymbol{HFI:FIRAS:gain:calibration:accuracy:statistical:217GHz}{}
\setsymbol{HFI:FIRAS:gain:calibration:accuracy:statistical:353GHz}{2.5\%}
\setsymbol{HFI:FIRAS:gain:calibration:accuracy:statistical:545GHz}{1\%}
\setsymbol{HFI:FIRAS:gain:calibration:accuracy:statistical:857GHz}{0.5\%}

\setsymbol{HFI:FIRAS:gain:calibration:accuracy:systematic:100GHz}{}
\setsymbol{HFI:FIRAS:gain:calibration:accuracy:systematic:143GHz}{}
\setsymbol{HFI:FIRAS:gain:calibration:accuracy:systematic:217GHz}{}
\setsymbol{HFI:FIRAS:gain:calibration:accuracy:systematic:353GHz}{}
\setsymbol{HFI:FIRAS:gain:calibration:accuracy:systematic:545GHz}{7\%}
\setsymbol{HFI:FIRAS:gain:calibration:accuracy:systematic:857GHz}{7\%}

\setsymbol{HFI:FIRAS:zero:point:accuracy:100GHz:units}{0.8\MJysr}
\setsymbol{HFI:FIRAS:zero:point:accuracy:143GHz:units}{}
\setsymbol{HFI:FIRAS:zero:point:accuracy:217GHz:units}{}
\setsymbol{HFI:FIRAS:zero:point:accuracy:353GHz:units}{1.4\MJysr}
\setsymbol{HFI:FIRAS:zero:point:accuracy:545GHz:units}{2.2\MJysr}
\setsymbol{HFI:FIRAS:zero:point:accuracy:857GHz:units}{1.7\MJysr}

\setsymbol{HFI:FIRAS:zero:point:accuracy:100GHz}{0.8}
\setsymbol{HFI:FIRAS:zero:point:accuracy:143GHz}{}
\setsymbol{HFI:FIRAS:zero:point:accuracy:217GHz}{}
\setsymbol{HFI:FIRAS:zero:point:accuracy:353GHz}{1.4}
\setsymbol{HFI:FIRAS:zero:point:accuracy:545GHz}{2.2}
\setsymbol{HFI:FIRAS:zero:point:accuracy:857GHz}{1.7}

\setsymbol{HFI:unit:conversion:100GHz:units}{0.2415\MJysrmK}
\setsymbol{HFI:unit:conversion:143GHz:units}{0.3694\MJysrmK}
\setsymbol{HFI:unit:conversion:217GHz:units}{0.4811\MJysrmK}
\setsymbol{HFI:unit:conversion:353GHz:units}{0.2883\MJysrmK}
\setsymbol{HFI:unit:conversion:545GHz:units}{0.05826\MJysrmK}
\setsymbol{HFI:unit:conversion:857GHz:units}{0.002238\MJysrmK}

\setsymbol{HFI:unit:conversion:100GHz}{0.2415}
\setsymbol{HFI:unit:conversion:143GHz}{0.3694}
\setsymbol{HFI:unit:conversion:217GHz}{0.4811}
\setsymbol{HFI:unit:conversion:353GHz}{0.2883}
\setsymbol{HFI:unit:conversion:545GHz}{0.05826}
\setsymbol{HFI:unit:conversion:857GHz}{0.002238}

\setsymbol{HFI:colour:correction:alpha=-2:V1.01:100GHz}{0.9893}
\setsymbol{HFI:colour:correction:alpha=-2:V1.01:143GHz}{0.9759}
\setsymbol{HFI:colour:correction:alpha=-2:V1.01:217GHz}{1.0007}
\setsymbol{HFI:colour:correction:alpha=-2:V1.01:353GHz}{1.0028}
\setsymbol{HFI:colour:correction:alpha=-2:V1.01:545GHz}{1.0019}
\setsymbol{HFI:colour:correction:alpha=-2:V1.01:857GHz}{0.9889}

\setsymbol{HFI:colour:correction:alpha=0:V1.01:100GHz}{1.0008}
\setsymbol{HFI:colour:correction:alpha=0:V1.01:143GHz}{1.0148}
\setsymbol{HFI:colour:correction:alpha=0:V1.01:217GHz}{0.9909}
\setsymbol{HFI:colour:correction:alpha=0:V1.01:353GHz}{0.9888}
\setsymbol{HFI:colour:correction:alpha=0:V1.01:545GHz}{0.9878}
\setsymbol{HFI:colour:correction:alpha=0:V1.01:857GHz}{1.0014}

\usepackage{txfonts}
\usepackage[dvips]{graphicx}
\usepackage{epsf}
\usepackage{natbib}
\usepackage{longtable, lscape}
\usepackage[figuresright]{rotating}

\def\planck{\textit{Planck\/}}

\newcommand{\swift}{\textit{Swift}}
\newcommand{\fermi}{\textit{Fermi}}

\bibpunct{(}{)}{;}{a}{}{,}
\begin{document}
\author{\small
Planck Collaboration:
J.~Aatrokoski\inst{1}
\and
P.~A.~R.~Ade\inst{82}
\and
N.~Aghanim\inst{53}
\and
H.~D.~Aller\inst{4}
\and
M.~F.~Aller\inst{4}
\and
E.~Angelakis\inst{75}
\and
M.~Arnaud\inst{69}
\and
M.~Ashdown\inst{66, 7}
\and
J.~Aumont\inst{53}
\and
C.~Baccigalupi\inst{80}
% ...
}
\institute{\small
Aalto University Mets\"{a}hovi Radio Observatory, Mets\"{a}hovintie 114, FIN-02540 Kylm\"{a}l\"{a}, Finland\\
\and
Agenzia Spaziale Italiana Science Data Center, c/o ESRIN, via Galileo Galilei, Frascati, Italy\\
\and
Agenzia Spaziale Italiana, Viale Liegi 26, Roma, Italy\\
\and
Astronomy Department, University of Michigan, 830 Dennison Building, 500 Church Street, Ann Arbor, Michigan 48109-1042, U.S.A.\\
\and
Astroparticle Physics Laboratory, NASA/Goddard Space Flight Center, Greenbelt, MD 20771, U.S.A.\\
\and
Astroparticule et Cosmologie, CNRS (UMR7164), Universit\'{e} Denis Diderot Paris 7, B\^{a}timent Condorcet, 10 rue A. Domon et L\'{e}onie Duquet, Paris, France\\
\and
Astrophysics Group, Cavendish Laboratory, University of Cambridge, J J Thomson Avenue, Cambridge CB3 0HE, U.K.\\
\and
Atacama Large Millimeter/submillimeter Array, ALMA Santiago Central Offices, Alonso de Cordova 3107, Vitacura, Casilla 763 0355, Santiago, Chile\\
\and
Australia Telescope National Facility, CSIRO, P.O. Box 76, Epping, NSW 1710, Australia\\
\and
CITA, University of Toronto, 60 St. George St., Toronto, ON M5S 3H8, Canada\\
\and
CNRS, IRAP, 9 Av. colonel Roche, BP 44346, F-31028 Toulouse cedex 4, France\\
\and
California Institute of Technology, Pasadena, California, U.S.A.\\
\and
DAMTP, University of Cambridge, Centre for Mathematical Sciences, Wilberforce Road, Cambridge CB3 0WA, U.K.\\
\and
DSM/Irfu/SPP, CEA-Saclay, F-91191 Gif-sur-Yvette Cedex, France\\
\and
DTU Space, National Space Institute, Juliane Mariesvej 30, Copenhagen, Denmark\\
\and
Departamento de F\'{\i}sica, Universidad de Oviedo, Avda. Calvo Sotelo s/n, Oviedo, Spain\\
\and
Department of Astronomy and Astrophysics, Pennsylvania State University, 525 Davey Lab, University Park, PA 16802, U.S.A.\\
\and
Department of Astronomy and Astrophysics, University of Toronto, 50 Saint George Street, Toronto, Ontario, Canada\\
\and
Department of Physics \& Astronomy, University of British Columbia, 6224 Agricultural Road, Vancouver, British Columbia, Canada\\
\and
Department of Physics, Gustaf H\"{a}llstr\"{o}min katu 2a, University of Helsinki, Helsinki, Finland\\
\and
Department of Physics, Purdue University, 525 Northwestern Avenue, West Lafayette, Indiana, U.S.A.\\
\and
Department of Physics, University of California, Berkeley, California, U.S.A.\\
\and
Department of Physics, University of California, One Shields Avenue, Davis, California, U.S.A.\\
\and
Department of Physics, University of California, Santa Barbara, California, U.S.A.\\
\and
Department of Physics, University of Illinois at Urbana-Champaign, 1110 West Green Street, Urbana, Illinois, U.S.A.\\
\and
Dipartimento di Fisica G. Galilei, Universit\`{a} degli Studi di Padova, via Marzolo 8, 35131 Padova, Italy\\
\and
Dipartimento di Fisica M. Merlin dell'Universit\`{a} e del Politecnico di Bari, 70126 Bari, Italy\\
\and
Dipartimento di Fisica, Universit\`a degli Studi di Perugia, 06123 Perugia, Italy\\
\and
Dipartimento di Fisica, Universit\`{a} La Sapienza, P. le A. Moro 2, Roma, Italy\\
\and
Dipartimento di Fisica, Universit\`{a} degli Studi di Milano, Via Celoria, 16, Milano, Italy\\
\and
Dipartimento di Fisica, Universit\`{a} degli Studi di Trieste, via A. Valerio 2, Trieste, Italy\\
\and
Dipartimento di Fisica, Universit\`{a} di Ferrara, Via Saragat 1, 44122 Ferrara, Italy\\
\and
Dipartimento di Fisica, Universit\`{a} di Roma Tor Vergata, Via della Ricerca Scientifica, 1, Roma, Italy\\
\and
Discovery Center, Niels Bohr Institute, Blegdamsvej 17, Copenhagen, Denmark\\
\and
Dpto. Astrof\'{i}sica, Universidad de La Laguna (ULL), E-38206 La Laguna, Tenerife, Spain\\
\and
European Southern Observatory, ESO Vitacura, Alonso de Cordova 3107, Vitacura, Casilla 19001, Santiago, Chile\\
\and
European Space Agency, ESAC, Planck Science Office, Camino bajo del Castillo, s/n, Urbanizaci\'{o}n Villafranca del Castillo, Villanueva de la Ca\~{n}ada, Madrid, Spain\\
\and
European Space Agency, ESTEC, Keplerlaan 1, 2201 AZ Noordwijk, The Netherlands\\
\and
Finnish Centre for Astronomy with ESO (FINCA), University of Turku, V\"{a}is\"{a}l\"{a}ntie 20, FIN-21500, Piikki\"{o}, Finland\\
\and
Haverford College Astronomy Department, 370 Lancaster Avenue, Haverford, Pennsylvania, U.S.A.\\
\and
Helsinki Institute of Physics, Gustaf H\"{a}llstr\"{o}min katu 2, University of Helsinki, Helsinki, Finland\\
\and
INAF - Osservatorio Astrofisico di Catania, Via S. Sofia 78, Catania, Italy\\
\and
INAF - Osservatorio Astronomico di Padova, Vicolo dell'Osservatorio 5, Padova, Italy\\
\and
INAF - Osservatorio Astronomico di Roma, via di Frascati 33, Monte Porzio Catone, Italy\\
\and
INAF - Osservatorio Astronomico di Trieste, Via G.B. Tiepolo 11, Trieste, Italy\\
\and
INAF/IASF Bologna, Via Gobetti 101, Bologna, Italy\\
\and
INAF/IASF Milano, Via E. Bassini 15, Milano, Italy\\
\and
INRIA, Laboratoire de Recherche en Informatique, Universit\'{e} Paris-Sud 11, B\^{a}timent 490, 91405 Orsay Cedex, France\\
\and
ISDC Data Centre for Astrophysics, University of Geneva, ch. d'Ecogia 16, Versoix, Switzerland\\
\and
Imperial College London, Astrophysics group, Blackett Laboratory, Prince Consort Road, London, SW7 2AZ, U.K.\\
\and
Infrared Processing and Analysis Center, California Institute of Technology, Pasadena, CA 91125, U.S.A.\\
\and
Institut N\'{e}el, CNRS, Universit\'{e} Joseph Fourier Grenoble I, 25 rue des Martyrs, Grenoble, France\\
\and
Institut d'Astrophysique Spatiale, CNRS (UMR8617) Universit\'{e} Paris-Sud 11, B\^{a}timent 121, Orsay, France\\
\and
Institut d'Astrophysique de Paris, CNRS UMR7095, Universit\'{e} Pierre \& Marie Curie, 98 bis boulevard Arago, Paris, France\\
\and
Institut de Radioastronomie Millim\'{e}trique (IRAM), Avenida Divina Pastora 7, Local 20, 18012 Granada, Spain\\
\and
Institute of Astronomy and Astrophysics, Academia Sinica, Taipei, Taiwan\\
\and
Institute of Astronomy, University of Cambridge, Madingley Road, Cambridge CB3 0HA, U.K.\\
\and
Institute of Theoretical Astrophysics, University of Oslo, Blindern, Oslo, Norway\\
\and
Instituto de Astrof\'{\i}sica de Canarias, C/V\'{\i}a L\'{a}ctea s/n, La Laguna, Tenerife, Spain\\
\and
Instituto de F\'{\i}sica de Cantabria (CSIC-Universidad de Cantabria), Avda. de los Castros s/n, Santander, Spain\\
\and
Istituto Nazionale di Fisica Nucleare, Sezione di Bari, 70126 Bari, Italy\\
\and
Istituto Nazionale di Fisica Nucleare, Sezione di Perugia, 06123 Perugia, Italy\\
\and
Istituto di Fisica del Plasma, CNR-ENEA-EURATOM Association, Via R. Cozzi 53, Milano, Italy\\
\and
Jet Propulsion Laboratory, California Institute of Technology, 4800 Oak Grove Drive, Pasadena, California, U.S.A.\\
\and
Jodrell Bank Centre for Astrophysics, Alan Turing Building, School of Physics and Astronomy, The University of Manchester, Oxford Road, Manchester, M13 9PL, U.K.\\
\and
Kavli Institute for Cosmology Cambridge, Madingley Road, Cambridge, CB3 0HA, U.K.\\
\and
Key Laboratory of Optical Astronomy, National Astronomical Observatories, Chinese Academy of Sciences, 20A Datun Road, Chaoyang District, Beijing 100012, China\\
\and
LERMA, CNRS, Observatoire de Paris, 61 Avenue de l'Observatoire, Paris, France\\
\and
Laboratoire AIM, IRFU/Service d'Astrophysique - CEA/DSM - CNRS - Universit\'{e} Paris Diderot, B\^{a}t. 709, CEA-Saclay, F-91191 Gif-sur-Yvette Cedex, France\\
\and
Laboratoire Traitement et Communication de l'Information, CNRS (UMR 5141) and T\'{e}l\'{e}com ParisTech, 46 rue Barrault F-75634 Paris Cedex 13, France\\
\and
Laboratoire de Physique Subatomique et de Cosmologie, CNRS/IN2P3, Universit\'{e} Joseph Fourier Grenoble I, Institut National Polytechnique de Grenoble, 53 rue des Martyrs, 38026 Grenoble cedex, France\\
\and
Laboratoire de l'Acc\'{e}l\'{e}rateur Lin\'{e}aire, Universit\'{e} Paris-Sud 11, CNRS/IN2P3, Orsay, France\\
\and
Lawrence Berkeley National Laboratory, Berkeley, California, U.S.A.\\
\and
Max-Planck-Institut f\"{u}r Astrophysik, Karl-Schwarzschild-Str. 1, 85741 Garching, Germany\\
\and
Max-Planck-Institut f\"{u}r Radioastronomie, Auf dem H\"{u}gel 69, 53121 Bonn, Germany\\
\and
MilliLab, VTT Technical Research Centre of Finland, Tietotie 3, Espoo, Finland\\
\and
National University of Ireland, Department of Experimental Physics, Maynooth, Co. Kildare, Ireland\\
\and
Niels Bohr Institute, Blegdamsvej 17, Copenhagen, Denmark\\
\and
Owens Valley Radio Observatory, Mail code 249-17, California Institute of Technology, 1200 E California Blvd, Pasadena, CA 91125, U.S.A.\\
\and
SISSA, Astrophysics Sector, via Bonomea 265, 34136, Trieste, Italy\\
\and
SUPA, Institute for Astronomy, University of Edinburgh, Royal Observatory, Blackford Hill, Edinburgh EH9 3HJ, U.K.\\
\and
School of Physics and Astronomy, Cardiff University, Queens Buildings, The Parade, Cardiff, CF24 3AA, U.K.\\
% ...
}

\title{\planck\ early results. XV. Spectral energy distributions and radio continuum spectra 
of northern extragalactic radio sources}

\abstract{ Spectral energy distributions (SEDs) and radio continuum
  spectra are presented for a northern sample of 104 extragalactic radio
  sources, based on the \planck\ Early Release Compact Source Catalogue
  (ERCSC) and simultaneous multifrequency data. The nine \planck\
  frequencies, from 30 to 857\,GHz, are complemented by a set of
  simultaneous observations ranging from radio to gamma-rays. This is
  the first extensive frequency coverage in the radio and millimetre
  domains for an essentially complete sample of extragalactic radio
  sources, and it shows how the individual shocks, each in their own
  phase of development, shape the radio spectra as they move in the 
  relativistic jet. The SEDs presented in this paper were fitted with
  second and third degree polynomials to estimate the frequencies of
  the synchrotron and inverse Compton (IC) peaks, and the spectral indices 
  of low and high 
  frequency radio data, including the \planck\ ERCSC data, were calculated.
  SED modelling methods are discussed, with an emphasis on proper,
  physical modelling of the synchrotron bump using multiple
  components. \planck\ ERCSC data also suggest that the
  original accelerated electron energy spectrum could be much harder
  than commonly thought, with power-law index around 1.5 instead of the canonical
  2.5. The implications of this are discussed for the acceleration mechanisms
  effective in blazar shock. Furthermore in many cases the \planck\ data 
  indicate that gamma-ray emission must originate in the same shocks that
  produce the radio emission.}

\keywords{galaxies: active -- BL Lacertae objects: general -- quasars: general -- radiation mechanisms: non-thermal -- radio continuum: galaxies}

\authorrunning{Planck Collaboration}
\titlerunning{SEDs and radio spectra of northern AGN}

\maketitle
\allearlypapers

\section{Introduction}

This paper is part of the first series of publications based on measurements made with the 
\planck\footnote{\Planck\ (http://www.esa.int/planck) is a project of the European Space
Agency (ESA) with instruments provided by two scientific consortia funded by ESA member
states (in particular the lead countries France and Italy), with contributions from NASA
(USA) and telescope reflectors provided by a collaboration between ESA and a scientific
consortium led and funded by Denmark.} satellite.  
\Planck\ \citep{tauber2010a, planck2011-1.1} is the third-generation space mission to measure 
the anisotropy of the cosmic microwave background (CMB). It observes the sky in nine frequency 
bands covering 30--857\,GHz with high sensitivity and angular resolution from 31\arcm\ to 5\arcm. 
The Low Frequency Instrument (LFI; \citealt{Mandolesi2010, Bersanelli2010, planck2011-1.4}) 
covers the 30, 44, and 70\,GHz bands with amplifiers cooled to 20\,\hbox{K}. The High Frequency 
Instrument (HFI; \citealt{Lamarre2010, planck2011-1.5}) covers the 100, 143, 217, 353, 545, and 
857\,GHz bands with bolometers cooled to 0.1\,\hbox{K}. Polarization is measured in all but 
the highest two bands \citep{Leahy2010, Rosset2010}. A combination of radiative cooling and 
three mechanical coolers produces the temperatures needed for the detectors and optics 
\citep{planck2011-1.3}. Two data processing centres (DPCs) check and calibrate the data and 
make maps of the sky \citep{planck2011-1.7, planck2011-1.6}. \Planck's sensitivity, angular 
resolution, and frequency coverage make it a powerful instrument for Galactic and extragalactic 
astrophysics as well as cosmology.

The paper uses data from the \planck\ Early Release Compact Source Catalogue 
\citep[ERCSC;][]{planck2011-1.10}. The ERCSC provides positions and
flux densities of compact sources found in each of the nine \planck\ frequency maps. 
The flux densities are calculated using aperture photometry,
with careful modelling of \planck's elliptical beams.
The ERCSC includes data from the first all-sky survey, taken between 13 August
2009 and 6 June 2010. 
This unique dataset offers the first glimpse of the previously unmapped millimetre and sub-millimetre sky. 
It is used here to create spectral energy distributions of 104 radio-bright, northern active galactic 
nuclei (AGN), with the most complete coverage in the radio to sub-millimetre frequencies to date. 

Radio-loud AGN host jets of relativistic matter emanating symmetrically from the core. These 
jets produce copious amounts of non-thermal radiation, which dominates the spectral energy 
distributions (SEDs) of such sources compared with any thermal emission from the nucleus, i.e., the accretion 
disk. The SEDs typically consist of two broad-band bumps, the one at lower frequencies attributed 
to synchrotron radiation, and the other at higher frequencies attributed to inverse Compton (IC) radiation. The 
peak frequencies of the two bumps vary from one object to another. The peak of the synchrotron component can be 
between the infrared and X-ray domains, and the IC peak can range from MeV to GeV energies. The sequence in the peak 
frequencies of the emitted energy, and the factors that create it, have been a hot topic in blazar 
research for more than a decade \citep[e.g.,][]{fossati97, ghisellini98, padovani07, ghisellini08, nieppola08, 
ghisellini08, sambruna10}.

Research on blazar SEDs has concentrated on two approaches. The first is fitting the SED with a 
simple function, usually a second or third order polynomial, to obtain the pivotal peak frequencies and 
luminosities of the radiation components in a straightforward manner. This approach is typically used when
studying large samples \citep{fossati98, nieppola06, sambruna06, nieppola08, abdo10_sed}. The second 
approach is detailed modelling of the SED, starting with the definition of initial parameters such as 
electron energy, magnetic field intensity, and Doppler factor. This method is more time-consuming and is
used mostly for individual sources \citep[e.g.,][]{acciari10, collmar10}. The standard model is a leptonic, 
homogeneous, one-zone 
model, where the emission originates in a single component (for a review of the blazar emission models, 
see \citealt{bottcher10}). One-zone models are useful as first-order approximations, but, in reality, AGN 
jets are rarely, if ever, dominated by a single source of radiation. The material in the jets flows 
through shocks in the jet, which locally enhance the radiation (the ``shock-in-jet'' model;
\citealt{marscher85, valtaoja92_moniIII}). There can be several of these shocks in the jet simultaneously, 
and adding these to the emission of the quiescent jet, we have several radiation components. Therefore, 
ideally, the SEDs and radio spectra should be modelled with more than one component. Such 
modelling is also necessary for the proper identification of the high frequency (IC) emission sites.

In this paper we use the nearly complete SEDs and well-covered radio spectra provided by \planck\ to look 
for signs of these multiple components contributing to the total radiation that we observe. The paper is 
structured as follows. In Sects.~\ref{sam} and \ref{data} we introduce our sample and summarize 
the multifrequency data used in our study. 
The spectral 
energy distributions and radio spectra are discussed in Sects.~\ref{seds} and \ref{spectra}, 
and their modelling is described in Sect.~\ref{model}. In Sect.~\ref{acceleration} we 
discuss the implications of our results for understanding the acceleration mechanisms in blazar jets, 
and in Sect.~\ref{con}  we summarize our conclusions. Throughout the paper we adopt the sign convention for 
spectral index, $\alpha$: $S_\nu \propto \nu^{\alpha}$. The errors of numerical values marked with a plus--minus sign correspond to one standard deviation.

\section{The sample}
\label{sam}

The complete sample presented in this paper consists of 104 northern and equatorial radio-loud AGN. 
It includes all AGN with 
declination $\geq-10^{\circ}$ that have a measured average radio flux density at 37\,GHz exceeding 1\,Jy. 
Most of the sample sources have been monitored at Mets\"ahovi Radio Observatory for many years, and the 
brightest sources have been observed for up to 30 years. 
The sample can be divided into subclasses as follows: 33 high-polarization quasars (HPQs), 
21 low-polarization quasars (LPQs), 21 BL Lacertae objects (BLOs), 19 normal quasars (QSOs), 9 radio galaxies 
(GALs), and one unclassified source (J184915+67064). (See, for example, \citet{hovatta08,hovatta09} for
additional information on the classification.) 
By high-polarized quasars we mean objects which have a measured optical polarization 
$\geq3\%$ at some epoch, while low-polarized quasars have a 
polarization $\leq3\%$. Normal quasars have no polarization measured, so they could be either HPQs 
or LPQs. Radio galaxies are non-quasar AGN. The full sample is listed in Table~\ref{sample}. Columns 1 and 2
give the name and J2000 name for the source,  and for some sources an alternative name is given in Col. 3. 
The coordinates of the sources are given in Cols. 4 and 5. The start dates of the \planck\ scans are listed 
in Cols. 6 and 7. The average 37\,GHz flux density from Mets\"ahovi observations is given in Col. 8. 
For Col. 9 onwards, see Sect.~\ref{seds}.

\addtocounter{table}{1}

\section{Multifrequency data}
\label{data}

The core of our data set is the \planck\ ERCSC. The construction and contents of the catalogue 
are described in \citet{planck2011-1.10}.
For most sources,
the ERCSC flux density values are averages of two scans, separated by about six months.
To enable extensive multifrequency studies with a simultaneous data set, the \planck\ Extragalactic Working 
Group has coordinated a programme in which ground-based and space-borne telescopes 
observe the sources in unison with \planck. In this paper we present SEDs based on the averaged ERCSC 
flux densities and on multifrequency supporting observations taken within two weeks
of the \planck\ scans.
In a later paper we will present single-epoch SEDs, constructed 
from all the available \planck\ and supplementary simultaneous multifrequency data.
This collaborative programme includes 12 observatories around the world (Table~\ref{observatories}).
Archival data have been obtained from the literature and from the
the search tool at the ASI (Agenzia Spaziale Italiana) Science Data Center (ASDC) 
web page\footnote{http://www.asdc.asi.it}.

\begin{table*}
\caption{Optical and radio observatories that participated in
the \planck\ multifrequency campaigns}
\label{observatories}
\centering
\begin{tabular}{l c}
\hline\hline
\noalign{\smallskip}
\textit{Radio} & \\
Observatory & Frequencies [GHz]\\
\hline
APEX, Chile & 345 \\
ATCA, Australia & 4.5 -- 40 \\
Effelsberg, Germany & 2.64 -- 43 \\
IRAM Pico Veleta, Spain & 86, 142 \\
Medicina, Italy & 5, 8.3 \\
Mets\"ahovi, Finland & 37 \\
OVRO, USA & 15 \\
RATAN-600, Russia & 1.1, 2.3, 4.8, 7.7, 11.2, 21.7 \\
UMRAO & 4.8, 8.0, 14.5 \\
VLA, USA & 5, 8, 22, 43 \\
\hline\hline
\noalign{\smallskip}
\textit{Optical} & \\
Observatory & Band \\
\hline
KVA, Spain & $R$ \\
Xinglong, China & $i$ \\
\hline
\end{tabular}
\end{table*}

\subsection{Radio and submillimetre data}

Centimetre-band observations were obtained with the University of Michigan
Radio Astronomy Observatory's (UMRAO)
26-m prime focus paraboloid equipped with radiometers operating at central
frequencies of 4.8, 8.0, and 14.5\,GHz. Observations at all three frequencies employed
rotating polarimeter systems permitting both total flux density and linear
polarization to be measured. A typical measurement consisted of 8--16 individual
measurements over a 20--40\,min  period. Frequent drift scans were made across
stronger sources to verify the telescope pointing correction curves, and
observations of programme sources were intermixed with observations of a grid of calibrator
sources to correct for temporal changes in the antenna aperture efficiency. The
flux scale was based on observations of Cassiopeia A \citep{baars77}.
Details of the calibration and analysis techniques are described in \citet{aller85}.

\begin{table*}
\caption{Parameters for the RATAN-600 receivers.}             
\label{ratan}
\centering                              
\begin{tabular}{c c c c c}       
\hline\hline  
\noalign{\smallskip} 
$f_{\rm c}$ [GHz] & $\Delta f$ [GHz] & $\Delta T$ [mK] & $T_{\rm phys}$ [K] & $T_{\rm sys}$ [K] \\ 
\noalign{\smallskip}  
\hline    
21.7 & 2.5 & 3.5 & 15 & 77 \\
11.2 & 1.4 & 3 & 15 & 65 \\
7.7 & 1.0 & 3 & 15 & 62 \\
4.8 & 0.9 & 2.2 & 15 & 39 \\
2.3 & 0.4 & 8 & 310 & 95 \\
1.1 & 0.12 & 15 & 310 & 105 \\
\hline                                   
\end{tabular}
\end{table*}

Six-frequency broadband radio spectra were obtained with the RATAN-600 radio telescope
in transit mode by observing simultaneously at 1.1, 2.3, 4.8, 7.7, 11.2, and 21.7\,GHz
\citep{parijskij93,berlin96}. 
The parameters of the receivers are listed in Table~\ref{ratan}, where $\nu_{\rm c}$ is the central frequency, 
$\Delta \nu$ is the bandwidth, $\Delta T$ is the sensitivity of the radiometer over a 1\,s 
integration, $T_{\rm phys}$ is the physical temperature of the front-end amplifier,
and $T_{\rm sys}$ is the noise temperature of the whole system at the given frequency. 
Data were reduced using the RATAN standard software \textit{FADPS} (Flexible Astronomical Data 
Processing System) reduction package \citep{verkhodanov97_data_reduction}. The flux 
density measurement procedure at RATAN-600 is described by \citet{aliakberov85}. 

The 37\,GHz observations were made with the 13.7-m Mets\"ahovi radio telescope
using a 1\,GHz bandwidth, dual-beam receiver centred at 
36.8\,GHz. The observations were ON--ON observations, alternating the source and 
the sky in each feed horn. The integration time used to obtain each flux
density data point typically ranged from 1200 to 1400 s. The detection limit of the telescope at
37\,GHz is $\sim 0.2$\,Jy under optimal conditions. Data points with
a signal-to-noise ratio less than four were handled as non-detections.
The flux density scale was set by observations of DR\,21. Sources NGC\,7027, 3C\,274, 
and 3C\,84 were used as secondary calibrators. A detailed description of the
data reduction and analysis is given in \citet{terasranta98}.
The error estimate in the flux density includes the contribution from the
measurement rms and the uncertainty of the absolute calibration.

Quasi-simultaneous centimetre/millimetre radio spectra for a large number of \planck\ 
blazars have been obtained within the framework of a \textit{Fermi}-GST 
related monitoring programme of gamma-ray blazars (the F-GAMMA programme, 
\citealt{fuhrmann07, angelakis08}). The frequency range spans 
2.64\,GHz to 142\,GHz using the Effelsberg 100-m and IRAM 30-m
telescopes.
The Effelsberg measurements were conducted with the secondary focus heterodyne
receivers at 2.64, 4.85, 8.35, 10.45, 14.60, 23.05, 32.00, and 43.00\,GHz.
The observations were performed quasi-simultaneously with cross-scans, 
i.e., slewing over the source position in the azimuth and elevation 
directions, with an adaptive number of sub-scans for reaching the desired 
sensitivity (for details, see \citealt{fuhrmann08, angelakis08}). 
Pointing offset correction, gain correction, 
atmospheric opacity correction, and sensitivity correction have been 
applied to the data. 

The Institut de Radioastronomie Millim\'etrique
(IRAM) observations were carried out with calibrated cross-scans 
using the EMIR horizontal and vertical polarisation receivers operating 
at 86.2 and 142.3\,GHz. The opacity-corrected intensities were converted 
to the standard temperature scale and corrected for small remaining 
pointing offsets and systematic gain-elevation effects. The conversion to 
the standard flux-density scale was done using the instantaneous conversion 
factors derived from frequently observed primary calibrators (Mars, Uranus) and secondary
calibrators (W3(OH), K3-50A, NGC\,7027).
From this programme, radio spectra measured
quasi-simultaneously with the \planck\ observations have been collected for a
total of 37 \planck\ blazars during the period August 2009 to June 2010.

Many of the sources in the sample were monitored at 15\,GHz using the
40-m telescope of the Owens Valley Radio Observatory (OVRO) as part of a larger
monitoring programme \citep{richards11}.  The flux density of each source
was measured approximately twice weekly, with occasional gaps due to poor
weather or instrumental problems.  
The telescope is equipped with a cooled
receiver installed at prime focus, with two symmetric off-axis corrugated
horn feeds sensitive to left circular polarization. The telescope and
receiver combination produces a pair of approximately Gaussian beams (157\arcs\
 FWHM), separated in azimuth by 12\parcm95. 
The receiver has a
centre frequency of 15.0\,GHz, a 3.0\,GHz bandwidth, and a noise-equivalent
reception bandwidth of 2.5\,GHz. 
Measurements were made using a
Dicke-switched dual-beam system, with a second level of switching in azimuth
to alternate source and sky in each of the two horns. 
Calibration was referred to 3C\,286, for which a flux density of 3.44\,Jy at 15\,GHz 
is assumed \citep{baars77}. 
Details of the observations, calibration, and analysis are given by
\citet{richards11}.

The Very Large Array (VLA) and (since spring 2010) the Expanded VLA 
also observed a subset of the sources as simultaneously as possible.
Most of the VLA and EVLA runs were brief 1--2 hour blocks of time. In a one-hour
 block of time, in addition to flux calibrators and phase calibrators,
typically 5--8 \planck\ sources were observed. In many cases, VLA flux density and
phase calibrators were themselves of interest, since they were bright enough
to be detected by \planck. For these bright sources, the integration times
could be extremely short. At 4.86\,GHz and 8.46\,GHz, each target was observed
for approximately 30\,s. The integration times were typically 100\,s at 22.46\,GHz
and 120\,s at 43.34\,GHz. All VLA/EVLA flux density measurements were 
calibrated using standard values for one or both of the primary calibrator 
sources used by NRAO (3C\,48 or 3C\,286), and the $uv$ data were flagged, calibrated 
and imaged using standard NRAO software, \textit{AIPS} or \textit{CASA}.
The VLA and EVLA were in different 
configurations at different times in the several months
duration of the observations. As a consequence, the angular resolution
changed. In addition, for a given configuration, the resolution was much
finer at higher frequencies. For that reason, sources that showed signs of 
resolution in any VLA configuration or at any VLA frequency have been 
carefully flagged.

The \planck-ATCA Co-eval Observations (PACO) project \citep{massardi11} consists 
of observations of a compilation 
of sources selected in the Australia Telescope 20\,GHz survey \citep[AT20G;][]{massardi10a} 
taken with the Australia Telescope Compact Array (ATCA) in the frequency range 
4.5--40\,GHz. The observations were carried out at several epochs close in time with 
\planck\ satellite observations covering July 2009 -- August 2010. 
The PACO sample is a complete, flux-density limited, and spectrally-selected sample 
over the whole southern sky, with the exception of the region with Galactic
latitude $|b|<5^\circ$.

The Simultaneous Medicina \planck\ Experiment (SiMPlE; \citealt{procopio11}) 
uses the 32-m Medicina single-dish antenna at 5 and 8.3\,GHz 
to observe the 263 sources of the New Extragalactic \textit{WMAP} Point Source
(NEWPS) sample \citep{massardi09} with $\delta>0^{\circ}$, and partially overlapping 
with the PACO observations for $-10^{\circ}<\delta<0^{\circ}$. 
The project began in June 2010. Because of the lack of simultaneity with the ERCSC, the data have
been used here to match the PACO observations in the overlapping region and
to add information for $\delta>70^{\circ}$, a region so far poorly
covered at 5\,GHz.

Twelve sources from our sample were observed in the
submillimetre domain with the 12-m Atacama Pathfinder
EXperiment (APEX) in Chile. The observations were made using the
LABOCA bolometer array centred at 345\,GHz. Data were taken at two
epochs in 2009: September 3--4 2009 and November 12 2009.  The data
were reduced using the script package 
\textit{minicrush}\footnote{http://www.submm.caltech.edu/$\sim$sharc/crush/}, version 30-Oct-2009, 
with Uranus used as calibrator.

\subsection{Optical data}

The optical observations were made with the 35-cm KVA (Kungliga
Vetenskapsakademien) telescope on La Palma, Canary islands. All
observations were made through the $R$-band filter ($\lambda_{\rm eff}$ = 640\,nm)
 using a Santa Barbara ST-8 CCD camera with a gain factor of 2.3
$e^-$/ADU and readout noise of 14 electrons. The binning of pixels by
$2 \times 2$ pixels resulted in a plate scale of $0\farcs98$ per pixel. We
obtained three to six exposures of 180\,s per target. The images were
reduced in the standard way of subtracting the bias and dark frames
and dividing by twilight flat-fields.

The flux densities of the target and three to five stars in the target field were
measured with aperture photometry and the magnitude difference between
the target and a primary reference star in the same field was
determined. Using differential mode makes the observations insensitive
to variations in atmospheric transparency and accurate measurements
can be obtained even in partially cloudy conditions.  The $R$-band
magnitude of the primary reference star was determined from
observations made on photometric nights, using comparison stars in
known blazar fields as calibrators \citep{fiorucci96, fiorucci98,
raiteri98, villata98_bvr, nilsson07} and
taking into account the color term of the $R$-band filter
employed. After the $R$-band magnitude of the primary reference star was
determined, the object magnitudes were computed from the magnitude
differences. At this phase we assumed $V-R = 0.5$ for the targets.
Several stars in the field were used to check the quality of the
photometry and stability of the primary reference.  The uncertainties in
the magnitudes include the contribution from both measurement and
calibration errors.

The monitoring at Xinglong Station, National Astronomical Observatories of China, 
was performed with a 60/90-cm f/3 Schmidt telescope. The telescope
is equipped with a $4096\times4096$ E2V CCD, which has a pixel size of
$12\,\mu\textrm{m}$ and a spatial resolution of $1\farcs3$ per pixel. The
observations were made with an $i$ filter. Its
central wavelength and passband width are 668.5 and 51.4\,nm, respectively.
The exposure times are mostly 120\,s but can range from 60 to 180\,s, 
depending on weather and moon phase.

\subsection{\swift}

The {\it Swift} Gamma-Ray-Burst (GRB) Explorer \citep{gehrels2004} is a 
multi-frequency space observatory devoted to the discovery and rapid follow up 
of GRBs. There are three instruments on board the spacecraft:
the UV and Optical Telescope (UVOT, \citealt{roming2005}), the X-Ray Telescope 
(XRT, \citealt{burrows2005}) sensitive in the $0.3-10.0$\,keV band, and the 
Burst Alert Telescope (BAT, \citealt{barthelmy2005}) sensitive in the 15 -- 150\,keV 
band. Although the primary scientific goal of the satellite is the observation
of GRBs, the wide frequency coverage is useful for AGN studies because it 
covers the region where the transition between synchrotron and inverse Compton 
emission usually occurs.

When not engaged in GRB observations, {\it Swift} is available for target of 
opportunity (ToO) requests. The {\it Swift} team decided to devote an average 
of three ToO per week to the simultaneous observations of \planck\ AGN.
The simultaneity of Swift observations within two weeks to the \planck\ first or 
second scan, or both, is shown in Table~\ref{simultaneity}, marked 
sim\_1st, sim\_2nd, or sim\_1st\_2nd, respectively.

\subsubsection{UVOT}

{\it Swift} UVOT observations were carried out using the ``filter of the day'', i.e., 
one of the lenticular filters ($V$, $B$, $U$, $UVW1$, $UVM2$, and $UVW2$), unless otherwise 
specified in the ToO request. Therefore images are not always available for all filters.

The photometry analysis of all our sources was performed using the standard UVOT
software distributed within the \textit{HEAsoft 6.8.0} package, and the calibration
included in the latest release of the ``Calibration Database'' (CALDB).
A specific procedure has been developed to process all the ToO observations requested
by the \planck\ project. 
Counts were extracted from an aperture of 5\arcsec\, radius for 
all filters and converted to flux densities using the standard zero points \citep{poole08}. The flux
densities were 
then de-reddened using the appropriate values of $E(B-V)$ for each source taken 
from \citet{schlegel1998}, with $A_{\lambda}/E(B-V)$ ratios calculated for UVOT 
filters using the mean galactic interstellar extinction curve from \citet{Fitzpatrick1999}. 
No variability was detected within single exposures in any filter.

\subsubsection{XRT}

The XRT is usually operated in ``Auto State'' mode which automatically adjusts the CCD read-out mode
to the source brightness, in an attempt to avoid pile-up (see \citealt{burrows2005, hill2004}
for details of the XRT observing modes). As a consequence, part of the data sample was collected
using the most sensitive Photon Counting (PC) mode while Windowed Timing (WT) mode
was used for bright sources.

The XRT data were processed with the \textit{XRTDAS} software package (v. 2.5.1, \citealt{capalbi2005})
developed at the ASDC and distributed by the NASA High Energy Astrophysics Archive
Research Center (HEASARC) within the \textit{HEASoft} package (v. 6.9). Event files were calibrated
and cleaned with standard filtering criteria with the \textit{xrtpipeline} task, using the latest calibration
files available in the Swift CALDB. Events in the energy range 0.3--10\,keV with grades 0--12 (PC
mode) and 0--2 (WT mode) were used for the analysis.

Events for the spectral analysis were selected within a circle of a 20-pixel ($\sim 47\arcsec$) radius,
centred on the source position, which encloses about 90\% of the point spread function (PSF) 
at $1.5$\,keV \citep{moretti2005}.
For PC mode data, when the source count rate is above $\sim 0.5$\,counts/s data are
significantly affected by pile-up in the inner part of the PSF. For such
cases, after comparing the observed PSF profile with the analytical model derived by \citet{moretti2005},
we removed pile-up effects by excluding events detected within a certain inner radius around the
source position, and used an outer radius of 30 pixels. The value of the inner radius was evaluated
individually for each observation affected by pile-up, depending on the observed source count rate.

Ancillary response files were generated with the \textit{xrtmkarf} task applying corrections for the PSF
losses and CCD defects. Source spectra were binned to ensure a minimum of 20 counts per bin to
allow the $\chi^2$ minimization fitting technique to be used.
We fitted the spectra adopting an absorbed power-law model with photon index $\Gamma_x$ . When
deviations from a single power-law model were found, we adopted a log-parabolic law of the form
$F (E) = KE^{-(a+b\log(E))}$ \citep{massaro2004} which has been shown to fit the X-ray spectrum
of blazars well (e.g., \citealt{giommi05, tramacere2009}). This spectral model is described by only two
parameters: $a$, the photon index at $1$\,keV, and $b$, the curvature of the parabola. For both models
the amount of hydrogen-equivalent column density ($N_{\rm H}$) was fixed to the Galactic value along the
line of sight \citep{kalberla2005}.

\subsection{{\fermi}-LAT Observations and Data Analysis}
\label{sec:Fermi}

The Large Area Telescope (LAT) onboard \fermi\ is an electron-positron pair-conversion telescope sensitive 
to gamma-rays of energies from 20\,MeV to above 300\,GeV. The \fermi-LAT consists of a high-resolution 
silicon microstrip tracker, a CsI hodoscopic electromagnetic calorimeter, and an anticoincidence detector 
for charged particles background identification. 
A full description of the instrument and its performance can be found in \citet{atwood2009}. 
The large field of view ($\sim$2.4\,sr) allows the LAT to observe the full sky in survey mode every 3~hours. 
The LAT PSF strongly depends on both the energy and the conversion point in the 
tracker, but less on the incidence angle. For 1\,GeV normal incidence conversions in the upper section of 
the tracker, the PSF 68$\%$ containment radius is 0.6$^{\circ}$ .

The \fermi-LAT data considered for this analysis cover the period from 4 August 2008 to 4 November 2010. 
They have been analyzed using the standard {\em Fermi}-LAT \textit{ScienceTools} software 
package\footnote{http://fermi.gsfc.nasa.gov/ssc/data/analysis/documentation/Cicerone/} (version v9r16) 
and selecting for each source only photons above 100\,MeV, belonging to the diffuse class 
(Pass6 V3 IRF, \citealt{atwood2009}) which have the lowest background contamination. 
For each source, we selected only photons within a 15$^{\circ}$ region of interest (RoI) centred 
around the source itself. In order to avoid background contamination from the bright Earth limb, 
time intervals where the Earth entered the 
LAT Field of View (FoV) were excluded from the data sample. In addition, we have excluded observations 
in which the source was viewed at zenith angles larger than 105$^{\circ}$, where Earth's 
albedo gamma-rays increase the background contamination.
The data were analyzed with a binned maximum likelihood technique \citep{mattox96} using the analysis 
software (\textit{gtlike}) developed by the LAT 
team$\footnote{http://fermi.gsfc.nasa.gov/ssc/data/analysis/documentation/Cicerone/ Cicerone\_Likelihood}$. 
A model accounting for the diffuse emission as well as the nearby gamma-ray sources is included 
in the fit. 

The diffuse foreground, including Galactic interstellar emission, extragalactic gamma-ray emission, 
and residual cosmic ray (CR) background, has been modelled using the 
models$\footnote{http://fermi.gsfc.nasa.gov/ssc/data/access/lat/ BackgroundModels.html}$ 
\textit{gll$\_$iem$\_$v02} for the Galactic diffuse emission and \textit{isotropic$\_$iem$\_$v02} 
for the extra-galactic isotropic emission. Each source has been fit with a power law function 
\begin{equation}
 \frac{dN}{dE} = \frac{N(\gamma +1)E^{\gamma}}{E_{\textrm{max}}^{\gamma +1}-E_{\textrm{min}}^{\gamma +1}}
\end{equation}
where both the normalization factor $N$ and the photon index $\gamma$ are left free in the model fit.
The model also includes all the sources within a 20$^{\circ}$ RoI included in the \fermi-LAT one year catalogue 
\citep{abdo10_1fgl}, modelled using power law functions. If a source included in the model is a 
pulsar belonging to the Fermi pulsar catalog 
\citep{abdopsrcat}, we have modelled the source using a power-law with exponential cut-off and the 
spectral parameters from the pulsar catalogue. 
For the evaluation of the gamma-ray SEDs, the whole energy range from 100\,MeV to 300\,GeV is 
divided into two equal logarithmically spaced bins per decade. In each energy bin the standard 
\textit{ gtlike} binned analysis has been applied 
assuming for all the point sources in the model a power law spectrum with photon index fixed to $-$2.0. 
Assuming that in each energy bin the spectral shape can be approximated by a power law, the flux density of the 
source in all selected energy bins was evaluated, requiring in each energy bin a test statistic 
(TS)\footnote{The test statistic (TS) is defined as ${\rm TS} = -2\ln(L_0/L_1)$ 
with $L_0$ the likelihood of the null-hypothesis model as compared to the likelihood of a competitive model, $L_1$} 
greater than 10. 
If the TS is lower than 10, an upper limit is evaluated in that energy bin. 
Only statistical errors for the fit parameters are shown in the plots. 
Systematic effects are mainly based on uncertainties in the LAT effective area derived from the on-orbit estimations, 
and are $<$5$\%$ near 1\,GeV, 10$\%$ below 0.10\,GeV, and 20$\%$ above 10\,GeV. 

The LAT gamma-ray spectra of all AGN detected by Fermi are studied in \citet{abdo2010a}, based on 11 months of
\fermi-LAT data. For this paper, we derived the gamma-ray SEDs of the 104 sources in the sample
in three time intervals, presented in Table~\ref{simultaneity}. In the first interval
the \fermi-LAT observations are simultaneous to the \planck\ first or second scan, or both, within two weeks
(marked sim\_1st, sim\_2nd, or sim\_1st\_2nd in Table~\ref{simultaneity}). The \fermi-LAT data have in this
case been integrated over two weeks.
In the second interval the gamma-ray 
data are quasi-simultaneous to the \planck\ first or second scan, or both, having been integrated over two months 
(2M\_1st, 2M\_2nd, or 2M\_1st\_2nd in Table~\ref{simultaneity}). In the third interval are sources for which
\fermi-LAT data have been averaged over 27 months due to their faintness (27m in Table~\ref{simultaneity}).

Note that eight sources (0306+102, 0355+508, 0804+499, 0945+408, J1130+3815, 1413+135, 1928+738, 2005+403) that appear as significant gamma-ray detections in the 27 month dataset presented in this work, have not been included in any \fermi\ catalogue published so far \citep{abdo10_1fgl}. They can be considered new gamma-ray emitting sources, taking into account the possibility that their association with the radio source is purely spatial.

\addtocounter{table}{1}

\section{Spectral energy distributions}
\label{seds}

The SEDs for the whole sample are shown in Figs.~\ref{sedsfig} --\ref{last} (in panels on the left). 
The \planck\ measurements and the simultaneous
auxiliary data are indicated by red circles. The grey points represent archival 
data obtained from the literature and using the ASDC search tool. 
The measurements were regarded as simultaneous 
if they were taken within two weeks (for radio frequencies) or five days (for optical and higher energies) 
of the \planck\ measurement. Previous studies of the radio variability of 
blazars \citep{hovatta08,nieppola09} have shown that the two-week simultaneity limit is appropriate, 
as large-scale radio flux-density changes on that timescale are quite rare. The SEDs were fitted with 
second and third degree polynomials, namely 
\begin{equation}
\label{fit1}
\log \nu F_\nu= c_1(\log \nu)^2+c_2(\log \nu)+c_3,
\end{equation}
 and 
\begin{equation}
\label{fit2}
\log \nu F_\nu= c_4(\log \nu)^3+c_5(\log \nu)^2+c_6(\log \nu)+c_7,
\end{equation} 
where $c_i$ ($i=1 \ldots 7$) 
are fit parameters. We calculated the peak frequencies of the components from the third degree fit by default and 
used the second degree fit only if it seemed by visual inspection to be more accurate. Typically, in 
these cases, the third degree fit would not yield a sensible result at all. In the SEDs in 
Figs.~\ref{sedsfig} --\ref{last}, third degree fits are marked with a dashed line, second degree fits 
with a dotted line. 

The fits were 
divided into two classes, A and B, according to the fit quality, A being superior. We emphasize that this division is the subjective view of the authors, and is only meant to illustrate the varying quality of the SED fits. The classes have not been used in calculating the correlation coefficients (see below). We were able to obtain meaningful 
synchrotron fits for 60 sources (15 class A fits), and IC fits for 30 sources (10 class A fits). For 21 
sources we were able to fit both components. The 
synchrotron and IC peak frequencies are listed in Table~\ref{sample}. 
Columns 9 and 10 give the peak frequency and the fit quality for the synchrotron component, 
and Cols.~11 and 12 give the peak frequency and the fit quality for the inverse Compton component. 

\begin{figure}
   \resizebox{\hsize}{!} {\includegraphics{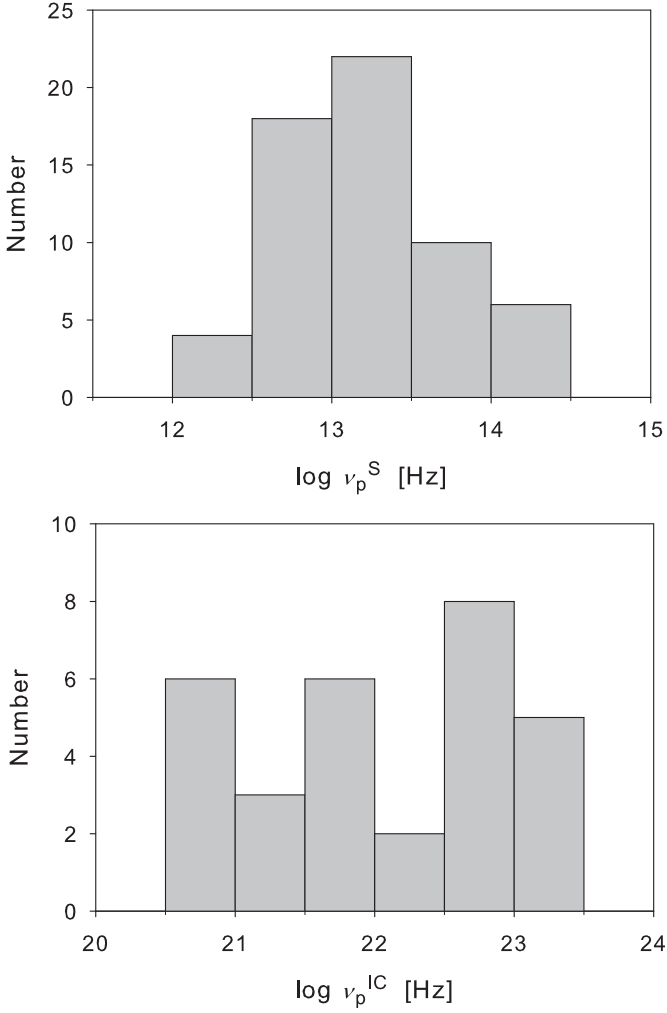}}
    \caption{The distributions of the synchrotron (top panel) and IC (bottom panel) peak frequencies.}
\label{nu_dist}
    \end{figure}

Figure~\ref{nu_dist} shows the distributions 
of the logarithms of the peak frequencies. The synchrotron peak frequencies, $\log\nu^{\rm S}_{\rm p}$, are 
typically very low, as we are 
dealing with bright radio sources. They range over two orders of magnitude, from 12.2 to 14.3, the average of the 
distribution being $13.2 \pm 0.5$. The source with the highest $\nu^{\rm S}_{\rm p}$ in the near-ultraviolet 
domain is 0716+714. 
The IC peak frequencies, $\log\nu^{\rm IC}_{\rm p}$, range over three orders of magnitude, 
from MeV to the high GeV gamma-ray region.
The distribution average is $22.0 \pm 0.9$. The source with the highest IC peak frequency is 1156+295, 
with $\log\nu^{\rm IC}_{\rm p} = 23.5$. The source with the lowest IC peak frequency, 
with $\log\nu^{\rm IC}_{\rm p} = 20.5$, is 
0836+710.

\begin{figure}
   \resizebox{\hsize}{!} {\includegraphics{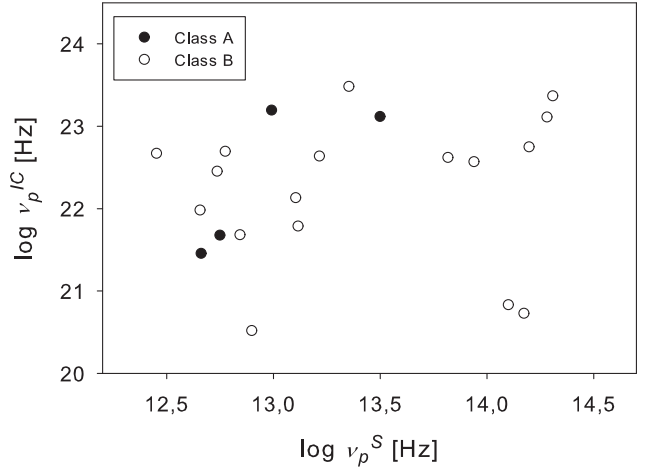}}
    \caption{The correlation of the peak frequencies of the synchrotron and the IC components. Solid circles 
denote class A fits, and open circles class B fits.}
\label{corr}
    \end{figure}

We have plotted the interdependence of the component peaks in Fig.~\ref{corr}. 
The two peak frequencies seem to have no significant correlation, 
which is confirmed by the Spearman rank correlation test ($\rho=0.301$, $P=0.092$, both class fits included). There is, 
however, a tendency for the sources with high $\nu^{\rm S}_{\rm p}$ to have a high $\nu^{\rm IC}_{\rm p}$. 
In the synchrotron-self-Compton (SSC) scenario, the 
separation of the component peaks depends on many factors, such as the electron Lorentz factors, magnetic 
field strengths, and particle densities. Therefore the lack of correlation as such does not allow any strong 
statements to be made about the origin of the IC radiation (SSC, or external Compton, EC). 

\begin{figure}
   \resizebox{\hsize}{!} {\includegraphics{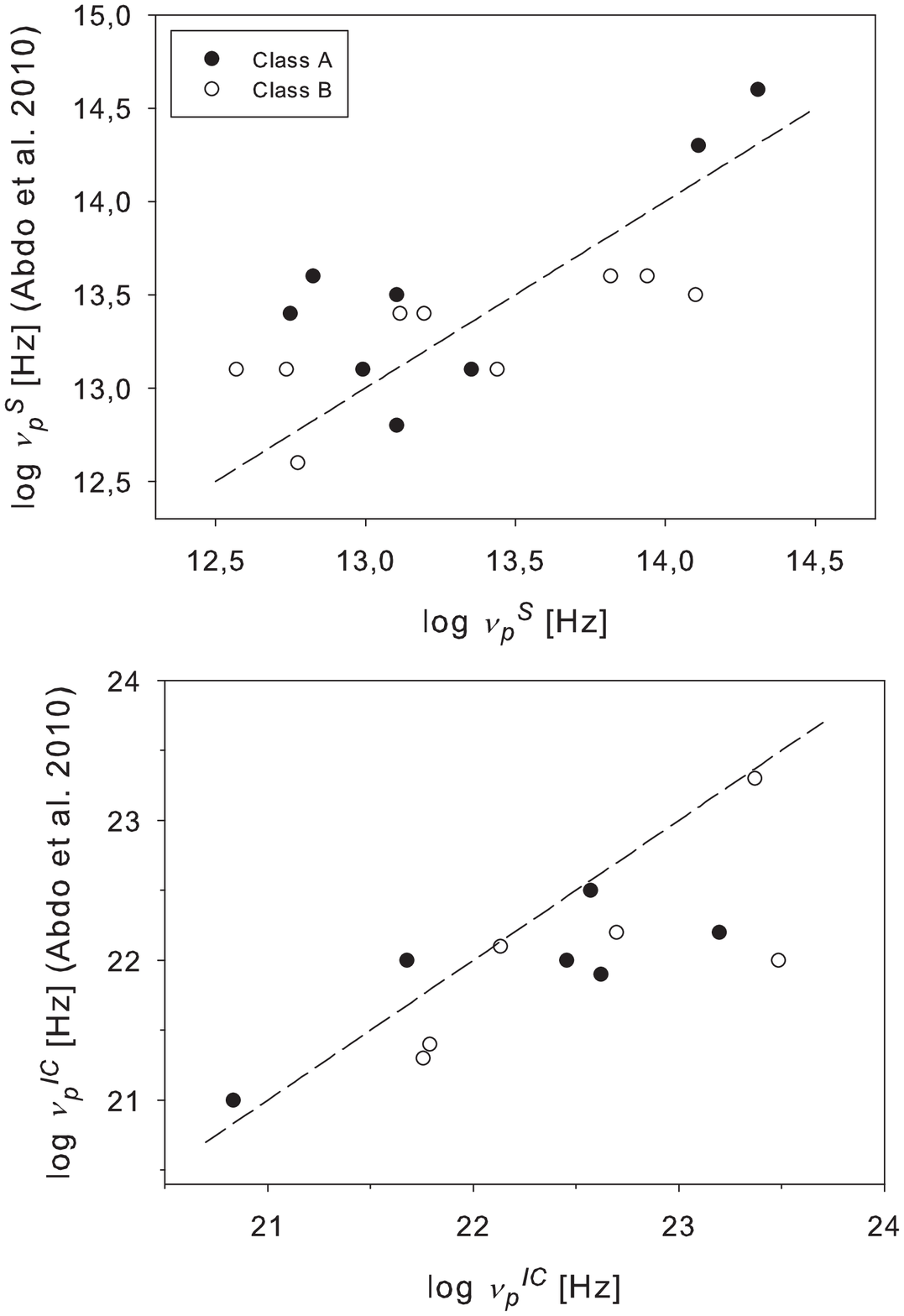}}
    \caption{Comparison between our synchrotron and IC peak frequencies and those of \citet{abdo10_sed}. 
The synchrotron peak frequencies are shown in the top panel and IC peak frequencies in the bottom panel. The 
dashed lines illustrate the one-to-one correspondence. Solid circles
denote class A fits, and open circles class B fits.}
\label{abdo_comp}
    \end{figure}

We also compared our peak frequencies with those of \citet{abdo10_sed}. These comparisons are illustrated in 
Fig.~\ref{abdo_comp}, where the one-to-one correspondence is shown with a dashed line. Both correlations are significant 
according to the Spearman test ($\rho=0.629$, $P=0.003$ for the synchrotron peaks 
and $\rho=0.660$, $P=0.010$ for the IC ones). 

\section{Radio spectra}
\label{spectra}

In addition to the spectral energy distributions, we have plotted the standard radio spectra 
for the sample sources. The aim was to look for signs of multiple 
components contributing to the synchrotron emission. The spectra are shown in Figs.~\ref{sedsfig}--\ref{last}
(in panels on the right). Red circles indicate low frequency (LF; $\leq 70$\,GHz) data simultaneous to \planck,
red stars indicate ERCSC LFI data, blue circles indicate high frequency (HF; $> 70$\,GHz) data simultaneous 
to \planck, and blue stars indicate ERCSC HFI data.
As can be seen, the \planck\ data and the supporting 
multifrequency data provide an excellent opportunity to find 
the signatures of possible multiple components or anomalous spectral shapes.
A statistical study of extragalactic radio sources in the ERCSC is presented in \citet{planck2011-6.1},
and SEDs of extreme radio sources, such as Gigahertz-Peaked Spectrum (GPS) sources in \citet{planck2011-6.2}.

\begin{figure}
\includegraphics[scale=0.8]{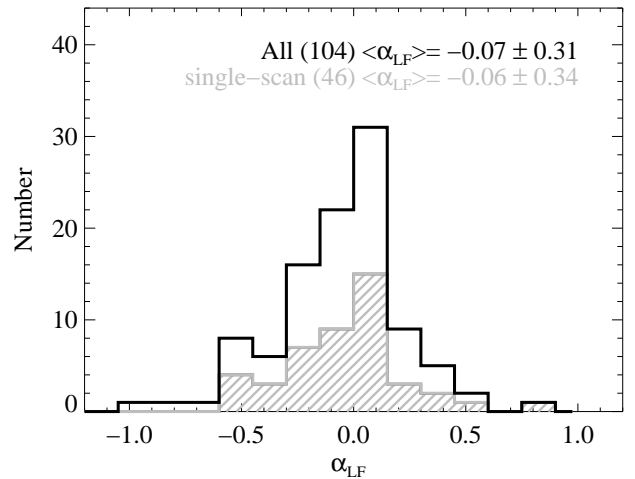}
 \caption{Distribution of LF spectral indices for the whole sample (104 sources).
The sources that have only been scanned once by \planck\ are shown hatched (46 sources).}
\label{lf}
\end{figure}

\begin{figure}
\includegraphics[scale=0.8]{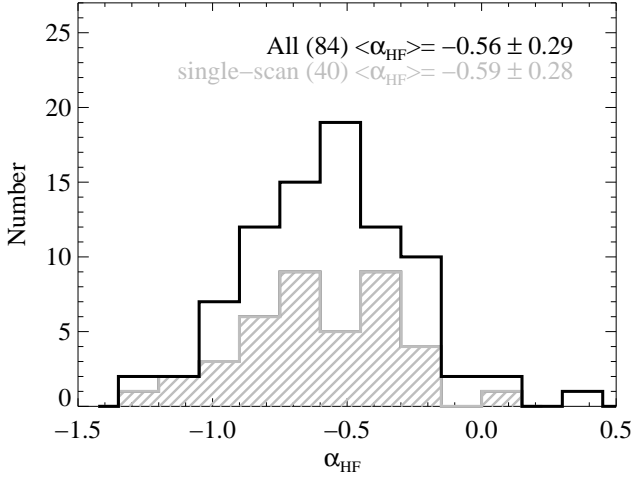}
 \caption{Distribution of HF spectral indices for the 84 sources that had three or more data points.
The sources that have only been scanned once by \planck\ are shown hatched (40 sources).}
\label{hf}
\end{figure}

\begin{figure}
\includegraphics[scale=0.8]{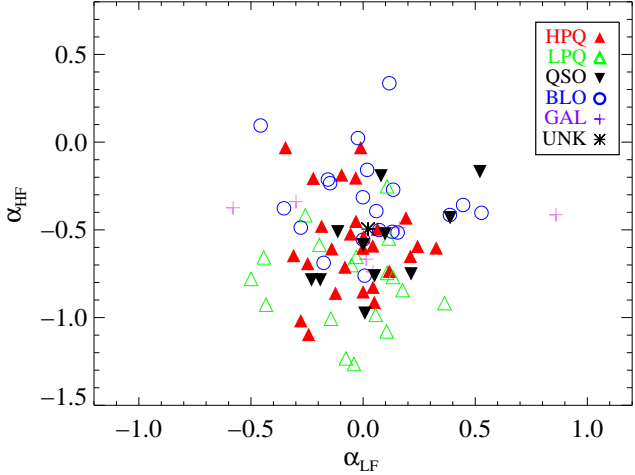}
 \caption{LF spectral indices $\alpha_{\rm LF}$ versus HF spectral indices $\alpha_{\rm HF}$ for the various AGN subclasses.}
\label{lf-hf}
\end{figure}

In Figs.~\ref{lf} and \ref{hf} we have plotted the distributions of the LF and HF spectral indices 
using only the Planck data for the latter.
When plotting the HF $\alpha$ values, we have only taken into account those sources
with three or more data points. This leaves us with 84 sources out of the 104. The sources
with only one \planck\ scan are shown hatched in Figs.~\ref{lf} and \ref{hf}.
Figure~\ref{lf-hf} shows $\alpha_{\rm LF}$ versus $\alpha_{\rm HF}$.
As expected, the LF indices are fairly flat, with an average of $-0.06$. Their distribution 
is narrow, with 91\% of the indices being in the range $\alpha_{\rm LF}=-0.5$--$0.5$. 
There are a couple of sources with remarkably steep LF spectra, namely 0552+398 and 2021+614, 
both having $\alpha_{\rm LF} \leq -0.8$. At the other end we have 0007+106 which has an inverted spectrum with 
$\alpha_{\rm LF}=0.86$. Also 1228+126 has a steep spectrum, although the fit value may be exaggerated in this case 
(see Fig.~\ref{1228}). For clarity, this source has been omitted from Figs.~\ref{lf} and \ref{lf-hf}.
  
The HF indices are concentrated around $-0.5$, having an average of $-0.56$. Rather similar distributions
for LF and HF spectral indices were found in \citet{planck2011-6.1}, especially in their Fig.~7. However, a large part 
of the $\alpha_{\rm HF}$ distribution is in the flat domain with $\alpha_{\rm HF} \geq -0.5$. 
Extreme examples are 1413+135 ($\alpha_{\rm HF}=0.02$) and 0954+658 ($\alpha_{\rm HF}=0.34$). 
It is unexpected that the spectral indices of blazars at sub-millimetre and FIR frequencies should be this flat. 
Only 28 of the 84 sources with HF spectral fits shown in Figs.~\ref{sedsfig}--\ref{last} have 
$\alpha < -0.7$, the ``canonical'' value for optically thin spectra, corresponding to an electron energy 
index $s$ of about $2.4$. This has two possible explanations: either the total HF spectra are defined by several 
underlying components, or the energy spectrum of the electron population producing the radiation is much 
harder than generally assumed. Although some HF spectra show evidence for several sub-components or even 
an upturn at the highest frequencies, many others appear to be perfectly straight and therefore perhaps originate 
from a single optically thin component. At the highest \planck\ frequencies the lifetimes of radiating 
electrons are short, and one would in most cases expect to see the optically 
thin spectra steepened by energy losses, with indices ranging from the original 
$\alpha_{\rm thin}$ to $\alpha_{\rm thin}-0.5$ and beyond. 
This makes spectral indices $> -0.7$ even more remarkable. 

To rule out contamination by dust in the host galaxies, we have estimated the possible contribution of an extremely bright submillimetre galaxy. We also estimated the flux densities of 2251+158 at several submillimetre bands in the case its luminosity were similar to the submillimetre galaxy. The comparison shows that if the dust component in 2251+158 were of the same brightness as in the extremely bright submillimetre galaxy, its emission would be completely swamped by the non-thermal emission from the jet.
The LF and HF spectral 
indices have been calculated using ERCSC data which for most sources is an average of two \planck\ scans
separated by about six months. However, the HF spectral index distribution for the sources
that have been observed only once by \planck\ is quite similar to that of the whole sample.
The implications of this result are discussed in more detail in Sect.~\ref{acceleration}.

As Fig.~\ref{lf-hf} shows, BLOs, HPQs and LPQs have different $\alpha_{\rm HF}$
distributions. In Table~\ref{av_indices} we have listed the average spectral indices for both frequency 
ranges for all subgroups. The LF spectra of radio galaxies are very steep on average, but the value is greatly 
influenced by the uncertain fit of 1228+126. If this source is left out of the calculations, the average $\alpha_{\rm LF}$ 
for galaxies is $0.10 \pm 0.52$. LPQs have the steepest spectra in the HF range. However, the standard deviations of 
the samples are substantial, which can also be seen 
as the large scatter in Fig.~\ref{lf-hf}. To quantify the possible differences between the AGN classes, 
we ran the Kruskal-Wallis ANOVA test. For the LF indices we could find no significant differences. 
However, in the case of the HF indices, the distributions of the subgroups differ significantly ($P<0.001$). Multiple-comparisons test with t distribution tells us that LPQs differ from all other subgroups ($P\leq0.031$). Also, BLO indices have a different distribution from all quasar subgroups ($P\leq0.019$).

\begin{table}
\caption{Average spectral indices for AGN subclasses. The errors correspond to one standard deviation.}
\label{av_indices}
\centering
\begin{tabular}{l c c c c}
\hline\hline
\noalign{\smallskip}
Class & number LF & number HF & average $\alpha_{\rm LF}$ & average $\alpha_{\rm HF}$ \\
\noalign{\smallskip}  
\hline
\noalign{\smallskip}  
BLO & 21 & 20 & $-0.01 \pm 0.25$ & $-0.34 \pm 0.26$\\
HPQ & 33 & 28 & $-0.06 \pm 0.22$ & $-0.58 \pm 0.27$ \\
LPQ & 21 & 18 & $-0.08 \pm 0.30$ & $-0.80 \pm 0.27$ \\
QSO & 19 & 10 & $-0.05 \pm 0.28$ & $-0.57 \pm 0.26$ \\
GAL & 9 & 7 & $-0.25 \pm 0.66$ & $-0.55 \pm 0.19$ \\
\hline
\end{tabular}
\end{table}

\section{Modelling the SEDs and spectra of blazars}
\label{model}

Spectral energy distributions rapidly became one of the main tools for understanding blazar 
physics after the EGRET instrument aboard the \textit{Compton Gamma-ray Observatory (CGRO)} 
satellite discovered strong gamma-ray 
emission from 3C\,279 \citep{hartman92} and, subsequently, from a large number of other radio-bright 
AGN. It was rapidly accepted that the double-peaked overall shape of the radio-to-gamma 
SED was due to synchrotron and inverse Compton radiation, but beyond this the agreement ended. 
The data simply lacked the sensitivity and the time coverage to sufficiently constrain the 
models. Even the most detailed and convincing effort for 3C\,279 \citep{hartman01_3C279} could be 
criticized for unrealistic physical assumptions \citep{lindfors05}.

With the coming of the \textit{Fermi} and \textit{Swift} satellites, ground-based TeV telescopes, and other 
satellite data, the inverse Compton high-energy SED of a large number of AGN can now be determined 
with hugely improved accuracy and time coverage. This has initiated a new era of AGN modelling 
(e.g., Proceedings of the Workshop ``Fermi meets Jansky: AGN in Radio and Gamma Rays''; \citealt{fermi-jansky10}). 
However, since the IC spectrum is created by the 
relativistic electrons scattering either the synchrotron photons (SSC) or 
external photons (EC), accurate knowledge of the shape of the synchrotron component 
is essential for all realistic modelling. Here the \planck\ data are invaluable, since
they provide the first simultaneous multifrequency coverage of the crucial radio-to-IR part of the 
synchrotron spectrum.

However, most of the ERCSC data are not suitable for detailed modelling, since they are averages 
of two spectra taken about six months apart. Considering the strong SED variability, accurate
modelling can be done only with the final \planck\ datasets and their several individual scans.
Here we limit ourselves to a short discussion of how \planck\ and supporting data can be used to improve our 
understanding of blazar physics.

The basic picture of a blazar is well established: a black hole surrounded by an accretion disk and 
a broad-line emitting region (BLR), and a relativistic jet which produces the lower-frequency bump of the 
SED through synchrotron emission. The jet is not totally stable; VLBI imaging shows how new components, 
identified as relativistic shocks in the jet, emerge from the radio core at intervals ranging from months 
to years. These growing and decaying shocks are also responsible for the total flux density variations 
in the radio-to-IR, and at least partly also in the optical regime. The important thing 
to note is that the 
synchrotron SED is rarely, if ever, the product of a single synchrotron-radiating component. The nature of the radio core is 
still being debated, as is the nature and the extent of the jet flow upstream of the radio core, between 
it and the black hole \citep[e.g.,][]{marscher10, marscher10_1510}. Shocked jet models 
\citep{marscher85, hughes89, valtaoja92_moniIII} 
provide a good account of the multifrequency variations in the total flux density. 
In the particular case of 3C\,279, \citet{lindfors06} showed that the optical variations can also be 
moderately well explained as the higher-frequency part of the shock emission.

The origin of the IC component is a matter of intense debate. In the \textit{CGRO} era, 
the favoured alternative was that the high-frequency radiation originates close to the black hole and 
the accretion disk, inside the BLR and well upstream of the radio core. With a strong radiation field 
external to the jet, external Compton is then likely to be the dominant emission mechanism. A minority view 
was that at least a part of the gamma-ray flaring comes from much farther out, at or downstream from the radio 
core. The evidence for this was that, statistically speaking, strong gamma-ray flaring tended to occur 
after the beginning of a millimetre flare \citep{valtaoja96_gamma, lahteenmaki03} and after the 
ejection of a new VLBI component \citep{jorstad01}. 
Occurring well outside the BLR, SSC seems in this case to be the only viable emission mechanism, but on the 
other hand simple SSC models have generally failed to explain the gamma-ray emission, requiring 
unrealistically low magnetic field strengths in other than the extreme TeV BL Lac objects. With new \textit{Fermi} data, 
the evidence for gamma-ray emission from the radio jet has strengthened 
\citep{kovalev09, pushkarev10, leontavares11, nieppola10, tornikoski10, valtaoja10}, 
but the source of the upscattered photons remains a puzzle. 

Most of the attempts to model the high-energy bump have adopted a ``reverse engineering'' method, starting 
from the secondary IC SED and not from the primary synchrotron SED. The original electron energy 
spectrum is considered to be rather freely adjustable, with parameters chosen to produce a good fit of 
the calculated IC spectrum to the observations within the assumed model. In most cases, the synchrotron 
spectrum is assumed to originate in a single homogeneous component, and often the spectrum is adjusted 
only to the optical data, without any attempt to explain the radio-to-IR part of the SED. There is not 
much observational or theoretical evidence for the actual physical existence of such an IR-to-optical 
component: it is simply assumed to be there, because it can produce the observed IC spectrum.

To some extent such an approach is justifiable if the gamma-rays originate upstream of the 
radio core, and are therefore not directly connected to the observed radio jet and shocks, and their 
lower-frequency synchrotron emission. However, even in this case physically more realistic models must 
take into account also the radio-to-IR part, for several reasons.
First, whatever the exact mechanism, 
the radio shocks also contribute to the IC spectrum, and their contribution must be accounted for in the 
model fits. Assuming reasonable physical parameters, the shocks can quite often produce part or all 
of the observed X-ray emission through the SSC mechanism \citep[e.g.,][]{lindfors06}. Indeed, a standard
method for estimating the physical parameters of VLBI components is to assume that the X-ray emission is 
produced through SSC \citep{ghisellini93,marscher87}. Secondly, the components 
seen in the radio, especially in the \planck\ HFI regime, contribute to the optical synchrotron emission, 
and this contribution must be subtracted from the putative IR-to-optical component, which, as explained 
above, in many models generates the whole IC spectrum. For example, in the present sample, at least 0235+164, 
0917+449, 0954+658, 2227$-$088, and especially 2251+158 show ongoing millimetre-to-submillimetre flaring, which leaves 
little room for any additional components between the highest \planck\ frequencies and the optical. 
This is the case in particular if the optically thin spectra are originally very flat, as we argue in 
Sects.~\ref{examples} and~\ref{acceleration}.

On the other hand, if gamma-rays do come from the shocks at, or downstream from, the radio core, as 
increasing evidence indicates, the obvious first step is to model the shocks themselves and their spectra 
as accurately as possible using \planck\ and other radio data, and only after that attempt to calculate the 
IC radiation they produce.

Finally, it is clear that the more realistic multicomponent models, whatever their exact details, will 
give rather different physical parameter values than the single-component models. As an example, if the 
X-ray emission and the gamma-ray emission are not produced by a single component but by two (or even more) 
components, the derived electron energy spectral parameters will be quite different.

\subsection{Some examples of different spectra and their implications}
\label{examples}

In this section we highlight five examples of different spectra of ERCSC sources which have been scanned
only once by \planck\ and therefore have true snapshot spectra presented in Figs.~\ref{sedsfig}--\ref{last}.
As stated before, our aim at this stage is not to model these sources, but to draw attention to some
features commonly seen in the spectral energy distributions, and to their implications for future
multi-epoch modelling of the \planck\ sources.
We have fitted the radio-to-optical spectra with one or several standard self-absorbed sychrotron components
(based on a power-law assumption for the electron energy spectrum). These are not model fits from a numerical,
physical code (except for Figs.~\ref{2251a} and \ref{2251b}), but rather are meant to guide the eye and to
illustrate the general shape of the radio-to-optical spectra. 

\begin{figure}
   \resizebox{\hsize}{!} {\includegraphics{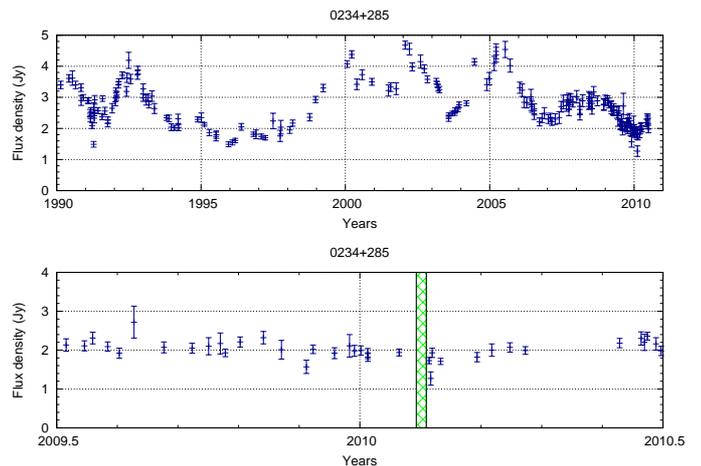}}
    \caption{The 37\,GHz lightcurve of 0234+285, measured at Mets\"ahovi Radio Observatory, showing
continuous variability. The period when \planck\ was observing the
source is indicated by the hatched region in the lower panel, which is a blow-up of the last year.}
\label{0234_lightcurve}
    \end{figure}

\begin{figure}
   \resizebox{\hsize}{!} {\includegraphics{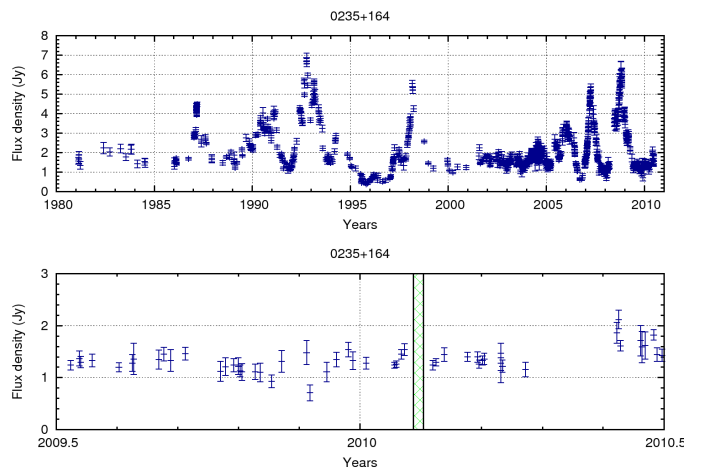}}
    \caption{The 37\,GHz lightcurve of 0235+164, details as in Fig.~\ref{0234_lightcurve}.}
\label{0235_lightcurve}
    \end{figure}

\begin{figure}
   \resizebox{\hsize}{!} {\includegraphics{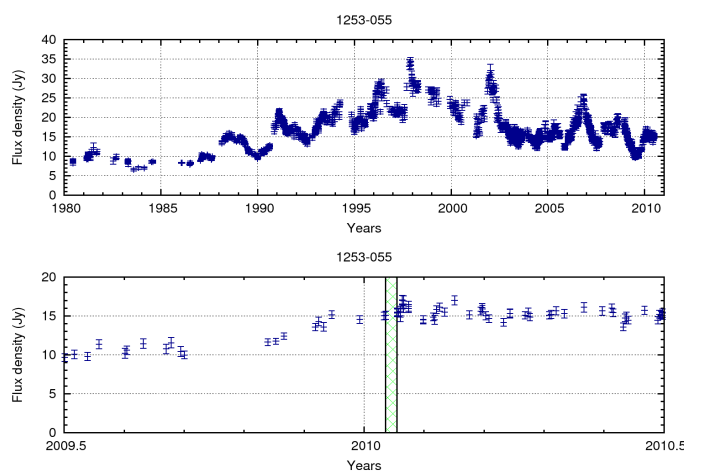}}
    \caption{The 37\,GHz lightcurve of 1253$-$055 (3C\,279), details as in Fig.~\ref{0234_lightcurve}.}
\label{1253_lightcurve}
    \end{figure}

\begin{figure}
   \resizebox{\hsize}{!} {\includegraphics{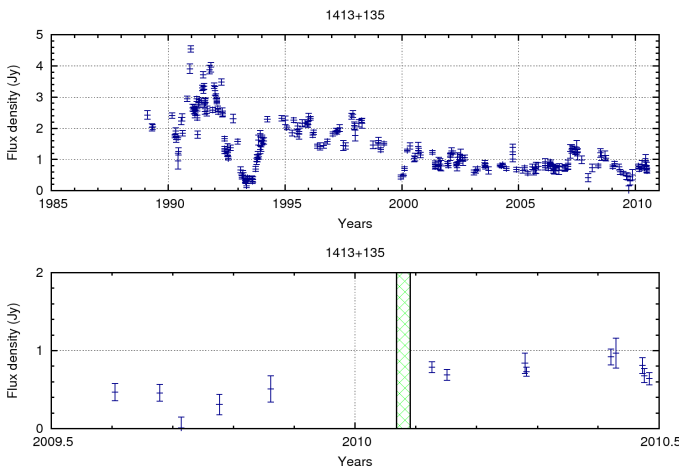}}
    \caption{The 37\,GHz lightcurve of 1413+135, details as in Fig.~\ref{0234_lightcurve}.}
\label{1413_lightcurve}
    \end{figure}

\begin{figure}
   \resizebox{\hsize}{!} {\includegraphics{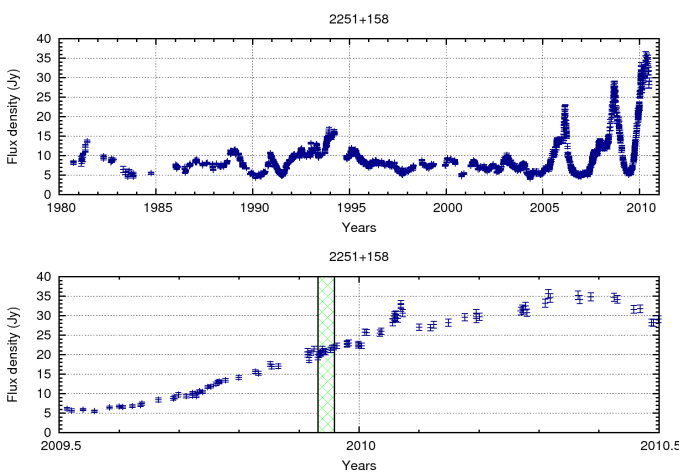}}
    \caption{The 37\,GHz lightcurve of 2251+158 (3C\,454.3), details as in Fig.~\ref{0234_lightcurve}.}
\label{2251_lightcurve}
    \end{figure}

We also use support observations to address the behaviour of the five example sources. 
In Figs.~\ref{0234_lightcurve}--\ref{2251_lightcurve} we have plotted the 37\,GHz
long-term flux density curves for the five sources.
The long-term data have been taken at Mets\"ahovi Radio Observatory. 

\begin{figure}
   \resizebox{\hsize}{!} {\includegraphics{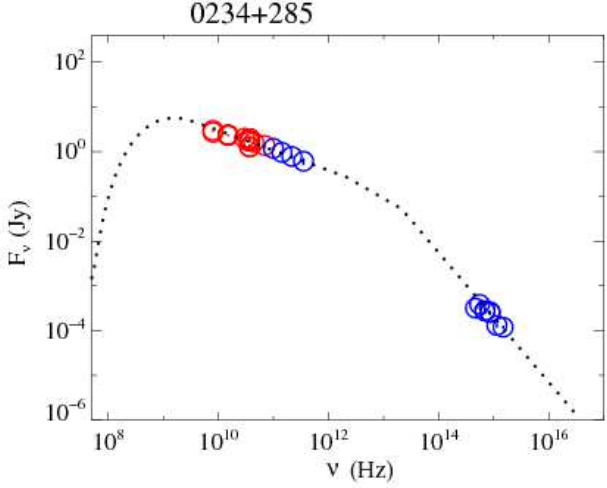}}
    \caption{A possible single-component spectrum for 0234+285 to illustrate the general shape. Red circles, 
LF data simultaneous to \planck; blue circles, HF data simultaneous to \planck.}
\label{0234}
    \end{figure}

\begin{figure}
   \resizebox{\hsize}{!} {\includegraphics{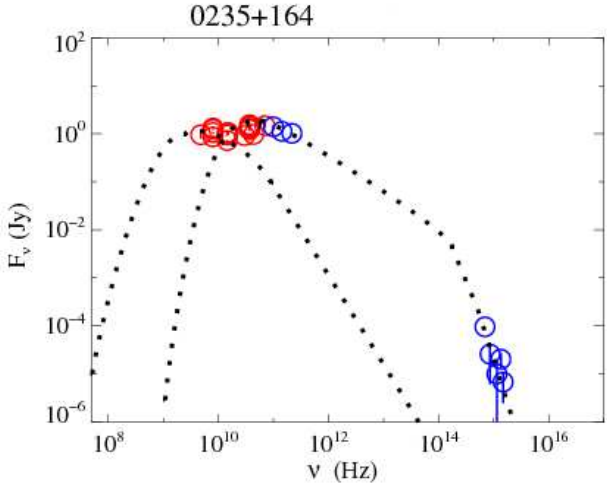}}
    \caption{A possible multiple-component spectrum for 0235+164 to illustrate the general shape. 
Symbols as in Fig.~\ref{0234}.}
\label{0235}
    \end{figure}

\begin{figure}
   \resizebox{\hsize}{!} {\includegraphics{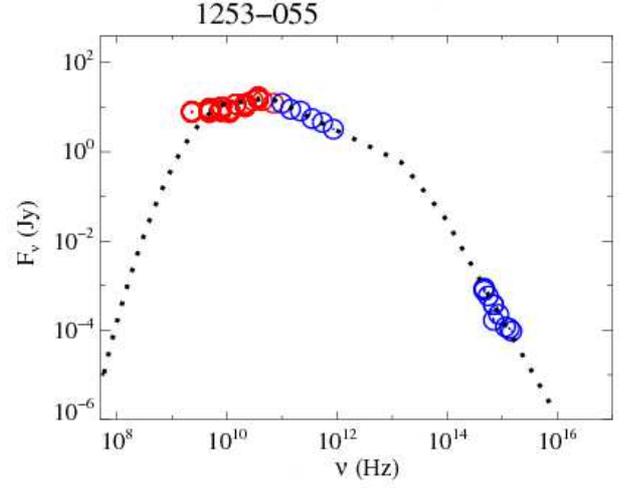}}
    \caption{The radio spectrum of 1253$-$055 shows that there is no room for
another component between radio and optical regimes. Symbols as in Fig.~\ref{0234}.}
\label{1253}
    \end{figure}

\begin{figure}
   \resizebox{\hsize}{!} {\includegraphics{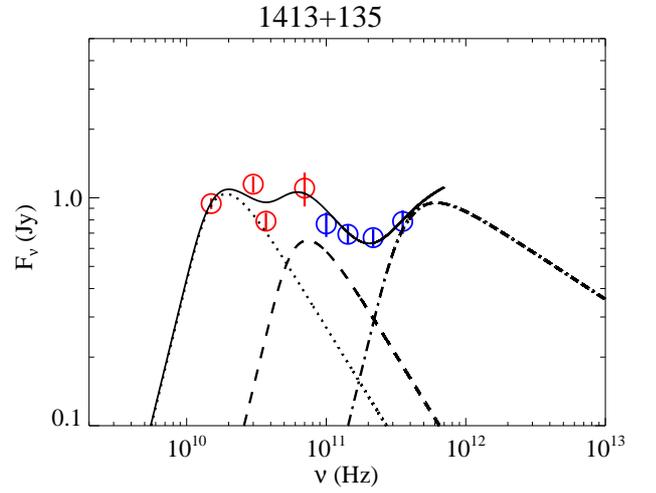}}
    \caption{The radio spectrum of 1413+135 is very flat and can be fitted with multiple components. 
Symbols as in Fig.~\ref{0234}.}
\label{1413}
    \end{figure}

\begin{figure}
  \resizebox{\hsize}{!} {\includegraphics{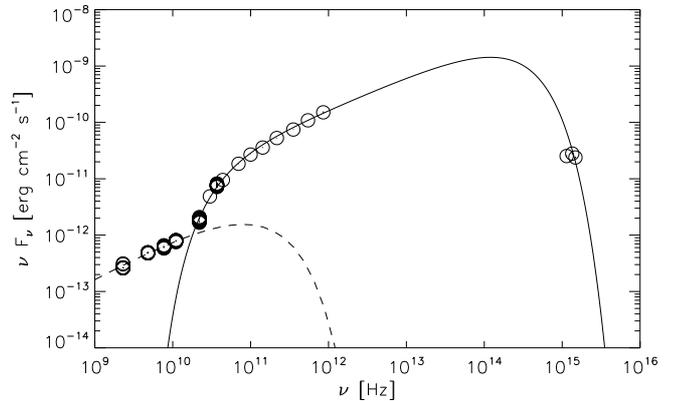}}
    \caption{The radio-optical SED of 2251+158 (3C\,454.3) fitted with one component and synchrotron
emission from the jet (dashed line) using a model by Tammi et al. (in preparation). The radio spectrum
is shown in Fig.~\ref{2251b}.}
\label{2251a}
    \end{figure}

\begin{figure}
  \resizebox{\hsize}{!} {\includegraphics{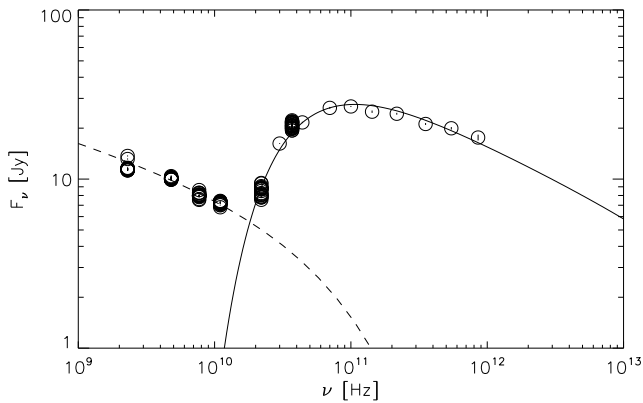}}
    \caption{A close-up of the radio spectrum and model fit of 2251+158 (3C\,454.3), details as in 
Fig.~\ref{2251a}.}
\label{2251b}
    \end{figure}

\begin{figure*}
\includegraphics[scale=0.8]{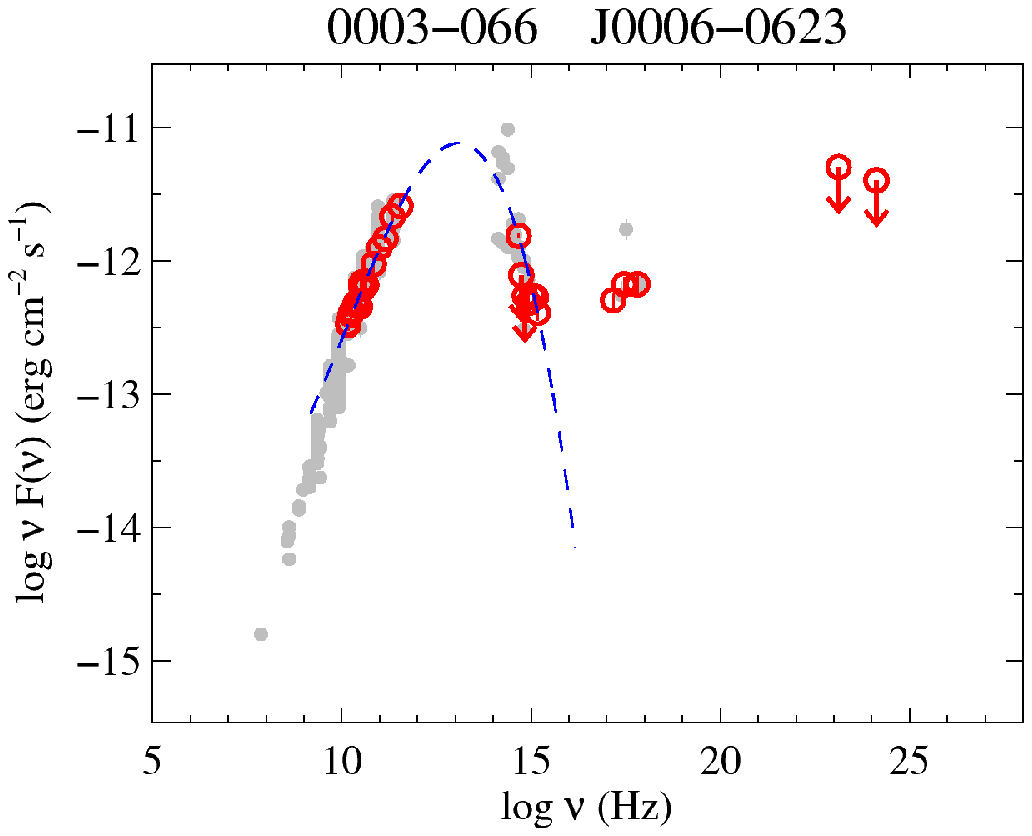}
\includegraphics[scale=0.8]{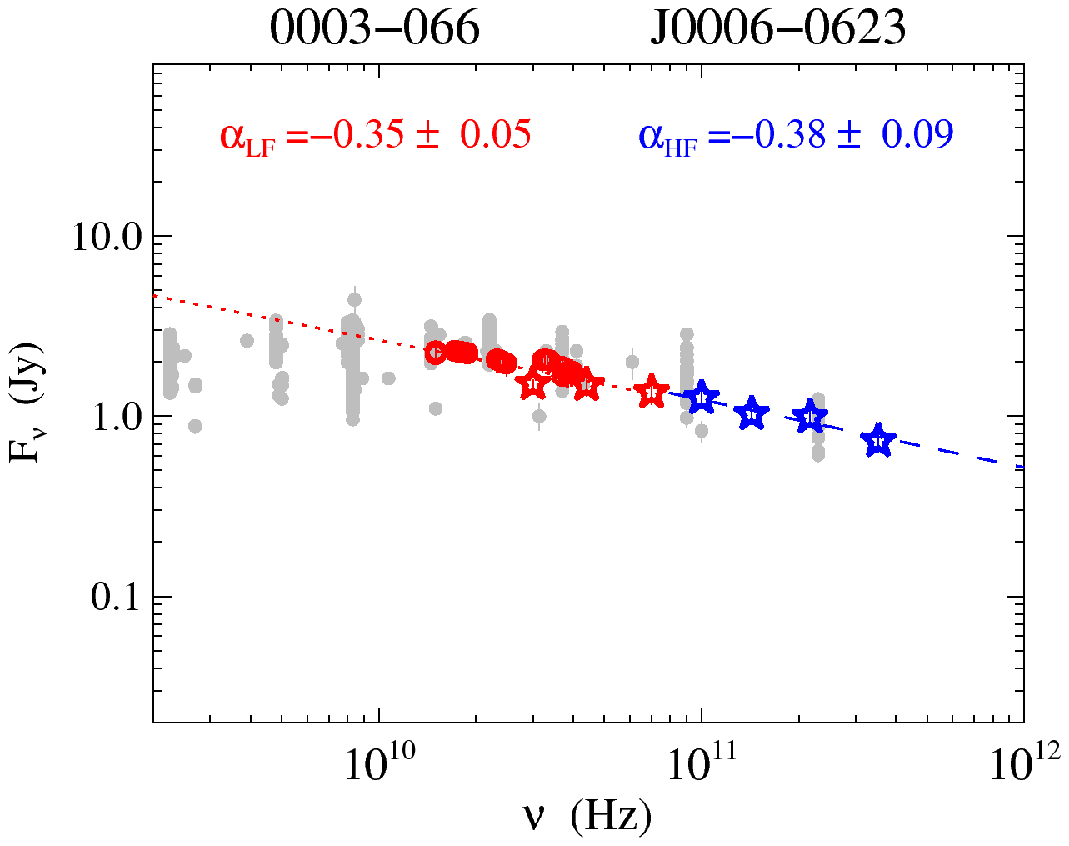}
 \caption{\emph{Left-panel:} The SED of the source 0003$-$066. Grey circles show the historical 
data. The red circles show data simultaneous to 
the \planck\ observations. The dotted and dashed lines show the second and third degree polynomials, 
respectively, fitted to the synchrotron and IC bumps in the SED.
\emph{Right-panel:} The radio spectrum of 0003$-$066. Red circles, LF data simultaneous to \planck; 
red stars, ERCSC LFI data; blue circles, HF data simultaneous to \planck;
blue stars, ERCSC HFI data.
The dashed and dotted lines are fits to simultaneous LF and HF data.}
\label{sedsfig}
\end{figure*}

\begin{figure*}
\includegraphics[scale=0.8]{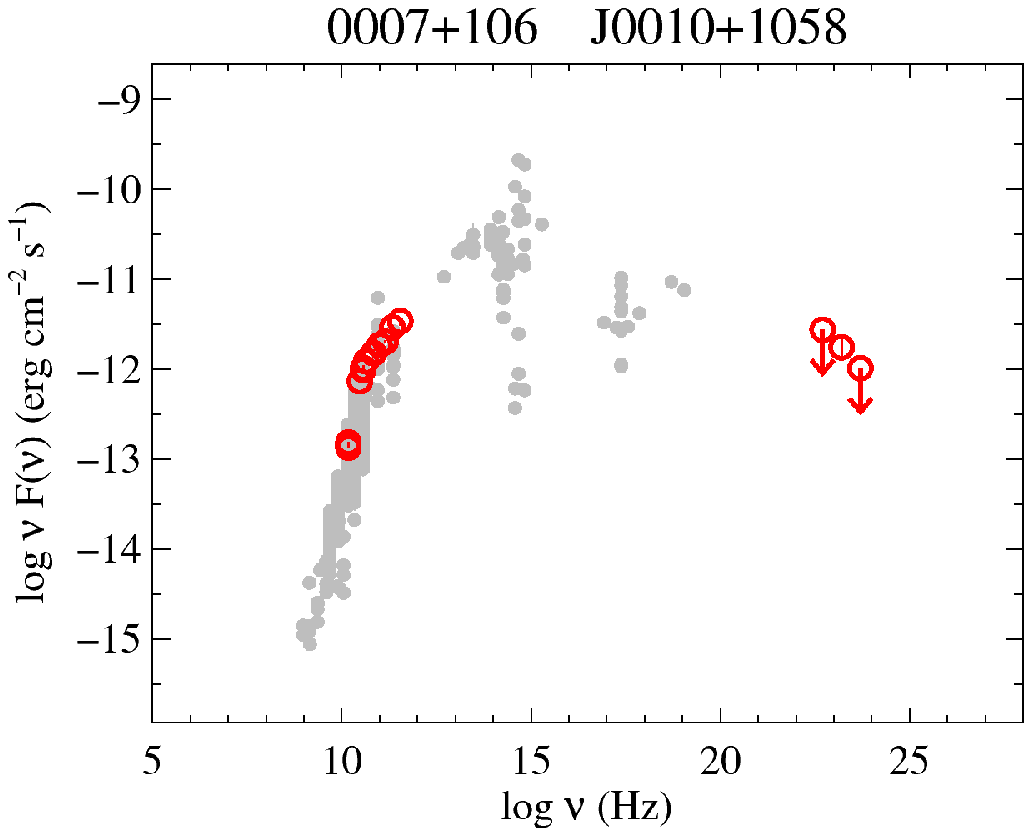}
\includegraphics[scale=0.8]{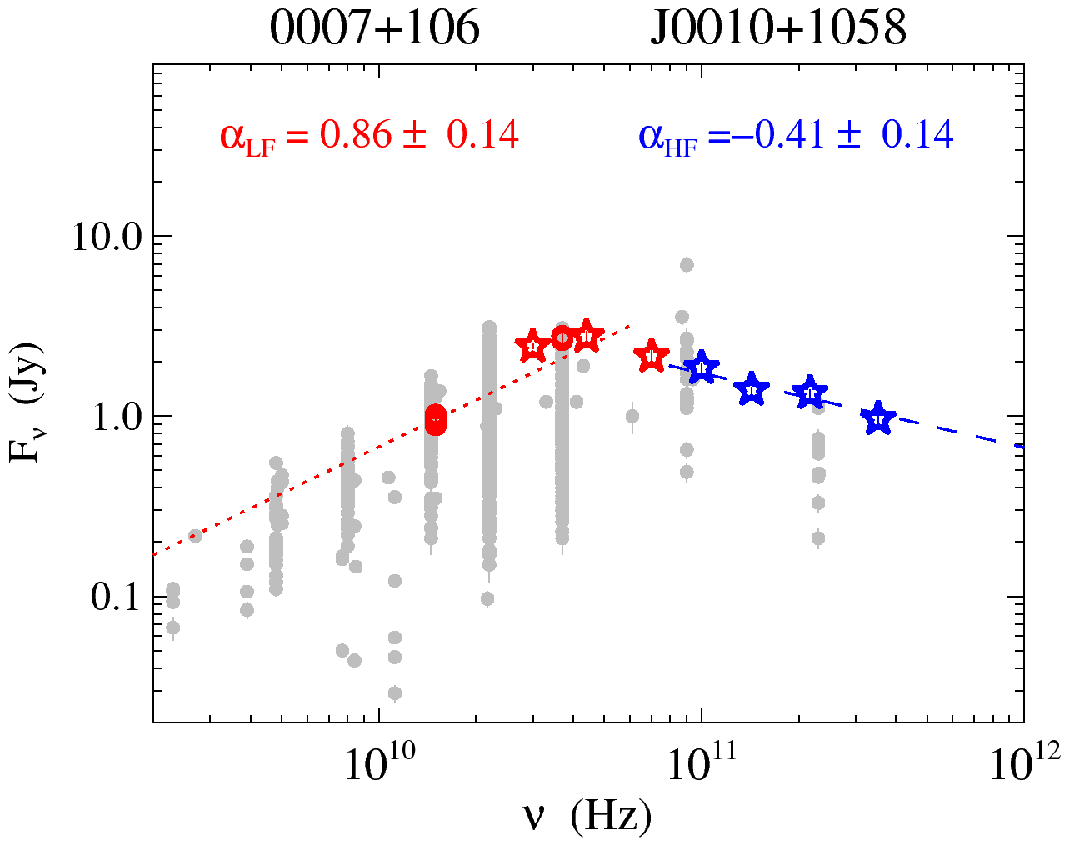}
 \caption{0007+106}
\label{0007_sed}
\end{figure*}

Even without numerical modelling we can address two important topics with these sources. The first one is the
spectral flatness of the optically thin radiation possibly hinting at non-standard acceleration processes.
The second one concerns the origin of the IC component. As Figs.~\ref{0234}--\ref{2251b} show there does not appear to be
room for additional synchrotron components between the \planck\ and the optical frequencies. The IC radiation
must therefore originate in the same shocks that also produce the radio emission, in contradiction to most theoretical models.

{\bf 0234+285} (Fig.~\ref{0234}).
This source was in a very uneventful state during the \planck\ observing period (Fig.~\ref{0234_lightcurve})
and shows a spectrum typical of sources in a quiescent stage. The radio spectrum can be modelled with a single 
synchrotron component, having $\alpha_{\rm thin} = -0.61$, which apparently steepens by $\Delta\alpha = -0.5$ 
and smoothly joins the optical spectrum (see also Fig.~\ref{0234_sed}). This synchrotron component can be 
identified with the latest significant flare (new shock), which peaked about two years earlier at 37\,GHz 
(Fig.~\ref{0234_lightcurve}). During low flux levels, in the absence of recent flares/shocks, the 
underlying, relatively stable jet may also contribute significantly to the synchrotron emission. Another 
similar example is the source 0007+106 (Fig.~\ref{0007_sed}). Lacking mid-IR data, we cannot prove that 
there is no additional synchrotron component, between the \planck\ frequencies and the optical, that  
might be the main source of the IC flux (Fig.~\ref{0234_sed}). However, the most straightforward 
interpretation of the data is that the IC flux also comes from the synchrotron component outlined in 
Fig.~\ref{0234}: from a rather old shock far downstream the jet, and far beyond the BLR.

{\bf 0235+164} (Fig.~\ref{0235}). 
The bumpiness of the radio spectrum shows clear evidence of multiple 
components (shocks) contributing to the observed total flux density. We have outlined two possible synchrotron 
components, which together could explain most of the observed radio spectrum. Although 0235+164 was 
relatively quiescent during the \planck\ observation (Fig.~\ref{0235_lightcurve}), there were at 
least four strong flares during the previous five years, and they all are likely to be still contributing 
significantly to the total radio flux density, at least around $10^{10}$\,Hz and below. Assuming reasonable physical 
parameters for these shocks, they can also produce significant amounts of X-rays through the SSC mechanism, 
which must be accounted for in a proper multicomponent modelling of the IC spectrum. As with 0234+285, 
the higher-turnover component with $\alpha_{\rm thin} = -0.4$ can be joined smoothly to the optical spectrum 
assuming a steepening by about $\Delta\alpha = -0.5$, leaving no room for additional mid-IR components. 
\textit{Fermi} measured only upper limits for the gamma-ray flux, which may be due to the fact that even the most 
recent shock, peaking around 100\,GHz, is already about two years old (Fig.~\ref{0235_lightcurve}).

{\bf 1253$-$055 (3C\,279).} (Fig.~\ref{1253}). 
Again, the highest-peaking radio component with $\alpha_{\rm thin} = -0.6$ 
can be made to smoothly join the steep optical spectrum. From Fig.~\ref{1253} it is rather obvious that 
there can be no additional mid-IR components which might provide a significant contribution to the IC 
spectrum. If we assume that there is another synchrotron component turning over somewhere above the \planck\ 
frequencies and joining smoothly to the optical spectrum, Fig.~\ref{1253} shows that its flux density cannot 
exceed the dashed line at any IR frequency between $10^{12}$ and $10^{14.5}$\,Hz. 
This is the case even if we make the assumption that the spectrum of the component seen at the highest 
\planck\ frequencies steepens very rapidly above 857\,GHz. 
Consequently, the \textit{Fermi} gamma-rays must originate in the latest shock, which started a few months 
before the \planck\ observation (Fig.~\ref{1253_lightcurve}).

{\bf 1413+135} (Fig.~\ref{1413}). This source exhibits a totally flat spectrum, with both the LF and the 
HF spectral indices approximately 0. Possible explanations are: an underlying electron energy index 
$s \approx 1$, which is rather unlikely; the superposition of many self-absorbed synchrotron components 
(shocks of various ages) with different turnover frequencies; or a continuous jet dominating the flux and 
having an integrated spectral index close to zero over a wide frequency range. As Fig.~\ref{1413_lightcurve} 
shows, the source has been rather inactive for over a decade, with no major ourbursts. The HFI 
spectrum may indicate an upturn, hinting at the presence of an IR component. We show a possible spectral 
decomposition, including an IR-peaking component. Further \planck\ observations are needed to ascertain the 
reality and the nature of this component, which may be either a synchrotron flare seen at an early 
developmental stage or, possibly, an infrared dust component. The correct alternative can be identified 
through variability information.

{\bf 2251+158 (3C\,454.3).} (Figs.~\ref{2251a} and \ref{2251b}). 
This source exhibits the most spectacular case of ongoing IR flaring, 
showing a very strong synchrotron component with a self-absorption turnover at about 80\,GHz and a very 
flat ($\alpha = -0.2$) optically thin spectrum. As Fig.~\ref{2251_lightcurve} shows, a series of 
spectacular flaring events has occurred in 2251+158 during the last five years, with \planck\ observing the 
source during the early stages of the strongest outburst ever seen in the source. In Fig.~\ref{2251a} 
we show a model fit to the radio-to-optical data, using a preliminary version of the code that we are developing 
(Tammi et al., in preparation). A close-up of the fit in the radio regime is shown in Fig.~\ref{2251b}.
As Fig.~\ref{2251_sed} also shows, it is impossible to squeeze in another 
component between the strong radio-IR flare and the optical regime. The gamma-ray emission must come from 
the growing shock that we see both in the total flux density monitoring curve (Fig.~\ref{2251_lightcurve})
and in the SED (Fig.~\ref{2251_sed}). Most of the theoretical model scenarios assume that gamma-rays 
originate close to the black hole and the accretion disk, well within the BLR and far upstream of the radio 
core. This is not a viable alternative for 2251+158, in which gamma-rays must come from the new shock 
component, parsecs downstream from the radio core. As Fig.~\ref{2251_lightcurve} shows, at the time of the 
\planck\ observation the shock had already been growing for about half a year, which, changed into the 
source frame, translates into a distance $L = D^2 c(1+z)^{-1} \Delta t_{\rm obs}$ (where $D$ is the Doppler
boosting factor, $c$ the speed of light, $z$ the redshift, and $\Delta t_{\rm obs}$ the time elapsed from 
the onset of the flare), of at least 10\,pc down the jet, 
assuming a modest value of $D \approx 10$, or 90\,pc using the
value of $D=33.2$  from \citet{hovatta09}. This is the distance that the shock 
has moved downstream after emerging from the radio core.

\section{The shape of the synchrotron spectrum: nonstandard acceleration mechanisms?}
\label{acceleration}

If the electron energy distribution is assumed to be a simple power-law with the form
$N(E) = K E^{-s}$, the synchrotron spectrum of a single homogeneous source has an
optically thick low-frequency part with a spectral index $\alpha = +2.5$,
 a turnover, and an optically thin part with a spectral index $\alpha$ related
to the electron energy index by $\alpha = -(s-1)/2$. Energy losses eventually cause the
spectrum to steepen to $\alpha - 0.5$.

Since the earliest times of AGN research, the canonical value for $\alpha$ has been
assumed to be about $-0.75$, corresponding to $s = 2.5$. Such an electron energy index
emerges naturally from the simplest first-order Fermi acceleration mechanism, and blazars
also tend to show slopes around this value at the high-frequency (10 to 100\,GHz) end of
the radio spectrum, where the sources are typically assumed to be optically thin.
However, \planck\ data now challenge this commonly accepted view. As Fig.~\ref{lf} shows,
the LF spectral slopes are, in the main, quite flat, and sometimes even rising. It is
likely that in most cases this results from the superposition of several distinct
synchrotron components, since the spectra are generally not smooth, and we also know from
VLBI observations that there are nearly always several components (shocks) present in any
source (see also Sect.~\ref{model}). The HF spectra (Fig.~\ref{hf}), on the other
hand, are often straight within the \planck\ flux density error bars, and give a 
definite impression of optically thin radiation. In some cases, such as 2251+158, the
spectrum even shows a  clear self-absorption turnover at high frequencies,
followed by a flat optically thin part with $\alpha = -0.2$ (Figs.~\ref{2251a} and \ref{2251b}).

The flatness of the HF spectra, shown in Fig.~\ref{hf}, is remarkable. Only a third 
of the sources have $\alpha$ steeper than $-0.7$, while 15 have $\alpha > -0.3$. While the 
present data are not sufficient to exclude the possibility that multiple components 
produce the flatness, and some sources such as 0954+658 do show evidence of an additional 
IR component (see also, e.g., \citealt{raiteri99}), we consider such an explanation unlikely 
for the whole sample. The errors in 
$\alpha_{\rm HF}$ cannot explain the very flat spectral indices, since they generally are smaller 
than $\Delta\alpha \approx 0.1$. 

\begin{table}
\caption{A list of the 10 sources scanned only once by \planck\ and with $\alpha_{\rm HF} > -0.5 + \Delta\alpha$.
S = straight HF spectrum; B = bumpy HF spectrum, possibly indicating multiple components.}
\label{straight}
\centering
\begin{tabular}{l c c}
\hline\hline
\noalign{\smallskip}
Source & $\alpha_{\rm HF}$ & Spectral shape\\
\noalign{\smallskip}
\hline
\noalign{\smallskip}
0003$-$066 & $-0.38 \pm 0.09$ & S\\
0059+581 & $-0.19 \pm 0.09$ & S\\
0235+164 & $-0.40 \pm 0.03$ & S\\
0430+052 & $-0.34 \pm 0.05$ & S\\
1308+326 & $-0.41 \pm 0.08$ & B?\\
1413+135 & $\phantom{+}0.02 \pm 0.12$ & B\\
1418+546 & $-0.16 \pm 0.04$ & S\\
1652+398 & $-0.21 \pm 0.04$ & B\\
1823+568 & $-0.31 \pm 0.13$ & S\\
2251+158 & $-0.20 \pm 0.02$ & S\\
\hline
\end{tabular}
\end{table}

In Table~\ref{straight} we list the 10 sources that fullfill the following criteria.
They have been scanned only once by \planck\ and they have $\alpha_{\rm HF} > -0.5 + \Delta\alpha$,
where $\Delta\alpha$ is the HF spectral fit error shown in Figs.~\ref{sedsfig} -- \ref{last}.
These form the most extreme end of the $\alpha_{\rm HF}$ distribution shown in Fig.~\ref{hf}. 
Three of them may show indications of multiple HF components in the spectra, but for
the seven other sources the HF spectra appear to be straight within the \planck\ errors.
If this is indeed the case, they cannot be explained with standard acceleration mechanisms
with $s \geq 2$.

The rise of a submillimetre flare is quite fast (see, for example, the JCMT monitoring data 
of \citealt{stevens94}); the self-absorption turnover of a fresh shock component  
passes through the \planck\ HFI frequencies rather rapidly, and one therefore sees it only 
rarely in the snapshot high frequency spectra. (In the present data only 2251+158 is clearly 
such a case, although 0235+164 and 1652+398 might be others.) 
Instead, in most cases we expect to see optically thin submillimetre spectra 
which have already experienced energy losses. Therefore the observed high-frequency 
spectral-index distribution of a sample of sources should show a range from $\alpha_{\rm thin} = -(s-1)/2$ to 
$\alpha_{\rm thin}-0.5$ and beyond, as the exponential steepening sets in, with spectral indices 
steeper than the initial $\alpha_{\rm thin}$ dominating. The observed  distribution in Fig.~\ref{hf} 
is thus incompatible with an initial $\alpha_{\rm thin}$ around $-0.7$, as we should then 
see a few values around $\alpha = -0.7$ and a distribution peaking towards $-1.2$, reflecting 
the $\Delta\alpha = -0.5$ steepening due to energy losses.

Instead, the spectral index distribution is compatible with an electron index $s \approx 1.5$, 
resulting in a distribution of $\alpha_{\rm thin}$ having smallest values around 
$-0.2$ to $-0.3$ and a maximum around $-0.7$.
Such a suggestion of a very hard original electron spectrum is not entirely new;
\citet{valtaoja88_moniII} studied the shock spectra and concluded that $\alpha$ is around
$-0.2$, and \citet{hughes91} also concluded that the flux density variations can best be
modelled with very hard spectra. \citet{gear94} also found that submillimetre spectra in
a sample of 48 sources tended to be flat. Indeed, the distribution they found (their figure 7)
 is quite similar to ours (Fig.~\ref{hf}).

As noted above, alternative explanations for the $\alpha_{\rm HF}$ distribution cannot be
totally excluded. However, we consider sources  such as 2251+158 
(Figs.~\ref{2251a} and \ref{2251b}) to be compelling evidence for the existence
of very hard non-standard electron spectra. The implications of this possibility are
discussed below.

If the original $\alpha_{\rm thin}$ really is around $-0.2$, the distribution shown in
Fig.~\ref{hf} becomes easy to understand: in many slightly older HFI components we see
the expected spectral steepening from $-0.2$ to $-0.7$ and beyond. One way to test this
hypothesis is to try to estimate the ages of the highest-frequency spectral components,
either from their turnover frequencies or from the total flux-density monitoring: the
older the component, the lower the turnover frequency and the steeper the spectra.
However, this analysis is beyond the scope of this paper and reliable conclusions require more
\planck\ data than presently exist.

An electron energy spectral index as hard as 1.5 has major implications for the
acceleration mechanisms dominating in blazar shocks. Particle acceleration efficiency is
strongly governed by the compression ratio of the flow, which in turn is confined by
well-known plasma physics, preventing the spectral index from becoming arbitrarily small.
For a relativistic shock the traditional first-order Fermi mechanism usually produces
spectral indices of 2.2 or so (and not 2.5, which follows from the nonrelativistic
theory), but not smaller. There are a few options, however, for bypassing the $s \approx
2.2$ limit. For example, power-law spectra with $s \gtrsim 1$ have been found in certain
kinds of oblique shocks \citep[e.g.,][]{kirk_heavens89}, as well as in parallel shocks in the
case of large-angle particle scattering due to very strong turbulence \citep[e.g.,][]{stecker_etal97},
 or when the compression ratio felt by the particles is
higher than that felt by the plasma \citep{ellison_etal90,virtanen05}. The last
alternative is an especially interesting one, because the combination of low density and
relatively strong magnetic field---conditions likely to be found in
Poynting-flux-dominated AGN jets---quickly leads to non-negligible Alfv\'en speeds and
enhancement of the compression ratio felt by the particles and, consequently, to very hard
particle spectra \citep[and references therein]{tammi08b,tammi08a}. High Alfv\'en
speeds and scattering centres that are not frozen into the plasma also enable 
\textit{second-order} Fermi acceleration, which can produce power-law spectra even with
$s<1$ in relativistic parallel shocks \citep{virtanen05} and can work on time-scales
comparable to the fastest blazar flares \citep{tammi_duffy09}.

As a simple order-of-magnitude reality check we can estimate the conditions required by
the standard first-order acceleration in a parallel step shock in the presence of an increased
scattering-centre compression ratio due to turbulence transmission \citep{vainio_etal03}.
Following the analysis of \citet{tammi08a}, we can estimate that in order
to achieve $s \approx 1.5$ in a shock with Lorentz factor $\Gamma = 10$, magnetic field
strength of the order of 0.02--2\,Gauss (depending on the composition and the density of the
plasma) is needed. This is in good agreement with the usual modelling parameters of these
sources, although we emphasise that the analysis used here is a simple one and may be
limited to parallel shocks and weak turbulence.

\section{Conclusions}
\label{con}

We have presented the averaged \planck\ ERCSC spectra together
with supporting ground and satellite observations obtained simultaneously
with the \planck\ scans. The 104 SEDs, supplemented with archival data,
demonstrate the usefulness of \planck\ data in determining the SED peak
frequencies and fluxes and in modelling the spectral energy distributions in
greater detail. In particular, the data demonstrate that the synchrotron
spectrum contains contributions from several physically distinct AGN
components: the LF and HF spectra are rarely smooth, except at the highest
radio frequencies where the source component with the highest turnover
becomes optically thin. The LF spectral indices cluster around $\alpha = 0$,
also indicating the superposition of many components with different turnover
frequencies. In modelling the synchrotron and  inverse-Compton SEDs, a
multicomponent approach is therefore necessary. While physical modelling is
beyond the scope of this paper, which uses averaged, rather than single-epoch, \planck\ 
ERCSC flux densities, we have shown some
examples of multicomponent decomposition of the SEDs and the
conclusions that can be drawn from the data.

The sources display remarkable variability, which must also be taken into
account in modelling the SEDs. We have presented some examples of total flux
density monitoring to demonstrate how sources
have been observed during \planck\ scans in different stages of activity.

The \planck\ HFI spectra are remarkably flat, with only a minority of sources
having spectral indices steeper than the ``canonical'' $\alpha = -0.7$, and many
apparently having $\alpha_{\rm thin}$ around $-0.2$ to $-0.4$. 
We suggest that the most likely interpretation for the very flat and straight
synchrotron spectra observed by \planck\ is a very hard original electron 
spectral energy index, clearly below $s = 2$. While this possiblity must be
confirmed by future data, an index $s = 1.5$ is compatible with the HF spectral 
index distribution seen in our sample. In a number of sources, energy losses 
steepen the HF spectra by $\Delta\alpha = 0.5$ from about $-0.3$ to about $-0.8$, 
and further to $-1$ and beyond, which can explain the observed distribution.
The hardness of the electron spectrum would also mean that the
synchrotron SED peak is in most cases related to the maximum electron
Lorentz factor achieved in the acceleration process. The hard spectra also
require a rethinking of the acceleration processes dominating in the
relativistic jets and shocks.

For some sources the HF spectra are flat or even rising, indicating the
presence of multiple synchrotron-emitting components peaking in the
gigahertz to terahertz regime, and ongoing flaring at still higher frequencies. However,
in many cases the \planck\ HFI spectrum appears to join the optical spectrum
smoothly, leaving little room for additional still higher-frequency
components. In these cases, the inverse-Compton gamma-rays must originate in
the synchrotron (i.e., shock) components seen in the \planck\ data.

\begin{acknowledgements}
The Planck Collaboration acknowledges the support of: ESA; CNES and
CNRS/INSU-IN2P3-INP (France); ASI, CNR, and INAF (Italy); NASA and DoE (USA); STFC and UKSA (UK); CSIC,
MICINN and JA (Spain); Tekes, AoF and CSC (Finland); DLR and MPG (Germany); CSA (Canada); DTU Space
(Denmark); SER/SSO (Switzerland); RCN (Norway); SFI (Ireland); FCT/MCTES (Portugal); and DEISA (EU).
A description of the Planck Collaboration and a list of its members, indicating
which technical or scientific activities they have been involved in, can be found via
http://www.rssd.esa.int/Planck.
The Mets\"ahovi and Tuorla observing projects are supported by the Academy
of Finland (grant numbers 212656, 210338, 121148, 127740 and 122352).
UMRAO is supported by a series of grants from the NSF and NASA, and by the
University of Michigan.
This publication is partly based on data acquired with the Atacama
Pathfinder Experiment (APEX). APEX is a collaboration between the
Max-Planck-Institut f\"ur Radioastronomie, the European Southern
Observatory, and the Onsala Space Observatory. This research is 
partly based on observations with the 100-m telescope of the MPIfR
(Max-Planck-Institut f\"ur Radioastronomie) at Effelsberg, the IRAM 30-m 
telescope, and the Medicina (Noto) telescope operated by INAF--Istituto di 
Radioastronomia. This paper makes use of observations obtained at the Very Large Array 
(VLA) which is an instrument of the National Radio Astronomy Observatory (NRAO). 
The NRAO is a facility of the National Science Foundation operated under cooperative agreement
by Associated Universities, Inc.
The observations at Xinglong station are supported by the Chinese National Natural Science 
Foundation grants 10633020, 10778714, and 11073032, and by the National Basic Research Program
of China (973 Program) No. 2007CB815403.
The OVRO 40-m monitoring program is supported in part by NASA.
The Australia Telescope is funded by the Commonwealth of Australia for
operation as a National Facility managed by CSIRO. 
The \textit{Fermi} LAT Collaboration acknowledges generous ongoing support from a number
of agencies and institutes that have supported both the development and the operation of
the LAT as well as scientific data analysis. These include the National Aeronautics and
Space Administration and the Department of Energy in the United States, the Commissariat
\`a l'Energie Atomique and the Centre National de la Recherche Scientifique / Institut
National de Physique Nucl\'eaire et de Physique des Particules in France, the Agenzia
Spaziale Italiana and the Istituto Nazionale di Fisica Nucleare in Italy, the Ministry of
Education, Culture, Sports, Science and Technology (MEXT), High Energy Accelerator
Research Organization (KEK) and Japan Aerospace Exploration Agency (JAXA) in Japan, and
the K.~A.~Wallenberg Foundation, the Swedish Research Council and the Swedish National
Space Board in Sweden.
Additional support for science analysis during the operations phase is gratefully
acknowledged from the Istituto Nazionale di Astrofisica in Italy and the Centre National
d'\'Etudes Spatiales in France.
Part of this work is based on archival data, software or on-line services provided by the
ASI Science Data Center ASDC.
We thank the \textit{Fermi} LAT team reviewers, S. Ciprini and M. Giroletti, for their effort
and valuable comments.

\end{acknowledgements}


\bibliographystyle{aa}
\bibliography{Planck_bib,enbib}

\Online

\longtab{1}{
\begin{landscape}
\begin{longtable}{l l l l r c c c c c c c}   
\caption{The complete 1\,Jy northern sample of AGN. Column 8 refers to the 37\,GHz observations of Mets\"ahovi Radio Observatory.}\label{sample}  \\              
\hline \hline 
\multicolumn{1}{c}{(1)} & \multicolumn{1}{c}{(2)} & \multicolumn{1}{c}{(3)} & \multicolumn{1}{c}{(4)} & \multicolumn{1}{c}{(5)} & (6) & (7) & (8) & (9) & (10) & (11) & (12)\\           
\multicolumn{1}{c}{Name} & \multicolumn{1}{c}{J2000 Name} & \multicolumn{1}{c}{Alias} & \multicolumn{1}{c}{R.A.(J2000)} & \multicolumn{1}{c}{Dec(J2000)} & 1st scan & 2nd scan & $S_{\textrm{ave}}$ [Jy] & $\log\nu_s$ & Syn quality & $\log\nu_{\rm IC}$ & IC quality\\  
\hline     
\endfirsthead        
\caption{continued.}\\
\hline\hline
\multicolumn{1}{c}{(1)} & \multicolumn{1}{c}{(2)} & \multicolumn{1}{c}{(3)} & \multicolumn{1}{c}{(4)} & \multicolumn{1}{c}{(5)} & (6) & (7) & (8) & (9) & (10) & (11) & (12)\\           
\multicolumn{1}{c}{Name} & \multicolumn{1}{c}{J2000 Name} & \multicolumn{1}{c}{Alias} & \multicolumn{1}{c}{R.A.(J2000)} & \multicolumn{1}{c}{Dec(J2000)} & 1st scan & 2nd scan & $S_{ave}$ [Jy] & $\log\nu_s$ & Syn quality & $\log\nu_{\rm IC}$ & IC quality\\       
\hline
\endhead  
\hline
\endfoot  
0003$-$066 & 0006$-$0623 & NRAO5 & 00:06:13.90 & $-$06:23:36.00 & 2009-12-12 & ... & 2.06 & 13.1 & A & ... & ... \\
0007+106 & 0010+1058 & IIIZW2 & 00:10:31.01 & 10:58:29.50 & 2009-12-25 & ... & 1.04 & ... & ... & ... & ... \\
0048$-$097 & 0050$-$0929 & PKS0048$-$097 & 00:50:41.20 & $-$09:29:06.00 & 2009-12-23 & ... & 1.34 & 14.1 & A & ... & ... \\
0059+581 & 0102+5824 & TXS 0059+581 & 01:02:45.76 & 58:24:11.14 & 2010-01-28 & ... & 2.97 & 13.6 & B & ... & ... \\
0106+013 & 0108+0135 & OC012 & 01:08:38.77 & 01:35:00.30 & 2010-01-04 & ... & 1.85 & 12.7 & B & ... & ... \\
J0125$-$0005 & 0125$-$0005 &  & 01:25:28.84 & $-$00:05:56.00 & 2010-01-08 & ... & 1.24 & 13.6 & B & ... & ... \\
0133+476 & 0136+4751 & DA55 & 01:36:58.59 & 47:51:29.10 & 2010-01-29 & ... & 3.02 & 12.8 & A & ... & ... \\
0149+218 & 0152+2207 &  & 01:52:18.06 & 22:07:07.70 & 2010-01-24 & ... & 1.06 & ... & ... & ... & ... \\
0202+149 & 0204+1514 & 4C15.05 & 02:04:50.41 & 15:14:11.00 & 2010-01-24 & ... & 1.64 & ... & ... & ... & ... \\
0212+735 & 0217+7349 &  & 02:17:30.81 & 73:49:32.60 & 2009-09-13 & 2010-02-11 & 2.05 & 12.4 & B & ... & ... \\
0224+671 & 0228+6721 &  & 02:28:50.05 & 67:21:03.00 & 2009-09-13 & 2010-02-10 & 1.28 & ... & ... & ... & ... \\
0234+285 & 0237+2848 & 4C28.07 & 02:37:52.41 & 28:48:09.00 & 2010-02-03 & ... & 2.86 & 13.1 & A & 22.1 & B \\
0235+164 & 0238+1636 &  & 02:38:38.80 & 16:36:59.00 & 2010-02-01 & ... & 2.33 & 13.1 & A & ... & ... \\
0238$-$084 & 0241$-$0815 &  & 02:41:04.80 & $-$08:15:20.75 & 2010-01-24 & ... & 1.33 & ... & ... & ... & ... \\
0306+102 & 0309+1029 & PKS0306+102 & 03:09:03.60 & 10:29:16.00 & 2010-02-06 & ... & 1.01 & 12.9 & B & ... & ... \\
0316+413 & 0319+4130 & 3C84 & 03:19:48.16 & 41:30:42.10 & 2010-02-13 & ... & 14.01 & ... & ... & 22.9 & B \\
0333+321 & 0336+3218 & NRAO140 & 03:36:30.11 & 32:18:29.30 & 2010-02-15 & ... & 1.49 & ... & ... & ... & ... \\
0336$-$019 & 0339$-$0146 & CTA026 & 03:39:30.94 & $-$01:46:36.00 & 2010-02-10 & ... & 2.21 & 13.1 & B & ... & ... \\
0355+508 & 0359+5057 & NRAO150 & 03:59:29.75 & 50:57:50.20 & 2009-09-13 & 2010-02-19 & 5.92 & ... & ... & ... & ... \\
0415+379 & 0418+3801 & 3C111 & 04:18:21.28 & 38:01:35.80 & 2010-02-22 & ... & 5.97 & ... & ... & ... & ... \\
0420$-$014 & 0423$-$0120 & OA129 & 04:23:15.80 & $-$01:20:33.10 & 2010-02-21 & ... & 7.27 & 13.2 & B & ... & ... \\
0430+052 & 0433+0521 & 3C120 & 04:33:11.10 & 05:21:15.60 & 2010-02-24 & ... & 2.65 & ... & ... & ... & ... \\
0446+112 & 0449+1121 & PKS0446+112 & 04:49:07.67 & 11:21:28.00 & 2010-02-27 & ... & 1.41 & ... & ... & ... & ... \\
0458$-$020 & 0501$-$0159 & PKS0458$-$020 & 05:01:12.81 & $-$01:59:14.00 & 2010-03-01 & ... & 1.86 & ... & ... & ... & ... \\
0507+179 & 0510+1800 &  & 05:10:02.37 & 18:00:41.58 & 2009-09-13 & 2010-03-03 & 1.14 & ... & ... & ... & ... \\
0528+134 & 0530+1331 & PKS0528+134 & 05:30:56.42 & 13:31:55.15 & 2009-09-14 & 2010-03-07 & 4.55 & ... & ... & 21.8 & B \\
0552+398 & 0555+3948 & DA193 & 05:55:30.81 & 39:48:49.17 & 2009-09-22 & 2010-03-09 & 3.18 & ... & ... & 21.4 & B \\
0605$-$085 & 0607$-$0834 & PKS0605$-$085 & 06:07:59.70 & $-$08:34:50.00 & 2009-09-19 & 2010-03-17 & 1.73 & 13.2 & B & ... & ... \\
0642+449 & 0646+4451 & OH471 & 06:46:32.03 & 44:51:16.59 & 2009-09-30 & 2010-03-16 & 2.36 & 12.5 & B & 22.7 & B \\
0716+714 & 0721+7120 &  & 07:21:53.30 & 71:20:36.00 & 2009-10-04 & 2010-03-10 & 2.12 & 14.3 & A & 23.4 & B \\
0723$-$008 & 0725$-$0054 & PKS0723$-$008 & 07:25:50.70 & $-$00:54:56.00 & 2009-10-09 & 2010-04-02 & 1.05 & 13.0 & A & ... & ... \\
0735+178 & 0738+1742 & PKS0735+17 & 07:38:07.40 & 17:42:19.00 & 2009-10-10 & 2010-03-31 & 1.93 & ... & ... & ... & ... \\
0736+017 & 0739+0137 &  & 07:39:18.03 & 01:37:04.60 & 2009-10-12 & 2010-04-04 & 1.85 & 14.0 & B & ... & ... \\
0748+126 & 0750+1231 &  & 07:50:52.05 & 12:31:04.83 & 2009-10-14 & 2010-04-04 & 2.42 & 12.7 & A & 21.5 & A \\
0754+100 & 0757+0956 & OI090.4 & 07:57:06.64 & 09:56:34.90 & 2009-10-15 & 2010-04-06 & 1.33 & 13.0 & B & ... & ... \\
0805$-$077 & 0808$-$0751 &  & 08:08:15.54 & $-$07:51:09.89 & 2009-10-22 & 2010-04-12 & 1.05 & 12.7 & B & ... & ... \\
0804+499 & 0808+4950 &  & 08:08:39.67 & 49:50:36.50 & 2009-10-12 & 2010-03-27 & 1.30 & 13.3 & B & ... & ... \\
0823+033 & 0825+0309 & PKS0823+033 & 08:25:50.30 & 03:09:24.00 & 2009-10-24 & 2010-04-14 & 1.48 & 13.7 & B & ... & ... \\
0827+243 & 0830+2411 & OJ248 & 08:30:52.08 & 24:10:60.00 & 2009-10-21 & 2010-04-09 & 1.48 & ... & ... & ... & ... \\
0836+710 & 0841+7053 & 4C71.07 & 08:41:24.37 & 70:53:42.20 & 2009-10-12 & 2010-03-17 & 1.80 & 12.9 & B & 20.5 & B \\
0851+202 & 0854+2006 & OJ287 & 08:54:48.80 & 20:06:30.00 & 2009-10-27 & 2010-04-15 & 3.90 & 13.1 & B & 21.8 & B \\
0906+430 & 0909+4253 & 3C216 & 09:09:33.50 & 42:53:46.08 & 2009-10-23 & 2010-04-09 & 1.02 & 13.2 & B & ... & ... \\
0917+449 & 0920+4441 &  & 09:20:58.46 & 44:41:53.99 & 2009-10-25 & 2010-04-11 & 1.27 & 12.7 & A & 21.7 & A \\
0923+392 & 0927+3902 & 4C39.25 & 09:27:03.01 & 39:02:20.90 & 2009-10-28 & 2010-04-15 & 7.06 & ... & ... & 22.7 & B \\
0945+408 & 0948+4039 & 4C40.24 & 09:48:55.34 & 40:39:44.60 & 2009-10-31 & 2010-04-18 & 1.32 & 13.1 & B & ... & ... \\
0953+254 & 0956+2515 &  & 09:56:49.88 & 25:15:16.05 & 2009-11-08 & 2010-04-27 & 1.13 & 13.1 & B & ... & ... \\
0954+658 & 0958+6533 & S40954+65 & 09:58:47.20 & 65:33:54.00 & 2009-10-21 & 2010-03-30 & 1.03 & 12.8 & B & 21.7 & B \\
1036+054 & 1038+0512 &  & 10:38:46.78 & 05:12:29.09 & 2009-11-29 & 2010-05-16 & 1.13 & ... & ... & ... & ... \\
TEX1040+244 & 1043+2408 &  & 10:43:09.00 & 24:08:35.00 & 2009-11-19 & 2010-05-10 & 1.02 & ... & ... & ... & ... \\
1055+018 & 1058+0133 & OL093 & 10:58:29.61 & 01:33:58.80 & 2009-12-07 & 2010-05-24 & 3.96 & 12.7 & B & 22.5 & A \\
J1130+3815 & 1130+3815 & QSO B1128+385? & 11:30:53.28 & 38:15:18.50 & 2009-11-21 & 2010-05-15 & 1.16 & 12.9 & B & ... & ... \\
1150+812 & 1153+8058 &  & 11:53:12.50 & 80:58:29.15 & 2009-10-16 & 2010-03-07 & 1.11 & ... & ... & ... & ... \\
1150+497 & 1153+4931 &  & 11:53:24.47 & 49:31:08.83 & 2009-11-16 & 2010-05-11 & 1.40 & ... & ... & ... & ... \\
1156+295 & 1159+2914 & 4C29.45 & 11:59:31.83 & 29:14:44.00 & 2009-12-04 & 2010-05-31 & 2.00 & 13.4 & A & 23.5 & B \\
1219+044 & 1222+0413 &  & 12:22:22.55 & 04:13:15.78 & 2009-12-29 & ... & 1.39 & ... & ... & ... & ... \\
1222+216 & 1224+2122 & PKS1222+216 & 12:24:54.51 & 21:22:47.00 & 2009-12-18 & ... & 1.16 & 14.2 & B & 22.7 & B \\
1226+023 & 1229+0203 & 3C273 & 12:29:06.69 & 02:03:08.60 & 2010-01-02 & ... & 24.37 & 14.1 & B & 20.8 & A \\
1228+126 & 1230+1223 & 3C274 & 12:30:49.42 & 12:23:28.00 & 2009-12-26 & ... & 14.04 & 13.4 & B & ... & ... \\
1253$-$055 & 1256$-$0547 & 3C279 & 12:56:11.17 & $-$05:47:21.50 & 2010-01-13 & ... & 17.20 & 12.8 & B & 22.7 & B \\
1308+326 & 1310+3220 & AUCVn & 13:10:28.66 & 32:20:43.80 & 2009-12-21 & ... & 2.26 & 12.6 & B & ... & ... \\
1324+224 & 1327+2210 &  & 13:27:00.86 & 22:10:50.16 & 2010-01-05 & ... & 1.15 & 12.8 & B & ... & ... \\
1413+135 & 1415+1320 &  & 14:15:58.80 & 13:20:24.00 & 2010-01-25 & ... & 1.51 & ... & ... & ... & ... \\
1418+546 & 1419+5423 & OQ530 & 14:19:46.60 & 54:23:14.00 & 2009-12-06 & ... & 1.16 & ... & ... & 20.6 & B \\
1502+106 & 1504+1029 & OR103 & 15:04:24.98 & 10:29:39.00 & 2010-02-08 & ... & 1.30 & ... & ... & ... & ... \\
1510$-$089 & 1512$-$0905 & PKS1510$-$089 & 15:12:50.53 & $-$09:05:59.00 & 2010-02-14 & ... & 2.48 & 13.4 & B & ... & ... \\
1546+027 & 1549+0237 &  & 15:49:29.44 & 02:37:01.20 & 2010-02-21 & ... & 2.42 & 13.8 & B & ... & ... \\
1548+056 & 1550+0527 &  & 15:50:35.27 & 05:27:10.45 & 2010-02-21 & ... & 2.04 & ... & ... & ... & ... \\
1606+106 & 1608+1029 & 4C10.45 & 16:08:46.20 & 10:29:07.80 & 2010-02-24 & ... & 1.65 & ... & ... & ... & ... \\
1611+343 & 1613+3412 & DA406 & 16:13:41.00 & 34:12:48.00 & 2010-02-23 & ... & 3.03 & 13.5 & B & ... & ... \\
1633+382 & 1635+3808 & 4C38.41 & 16:35:15.49 & 38:08:04.50 & 2010-03-02 & ... & 2.90 & ... & ... & 21.6 & B \\
1638+398 & 1640+3946 &  & 16:40:29.63 & 39:46:46.03 & 2010-03-04 & ... & 1.18 & ... & ... & ... & ... \\
1642+690 & 1642+6856 &  & 16:42:07.85 & 68:56:39.76 & 2009-11-06 & 2010-01-17 & 1.74 & ... & ... & ... & ... \\
1641+399 & 1642+3948 & 3C345 & 16:42:58.81 & 39:48:37.00 & 2010-03-05 & ... & 7.89 & 12.2 & A & ... & ... \\
1652+398 & 1653+3945 & MARK501 & 16:53:52.20 & 39:45:36.00 & 2010-03-09 & ... & 1.00 & ... & ... & ... & ... \\
1739+522 & 1740+5211 & S41739+52 & 17:40:36.98 & 52:11:43.00 & 2010-03-30 & ... & 1.45 & ... & ... & ... & ... \\
1741$-$038 & 1743$-$0350 & PKS1741$-$038 & 17:43:58.86 & $-$03:50:04.60 & 2009-09-13 & 2010-03-18 & 4.39 & ... & ... & ... & ... \\
1749+096 & 1751+0939 & PKS1749+096 & 17:51:32.70 & 09:39:01.00 & 2009-09-13 & 2010-03-21 & 4.50 & 13.0 & A & 23.2 & A \\
1803+784 & 1800+7828 & S51803+784 & 18:00:45.40 & 78:28:04.00 & 2009-10-11 & 2010-02-02 & 1.90 & 13.5 & A & 23.1 & A \\
1807+698 & 1806+6949 & 3C371.0 & 18:06:50.70 & 69:49:28.00 & 2009-10-21 & 2010-01-07 & 1.37 & 14.3 & A & 23.1 & B \\
1823+568 & 1824+5651 & 4C56.27 & 18:24:07.07 & 56:51:01.50 & 2010-04-17 & ... & 1.60 & 13.0 & B & ... & ... \\
1828+487 & 1829+4844 & 3C380 & 18:29:31.80 & 48:44:46.62 & 2010-04-11 & ... & 2.38 & 14.0 & B & ... & ... \\
J184915+67064 & 1849+6705 &  & 18:49:15.89 & 67:06:40.90 & 2009-10-12 & 2009-12-25 & 1.82 & 12.6 & A & ... & ... \\
1928+738 & 1927+7358 & 4C73.18 & 19:27:48.50 & 73:58:01.60 & 2009-09-28 & 2010-01-18 & 2.61 & ... & ... & 21.0 & B \\
1954+513 & 1955+5131 &  & 19:55:42.74 & 51:31:48.55 & 2009-10-09 & 2010-05-06 & 1.35 & 13.7 & B & ... & ... \\
2007+776 & 2005+7752 & S52007+77 & 20:05:31.10 & 77:52:43.00 & 2009-09-25 & 2010-01-27 & 1.40 & 13.1 & B & ... & ... \\
2005+403 & 2007+4029 &  & 20:07:44.94 & 40:29:48.60 & 2009-10-17 & 2010-05-02 & 2.04 & ... & ... & ... & ... \\
2021+614 & 2022+6136 & OW637 & 20:22:06.68 & 61:36:58.80 & 2009-12-20 & 2010-05-26 & 1.25 & ... & ... & ... & ... \\
2037+511 & 2038+5119 &  & 20:38:37.04 & 51:19:12.66 & 2009-11-25 & 2010-05-18 & 2.46 & 13.2 & B & ... & ... \\
2121+053 & 2123+0535 &  & 21:23:44.52 & 05:35:22.09 & 2009-11-02 & 2010-05-08 & 1.74 & ... & ... & ... & ... \\
2131$-$021 & 2134$-$0153 & PKS2131$-$021 & 21:34:10.31 & $-$01:53:17.24 & 2009-11-02 & 2010-05-09 & 1.46 & ... & ... & ... & ... \\
2134+004 & 2136+0041 & OX057 & 21:36:38.59 & 00:41:54.21 & 2009-11-04 & 2010-05-10 & 3.34 & ... & ... & 21.0 & A \\
2136+141 & 2139+1423 &  & 21:39:01.31 & 14:23:36.00 & 2009-11-11 & 2010-05-16 & 1.29 & 12.5 & B & ... & ... \\
2145+067 & 2148+0657 &  & 21:48:05.46 & 06:57:38.60 & 2009-11-10 & 2010-05-16 & 7.37 & ... & ... & ... & ... \\
2200+420 & 2202+4216 & BLLAC & 22:02:43.30 & 42:16:39.00 & 2009-12-13 & 2010-06-06 & 3.35 & 13.8 & B & 22.6 & A \\
2201+315 & 2203+3145 & 4C31.63 & 22:03:14.98 & 31:45:38.30 & 2009-12-02 & 2010-05-30 & 2.53 & 14.2 & B & 20.7 & B \\
2201+171 & 2203+1725 &  & 22:03:26.89 & 17:25:48.20 & 2009-11-21 & 2010-05-24 & 1.21 & 13.5 & B & ... & ... \\
2216$-$038 & 2218$-$0335 &  & 22:18:52.04 & $-$03:35:36.90 & 2009-11-14 & 2010-05-21 & 1.41 & 14.0 & B & ... & ... \\
2223$-$052 & 2225$-$0457 & 3C446 & 22:25:47.26 & $-$04:57:01.40 & 2009-11-15 & 2010-05-23 & 5.58 & 12.7 & B & 22.0 & A \\
2227$-$088 & 2229$-$0832 &  & 22:29:40.09 & $-$08:32:54.50 & 2009-11-14 & 2010-05-23 & 1.94 & ... & ... & 20.7 & B \\
2230+114 & 2232+1143 & CTA102 & 22:32:36.41 & 11:43:50.90 & 2009-11-27 & 2010-05-31 & 3.84 & 13.2 & B & 22.6 & B \\
2234+282 & 2236+2828 &  & 22:36:22.47 & 28:28:57.41 & 2009-12-10 & ... & 1.03 & 13.0 & B & ... & ... \\
2251+158 & 2253+1608 & 3C454.3 & 22:53:57.75 & 16:08:53.60 & 2009-12-06 & ... & 9.42 & 13.9 & B & 22.6 & A \\
4C45.51 & 2354+4553 &  & 23:54:21.68 & 45:53:04.00 & 2010-01-11 & ... & 1.12 & ... & ... & ... & ... \\
2353+816 & 2356+8152 &  & 23:56:22.79 & 81:52:52.26 & 2009-09-19 & 2010-02-08 & 1.59 & ... & ... & ... & ... \\
\end{longtable}
\end{landscape}
}

\longtab{4}{
\begin{longtable}{l l c c c}
\caption{Simultaneity of high energy data to \planck\ observations displayed in the SEDs.}\label{simultaneity}\\
\hline\hline
B1950 Name & J2000 Name & UVOT & SWIFT-XRT & FERMI\\
\hline
\endfirsthead
\caption{continued.}\\
\hline\hline
B1950 Name & J2000 Name & UVOT & SWIFT-XRT & FERMI\\
\hline
\endhead
\hline
\endfoot
  0003$-$066 & 0006$-$0623 & sim\_1st & sim\_1st & sim\_1st\\
  0007+106 & 0010+1058 & ... & ... & 27m\\
  0048$-$097 & 0050$-$0929 & sim\_1st & sim\_1st & sim\_1st\\
  0059+581 & 0102+5824 & ... & ... & sim\_1st\\
  0106+013 & 0108+0135 & ... & ... & sim\_1st\\
  J0125$-$0005 & 0125$-$0005 & ... & ... & sim\_1st\\
  0133+476 & 0136+4751 & sim\_1st & sim\_1st & sim\_1st\\
  0149+218 & 0152+2207 & ... & ... & sim\_1st\\
  0202+149 & 0204+1514 & ... & ... & sim\_1st\\
  0212+735 & 0217+7349 & ... & ... & 27m\\
  0224+671 & 0228+6721 & ... & ... & sim\_1st\\
  0234+285 & 0237+2848 & sim\_1st & sim\_1st & sim\_1st\\
  0235+164 & 0238+1636 & sim\_1st & sim\_1st & sim\_1st\\
  0238$-$084 & 0241$-$0815 & ... & ... & sim\_1st\\
  0306+102 & 0309+1029 & ... & ... & 27m\\
  0316+413 & 0319+4130 & ... & ... & 27m\\
  0333+321 & 0336+3218 & ... & ... & 27m\\
  0336$-$019 & 0339$-$0146 & ... & ... & sim\_1st\\
  0355+508 & 0359+5057 & ... & ... & sim\_1st\\
  0415+379 & 0418+3801 & ... & ... & sim\_1st\\
  0420$-$014 & 0423$-$0120 & sim\_1st & sim\_1st & sim\_1st\\
  0430+052 & 0433+0521 & sim\_1st & sim\_1st & sim\_1st\\
  0446+112 & 0449+1121 & ... & ... & 27m\\
  0458$-$020 & 0501$-$0159 & ... & ... & 27m\\
  0507+179 & 0510+1800 & ... & ... & sim\_1st\\
  0528+134 & 0530+1331 & sim\_2nd & sim\_1st\_2nd & sim\_1st\\
  0552+398 & 0555+3948 & sim\_1st & sim\_1st & sim\_1st\\
  0605$-$085 & 0607$-$0834 & ... & ... & sim\_1st\\
  0642+449 & 0646+4451 & sim\_1st & sim\_1st & sim\_1st\\
  0716+714 & 0721+7120 & sim\_1st\_2nd & sim\_1st\_2nd & sim\_1st\_2nd\\
  0723$-$008 & 0725$-$0054 & ... & ... & ...\\
  0735+178 & 0738+1742 & ... & ... & 27m\\
  0736+017 & 0739+0137 & sim\_1st & sim\_1st & sim\_1st\\
  0748+126 & 0750+1231 & sim\_2nd & sim\_2nd & 2M\_2nd\\
  0754+100 & 0757+0956 & ... & ... & 2M\_1st\\
  0805$-$077 & 0808$-$0751 & ... & ... & sim\_1st\\
  0804+499 & 0808+4950 & sim\_1st & sim\_1st & 2M\_1st\\
  0823+033 & 0825+0309 & ... & ... & 27m\\
  0827+243 & 0830+2411 & ... & ... & 2M\_1st\\
  0836+710 & 0841+7053 & sim\_2nd & sim\_2nd & 2M\_2nd\\
  0851+202 & 0854+2006 & sim\_2nd & sim\_2nd & 2M\_1st\_2nd\\
  0906+430 & 0909+4253 & ... & ... & ...\\
  0917+449 & 0920+4441 & sim\_1st & sim\_1st & 2M\_1st\\
  0923+392 & 0927+3902 & sim\_2nd & sim\_2nd & sim\_2nd\\
  0945+408 & 0948+4039 & ... & ... & 2M\_1st\\
  0953+254 & 0956+2515 & ... & ... & 2M\_1st\\
  0954+658 & 0958+6533 & sim\_1st & sim\_1st & 2M\_1st\\
  1036+054 & 1038+0512 & ... & ... & 2M\_1st\\
  TEX1040+244 & 1043+2408 & ... & ... & 2M\_1st\\
  1055+018 & 1058+0133 & sim\_1st & sim\_1st & 2M\_1st\\
  J1130+3815 & 1130+3815 & ... & ... & 2M\_1st\\
  1150+812 & 1153+8058 & ... & ... & sim\_1st\\
  1150+497 & 1153+4931 & sim\_1st & sim\_1st & 2M\_1st\\
  1156+295 & 1159+2914 & sim\_1st\_2nd & sim\_1st\_2nd & sim\_1st\_2nd\\
  1219+044 & 1222+0413 & ... & ... & 27m\\
  1222+216 & 1224+2122 & sim\_1st & sim\_1st & sim\_1st\\
  1226+023 & 1229+0203 & sim\_1st & sim\_1st & sim\_1st\\
  1228+126 & 1230+1223 & sim\_1st & sim\_1st & sim\_1st\\
  1253$-$055 & 1256$-$0547 & sim\_1st & sim\_1st & sim\_1st\\
  1308+326 & 1310+3220 & sim\_1st & sim\_1st & sim\_1st\\
  1324+224 & 1327+2210 & ... & ... & sim\_1st\\
  1413+135 & 1415+1320 & ... & ... & sim\_1st\\
  1418+546 & 1419+5423 & ... & sim\_1st & 27m\\
  1502+106 & 1504+1029 & ... & ... & 27m\\
  1510$-$089 & 1512$-$0905 & sim\_1st & ... & sim\_1st\\
  1546+027 & 1549+0237 & sim\_1st & sim\_1st & sim\_1st\\
  1548+056 & 1550+0527 & ... & ... & sim\_1st\\
  1606+106 & 1608+1029 & ... & ... & sim\_1st\\
  1611+343 & 1613+3412 & sim\_1st & sim\_1st & sim\_1st\\
  1633+382 & 1635+3808 & ... & sim\_1st & sim\_1st\\
  1638+398 & 1640+3946 & ... & ... & 27m\\
  1642+690 & 1642+6856 & ... & ... & 2M\_1st\\
  1641+399 & 1642+3948 & sim\_1st & sim\_1st & sim\_1st\\
  1652+398 & 1653+3945 & ... & ... & 27m\\
  1739+522 & 1740+5211 & ... & ... & 27m\\
  1741$-$038 & 1743$-$0350 & ... & ... & sim\_1st\\
  1749+096 & 1751+0939 & sim\_2nd & sim\_2nd & sim\_2nd\\
  1803+784 & 1800+7828 & sim\_1st & sim\_1st & sim\_1st\\
  1807+698 & 1806+6949 & sim\_1st & sim\_1st & 2M\_1st\\
  1823+568 & 1824+5651 & ... & ... & 27m\\
  1828+487 & 1829+4844 & ... & ... & 27m\\
  J184915+670 & 1849+6705 & sim\_2nd & sim\_2nd & ...\\
  1928+738 & 1927+7358 & sim\_1st & sim\_1st & 2M\_1st\\
  1954+513 & 1955+5131 & ... & ... & ...\\
  2007+776 & 2005+7752 & ... & ... & 27m\\
  2005+403 & 2007+4029 & ... & ... & 2M\_1st\\
  2021+614 & 2022+6136 & ... & ... & 2M\_1st\\
  2037+511 & 2038+5119 & ... & ... & 2M\_1st\\
  2121+053 & 2123+0535 & ... & ... & 2M\_1st\\
  2131$-$021 & 2134$-$0153 & ... & ... & 2M\_1st\\
  2134+004 & 2136+0041 & sim\_2nd & sim\_2nd & 2M\_2nd\\
  2136+141 & 2139+1423 & ... & ... & 2M\_1st\\
  2145+067 & 2148+0657 & sim\_1st & sim\_1st & sim\_1st\\
  2200+420 & 2202+4216 & sim\_1st\_2nd & sim\_1st\_2nd & sim\_1st\_2nd\\
  2201+315 & 2203+3145 & sim\_1st & sim\_1st & sim\_1st\\
  2201+171 & 2203+1725 & ... & ... & 2M\_1st\\
  2216$-$038 & 2218$-$0335 & ... & ... & sim\_1st\\
  2223$-$052 & 2225$-$0457 & sim\_2nd & sim\_2nd & 2M\_2nd\\
  2227$-$088 & 2229$-$0832 & ... & sim\_1st & 2M\_1st\\
  2230+114 & 2232+1143 & sim\_1st & sim\_1st & sim\_1st\\
  2234+282 & 2236+2828 & ... & ... & sim\_1st\\
  2251+158 & 2253+1608 & sim\_1st & sim\_1st & sim\_1st\\
  4C45.51 & 2354+4553 & ... & ... & sim\_1st\\
  2353+816 & 2356+8152 & ... & ... & 27m\\
\end{longtable}
}

\begin{figure*}
\includegraphics[scale=0.8]{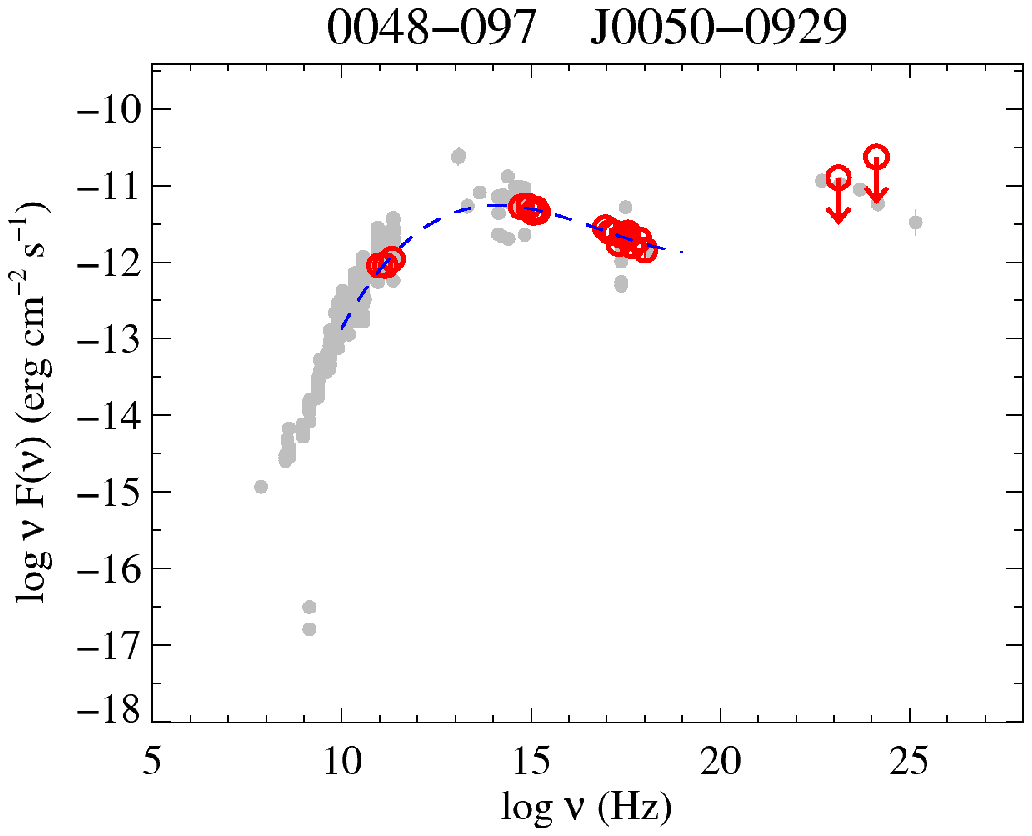}
\includegraphics[scale=0.8]{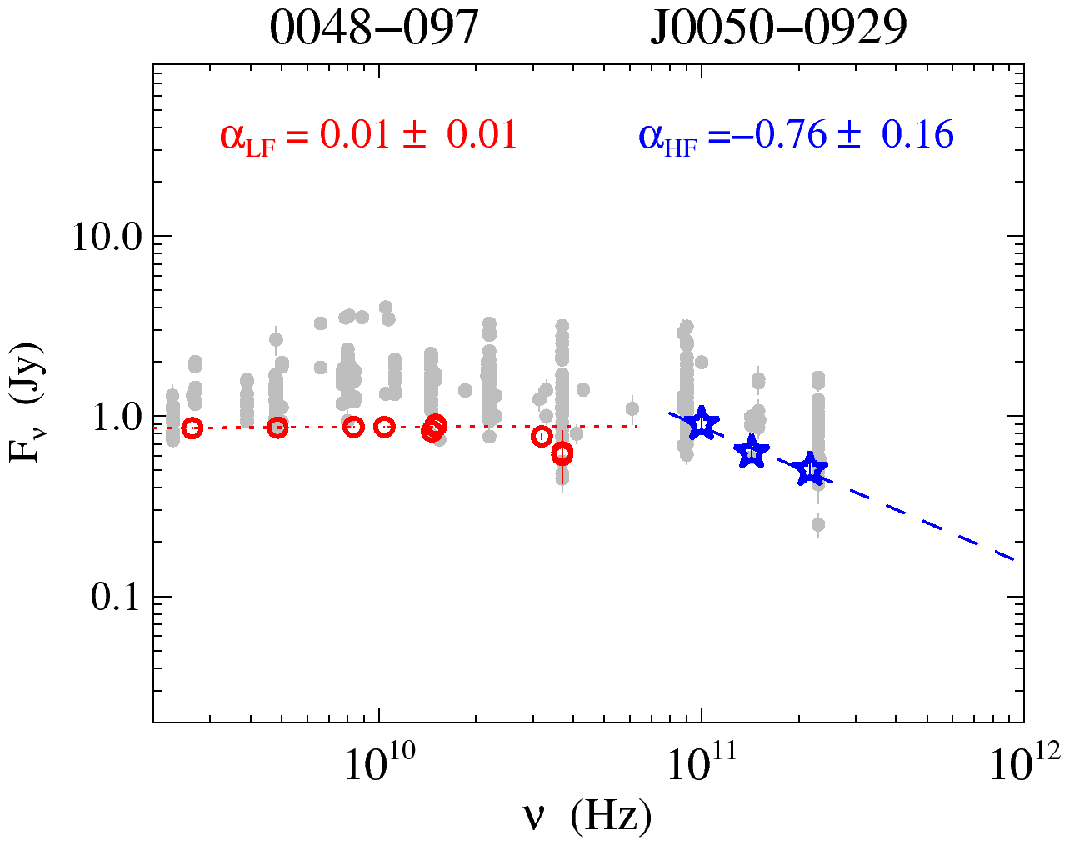}
 \caption{0048$-$097}
\end{figure*}
 
 \clearpage
 
\begin{figure*}
\includegraphics[scale=0.8]{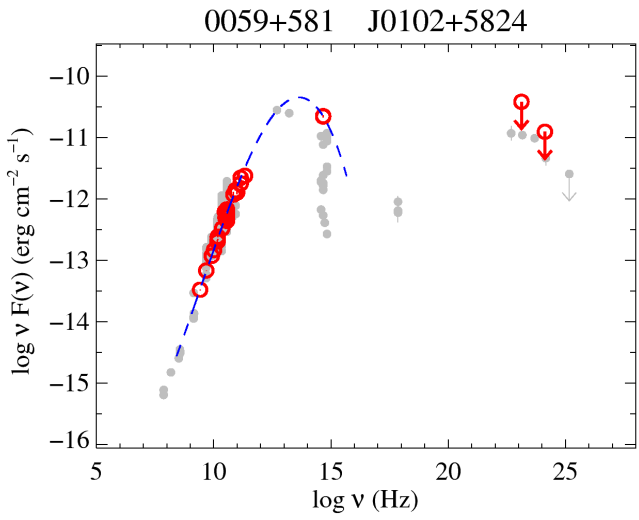}
\includegraphics[scale=0.8]{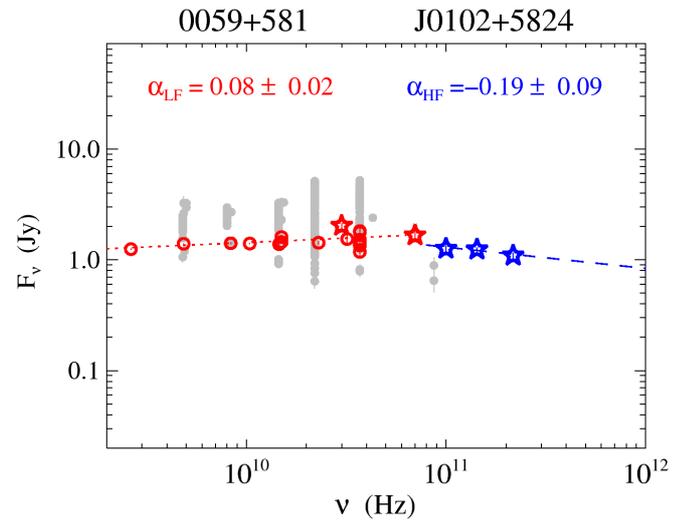}
 \caption{0059+581}
\end{figure*}

\begin{figure*}
\includegraphics[scale=0.8]{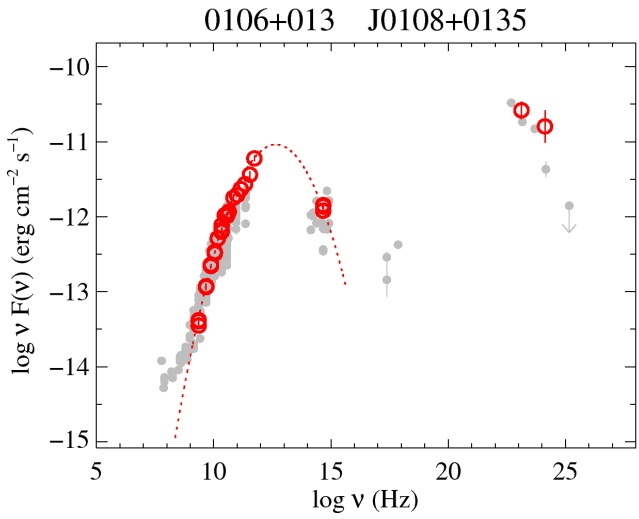}
\includegraphics[scale=0.8]{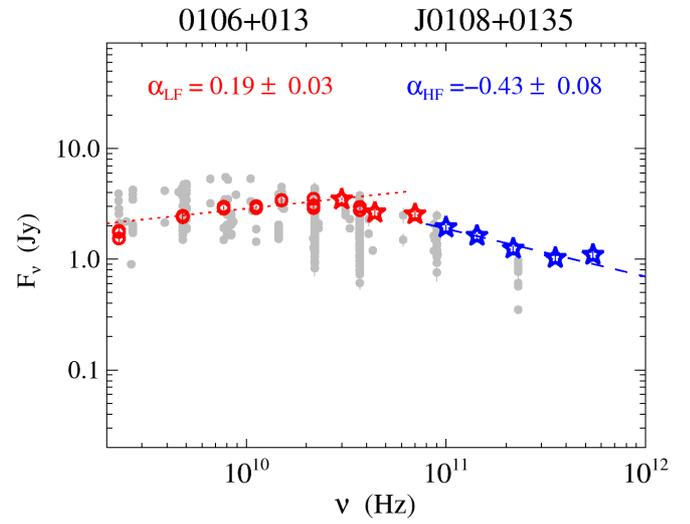}
 \caption{0106+013}
\end{figure*}

\begin{figure*}
\includegraphics[scale=0.8]{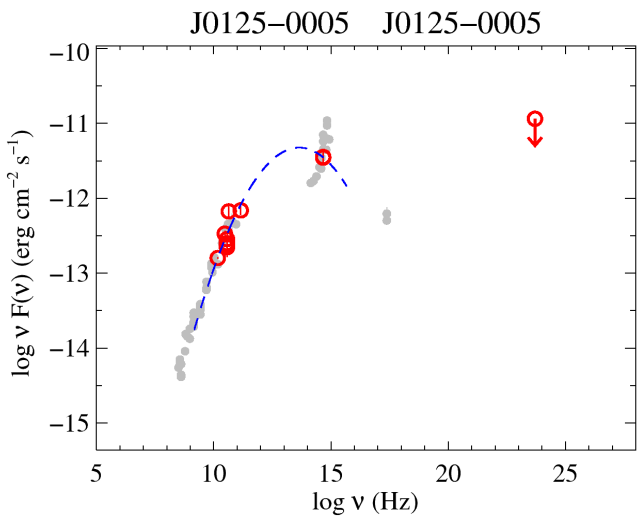}
\includegraphics[scale=0.8]{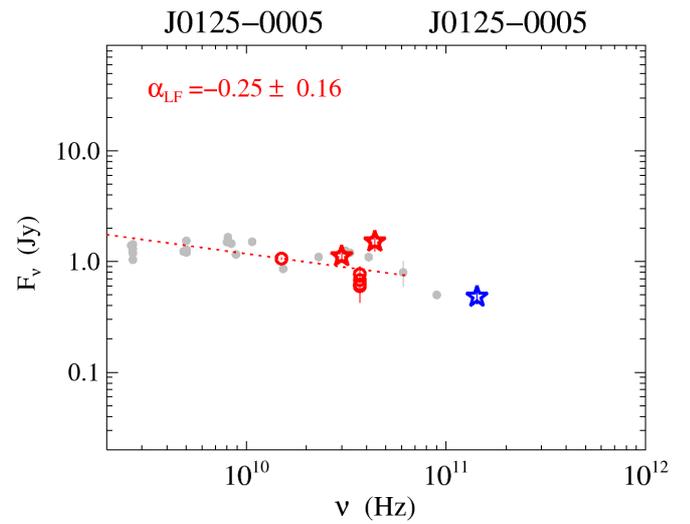}
 \caption{J0125$-$0005}
\end{figure*}
 
 \clearpage
 
\begin{figure*}
\includegraphics[scale=0.8]{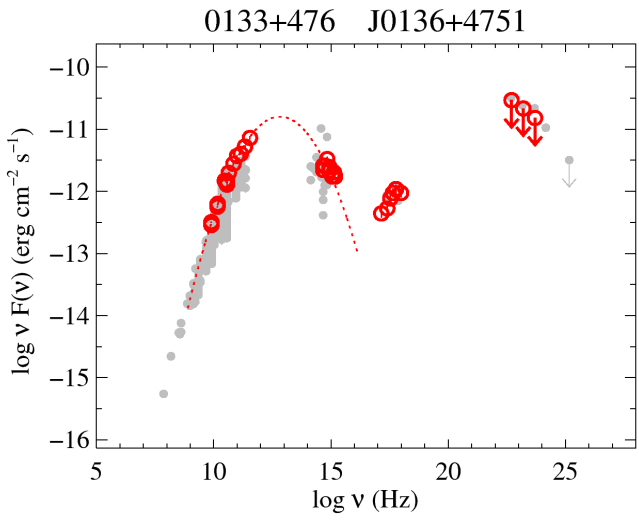}
\includegraphics[scale=0.8]{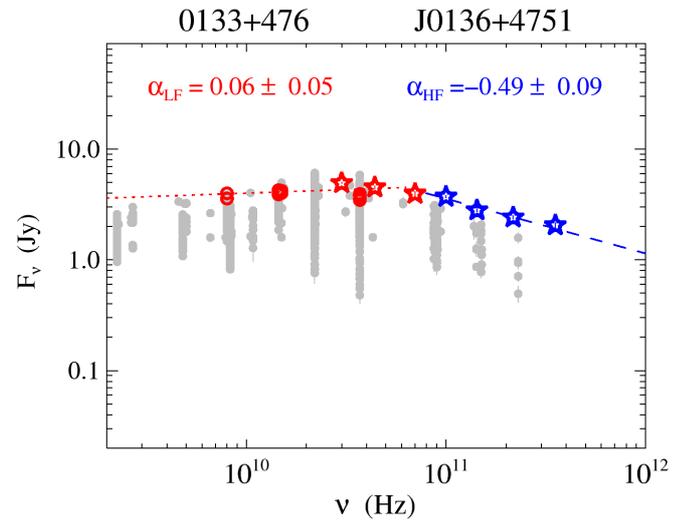}
 \caption{0133+476}
\end{figure*}

\begin{figure*}
\includegraphics[scale=0.8]{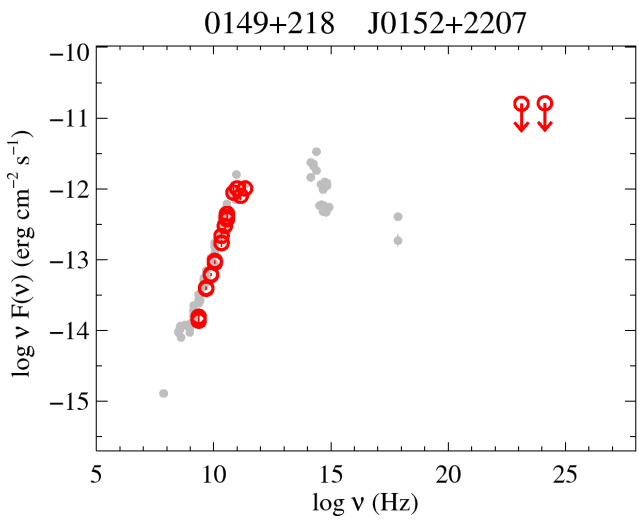}
\includegraphics[scale=0.8]{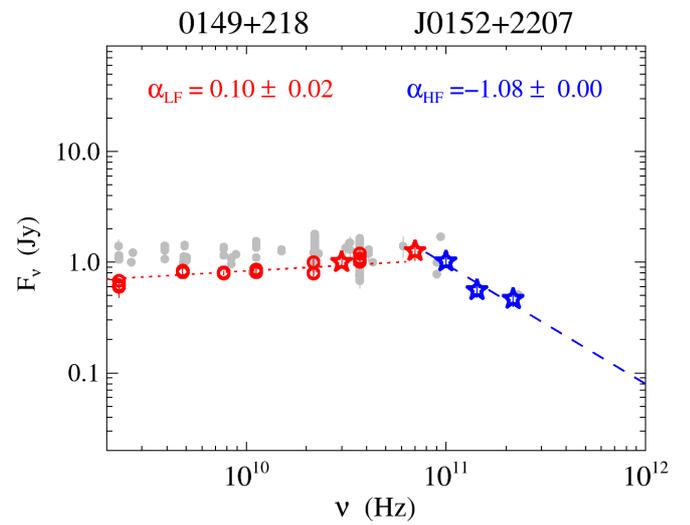}
 \caption{0149+218}
\end{figure*}

\begin{figure*}
\includegraphics[scale=0.8]{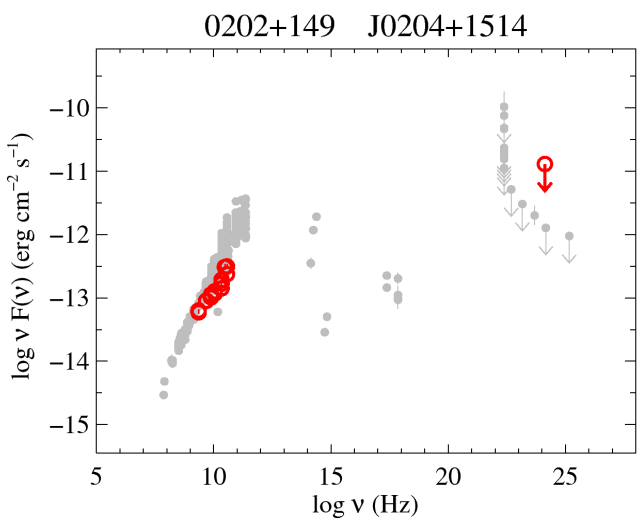}
\includegraphics[scale=0.8]{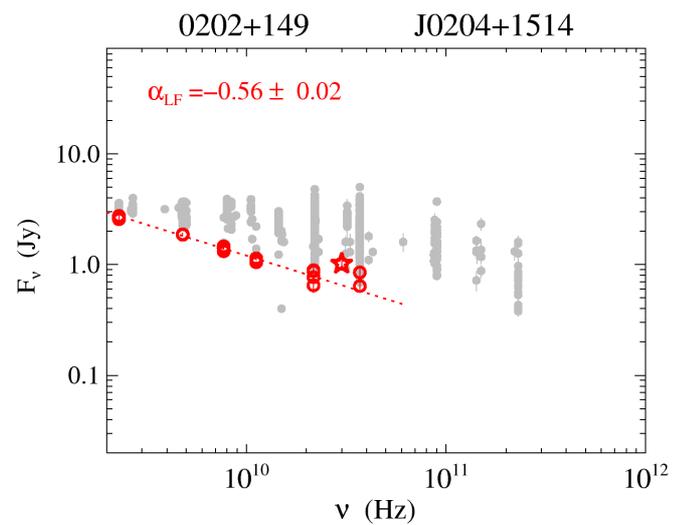}
 \caption{0202+149}
\end{figure*}
 
 \clearpage
 
\begin{figure*}
\includegraphics[scale=0.8]{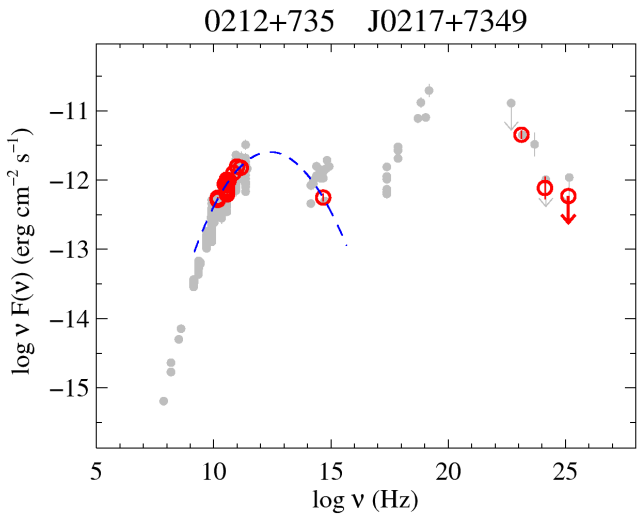}
\includegraphics[scale=0.8]{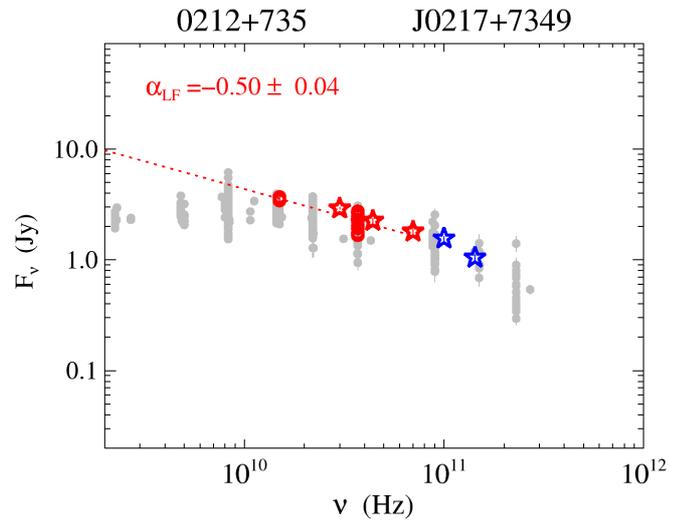}
 \caption{0212+735}
\end{figure*}

\begin{figure*}
\includegraphics[scale=0.8]{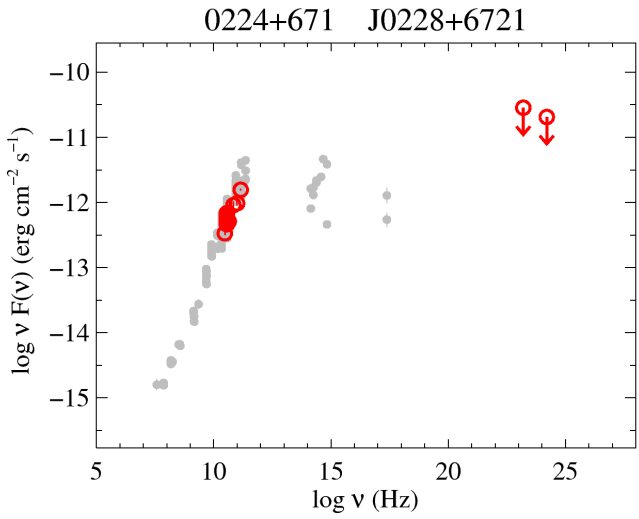}
\includegraphics[scale=0.8]{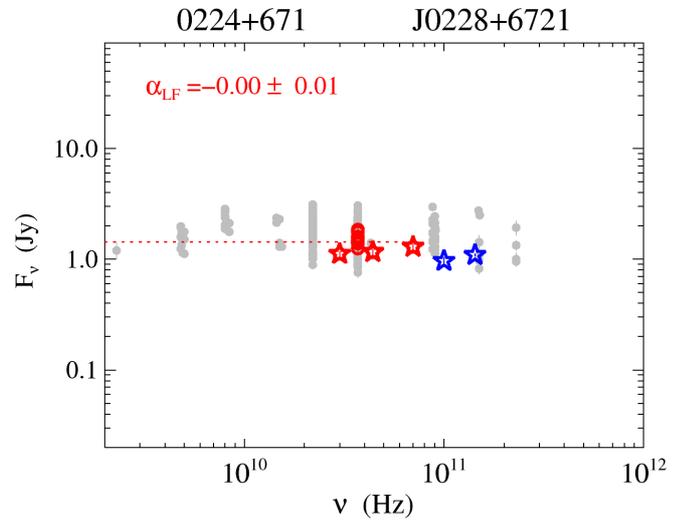}
 \caption{0224+671}
\end{figure*}

\begin{figure*}
\includegraphics[scale=0.8]{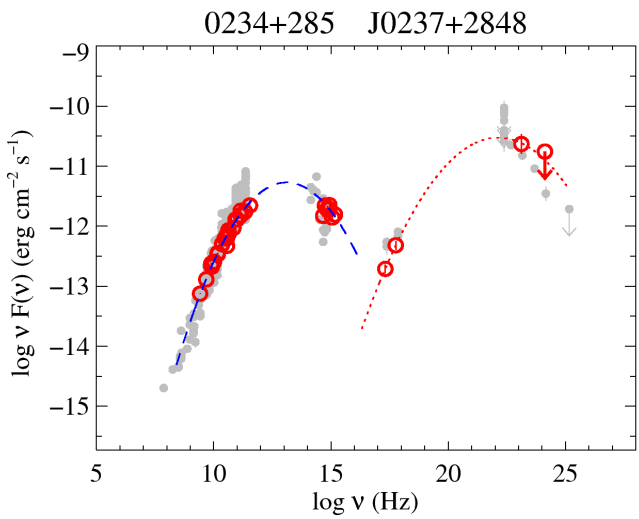}
\includegraphics[scale=0.8]{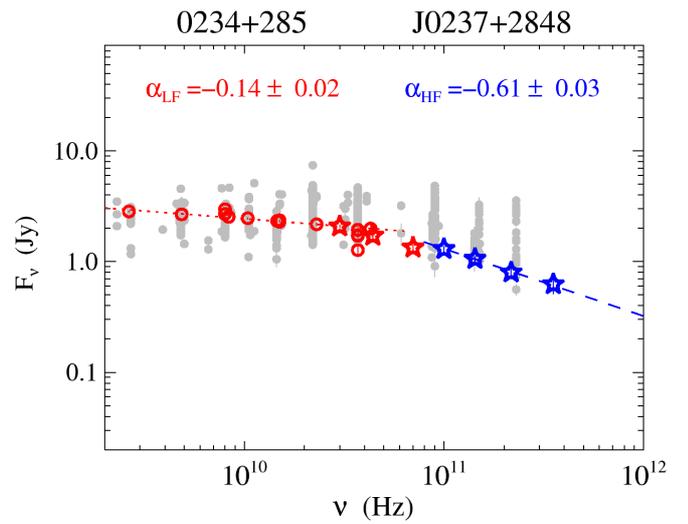}
 \caption{0234+285}
\label{0234_sed}
\end{figure*}
 
 \clearpage
 
\begin{figure*}
\includegraphics[scale=0.8]{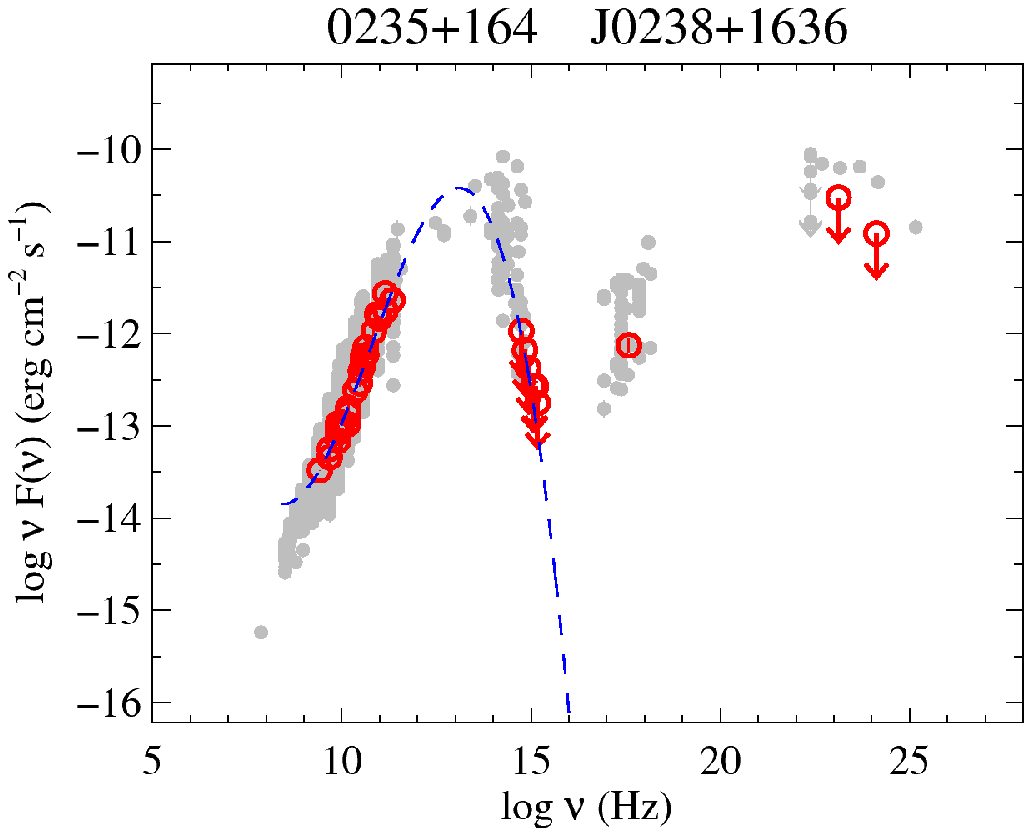}
\includegraphics[scale=0.8]{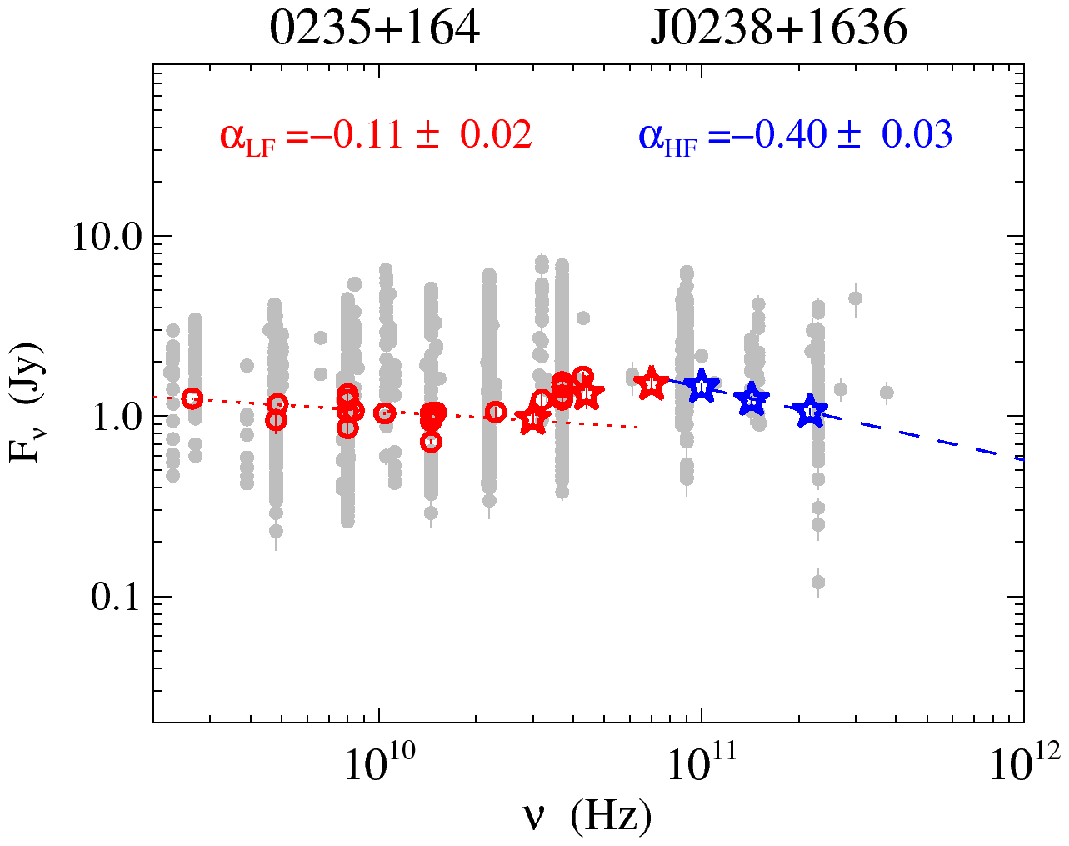}
 \caption{0235+164}
\end{figure*}

\begin{figure*}
\includegraphics[scale=0.8]{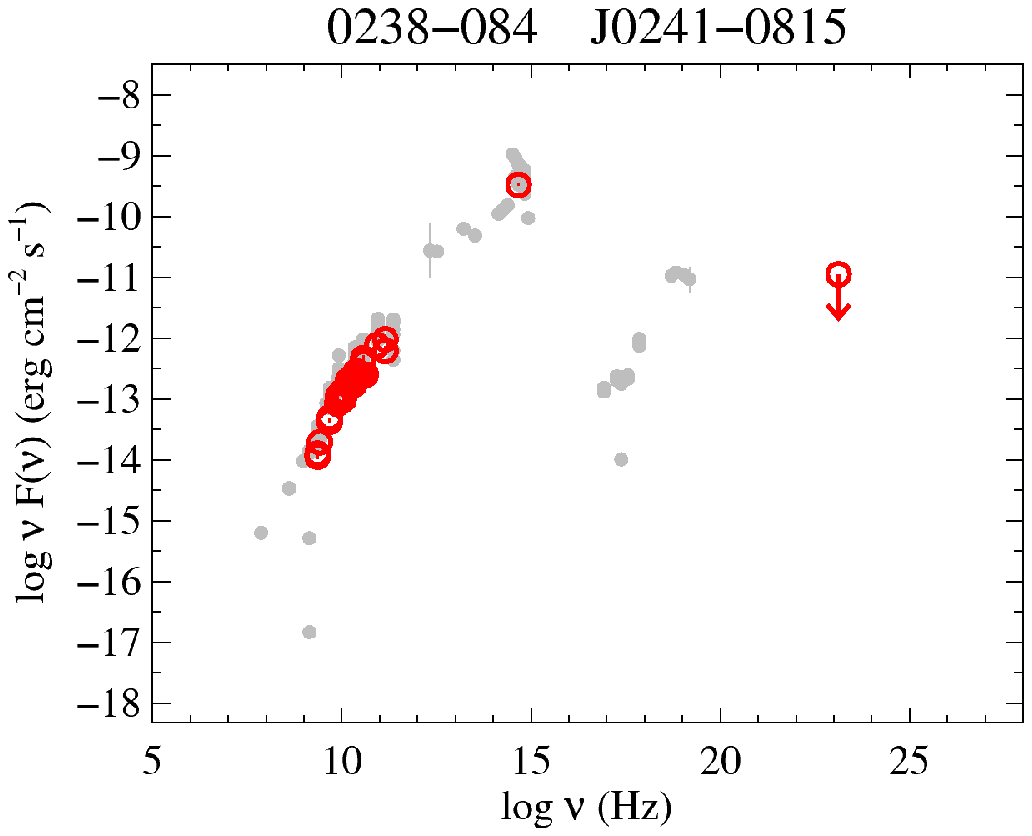}
\includegraphics[scale=0.8]{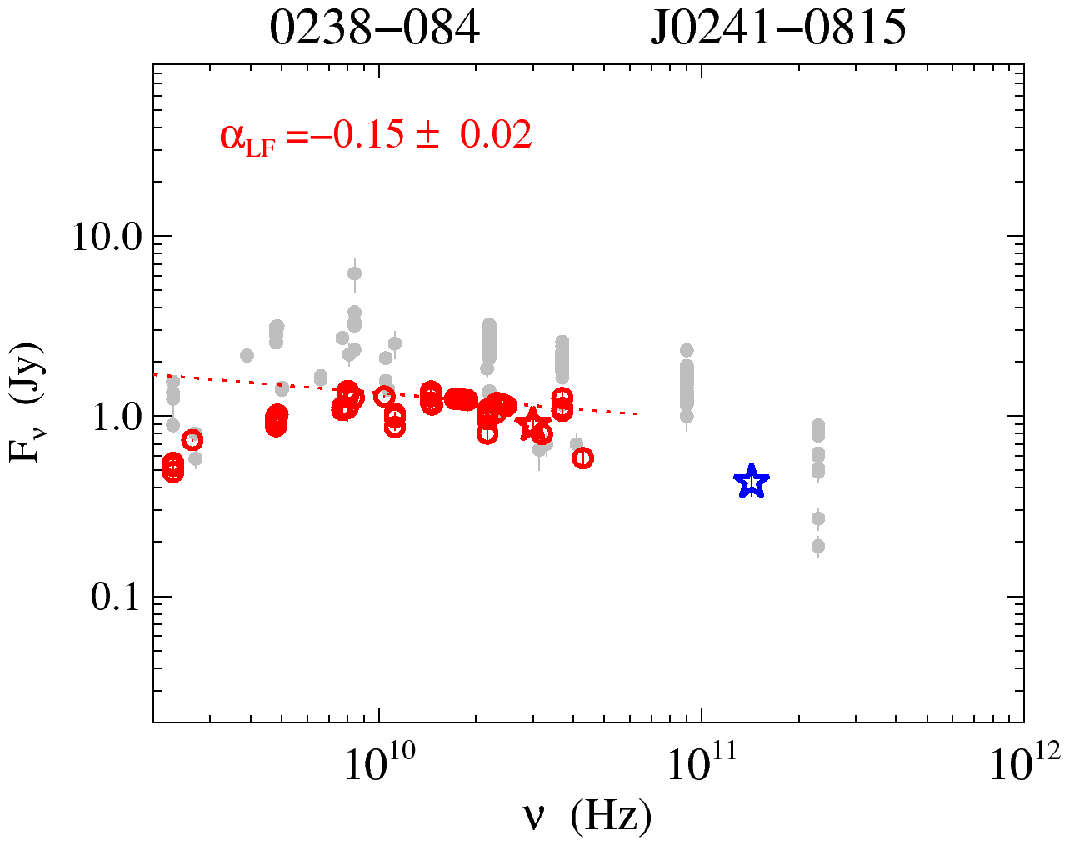}
 \caption{0238$-$084}
\end{figure*}

\begin{figure*}
\includegraphics[scale=0.8]{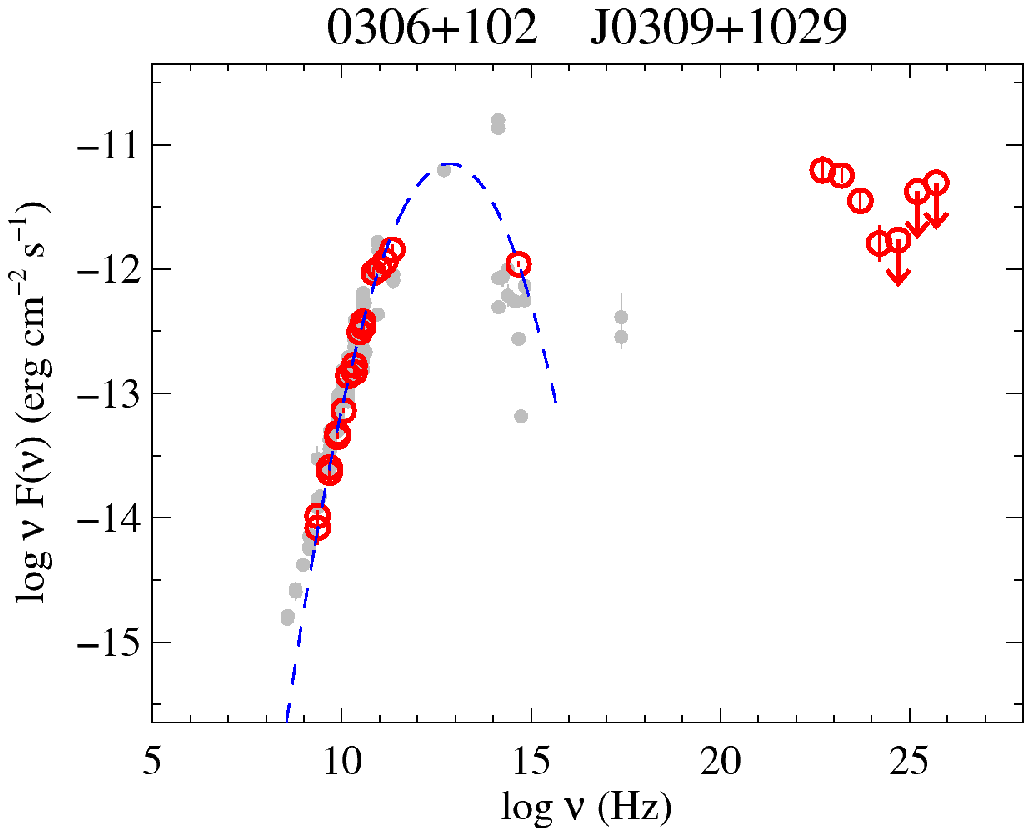}
\includegraphics[scale=0.8]{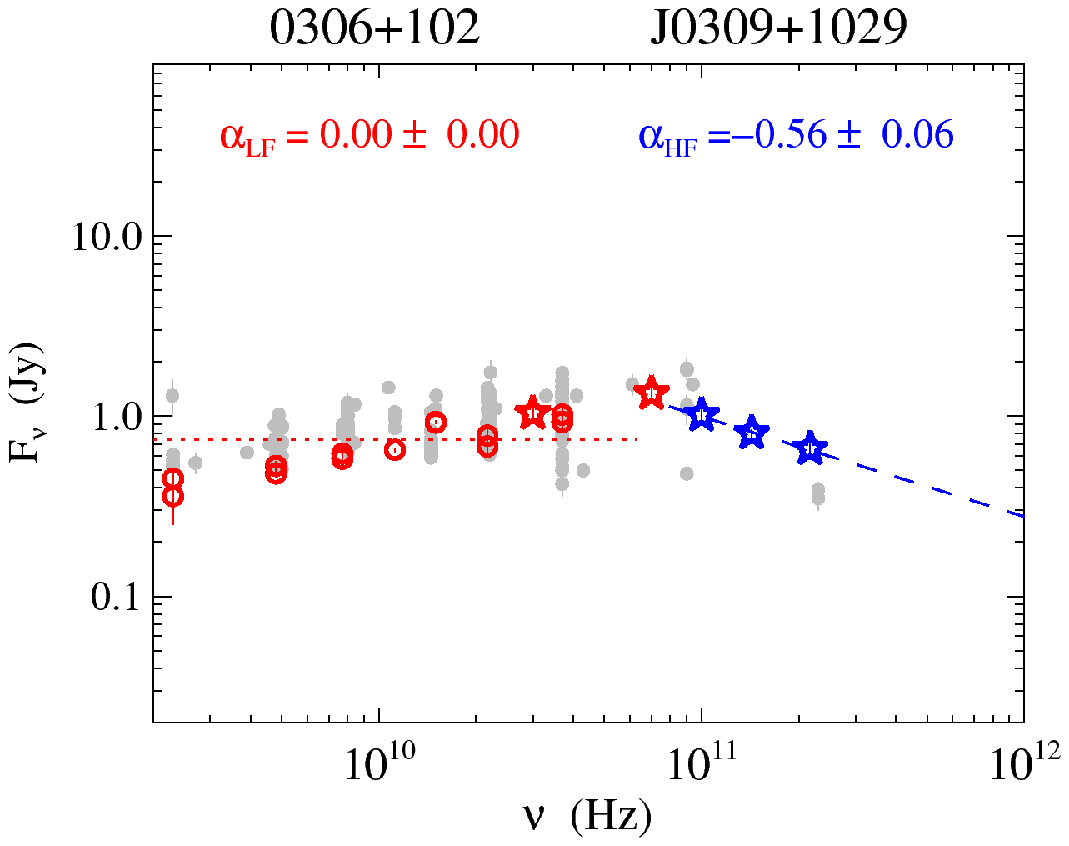}
 \caption{0306+102}
\end{figure*}
 
 \clearpage
 
\begin{figure*}
\includegraphics[scale=0.8]{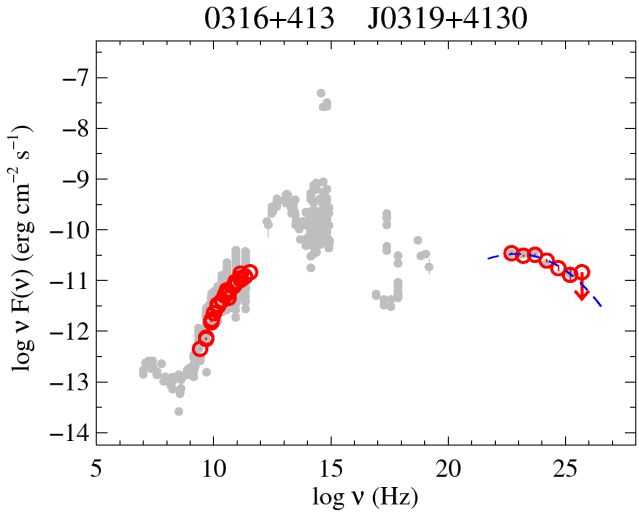}
\includegraphics[scale=0.8]{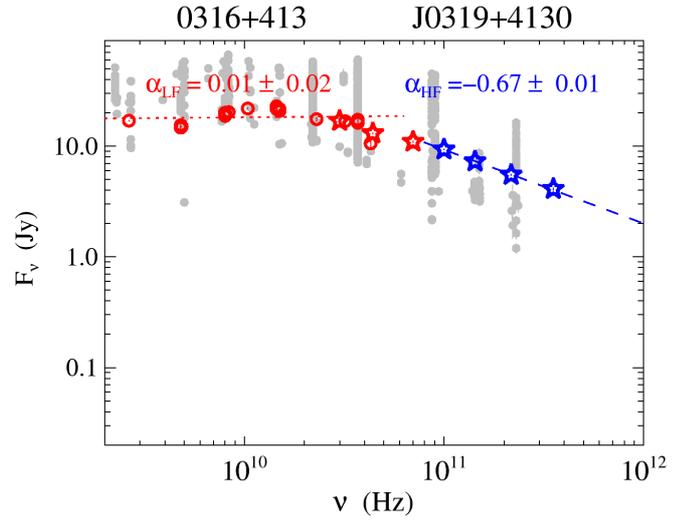}
 \caption{0316+413}
\end{figure*}

\begin{figure*}
\includegraphics[scale=0.8]{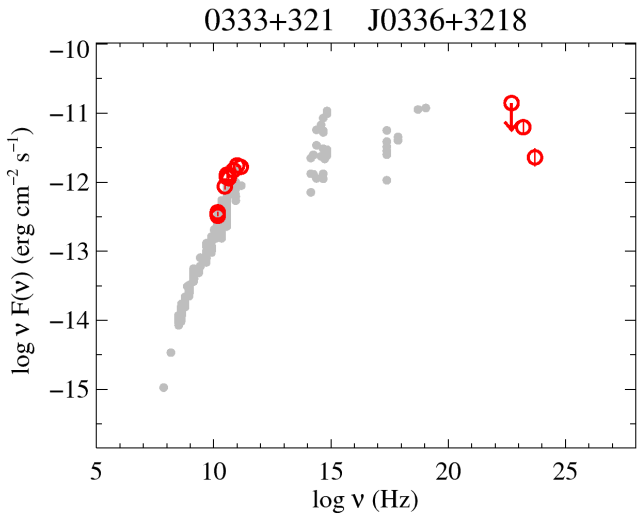}
\includegraphics[scale=0.8]{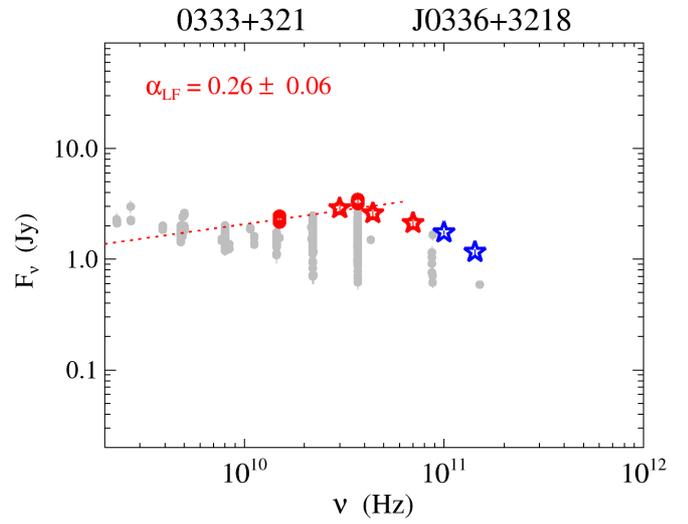}
 \caption{0333+321}
\end{figure*}

\begin{figure*}
\includegraphics[scale=0.8]{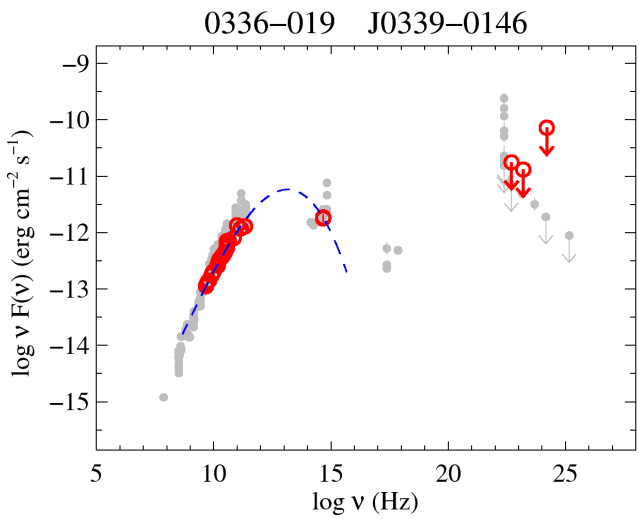}
\includegraphics[scale=0.8]{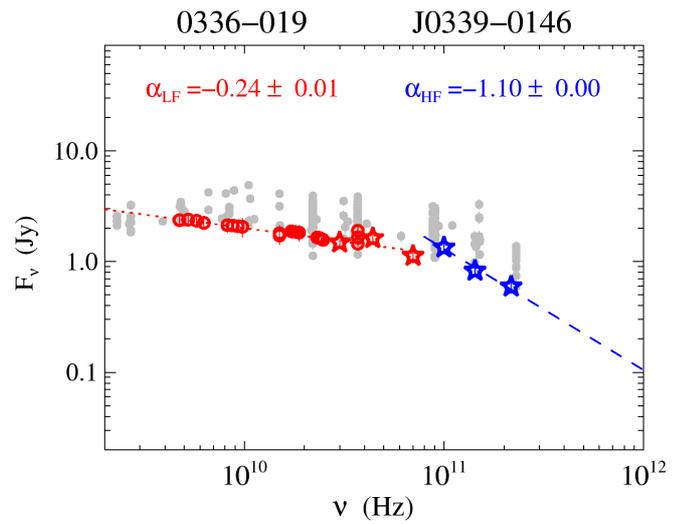}
 \caption{033$6-$019}
\end{figure*}
 
 \clearpage
 
\begin{figure*}
\includegraphics[scale=0.8]{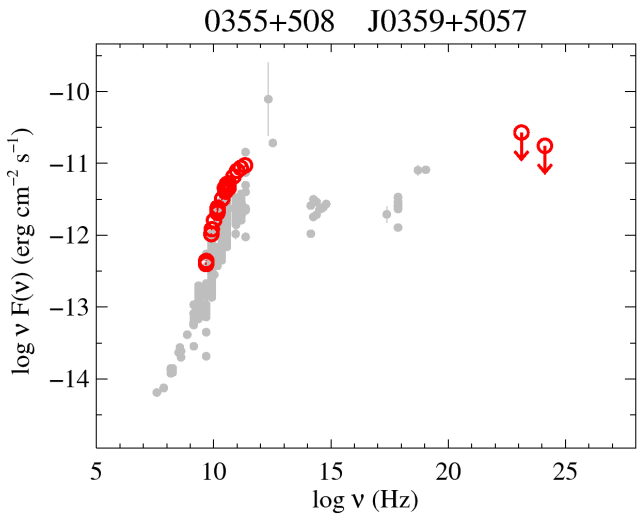}
\includegraphics[scale=0.8]{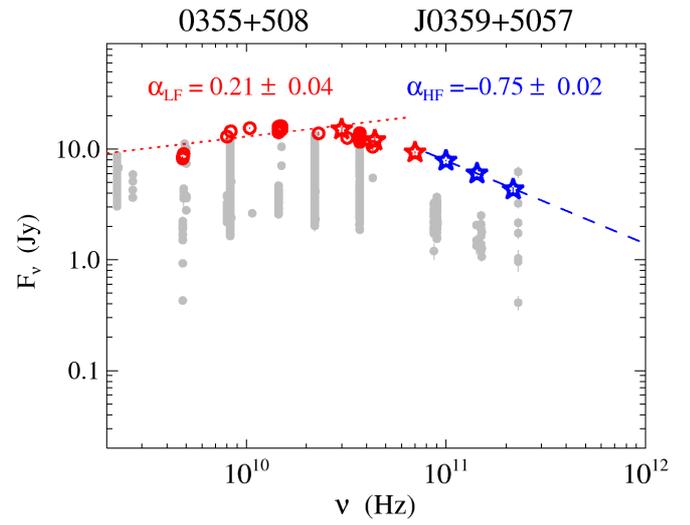}
 \caption{0355+508}
\end{figure*}

\begin{figure*}
\includegraphics[scale=0.8]{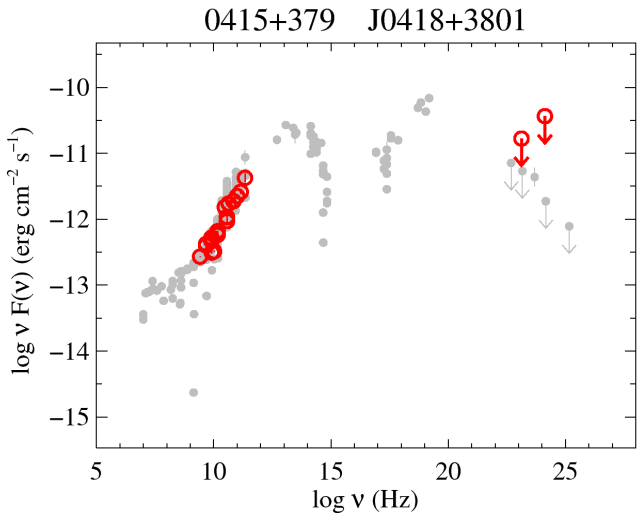}
\includegraphics[scale=0.8]{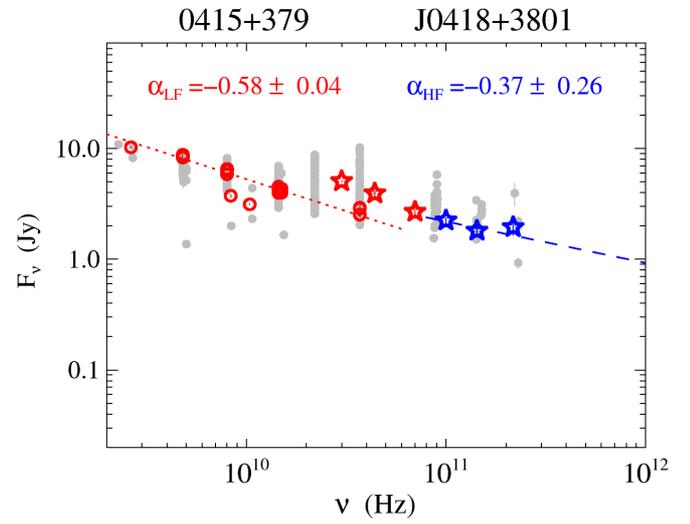}
 \caption{0415+379}
\end{figure*}

\begin{figure*}
\includegraphics[scale=0.8]{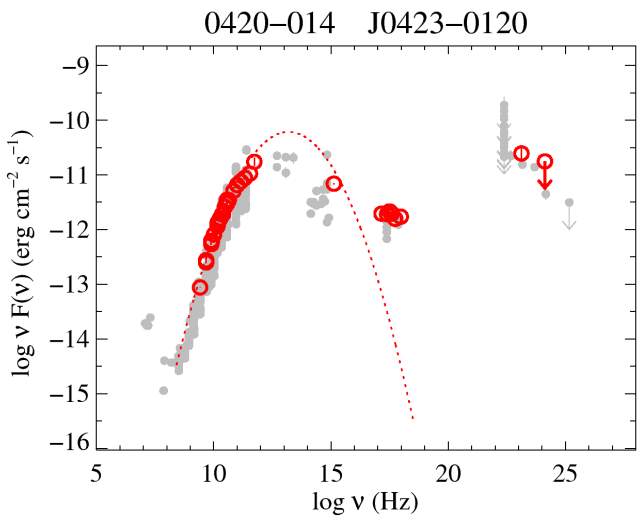}
\includegraphics[scale=0.8]{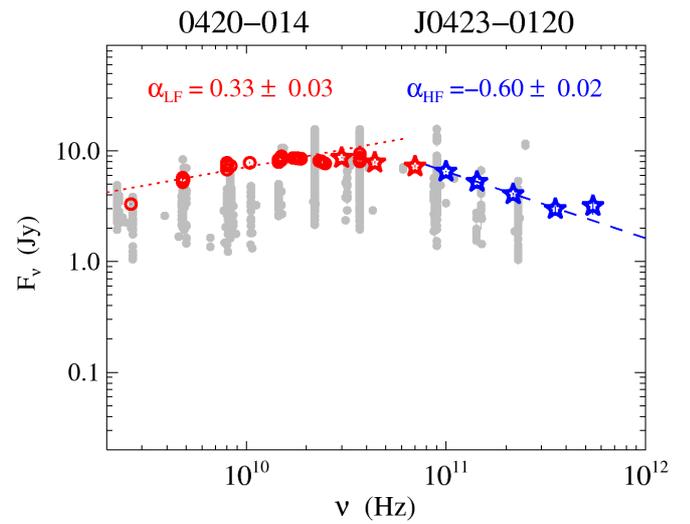}
 \caption{0420$-$014}
\end{figure*}
 
 \clearpage
 
\begin{figure*}
\includegraphics[scale=0.8]{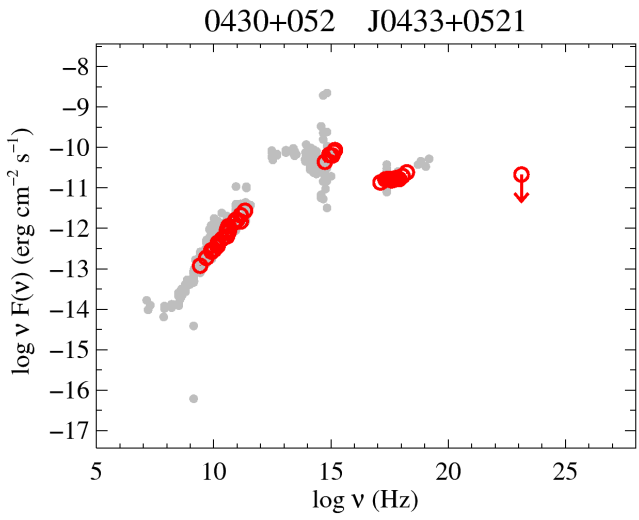}
\includegraphics[scale=0.8]{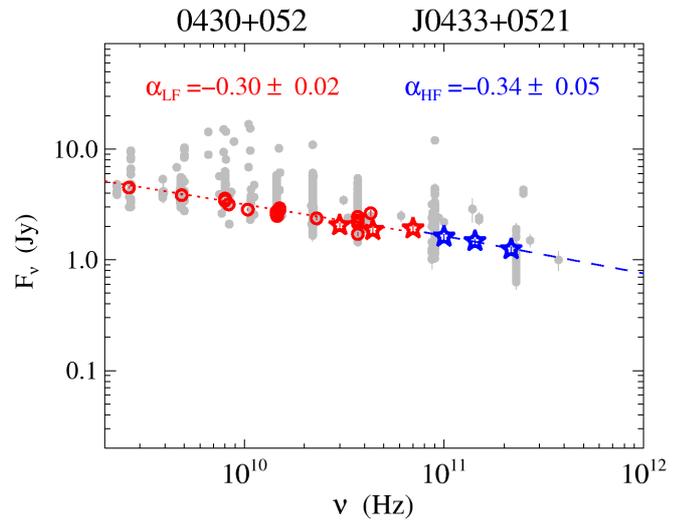}
 \caption{0430+052}
\end{figure*}

\begin{figure*}
\includegraphics[scale=0.8]{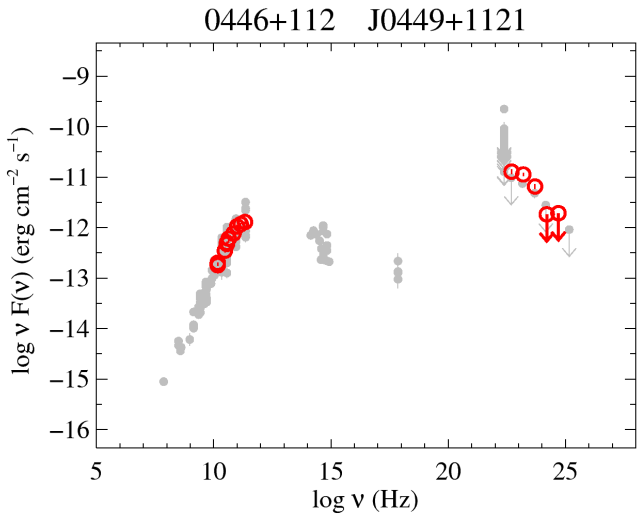}
\includegraphics[scale=0.8]{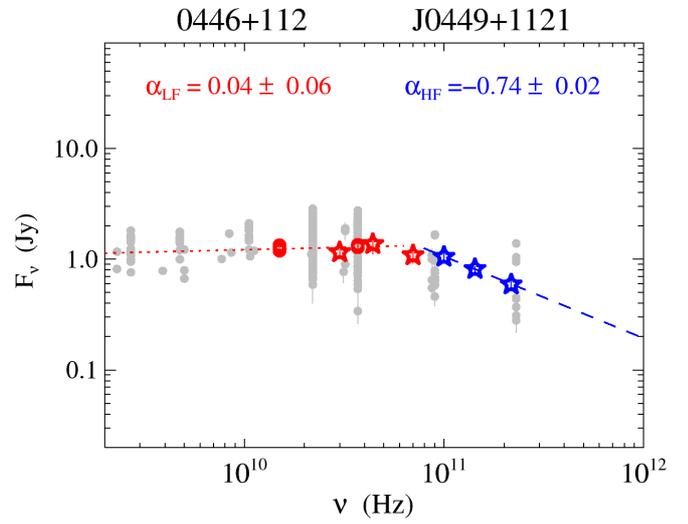}
 \caption{0446+112}
\end{figure*}

\begin{figure*}
\includegraphics[scale=0.8]{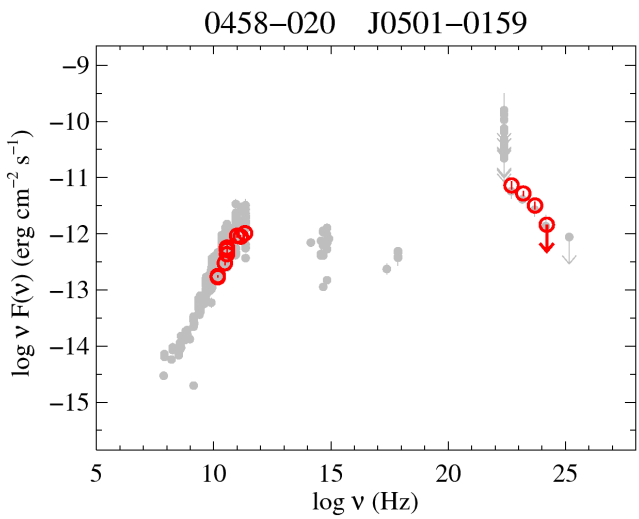}
\includegraphics[scale=0.8]{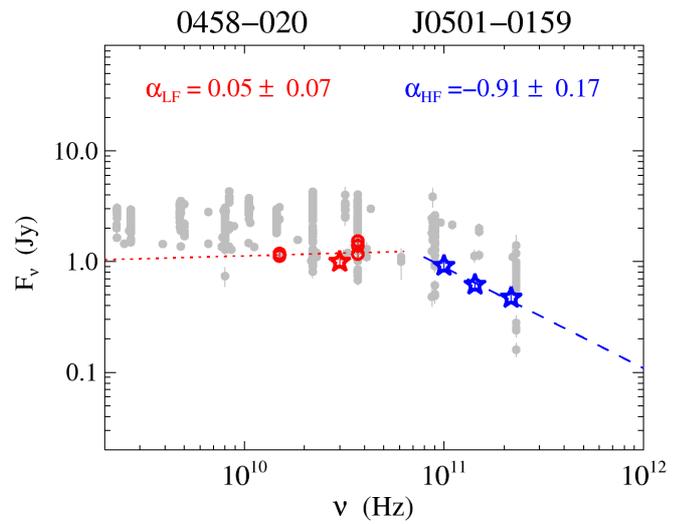}
 \caption{0458$-$020}
\end{figure*}
 
 \clearpage
 
\begin{figure*}
\includegraphics[scale=0.8]{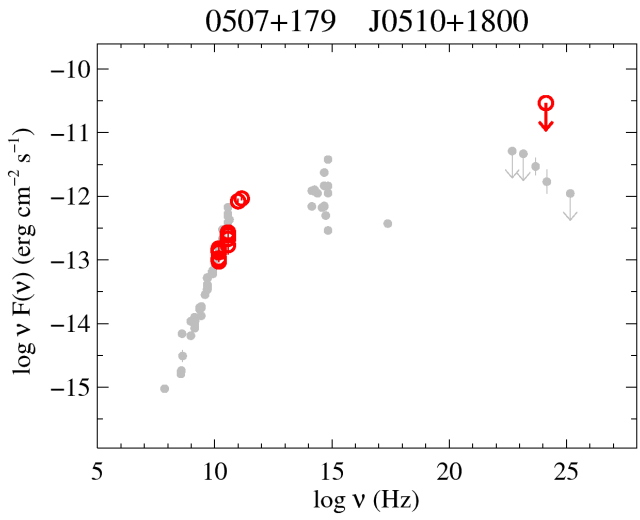}
\includegraphics[scale=0.8]{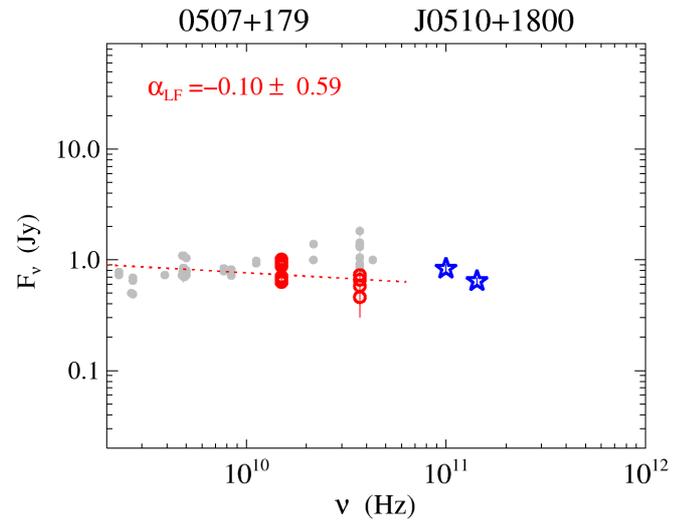}
 \caption{0507+179}
\end{figure*}

\begin{figure*}
\includegraphics[scale=0.8]{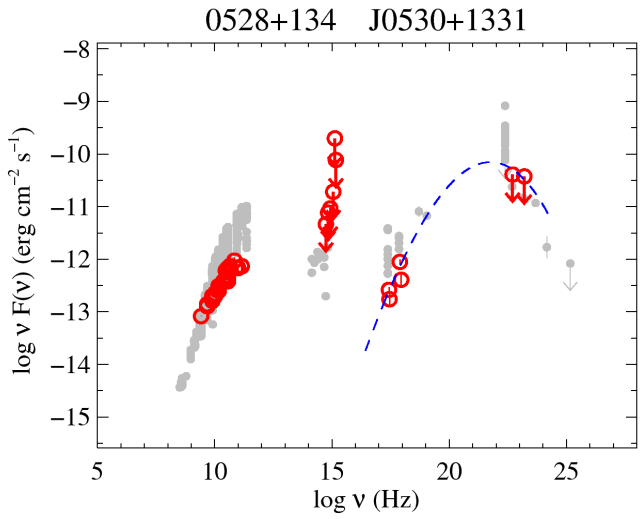}
\includegraphics[scale=0.8]{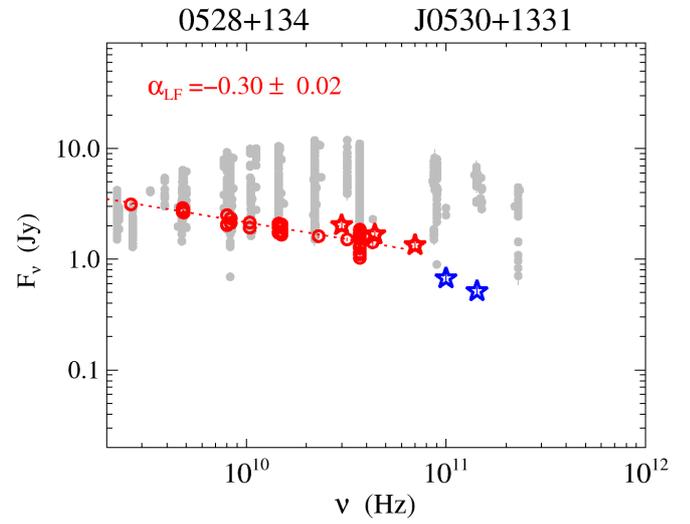}
 \caption{0528+134}
\end{figure*}

\begin{figure*}
\includegraphics[scale=0.8]{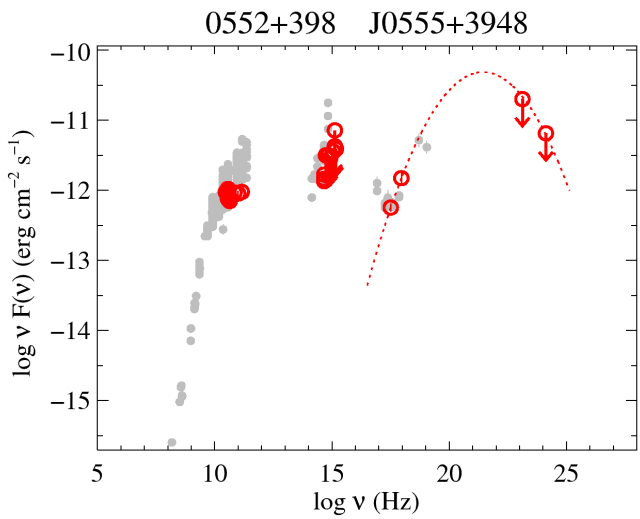}
\includegraphics[scale=0.8]{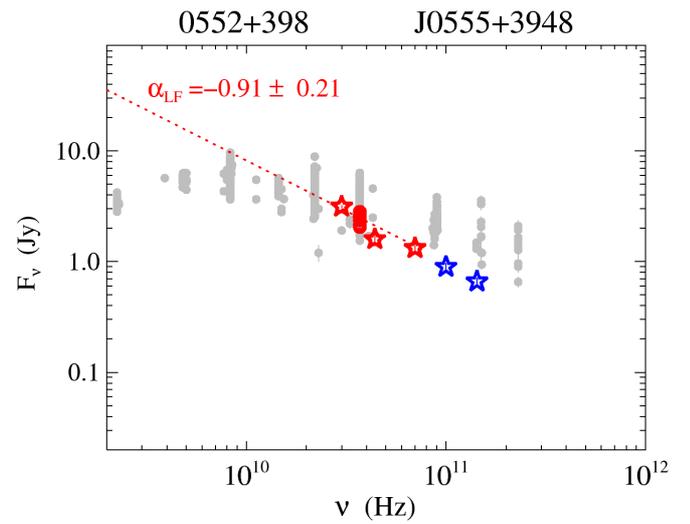}
 \caption{0552+398}
\end{figure*}
 
 \clearpage
 
\begin{figure*}
\includegraphics[scale=0.8]{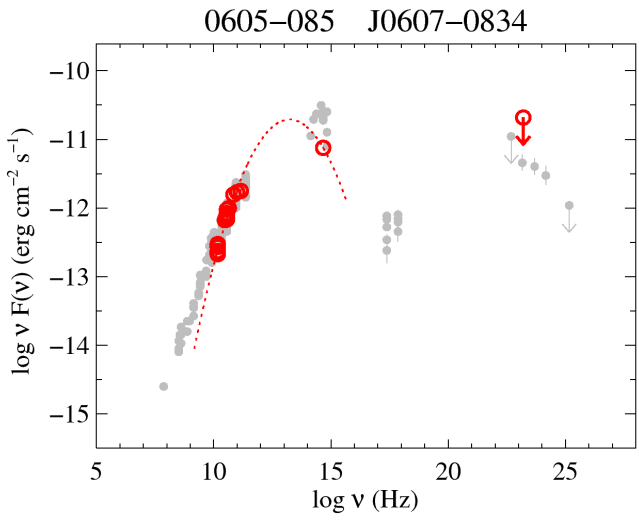}
\includegraphics[scale=0.8]{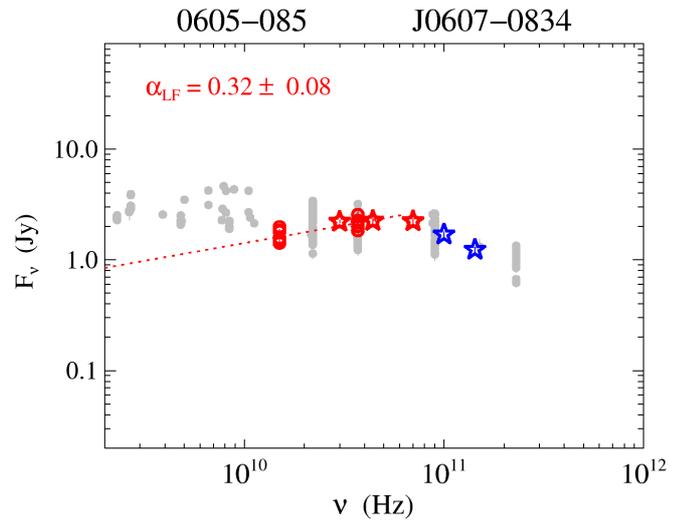}
 \caption{0605$-$085}
\end{figure*}

\begin{figure*}
\includegraphics[scale=0.8]{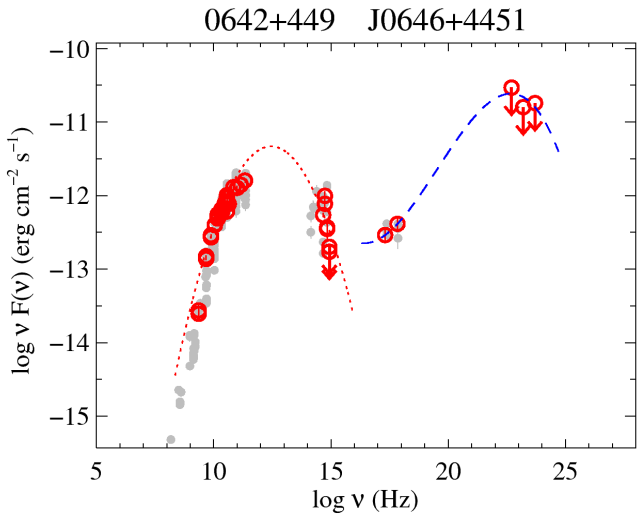}
\includegraphics[scale=0.8]{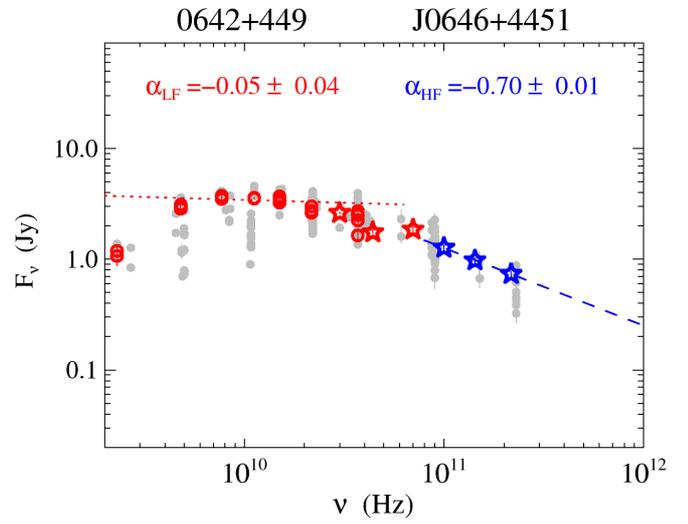}
 \caption{0642+449}
\end{figure*}

\begin{figure*}
\includegraphics[scale=0.8]{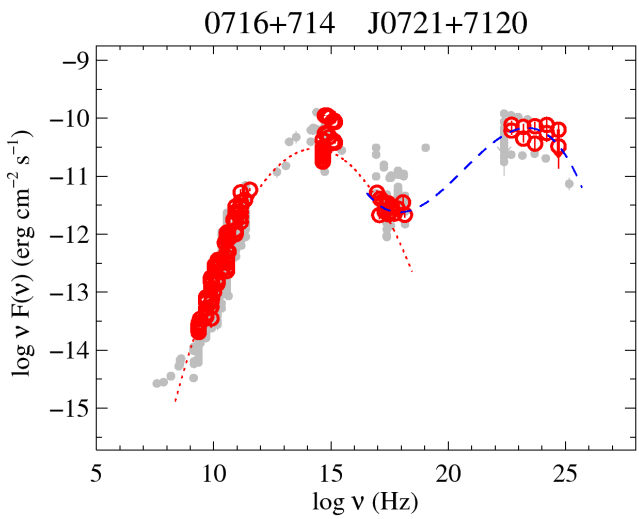}
\includegraphics[scale=0.8]{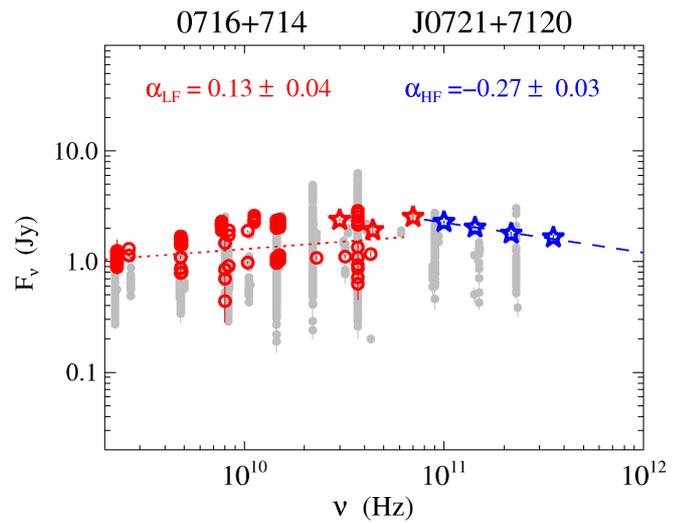}
 \caption{0716+714}
\end{figure*}
 
 \clearpage
 
\begin{figure*}
\includegraphics[scale=0.8]{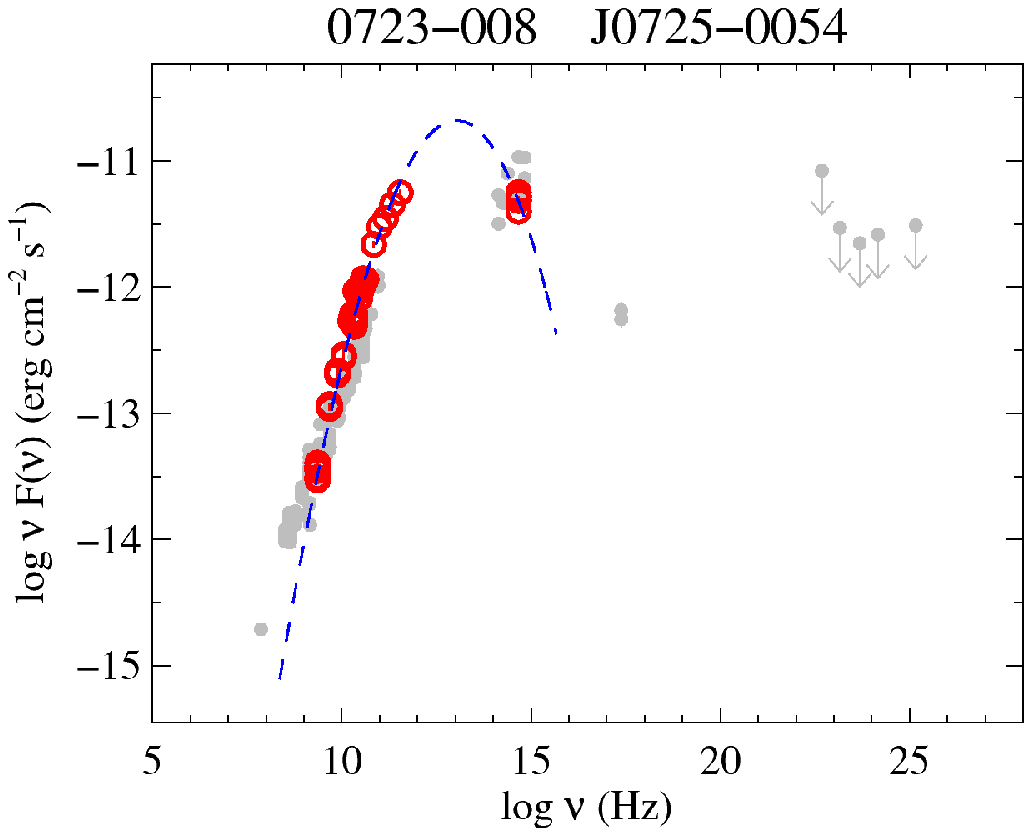}
\includegraphics[scale=0.8]{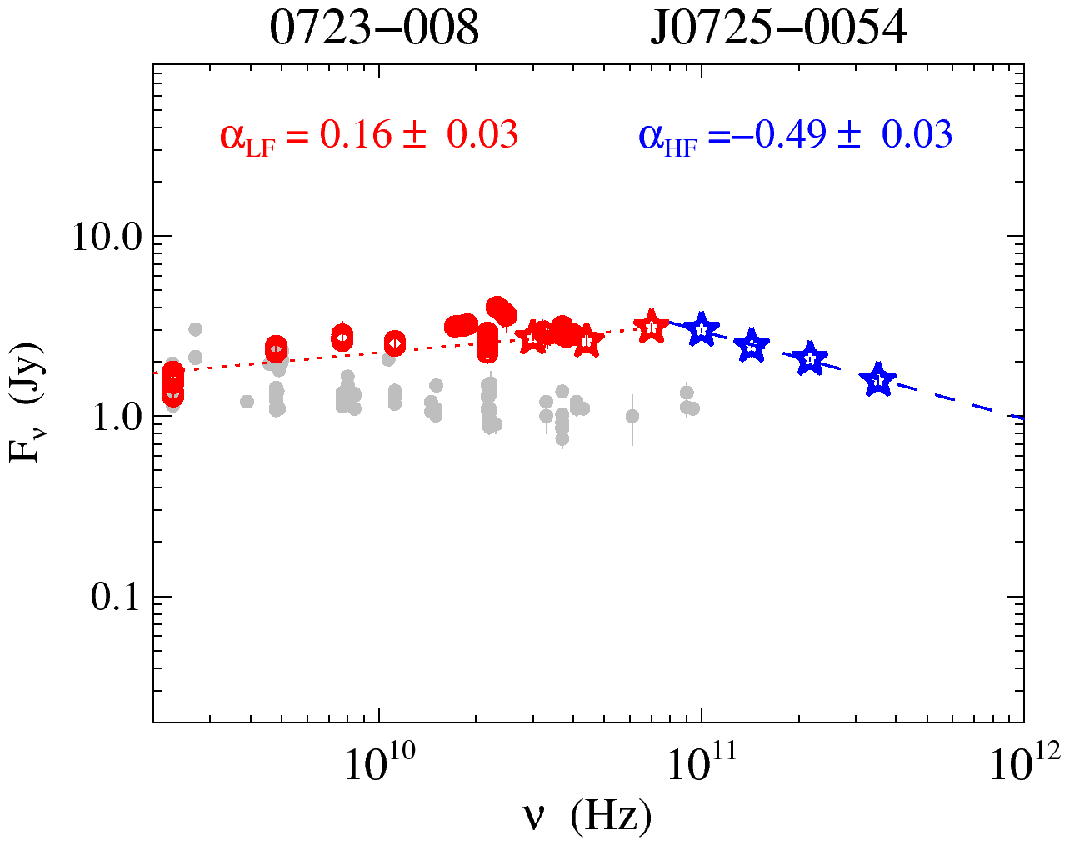}
 \caption{0723$-$008}
\end{figure*}

\begin{figure*}
\includegraphics[scale=0.8]{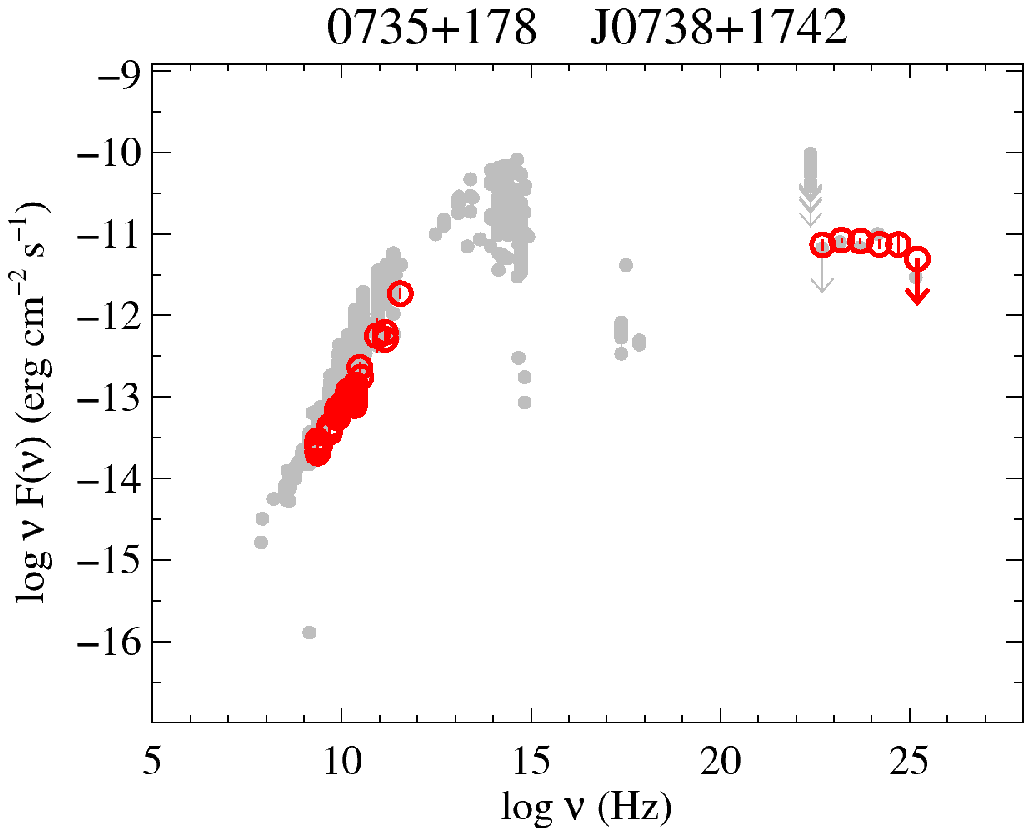}
\includegraphics[scale=0.8]{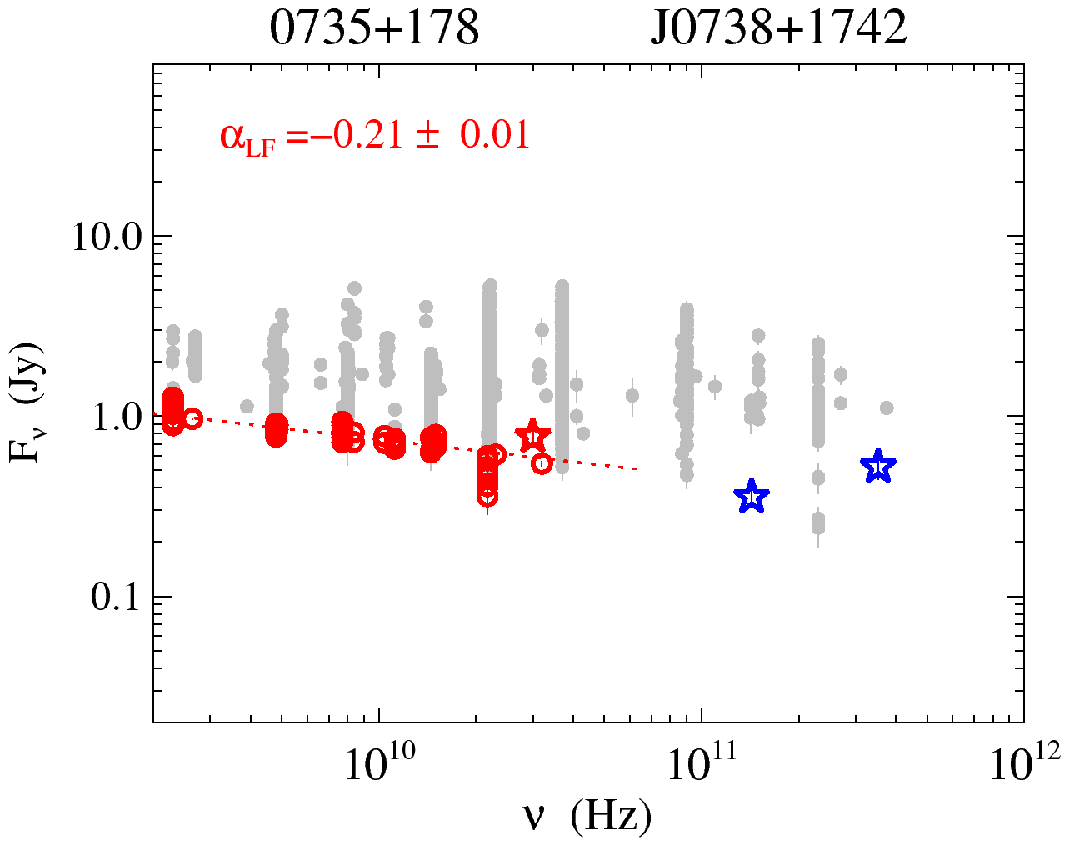}
 \caption{0735+178}
\end{figure*}

\begin{figure*}
\includegraphics[scale=0.8]{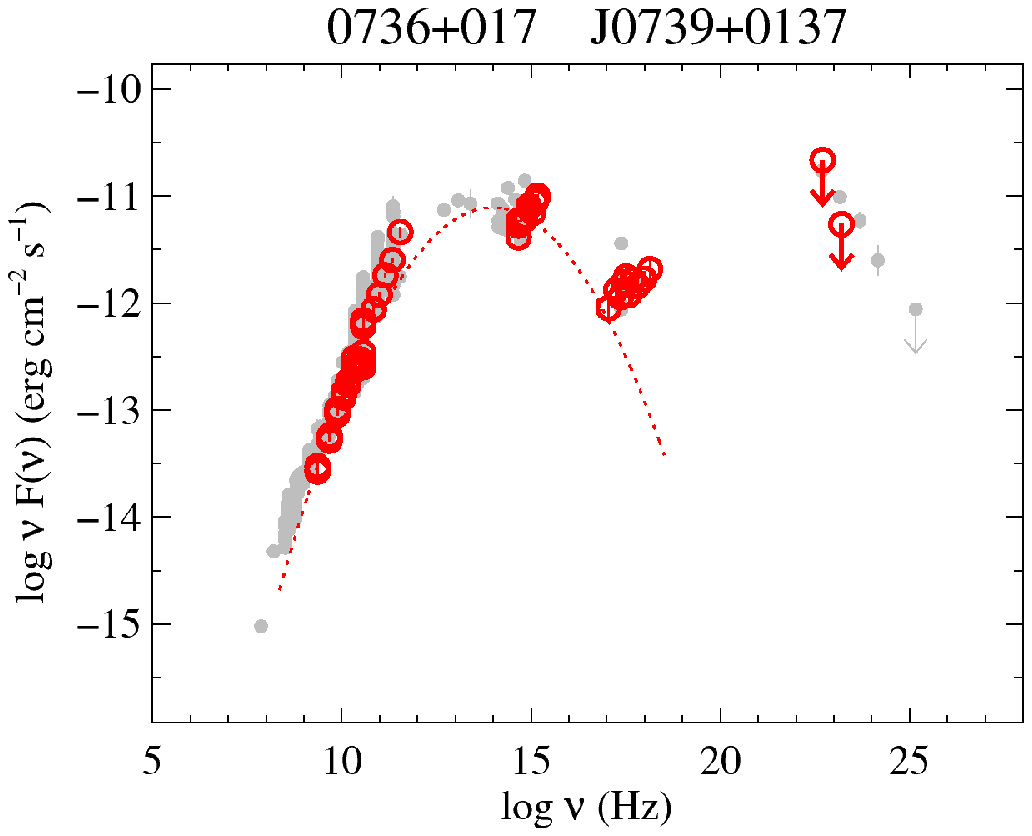}
\includegraphics[scale=0.8]{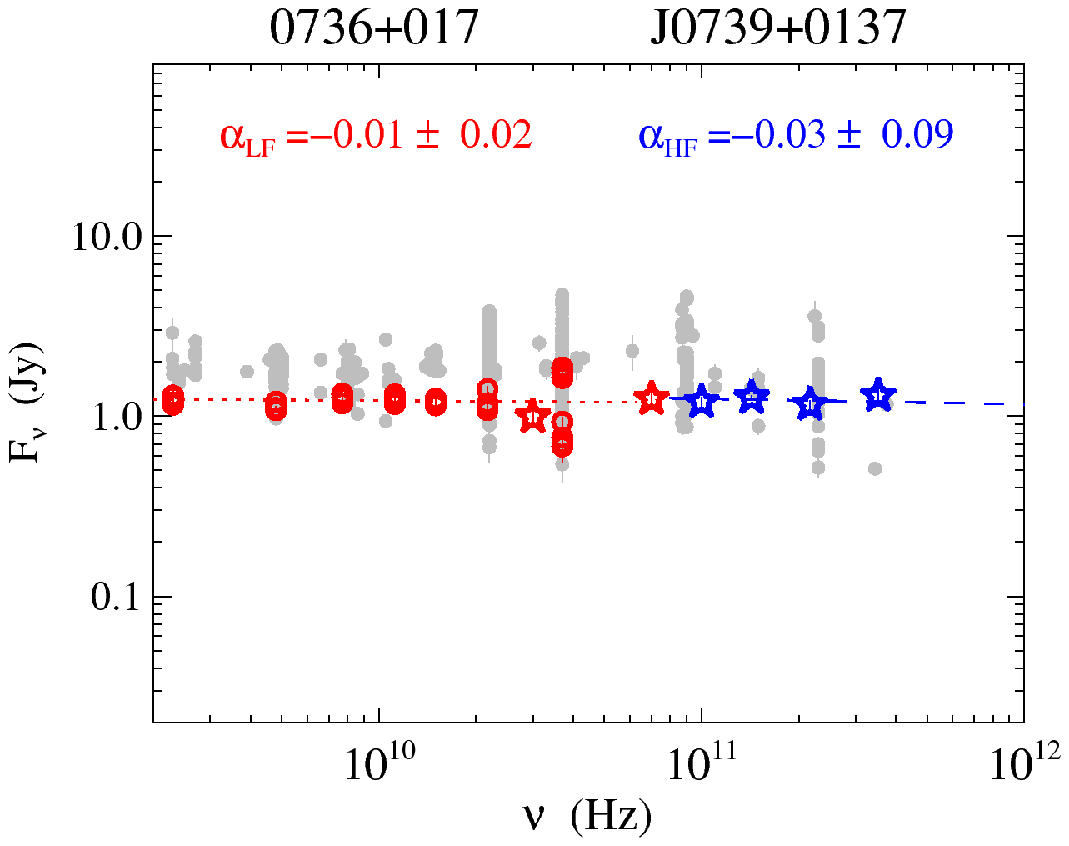}
 \caption{0736+017}
\end{figure*}
 
 \clearpage
 
\begin{figure*}
\includegraphics[scale=0.8]{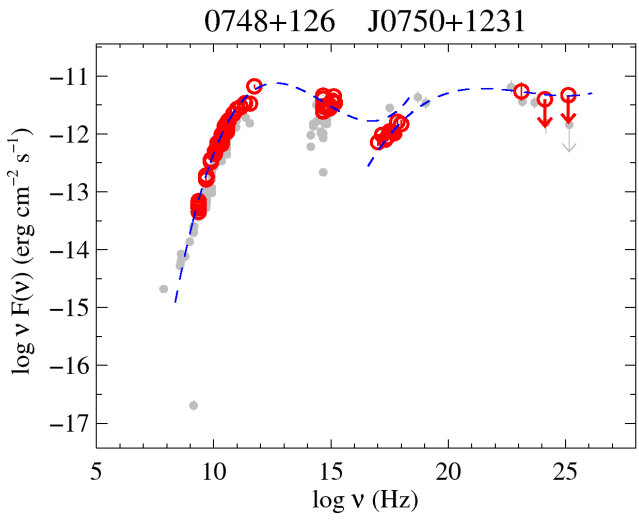}
\includegraphics[scale=0.8]{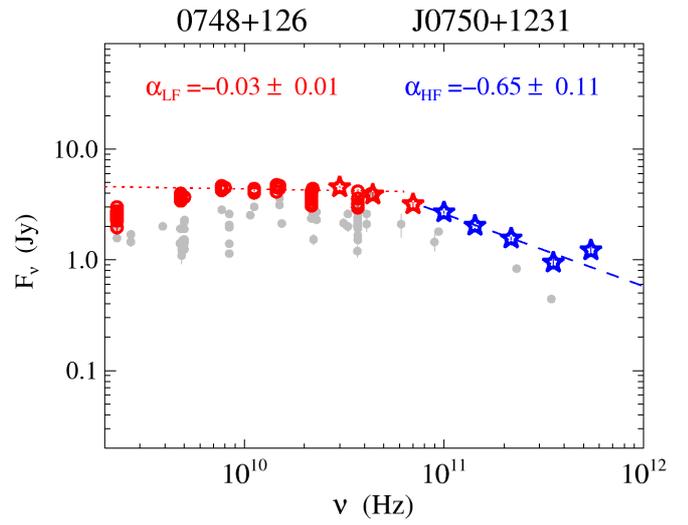}
 \caption{0748+126}
\end{figure*}

\begin{figure*}
\includegraphics[scale=0.8]{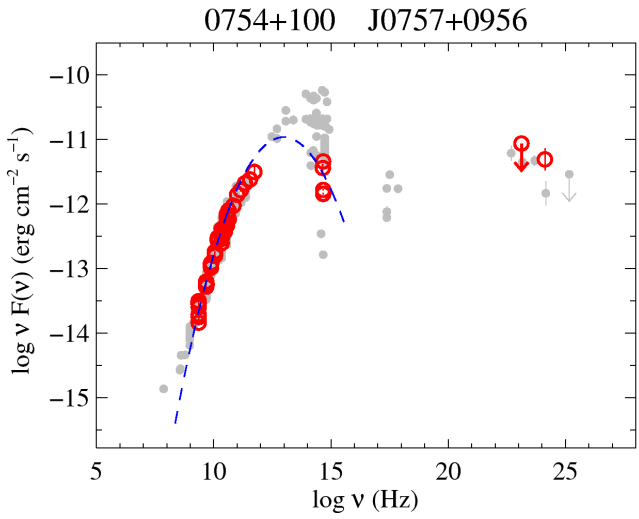}
\includegraphics[scale=0.8]{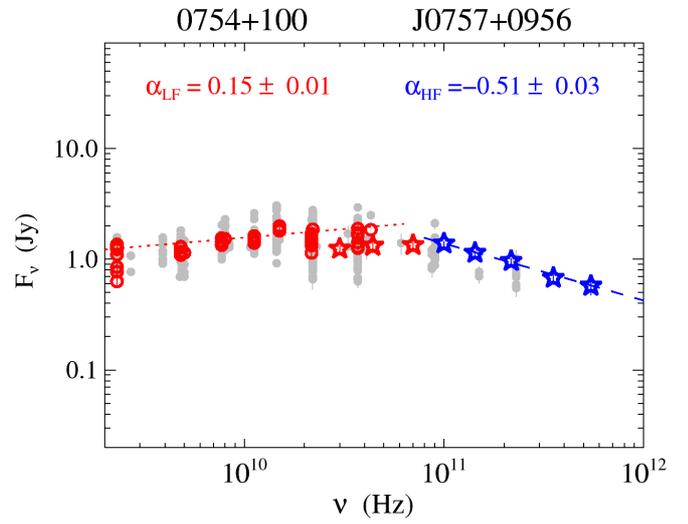}
 \caption{0754+100}
\end{figure*}

\begin{figure*}
\includegraphics[scale=0.8]{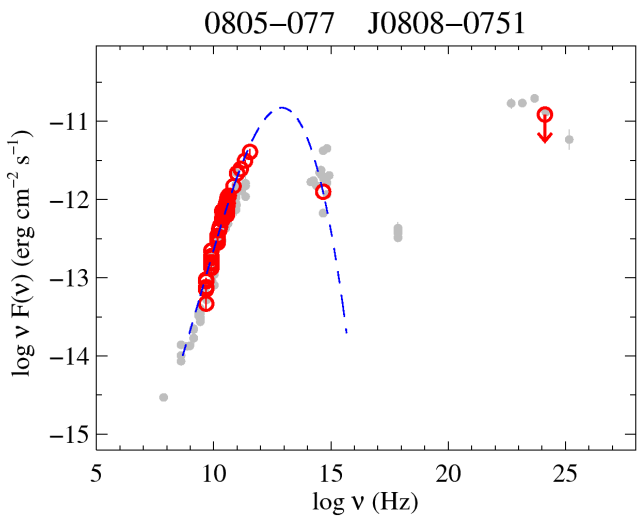}
\includegraphics[scale=0.8]{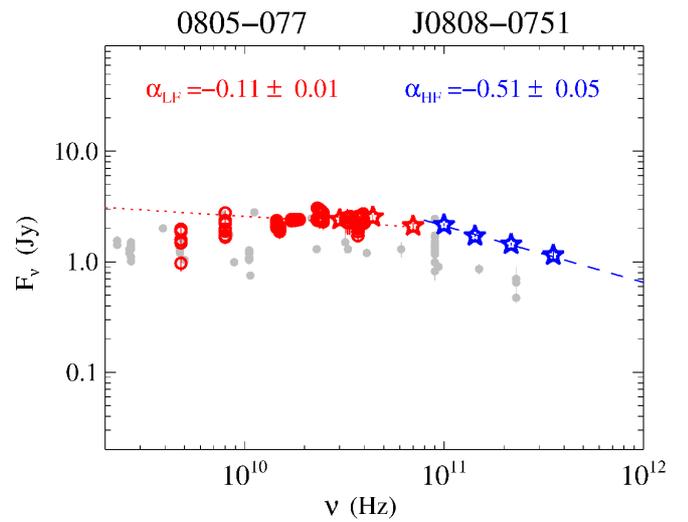}
 \caption{0805$-$077}
\end{figure*}
 
 \clearpage
 
\begin{figure*}
\includegraphics[scale=0.8]{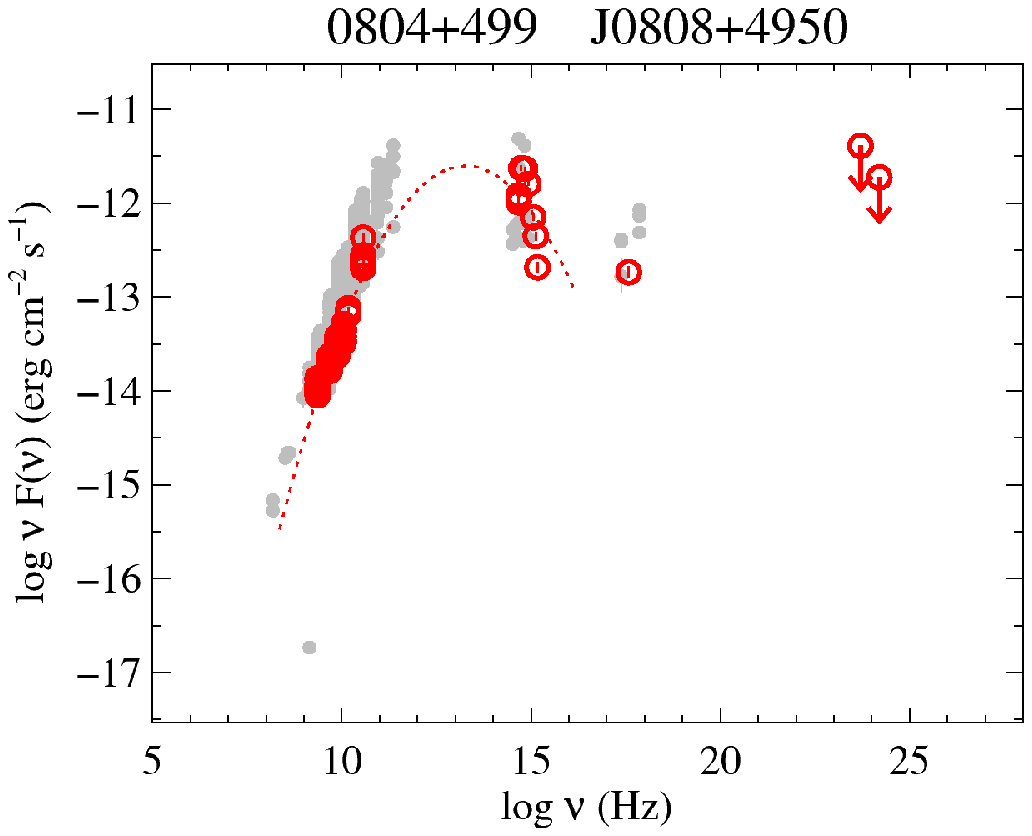}
\includegraphics[scale=0.8]{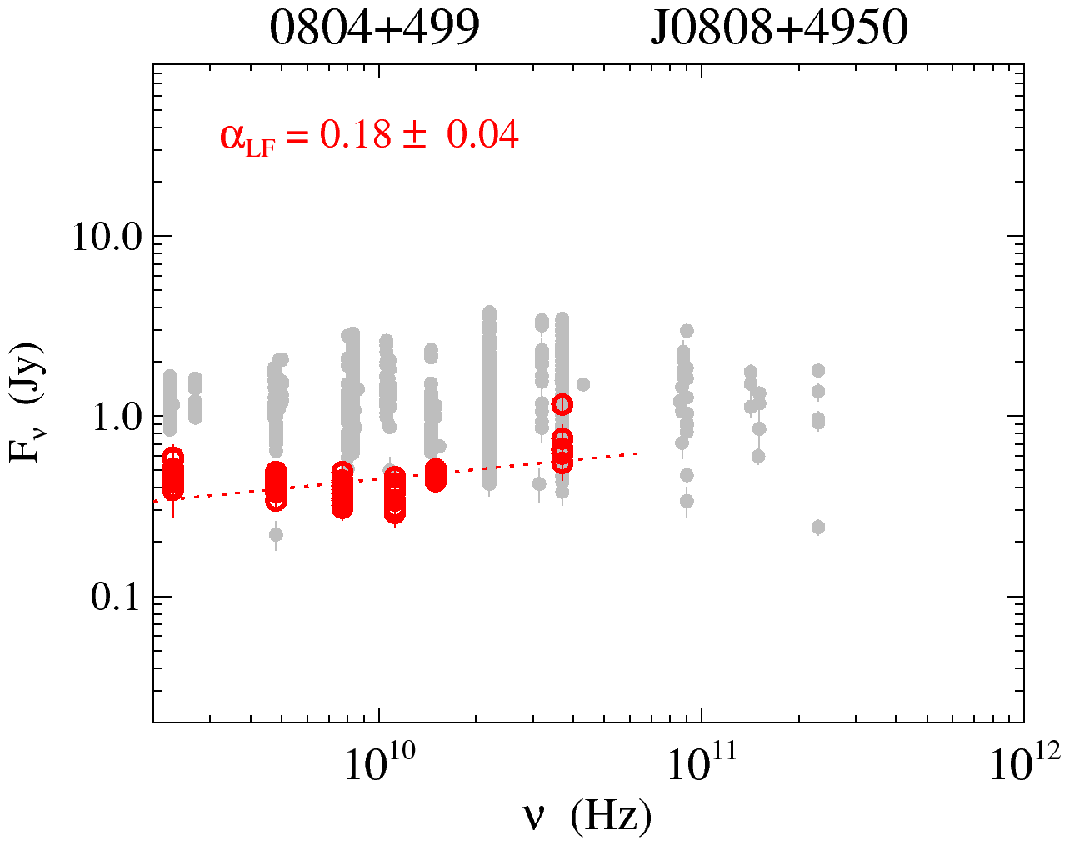}
 \caption{0804+499}
\end{figure*}

\begin{figure*}
\includegraphics[scale=0.8]{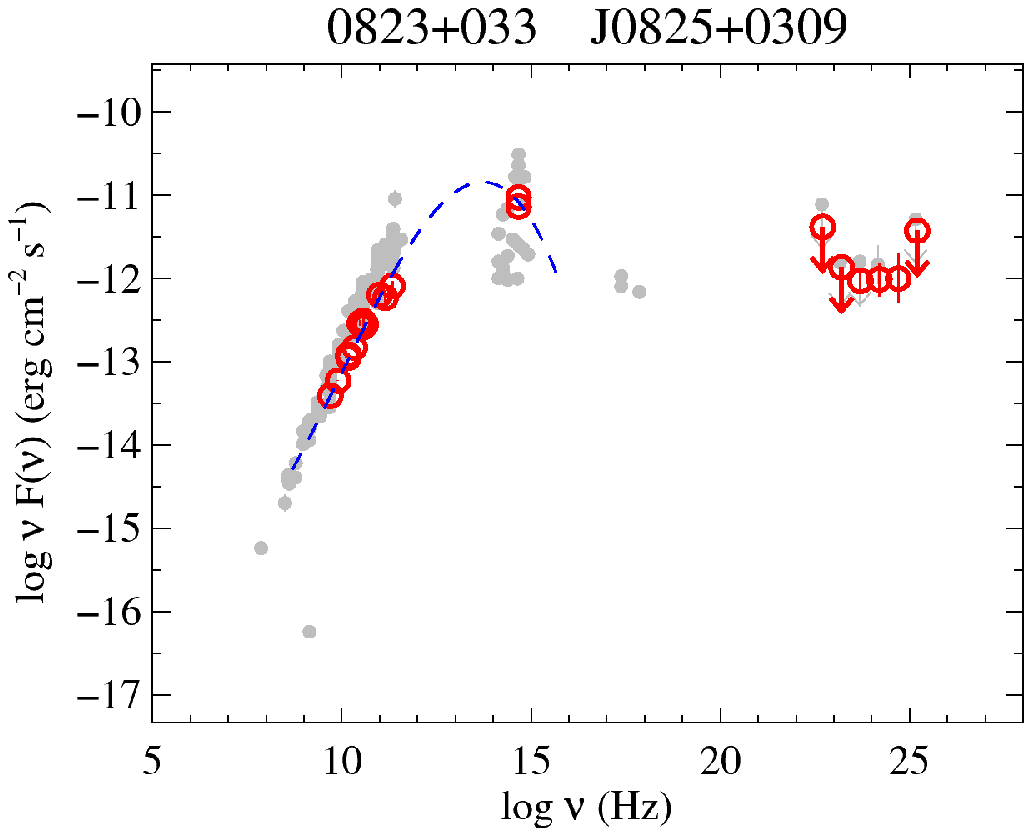}
\includegraphics[scale=0.8]{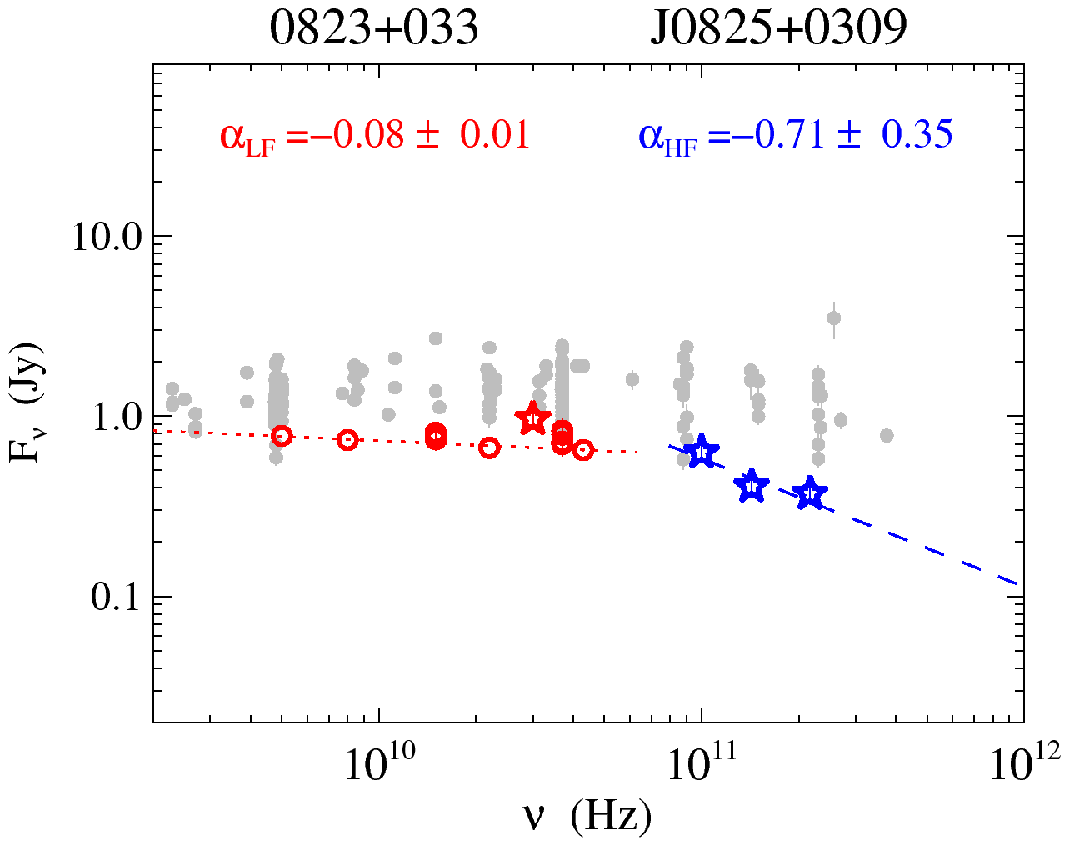}
 \caption{0823+033}
\end{figure*}

\begin{figure*}
\includegraphics[scale=0.8]{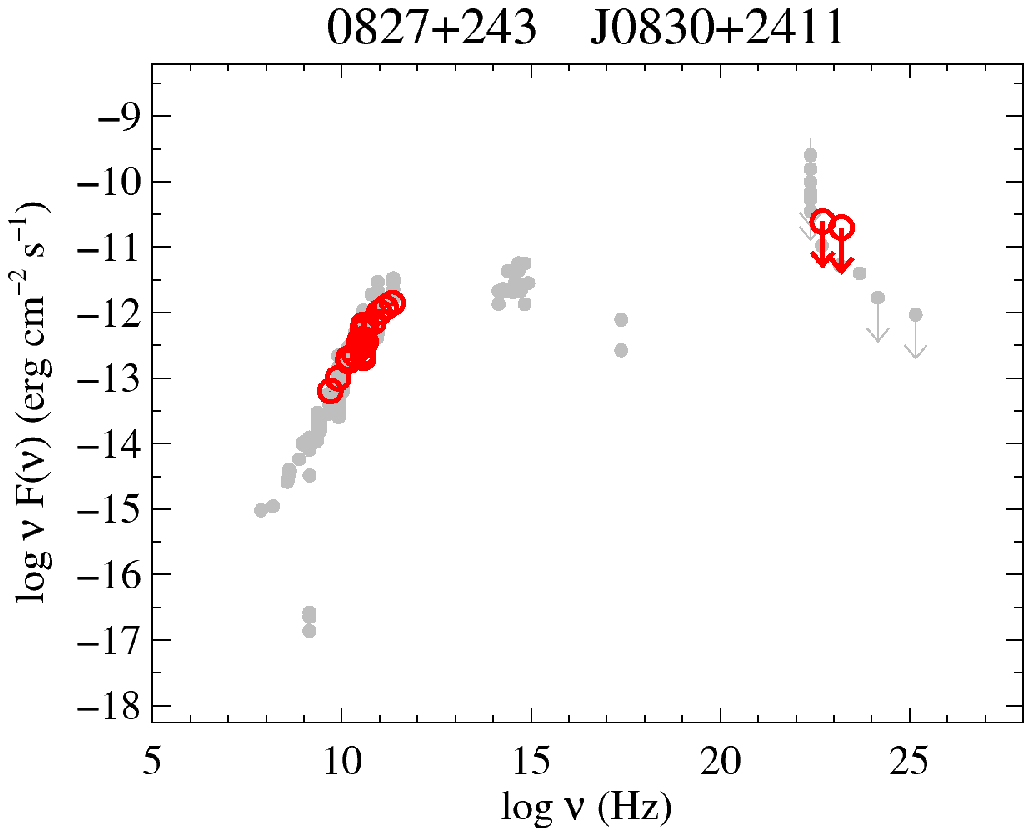}
\includegraphics[scale=0.8]{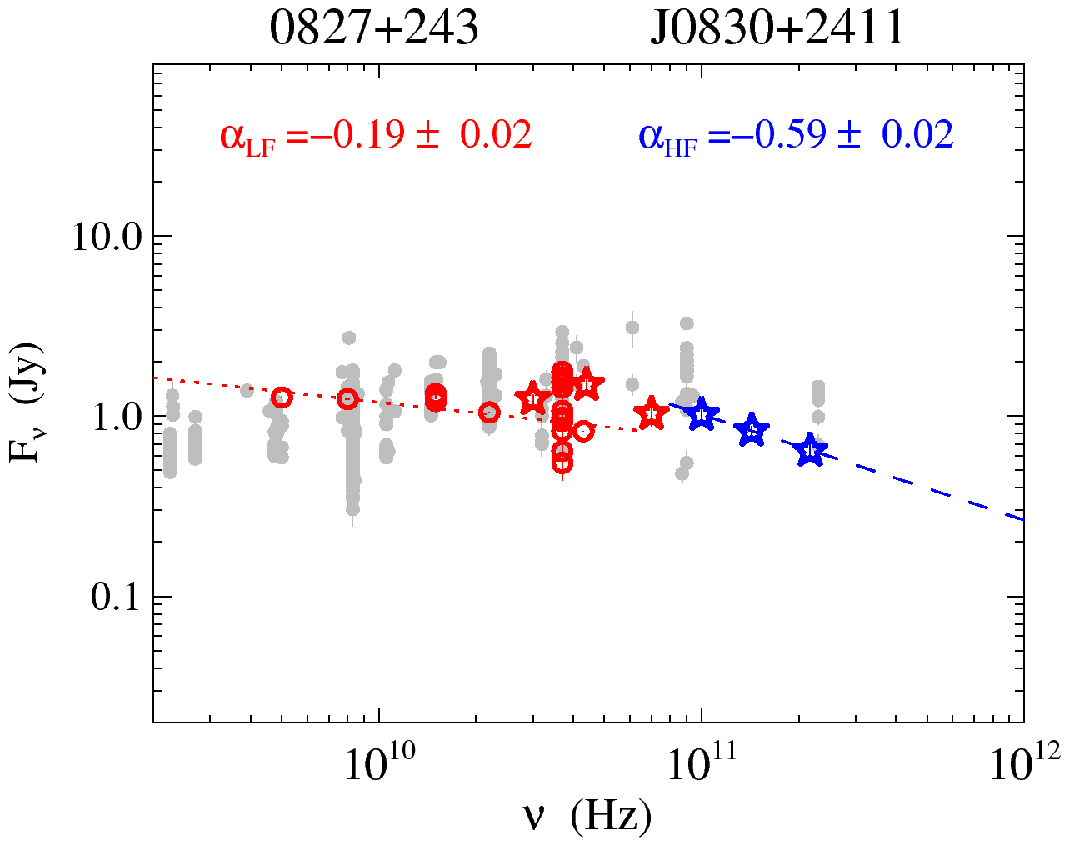}
 \caption{0827+243}
\end{figure*}
 
 \clearpage
 
\begin{figure*}
\includegraphics[scale=0.8]{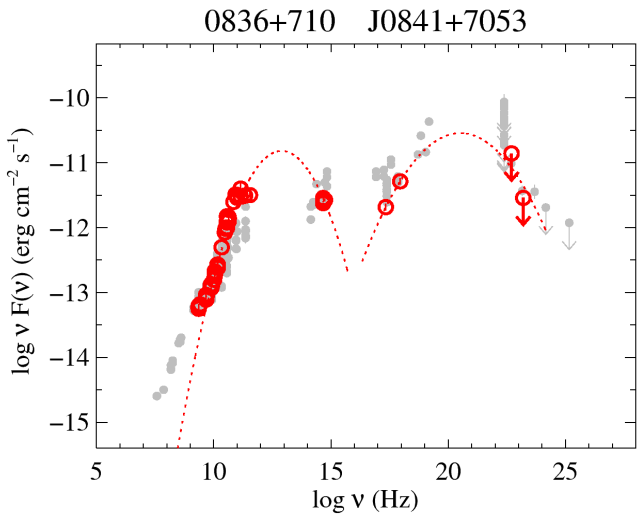}
\includegraphics[scale=0.8]{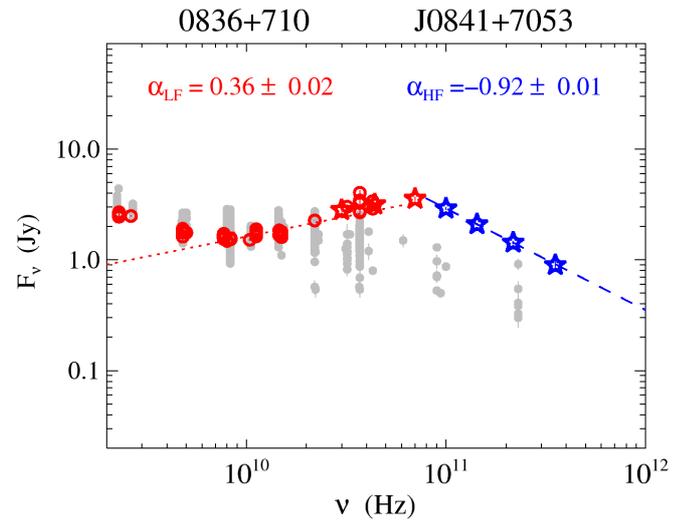}
 \caption{0836+710}
\end{figure*}

\begin{figure*}
\includegraphics[scale=0.8]{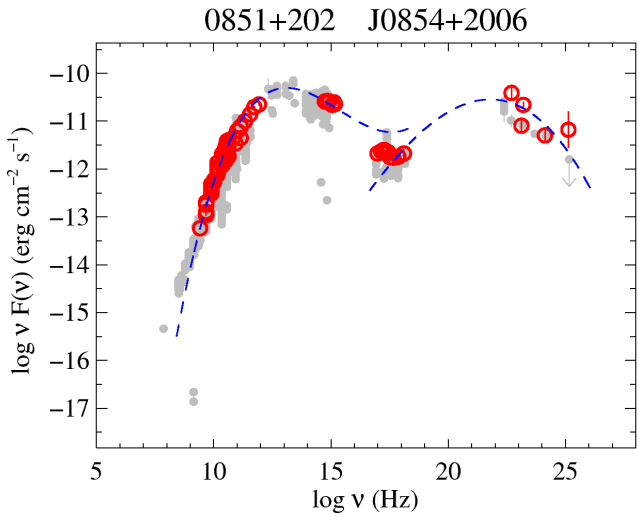}
\includegraphics[scale=0.8]{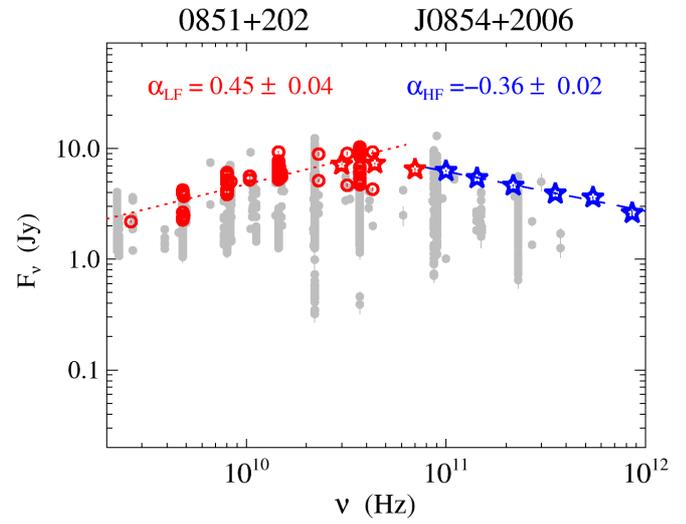}
 \caption{0851+202}
\end{figure*}

\begin{figure*}
\includegraphics[scale=0.8]{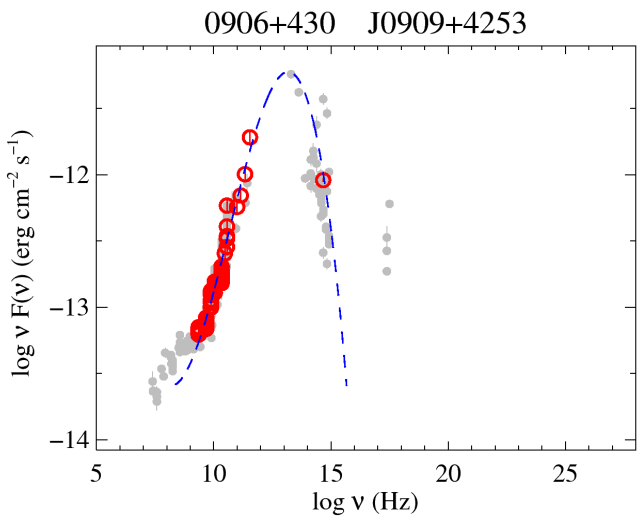}
\includegraphics[scale=0.8]{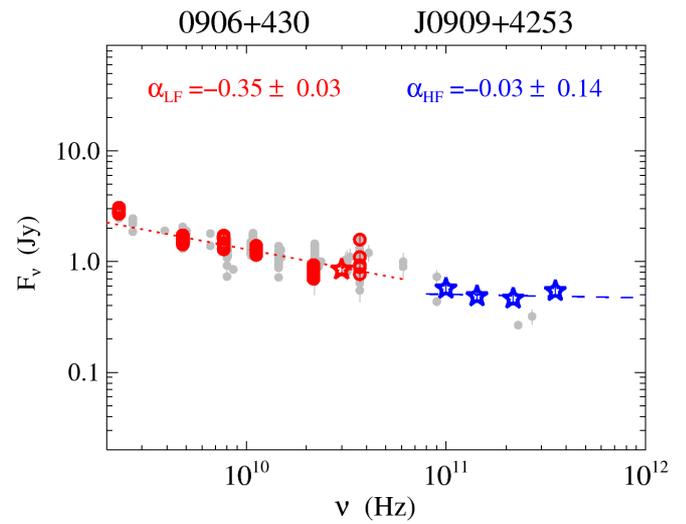}
 \caption{0906+430}
\end{figure*}
 
 \clearpage
 
\begin{figure*}
\includegraphics[scale=0.8]{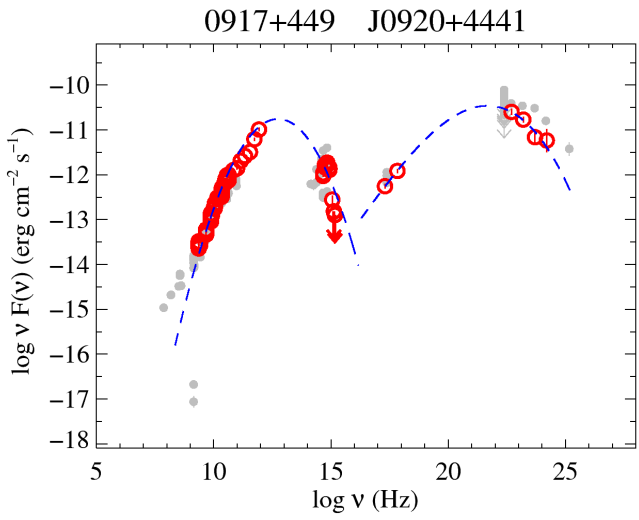}
\includegraphics[scale=0.8]{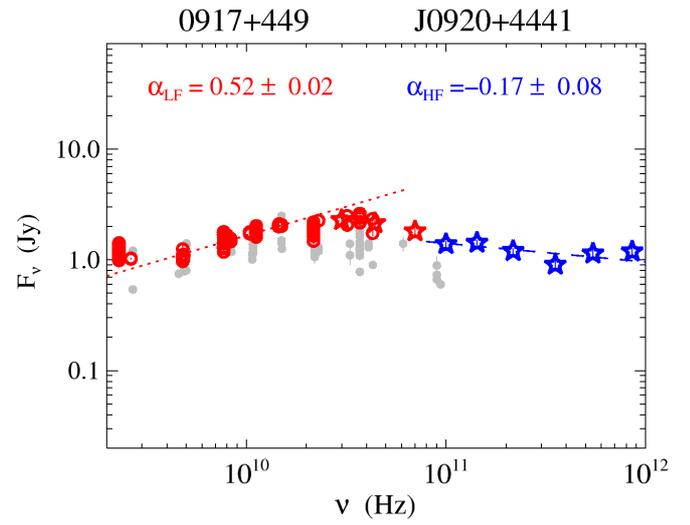}
 \caption{0917+449}
\end{figure*}

\begin{figure*}
\includegraphics[scale=0.8]{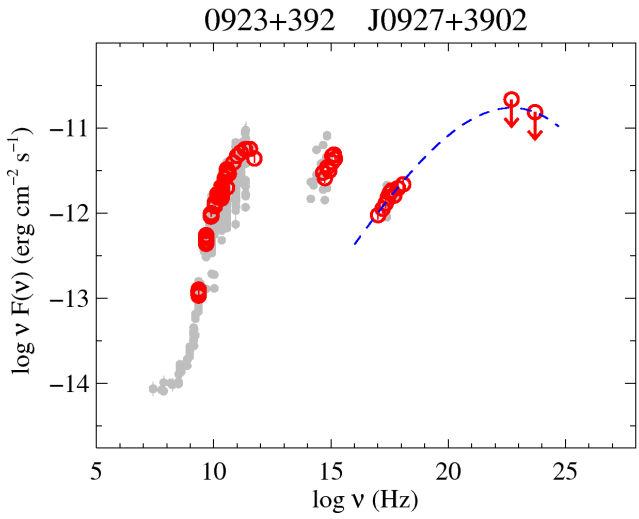}
\includegraphics[scale=0.8]{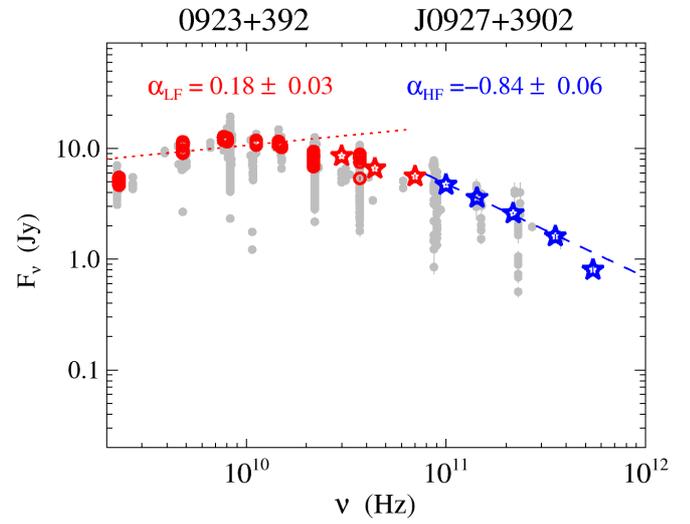}
 \caption{0923+392}
\end{figure*}

\begin{figure*}
\includegraphics[scale=0.8]{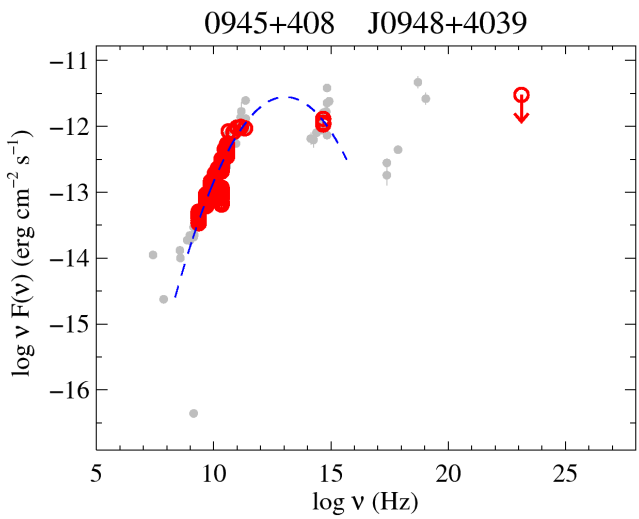}
\includegraphics[scale=0.8]{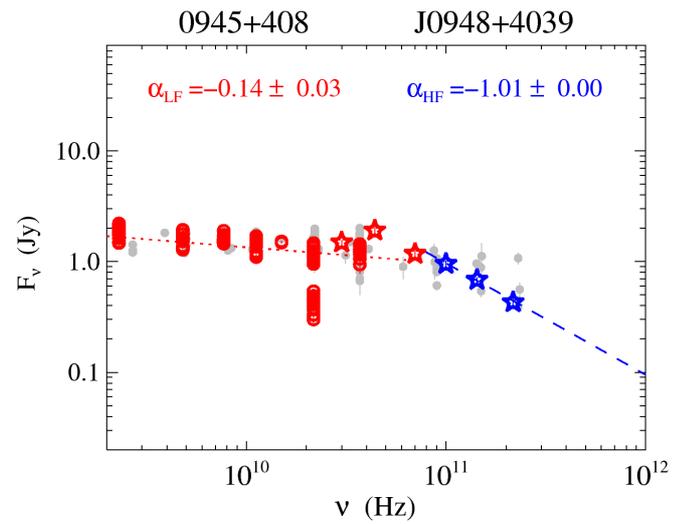}
 \caption{0945+408}
\end{figure*}
 
 \clearpage
 
\begin{figure*}
\includegraphics[scale=0.8]{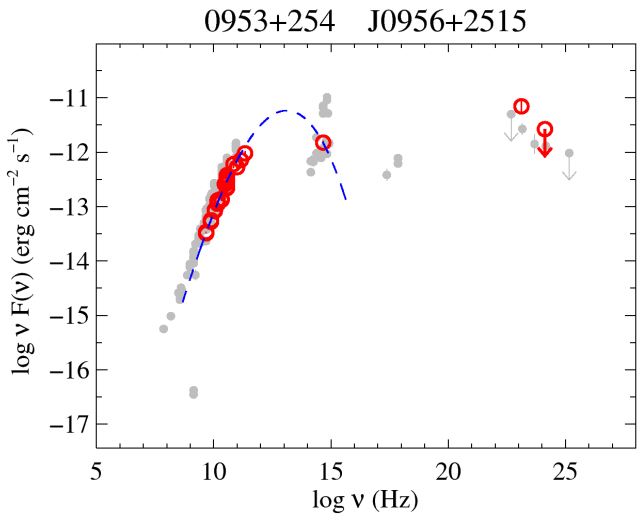}
\includegraphics[scale=0.8]{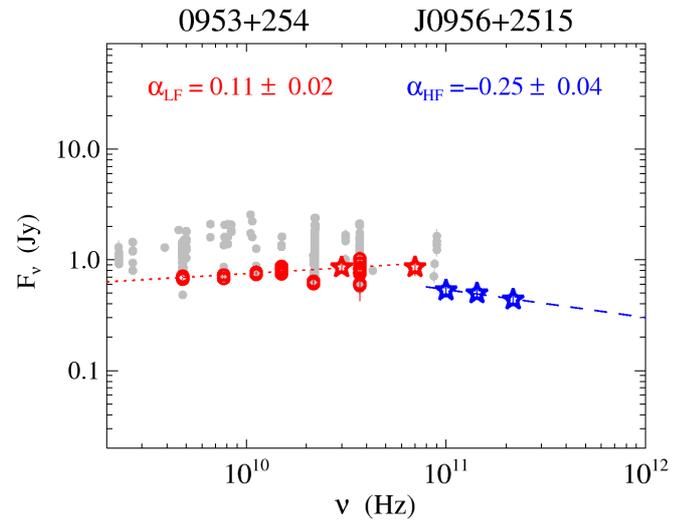}
 \caption{0953+254}
\end{figure*}

\begin{figure*}
\includegraphics[scale=0.8]{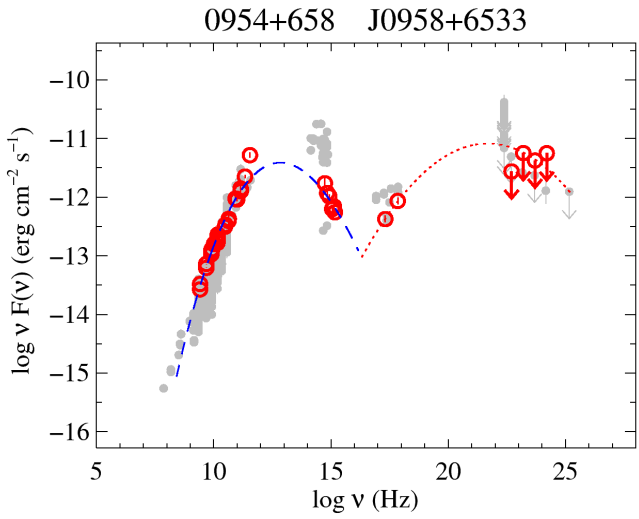}
\includegraphics[scale=0.8]{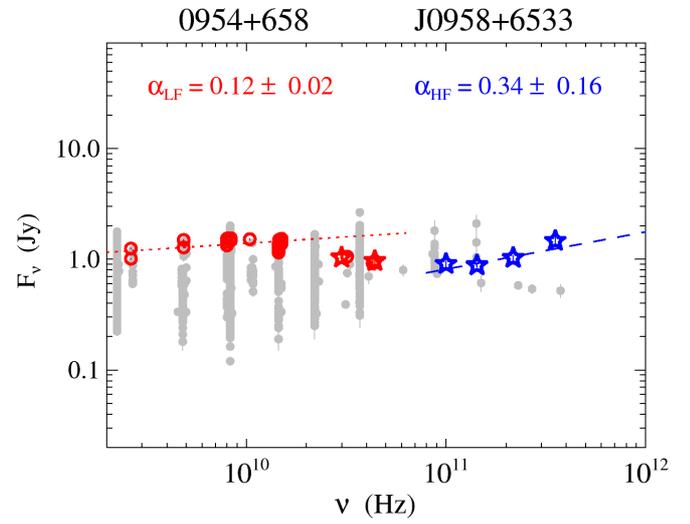}
 \caption{0954+658}
\end{figure*}

\begin{figure*}
\includegraphics[scale=0.8]{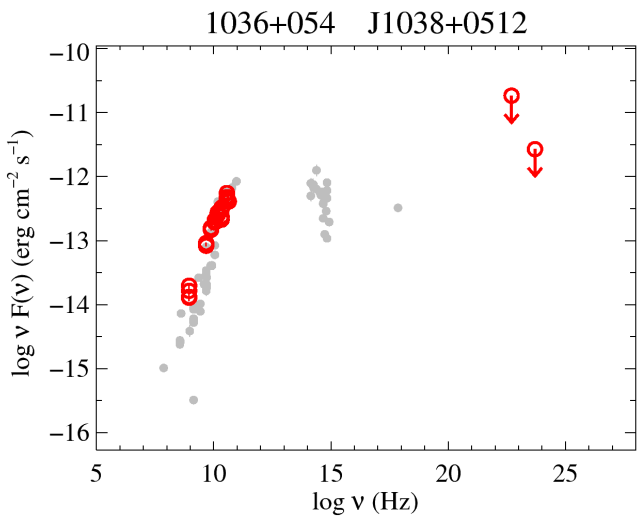}
\includegraphics[scale=0.8]{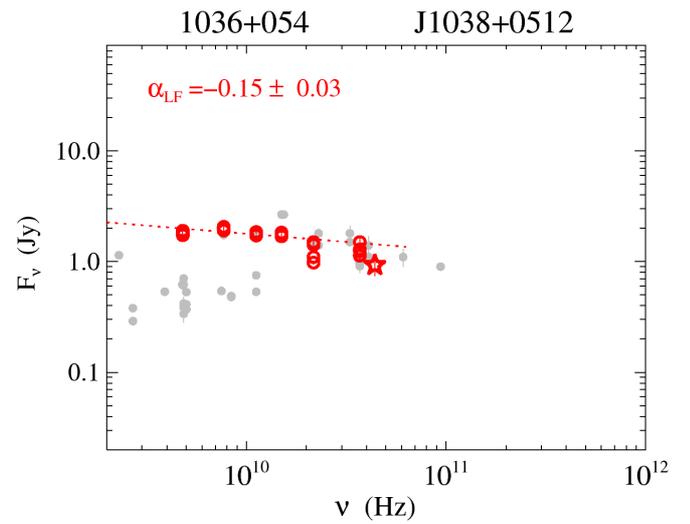}
 \caption{1036+054}
\end{figure*}
 
 \clearpage
 
\begin{figure*}
\includegraphics[scale=0.8]{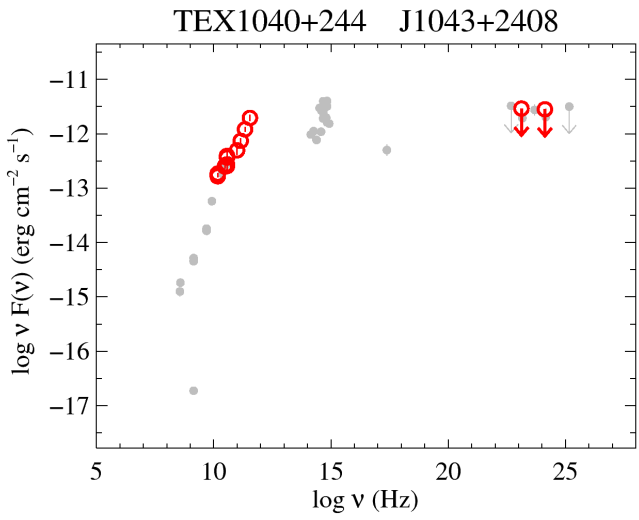}
\includegraphics[scale=0.8]{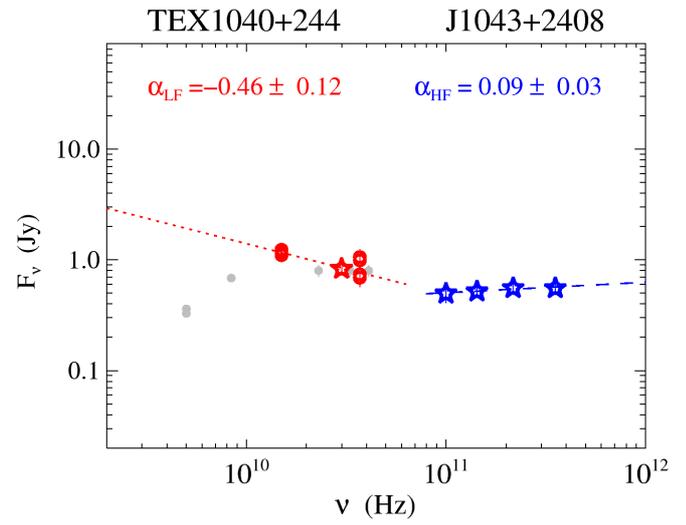}
 \caption{TEX1040+244}
\end{figure*}

\begin{figure*}
\includegraphics[scale=0.8]{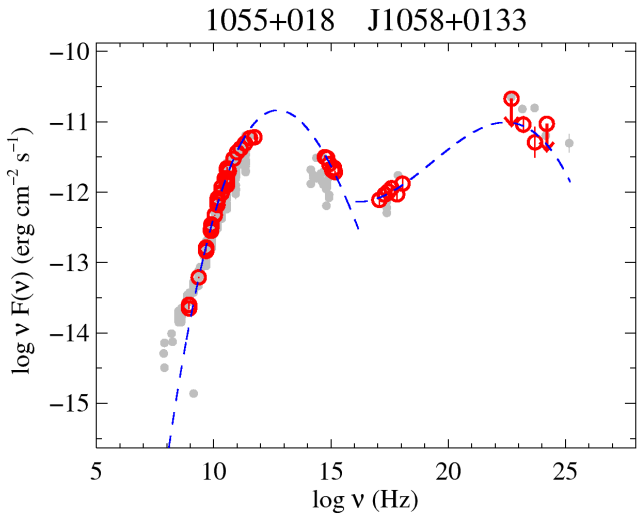}
\includegraphics[scale=0.8]{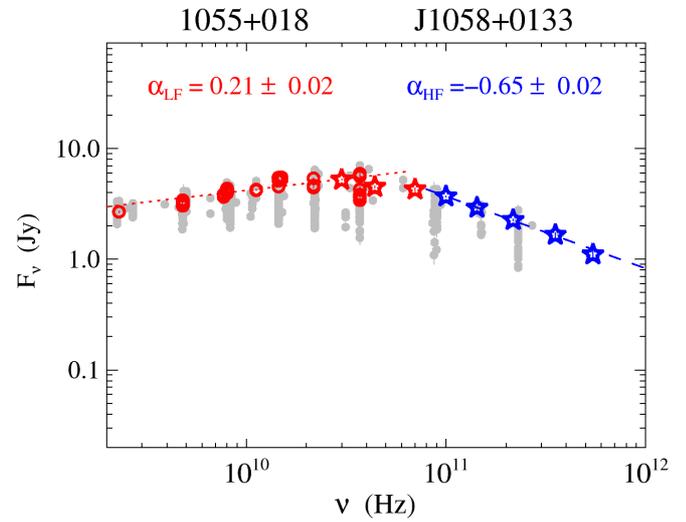}
 \caption{1055+018}
\end{figure*}

\begin{figure*}
\includegraphics[scale=0.8]{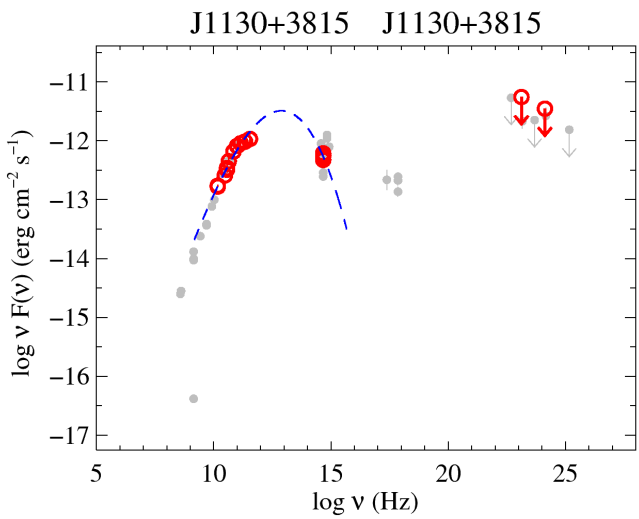}
\includegraphics[scale=0.8]{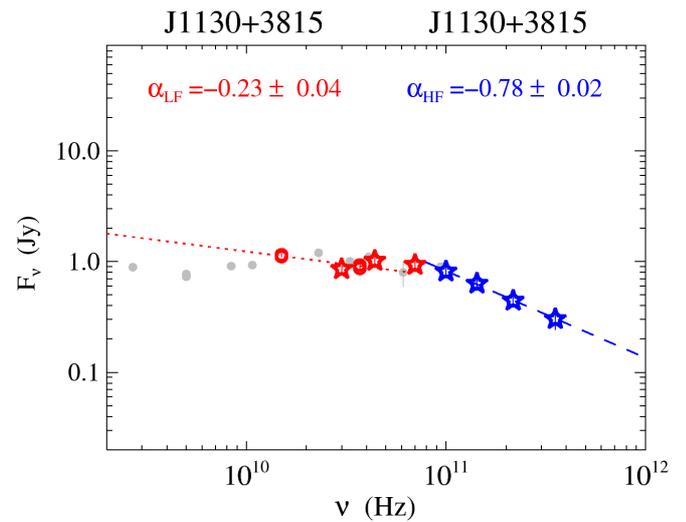}
 \caption{J1130+3815}
\end{figure*}
 
 \clearpage
 
\begin{figure*}
\includegraphics[scale=0.8]{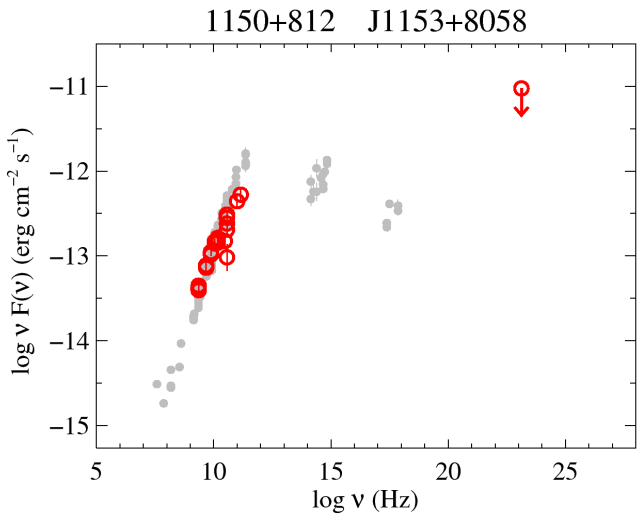}
\includegraphics[scale=0.8]{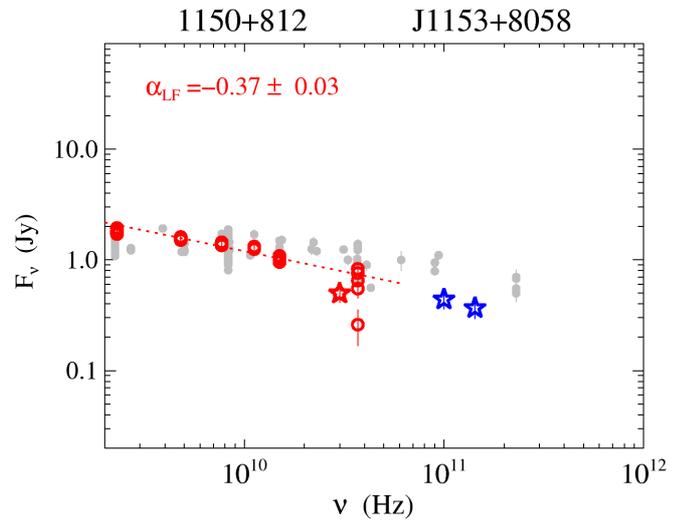}
 \caption{1150+812}
\end{figure*}

\begin{figure*}
\includegraphics[scale=0.8]{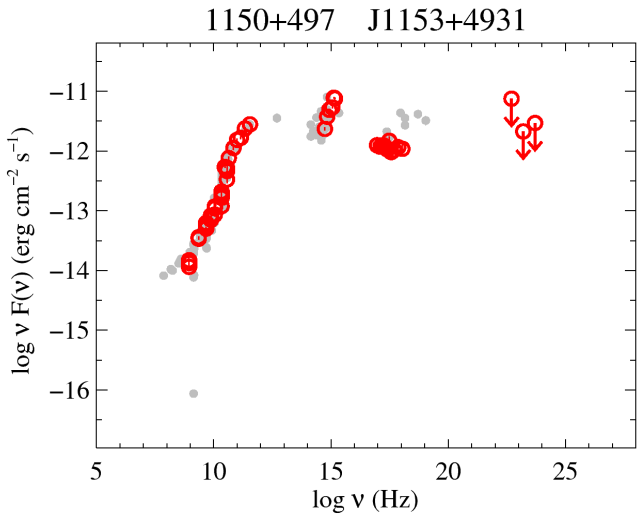}
\includegraphics[scale=0.8]{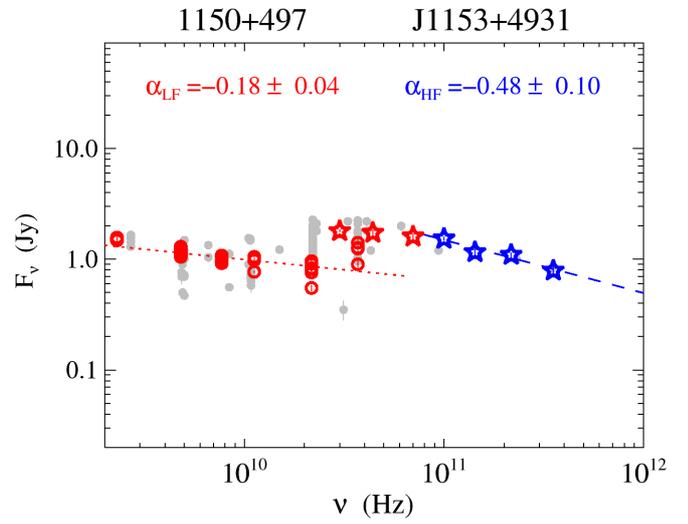}
 \caption{1150+497}
\end{figure*}

\begin{figure*}
\includegraphics[scale=0.8]{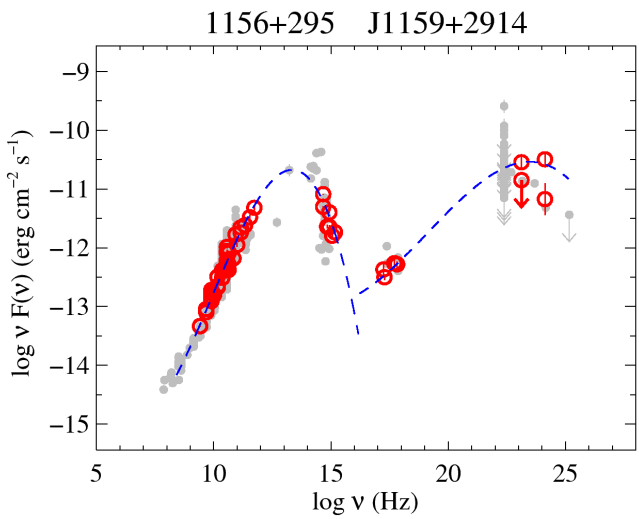}
\includegraphics[scale=0.8]{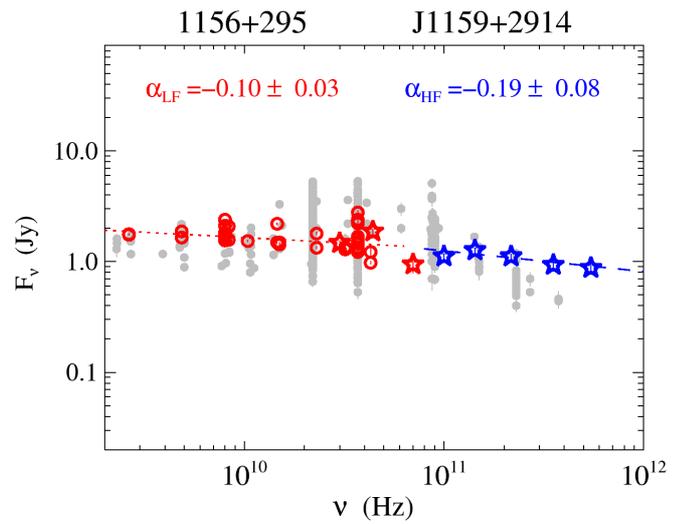}
 \caption{1156+295}
\end{figure*}
 
 \clearpage
 
\begin{figure*}
\includegraphics[scale=0.8]{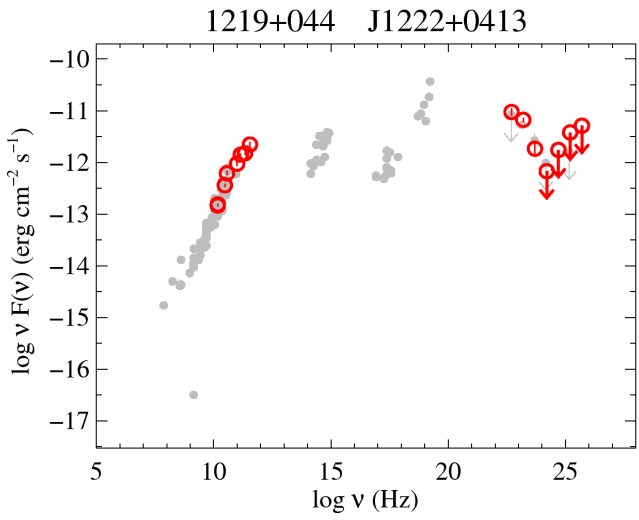}
\includegraphics[scale=0.8]{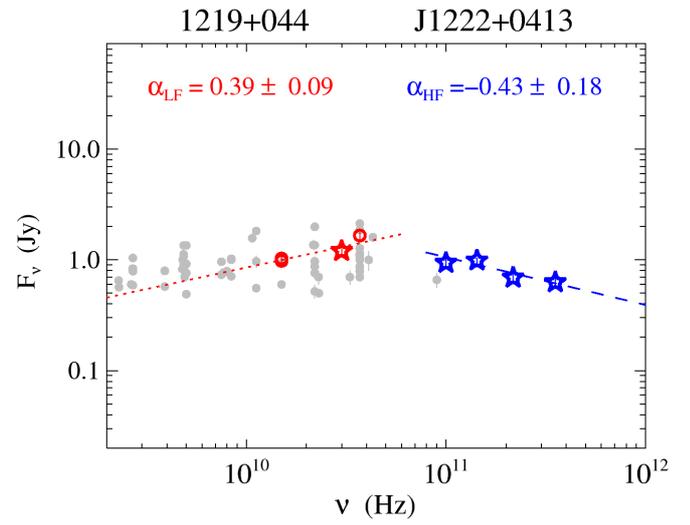}
 \caption{1219+044}
\end{figure*}

\begin{figure*}
\includegraphics[scale=0.8]{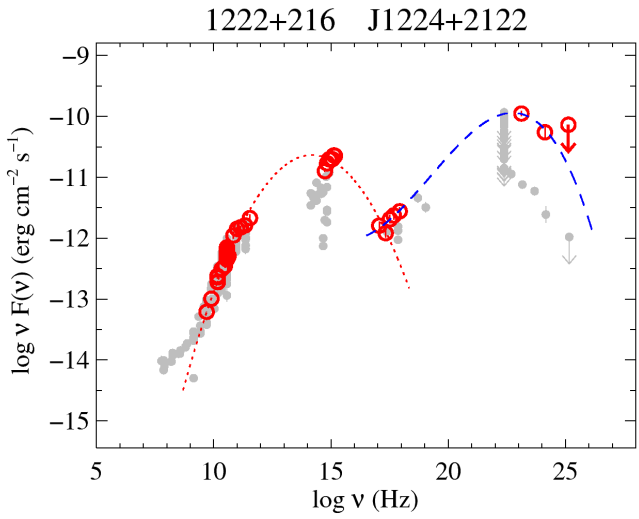}
\includegraphics[scale=0.8]{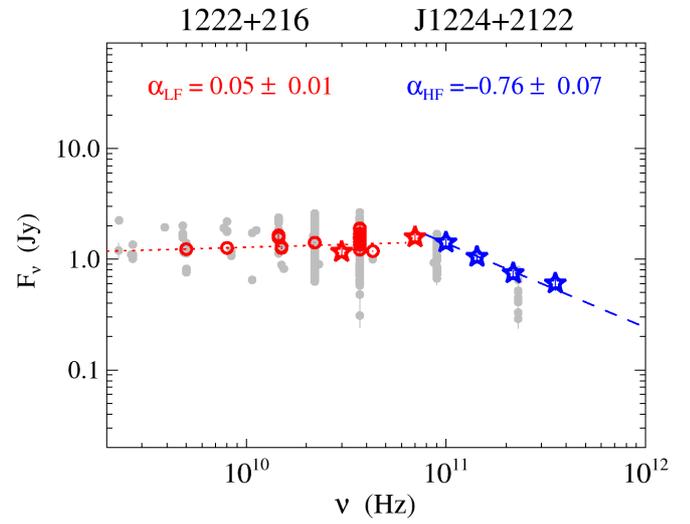}
 \caption{1222+216}
\end{figure*}

\begin{figure*}
\includegraphics[scale=0.8]{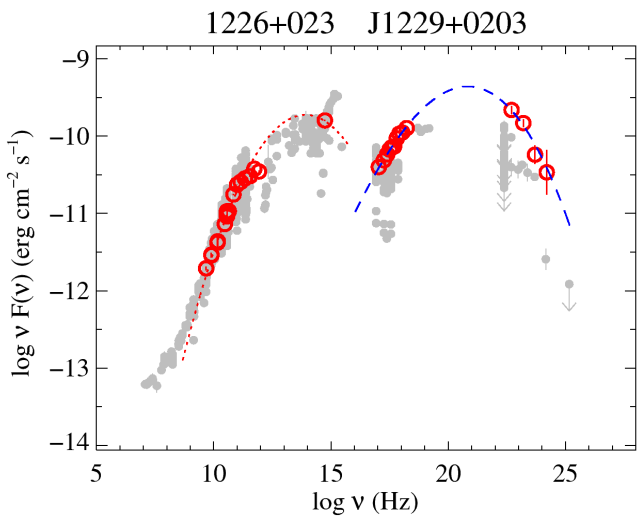}
\includegraphics[scale=0.8]{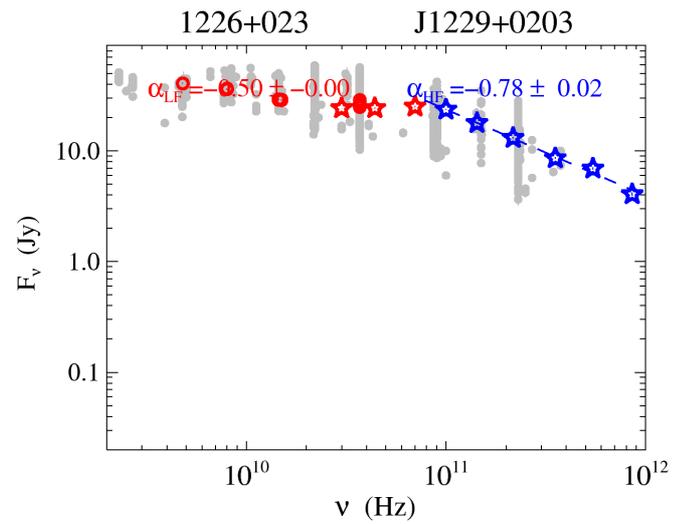}
 \caption{1226+023}
\end{figure*}
 
 \clearpage
 
\begin{figure*}
\includegraphics[scale=0.8]{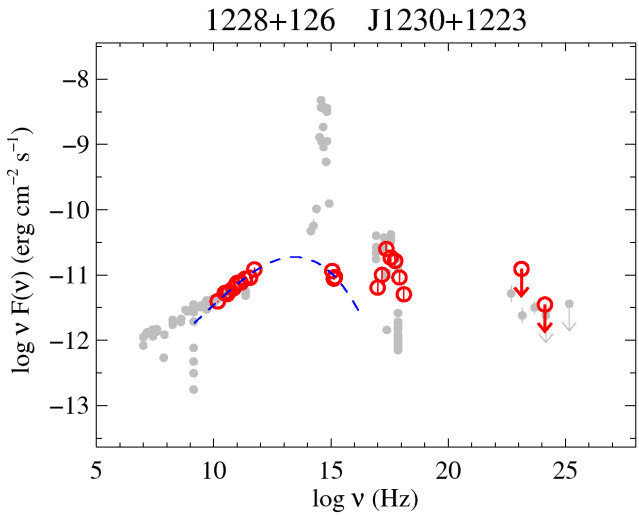}
\includegraphics[scale=0.8]{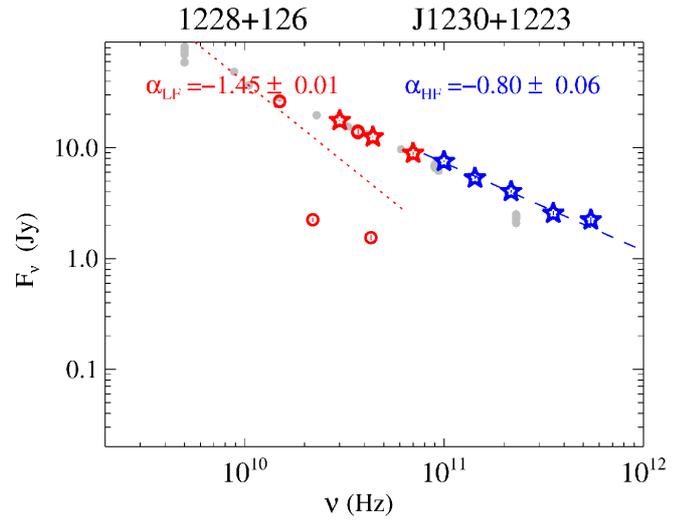}
 \caption{1228+126}\label{1228}
\end{figure*}

\begin{figure*}
\includegraphics[scale=0.8]{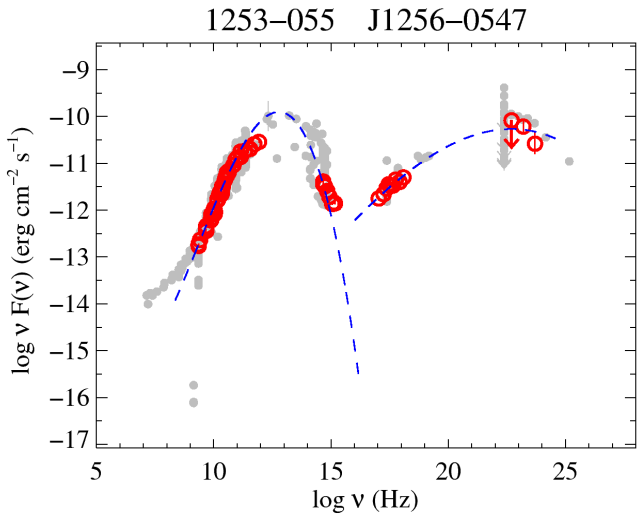}
\includegraphics[scale=0.8]{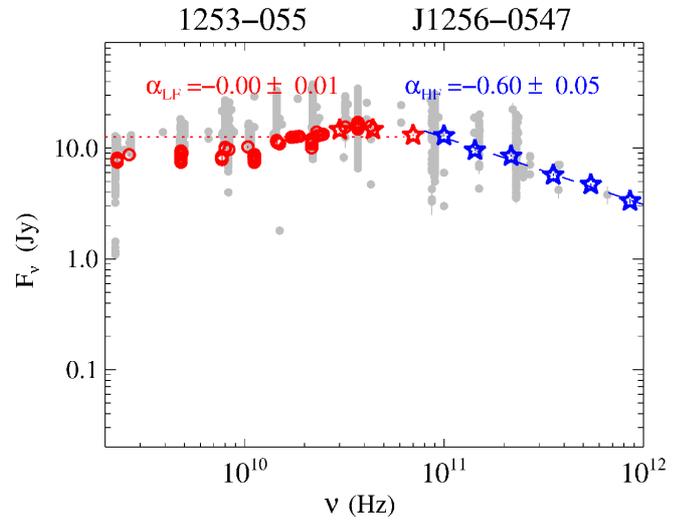}
 \caption{1253$-$055}
\end{figure*}

\begin{figure*}
\includegraphics[scale=0.8]{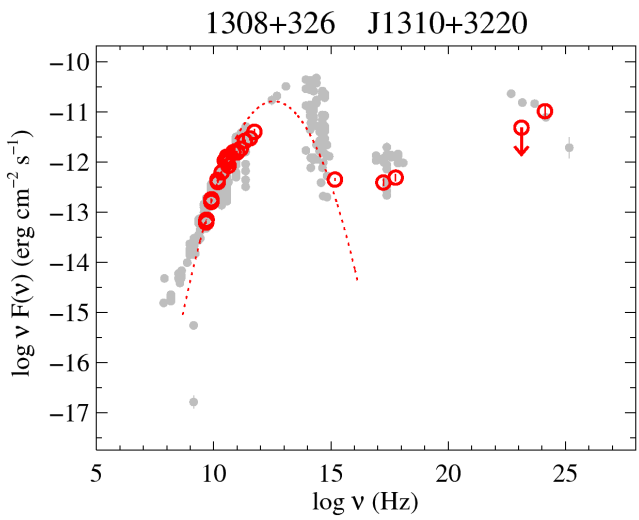}
\includegraphics[scale=0.8]{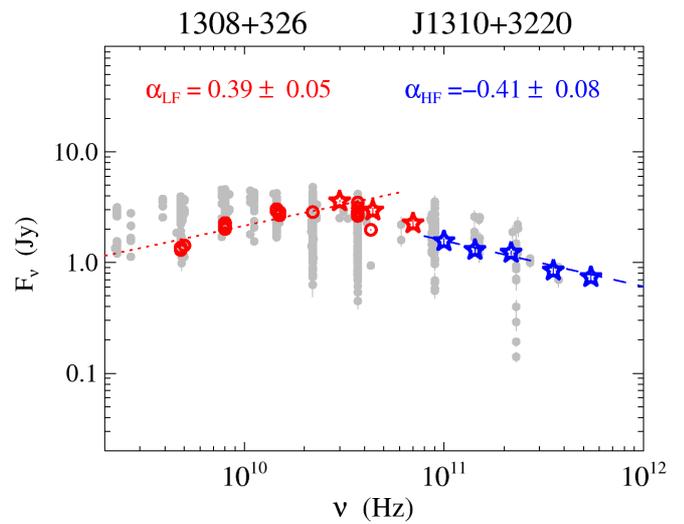}
 \caption{1308+326}
\end{figure*}
 
 \clearpage
 
\begin{figure*}
\includegraphics[scale=0.8]{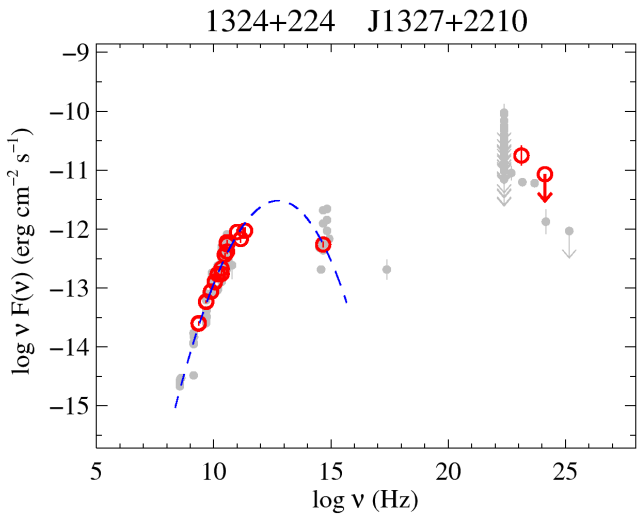}
\includegraphics[scale=0.8]{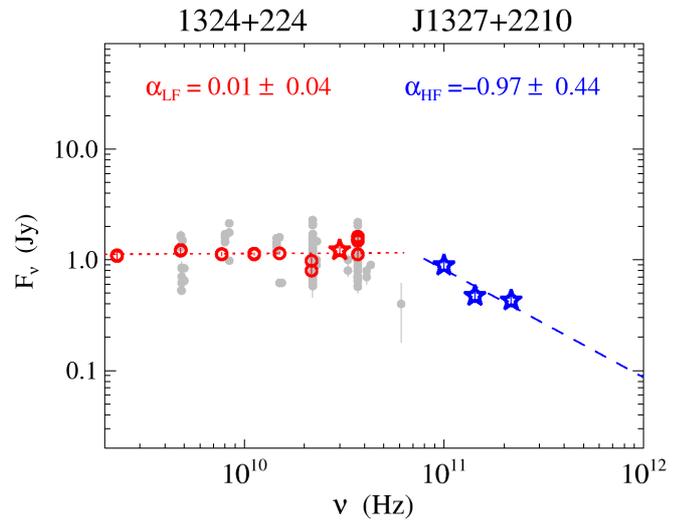}
 \caption{1324+224}
\end{figure*}

\begin{figure*}
\includegraphics[scale=0.8]{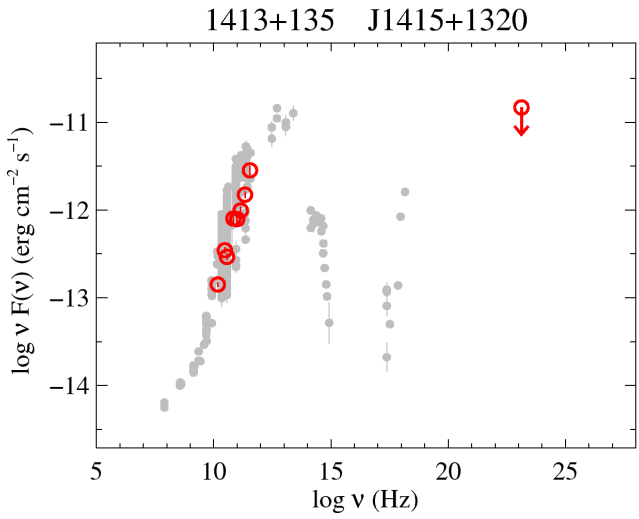}
\includegraphics[scale=0.8]{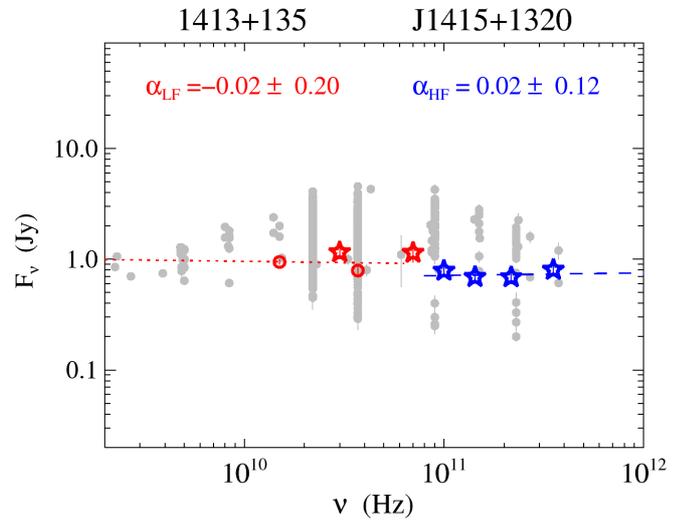}
 \caption{1413+135}
\end{figure*}

\begin{figure*}
\includegraphics[scale=0.8]{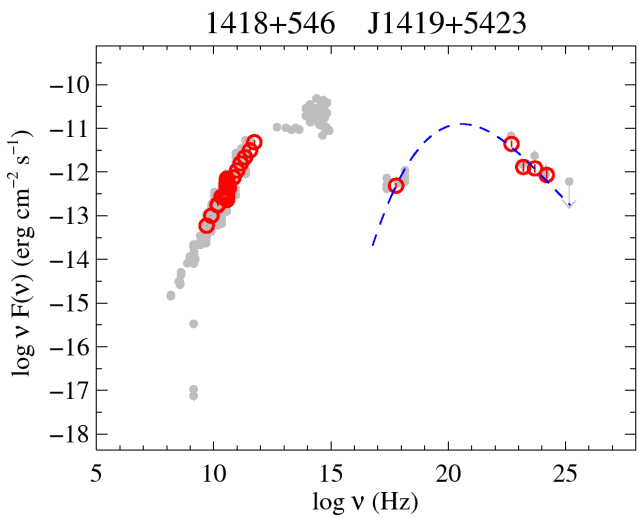}
\includegraphics[scale=0.8]{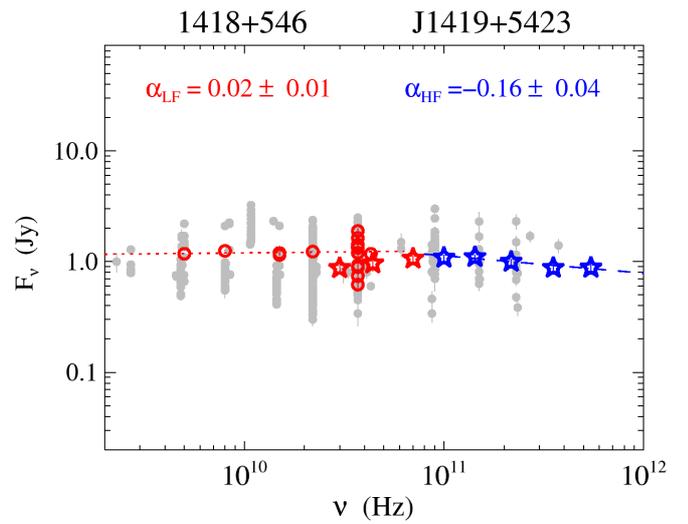}
 \caption{1418+546}
\end{figure*}
 
 \clearpage
 
\begin{figure*}
\includegraphics[scale=0.8]{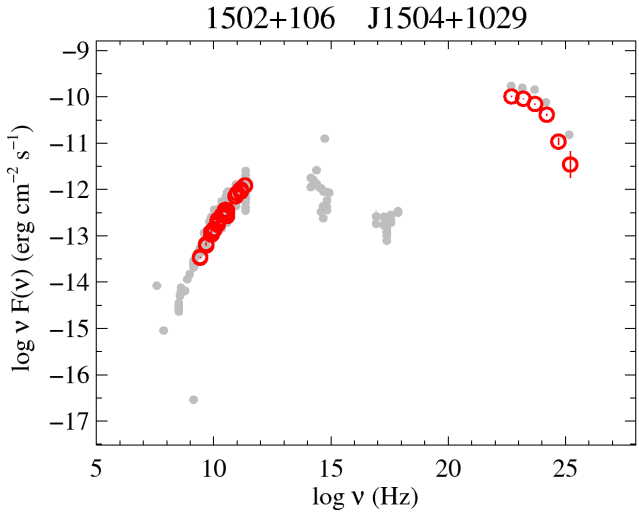}
\includegraphics[scale=0.8]{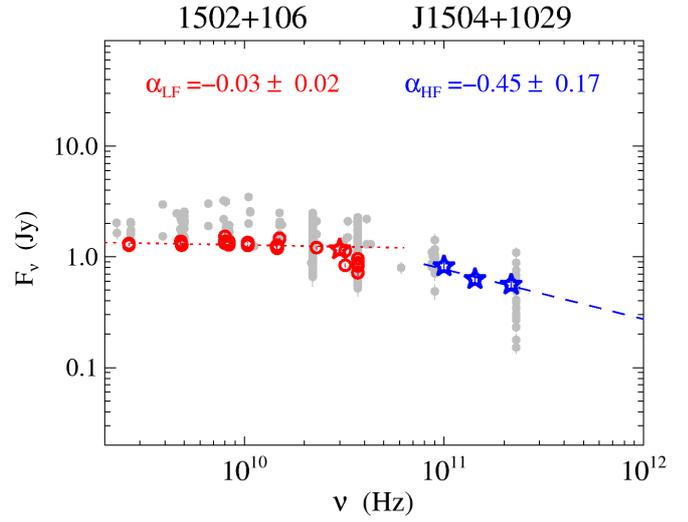}
 \caption{1502+106}
\end{figure*}

\begin{figure*}
\includegraphics[scale=0.8]{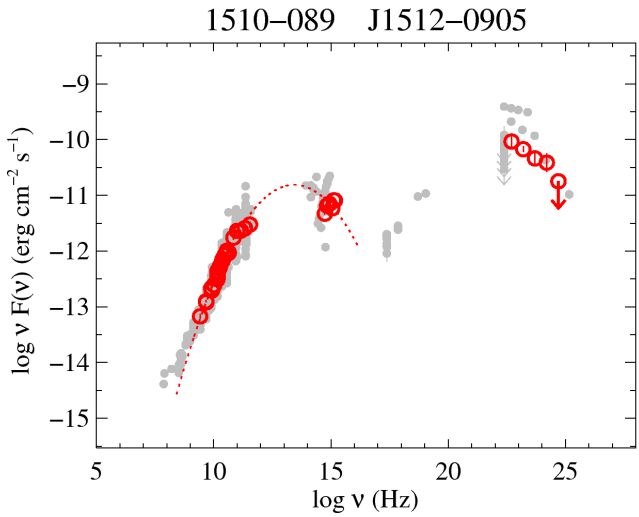}
\includegraphics[scale=0.8]{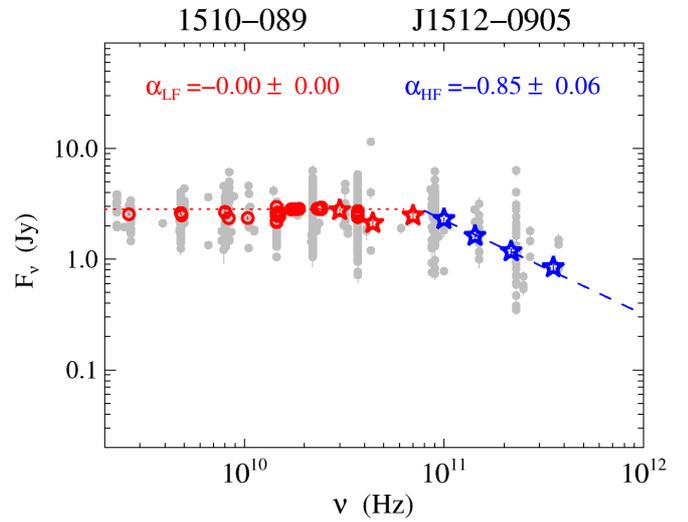}
 \caption{1510$-$089}
\end{figure*}

\begin{figure*}
\includegraphics[scale=0.8]{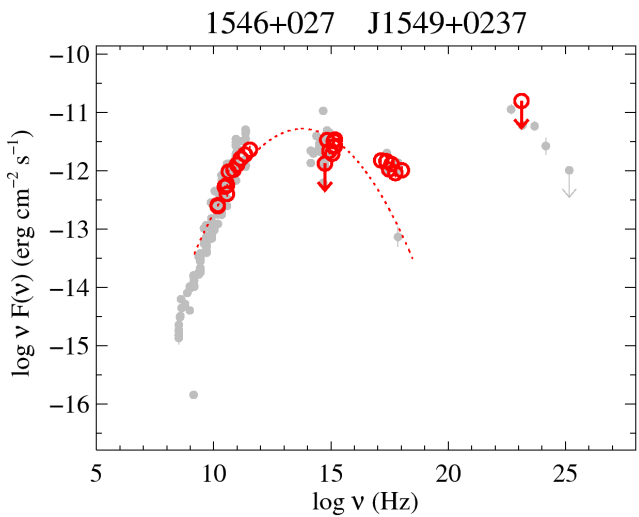}
\includegraphics[scale=0.8]{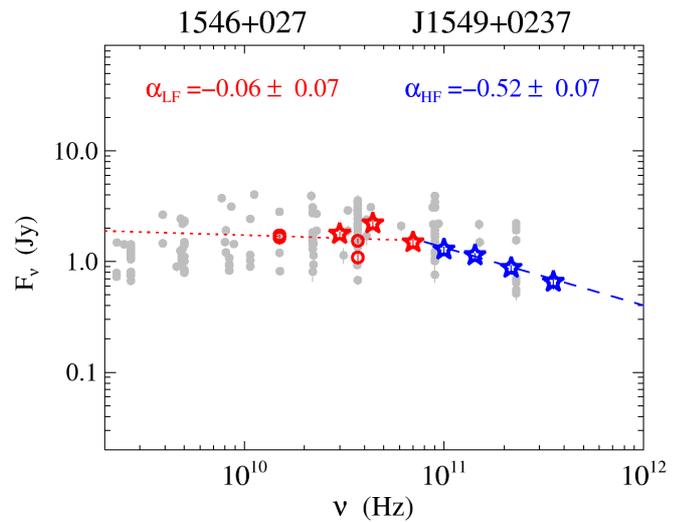}
 \caption{1546+027}
\end{figure*}
 
 \clearpage
 
\begin{figure*}
\includegraphics[scale=0.8]{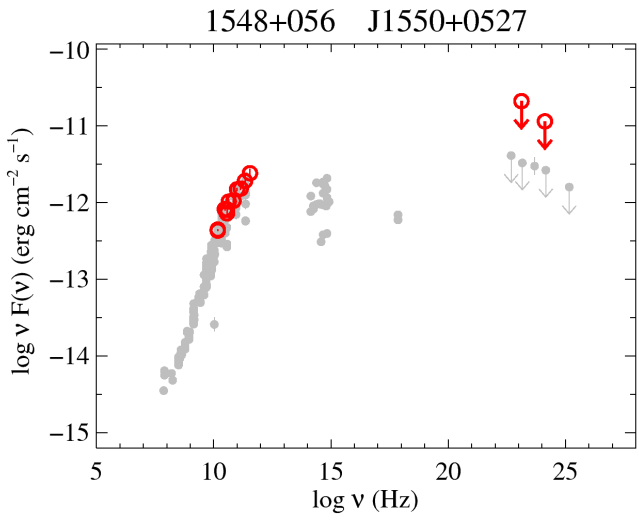}
\includegraphics[scale=0.8]{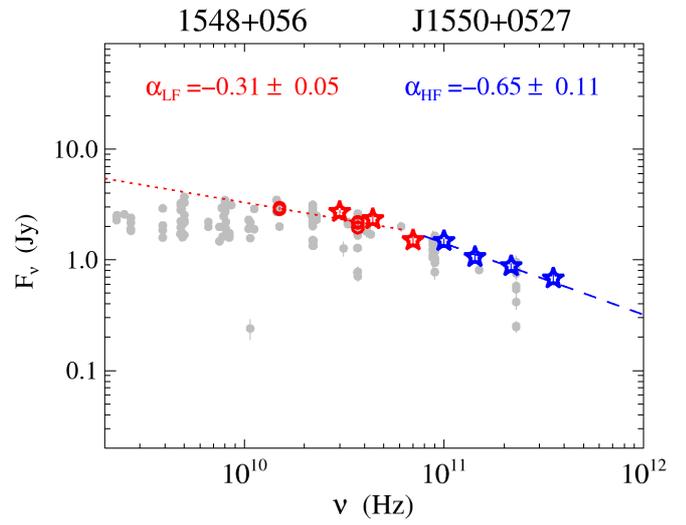}
 \caption{1548+056}
\end{figure*}

\begin{figure*}
\includegraphics[scale=0.8]{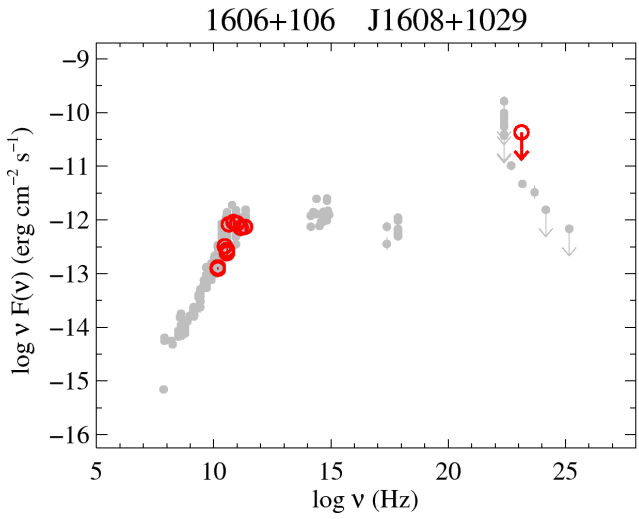}
\includegraphics[scale=0.8]{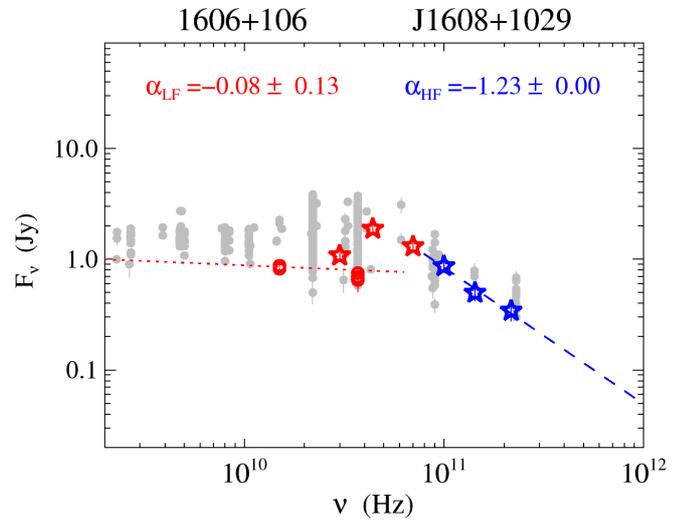}
 \caption{1606+106}
\end{figure*}

\begin{figure*}
\includegraphics[scale=0.8]{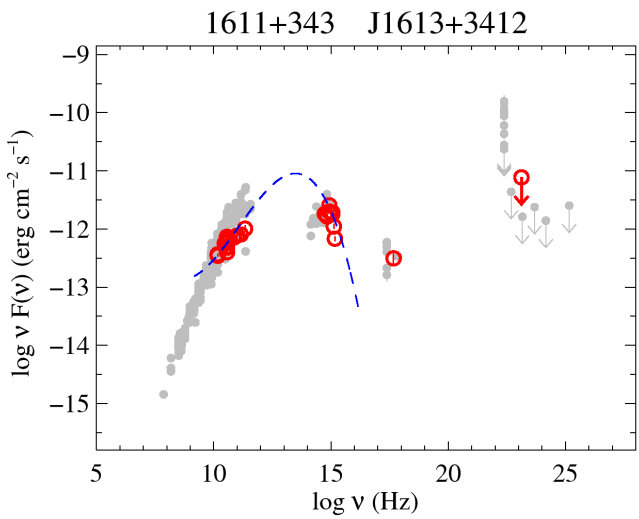}
\includegraphics[scale=0.8]{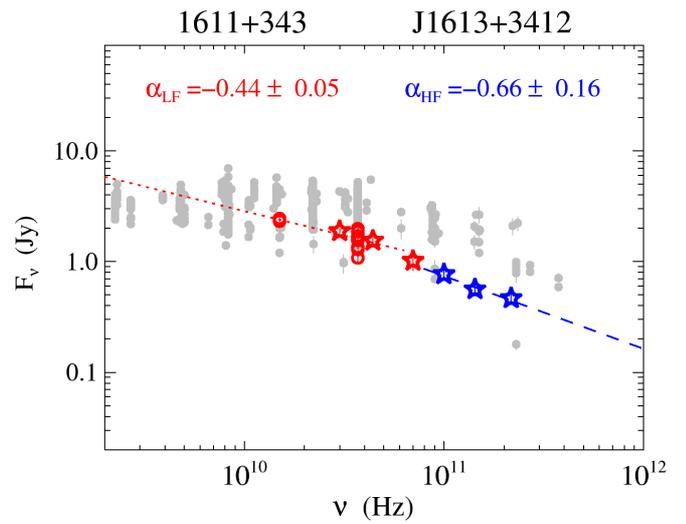}
 \caption{1611+343}
\end{figure*}
 
 \clearpage
 
\begin{figure*}
\includegraphics[scale=0.8]{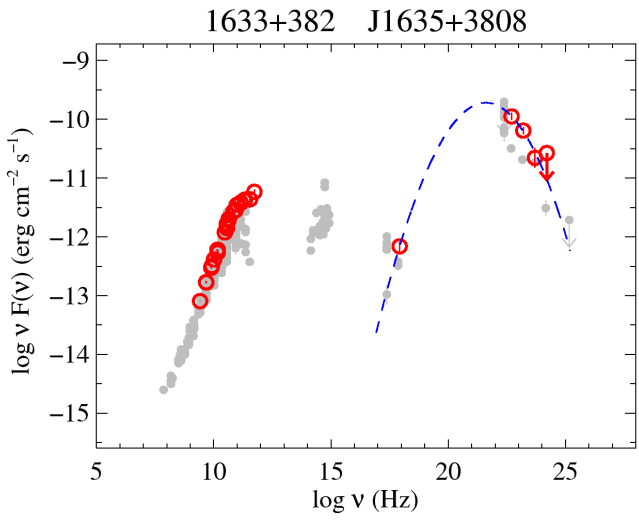}
\includegraphics[scale=0.8]{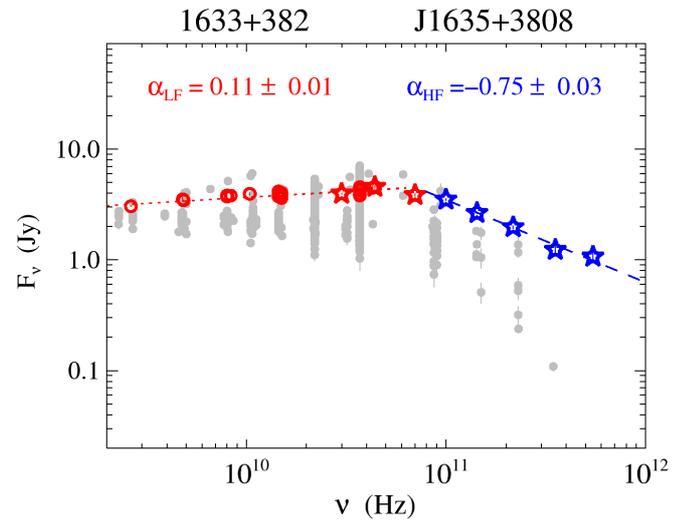}
 \caption{1633+382}
\end{figure*}

\begin{figure*}
\includegraphics[scale=0.8]{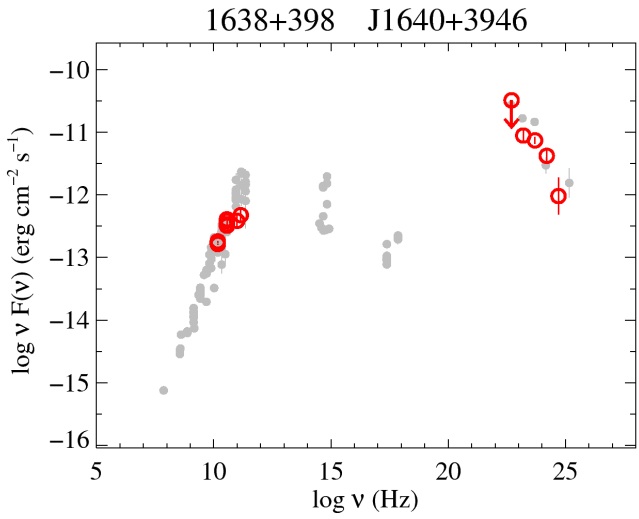}
\includegraphics[scale=0.8]{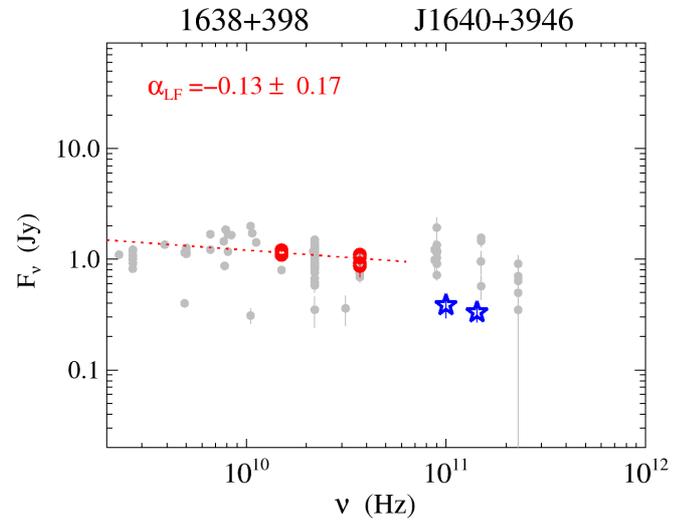}
 \caption{1638+398}
\end{figure*}

\begin{figure*}
\includegraphics[scale=0.8]{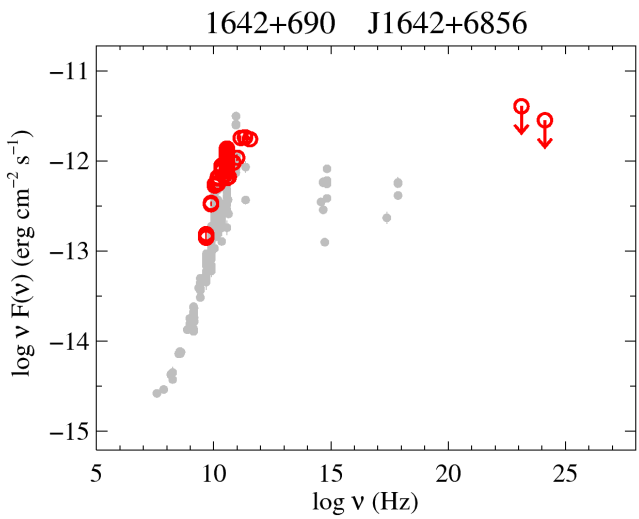}
\includegraphics[scale=0.8]{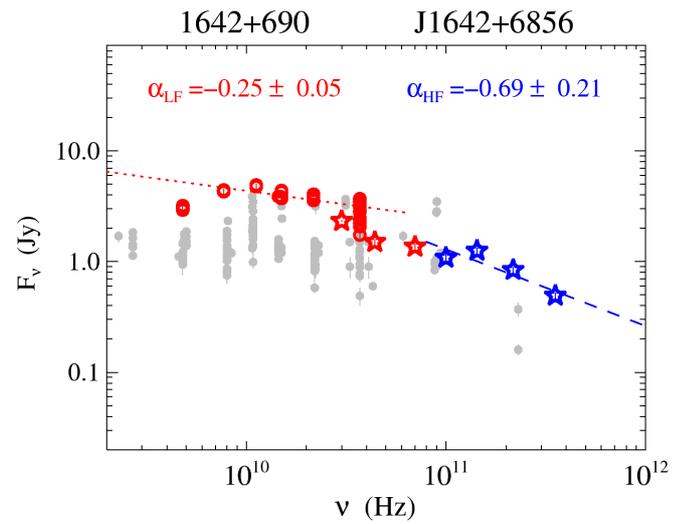}
 \caption{1642+690}
\end{figure*}
 
 \clearpage
 
\begin{figure*}
\includegraphics[scale=0.8]{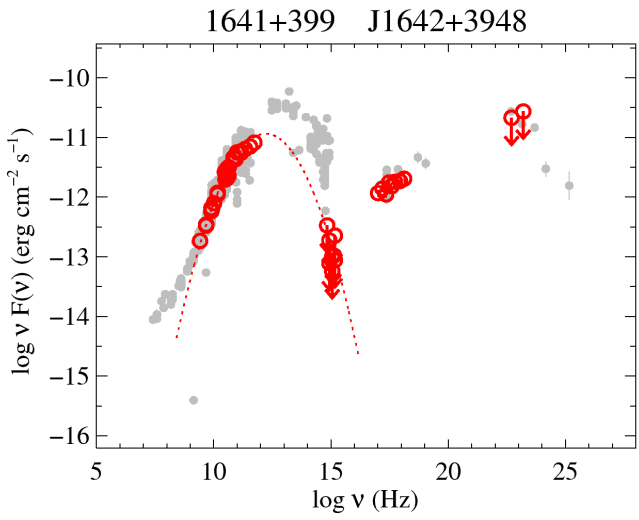}
\includegraphics[scale=0.8]{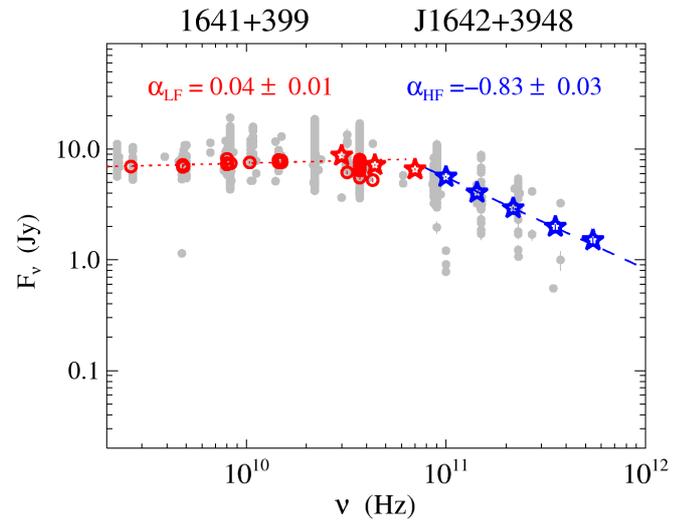}
 \caption{1641+399}
\end{figure*}

\begin{figure*}
\includegraphics[scale=0.8]{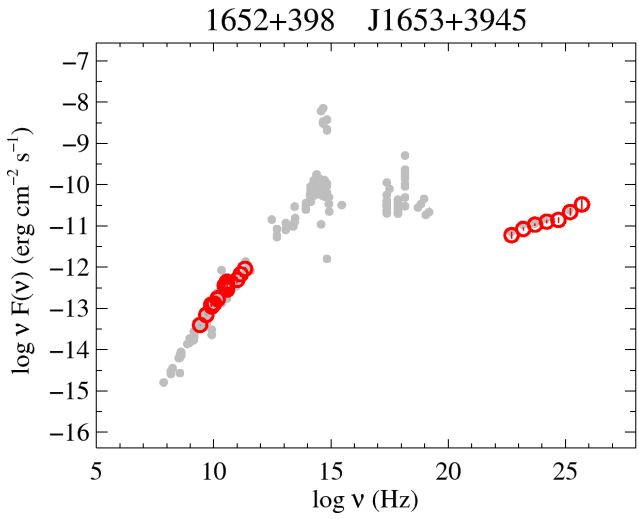}
\includegraphics[scale=0.8]{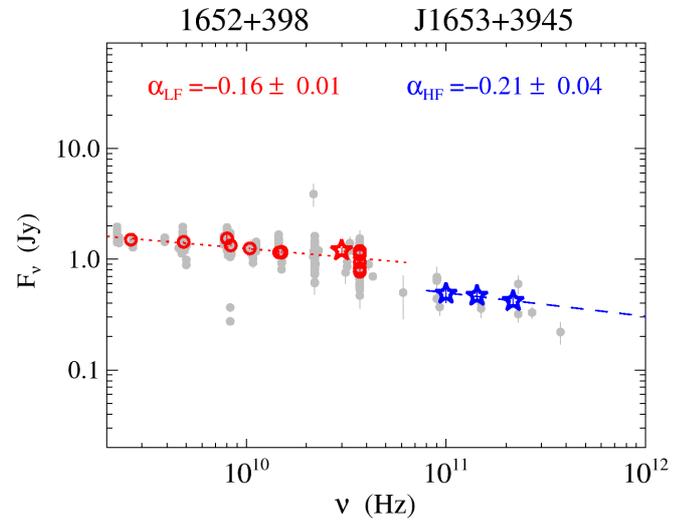}
 \caption{1652+398}
\end{figure*}

\begin{figure*}
\includegraphics[scale=0.8]{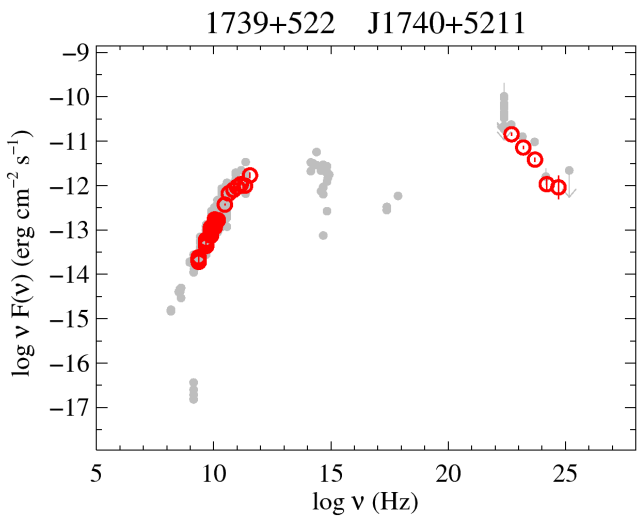}
\includegraphics[scale=0.8]{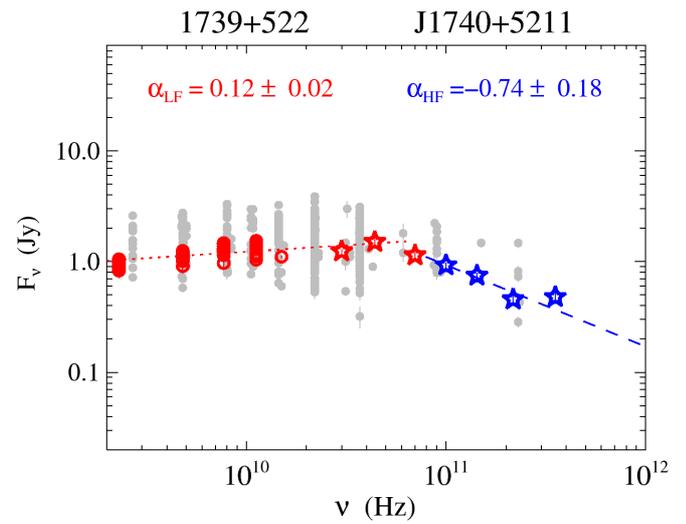}
 \caption{1739+522}
\end{figure*}
 
 \clearpage
 
\begin{figure*}
\includegraphics[scale=0.8]{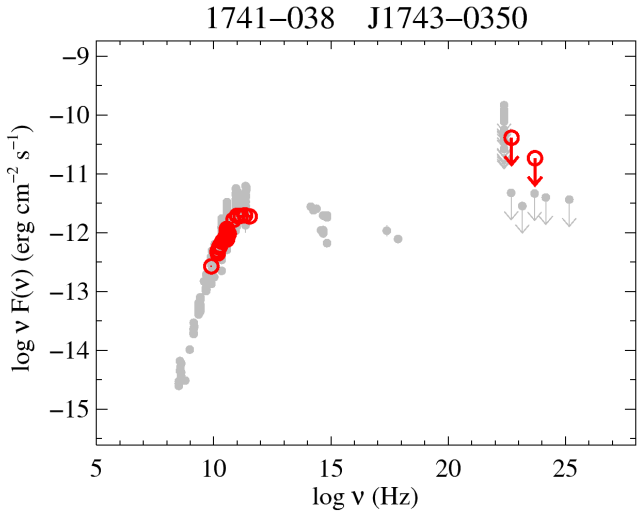}
\includegraphics[scale=0.8]{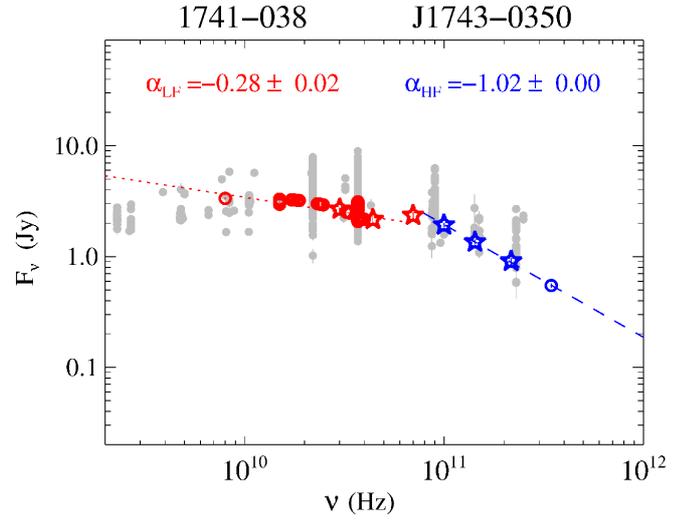}
 \caption{1741$-$038}
\end{figure*}

\begin{figure*}
\includegraphics[scale=0.8]{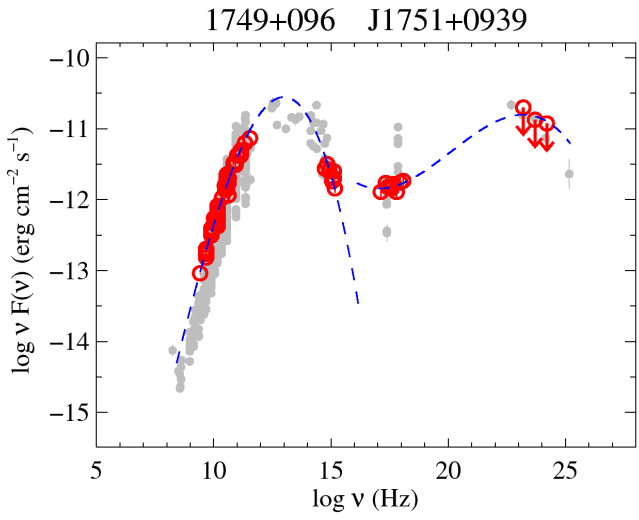}
\includegraphics[scale=0.8]{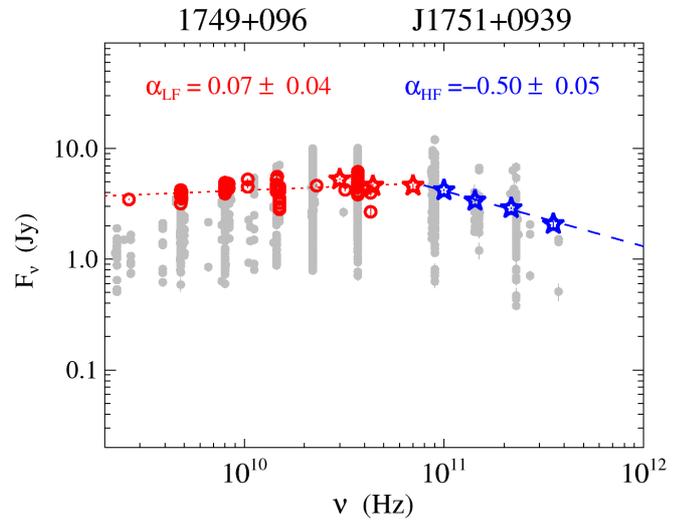}
 \caption{1749+096}
\end{figure*}

\begin{figure*}
\includegraphics[scale=0.8]{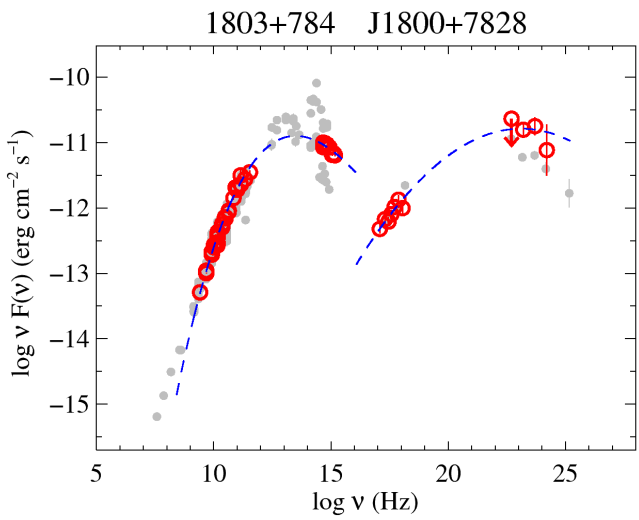}
\includegraphics[scale=0.8]{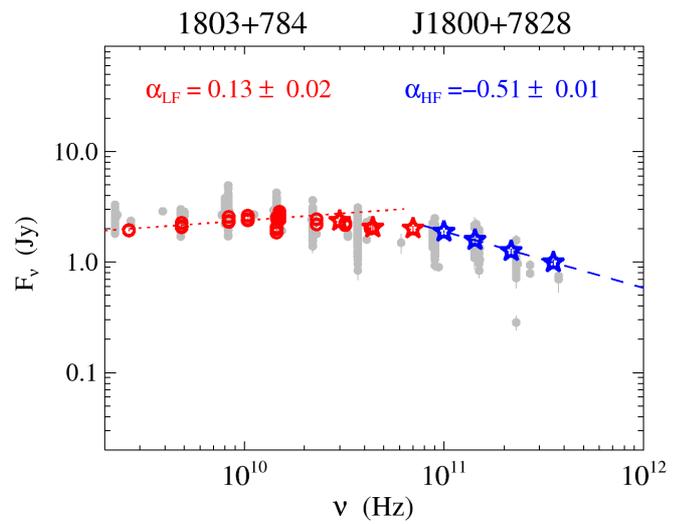}
 \caption{1803+784}
\end{figure*}
 
 \clearpage
 
\begin{figure*}
\includegraphics[scale=0.8]{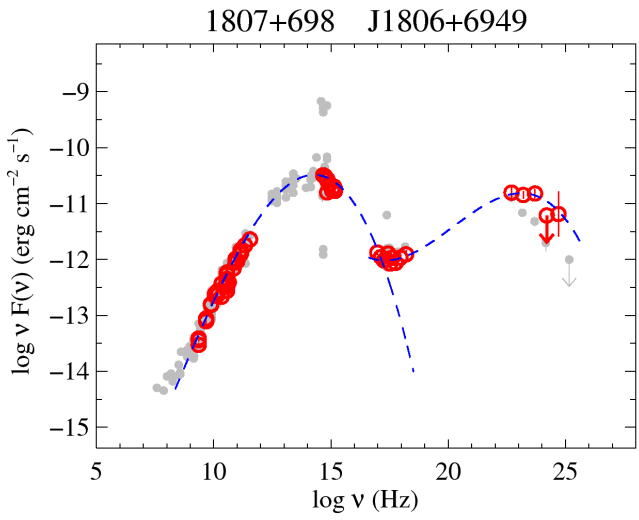}
\includegraphics[scale=0.8]{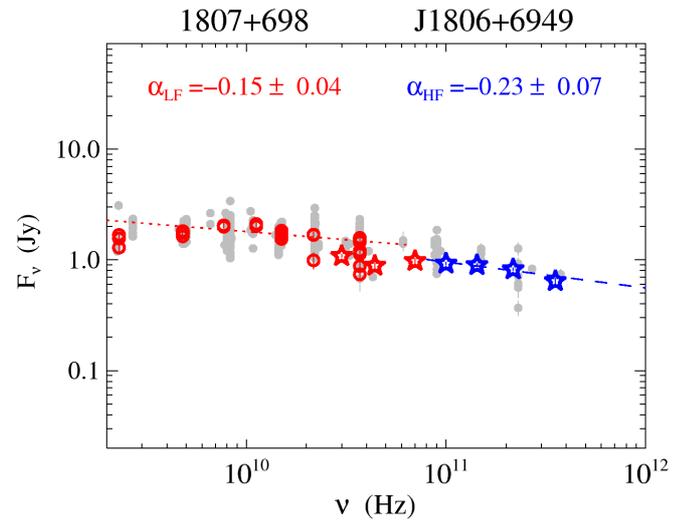}
 \caption{1807+698}
\end{figure*}

\begin{figure*}
\includegraphics[scale=0.8]{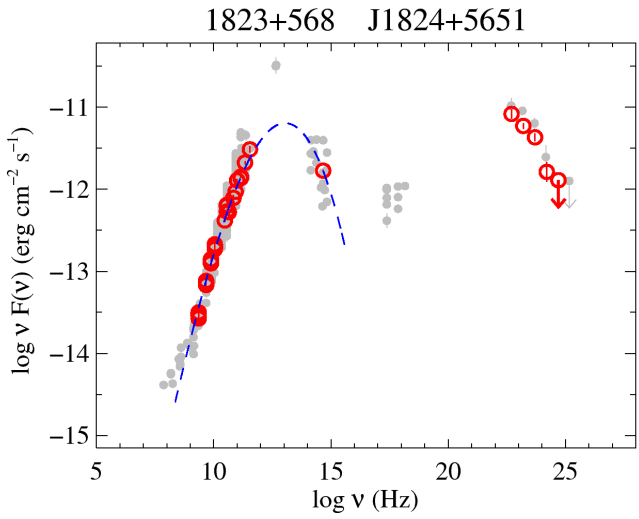}
\includegraphics[scale=0.8]{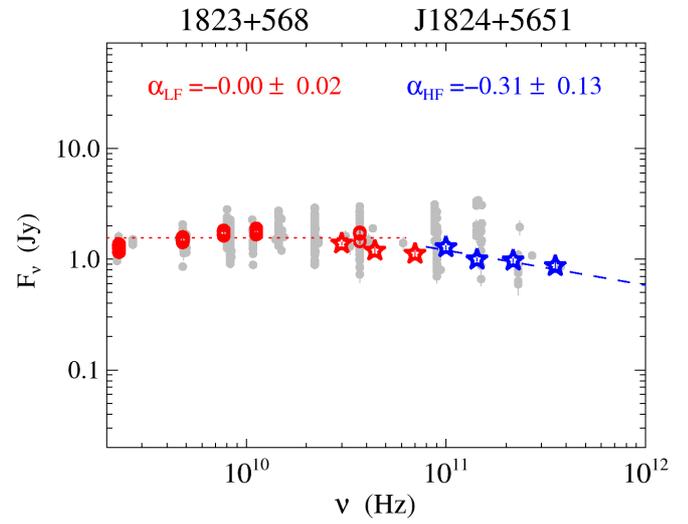}
 \caption{1823+568}
\end{figure*}

\begin{figure*}
\includegraphics[scale=0.8]{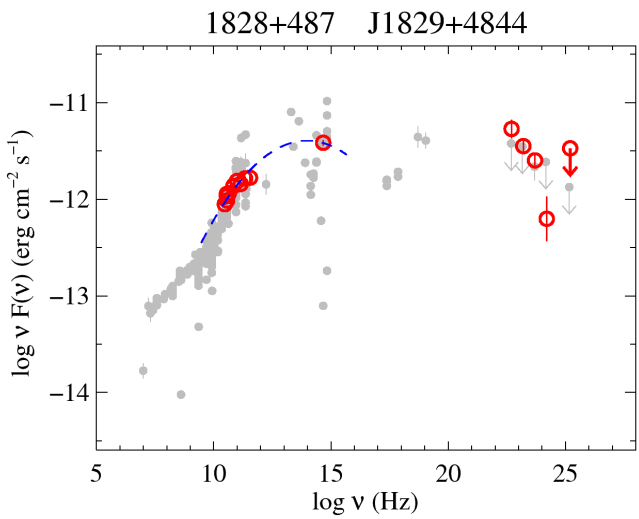}
\includegraphics[scale=0.8]{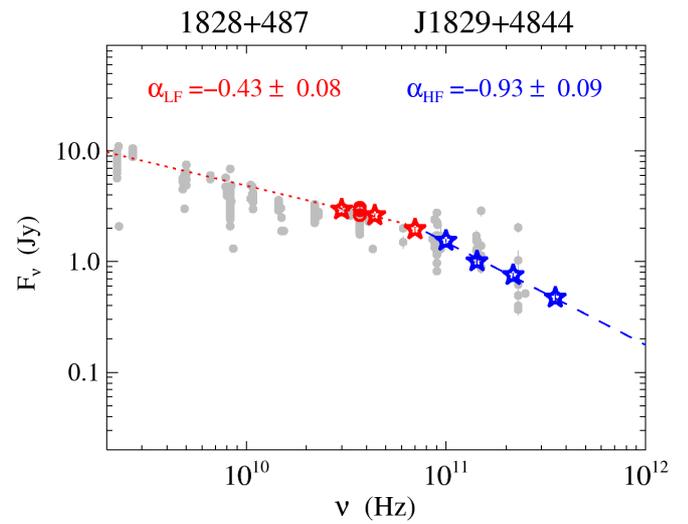}
 \caption{1828+487}
\end{figure*}
 
 \clearpage
 
\begin{figure*}
\includegraphics[scale=0.8]{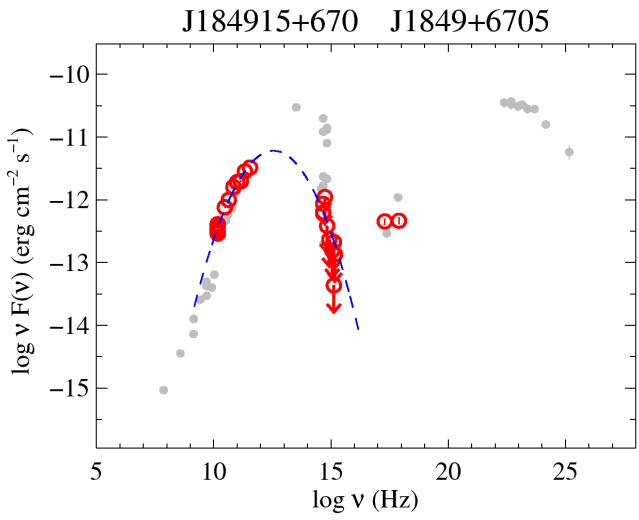}
\includegraphics[scale=0.8]{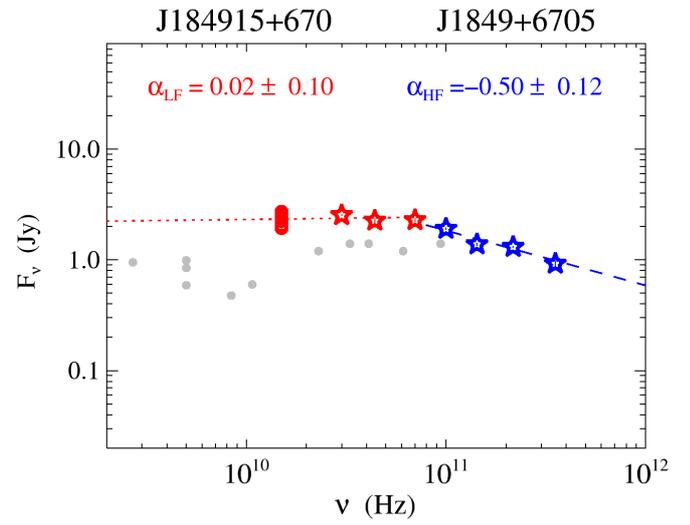}
 \caption{J184915+670}
\end{figure*}

\begin{figure*}
\includegraphics[scale=0.8]{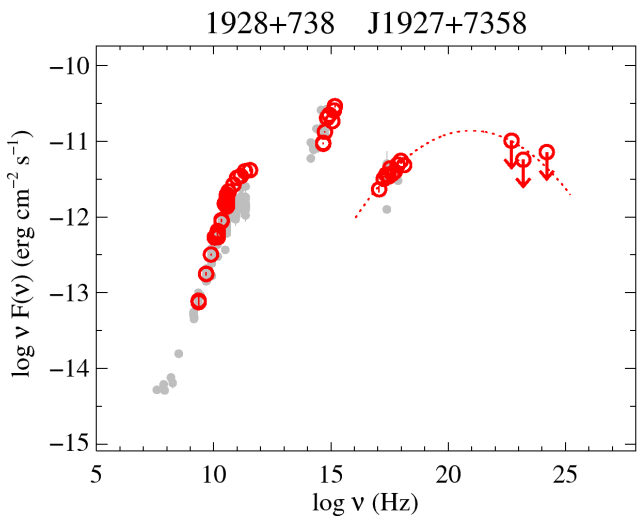}
\includegraphics[scale=0.8]{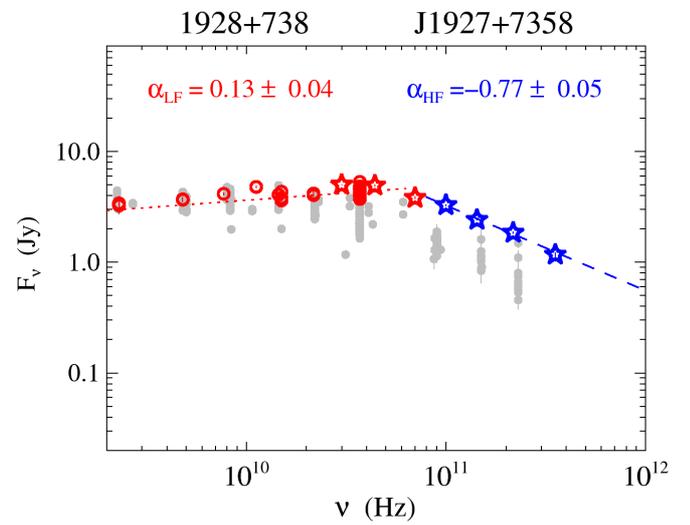}
 \caption{1928+738}
\end{figure*}

\begin{figure*}
\includegraphics[scale=0.8]{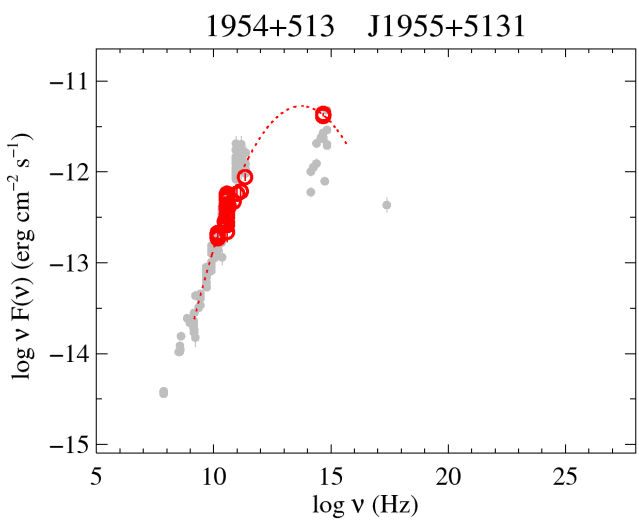}
\includegraphics[scale=0.8]{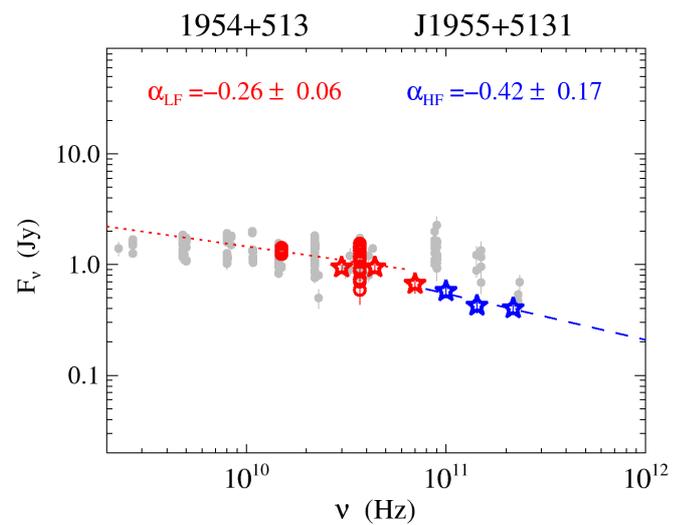}
 \caption{1954+513}
\end{figure*}
 
 \clearpage
 
\begin{figure*}
\includegraphics[scale=0.8]{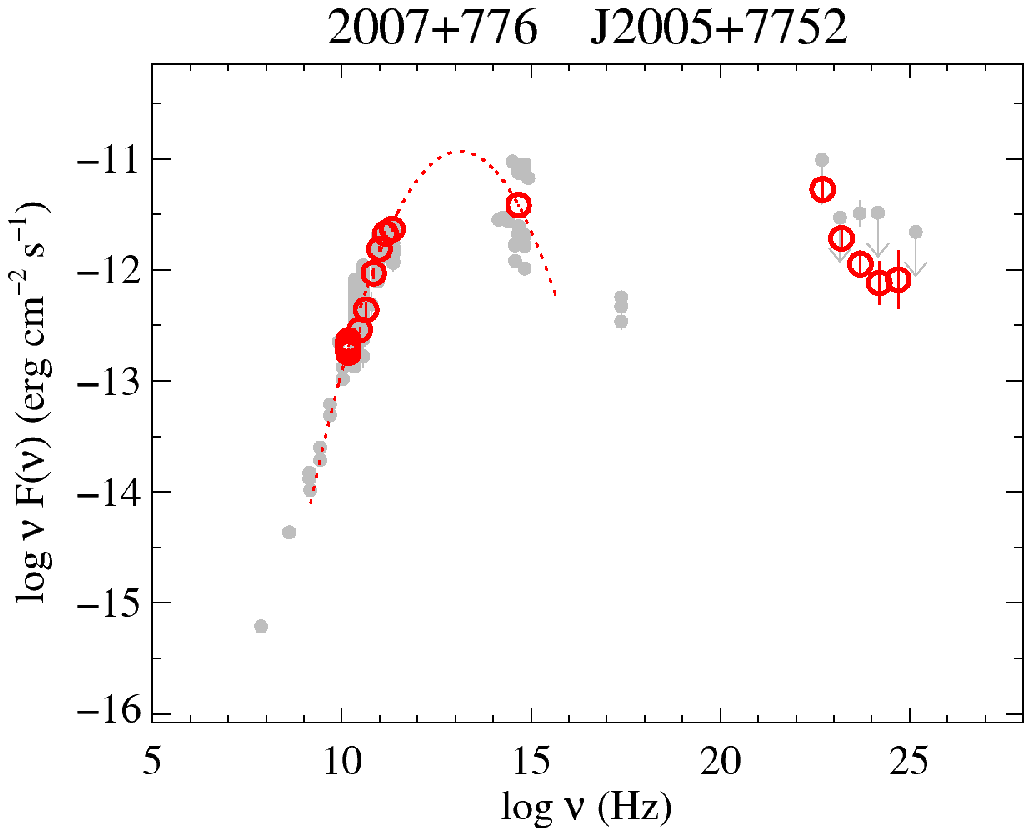}
\includegraphics[scale=0.8]{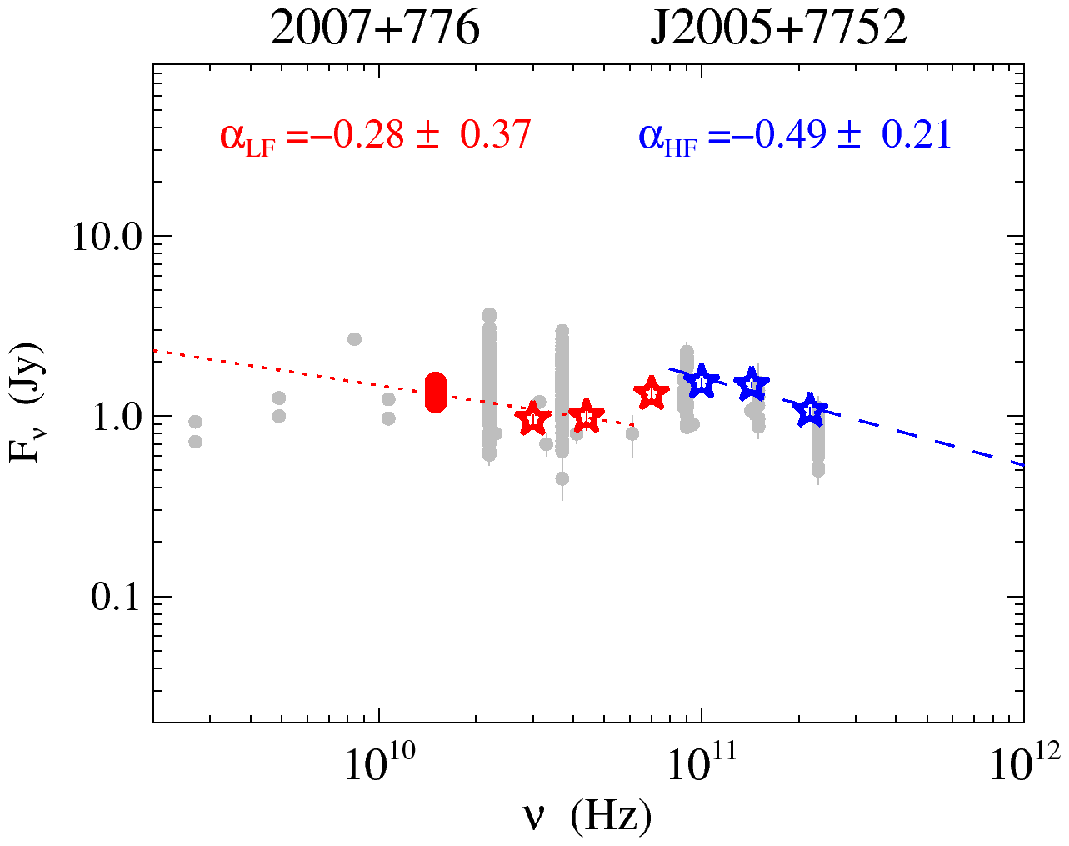}
 \caption{2007+776}
\end{figure*}

\begin{figure*}
\includegraphics[scale=0.8]{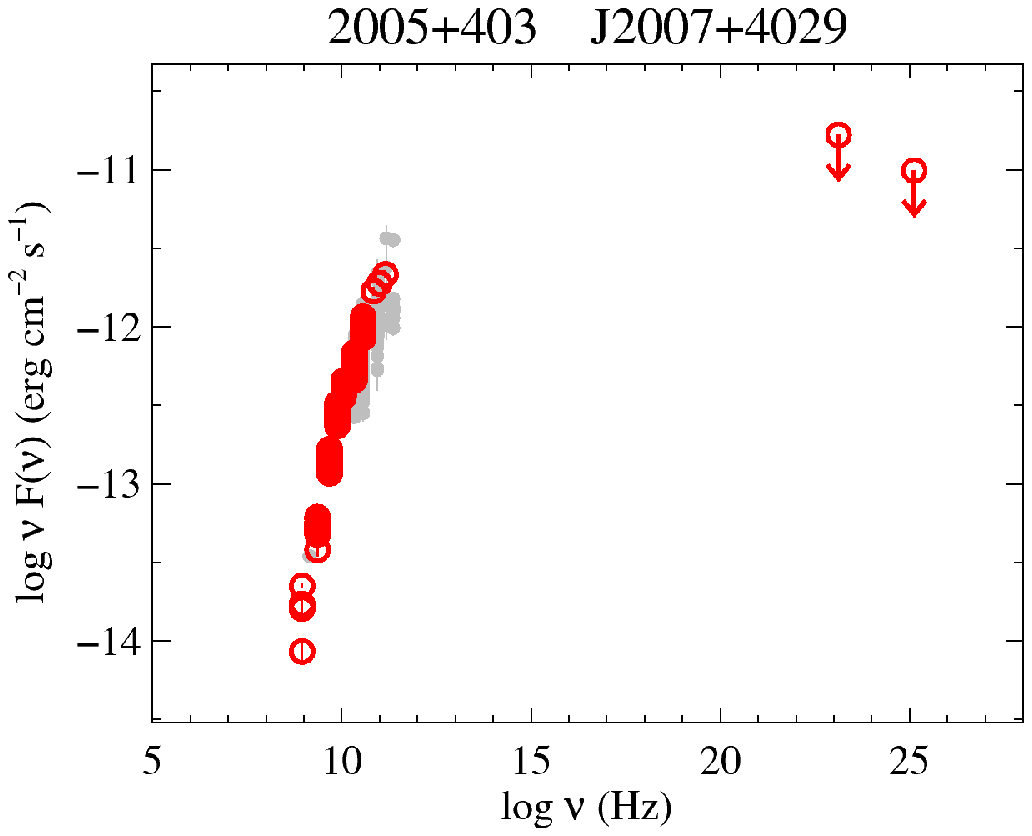}
\includegraphics[scale=0.8]{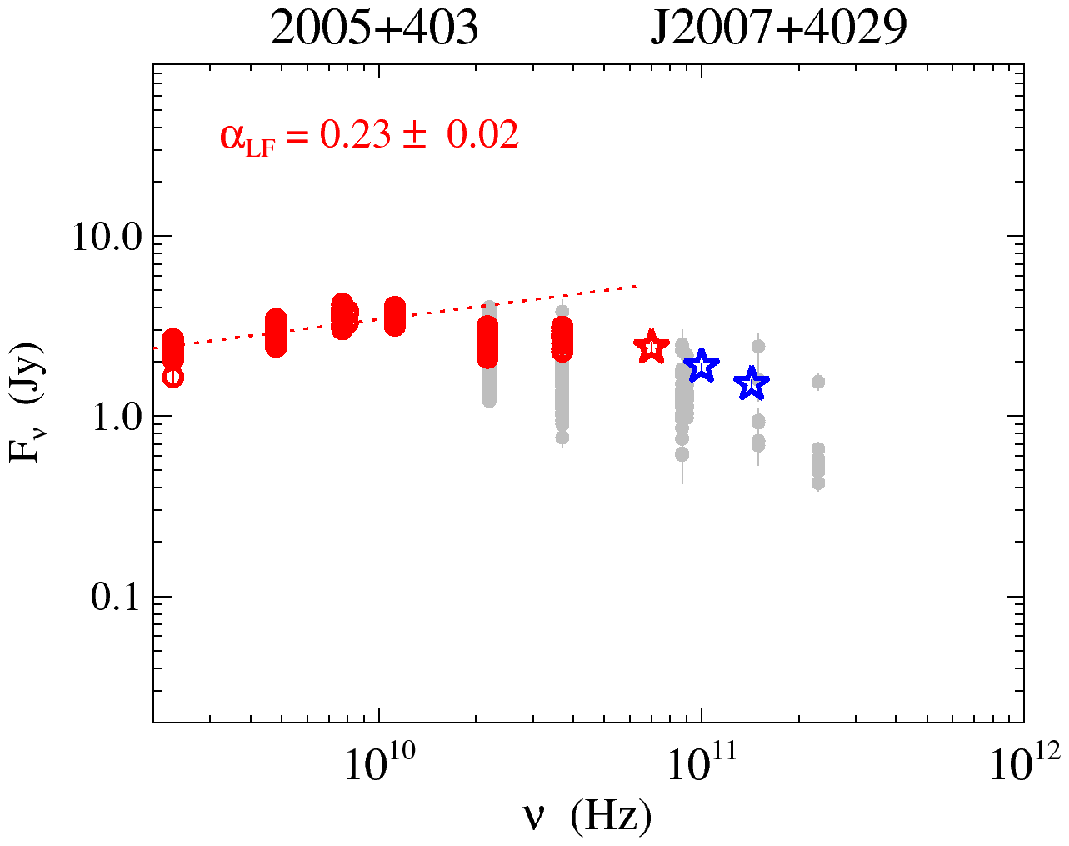}
 \caption{2005+403}
\end{figure*}

\begin{figure*}
\includegraphics[scale=0.8]{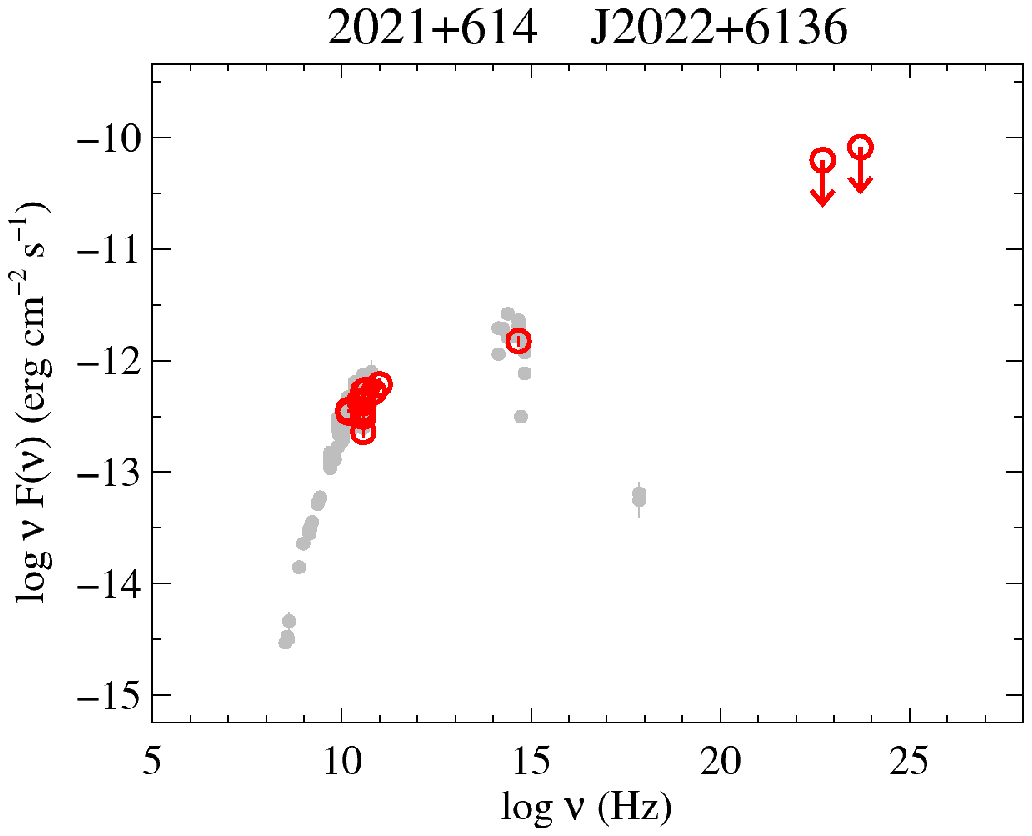}
\includegraphics[scale=0.8]{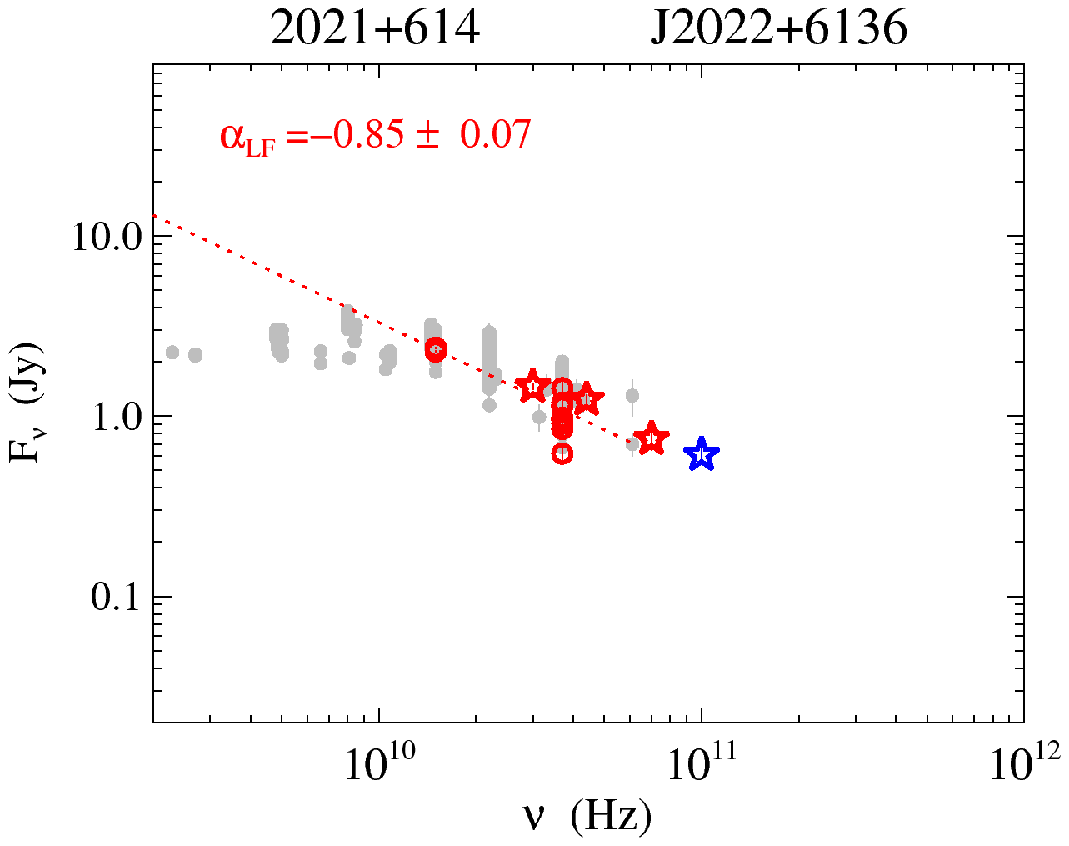}
 \caption{2021+614}
\end{figure*}
 
 \clearpage
 
\begin{figure*}
\includegraphics[scale=0.8]{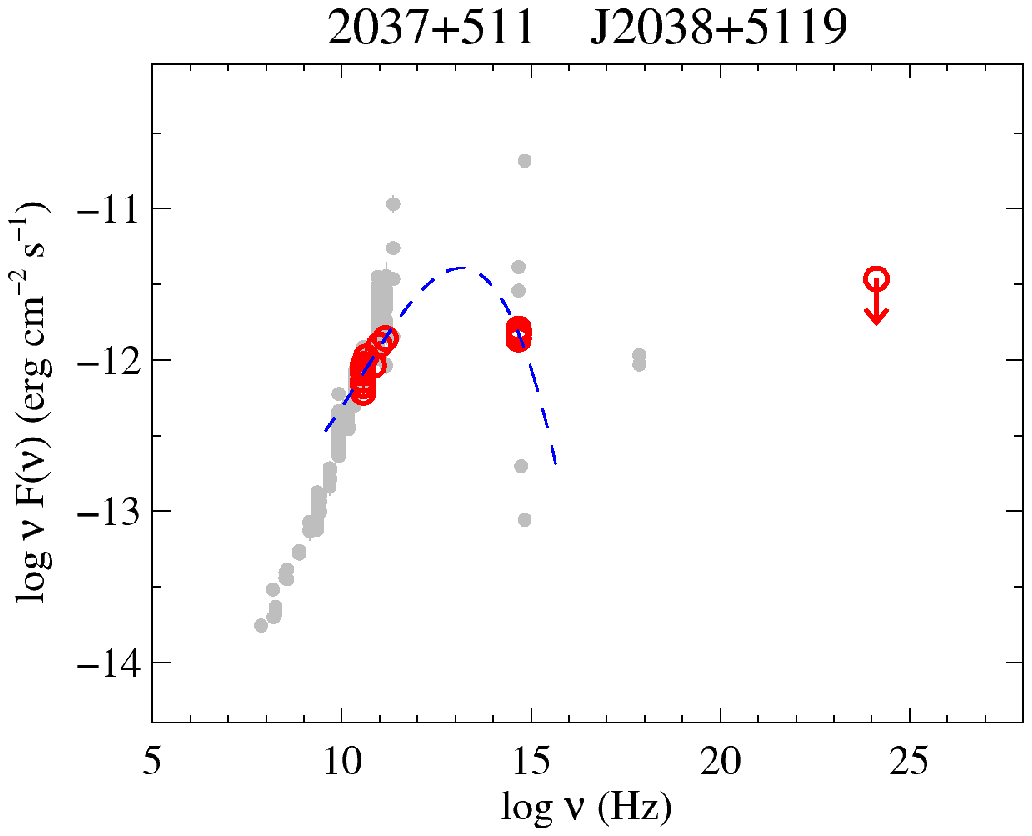}
\includegraphics[scale=0.8]{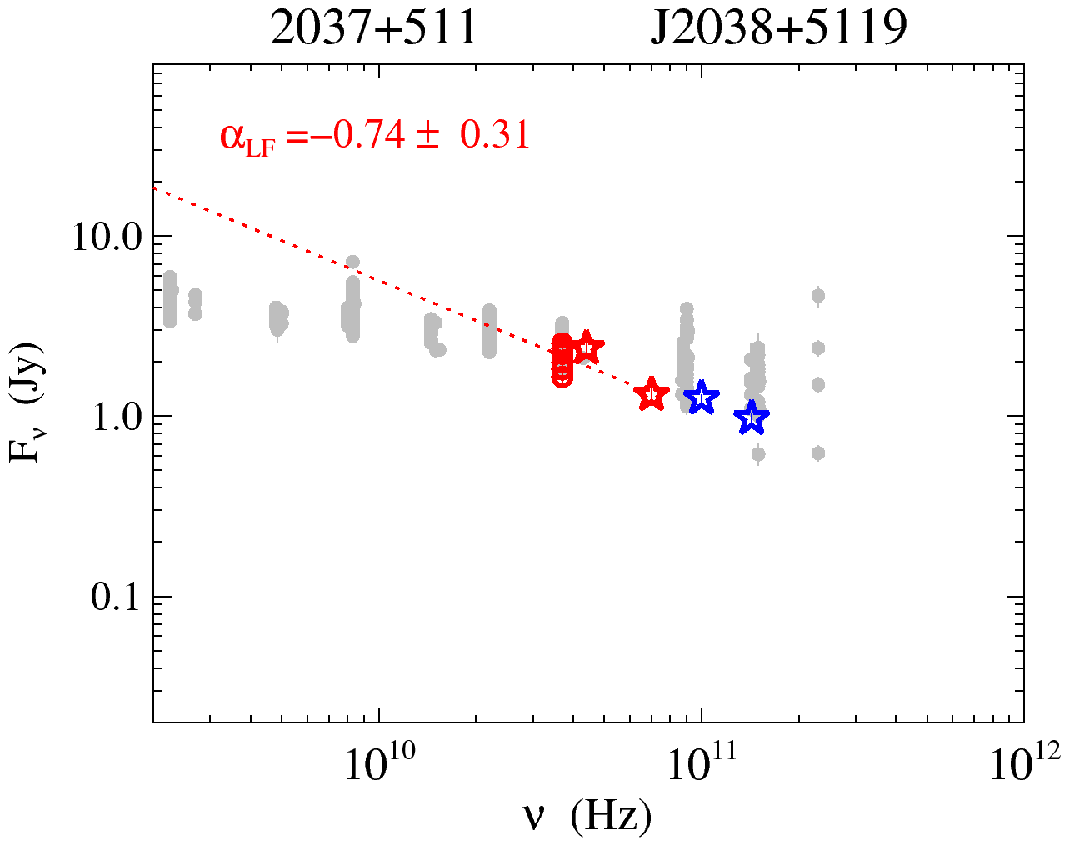}
 \caption{2037+511}
\end{figure*}

\begin{figure*}
\includegraphics[scale=0.8]{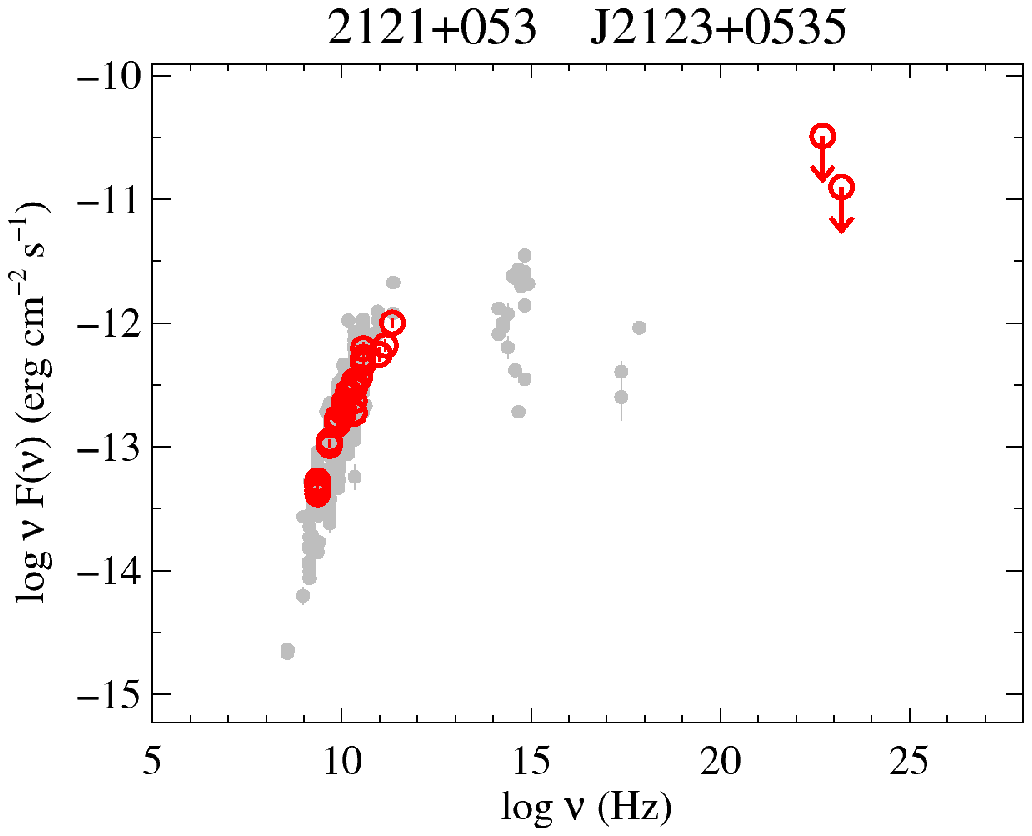}
\includegraphics[scale=0.8]{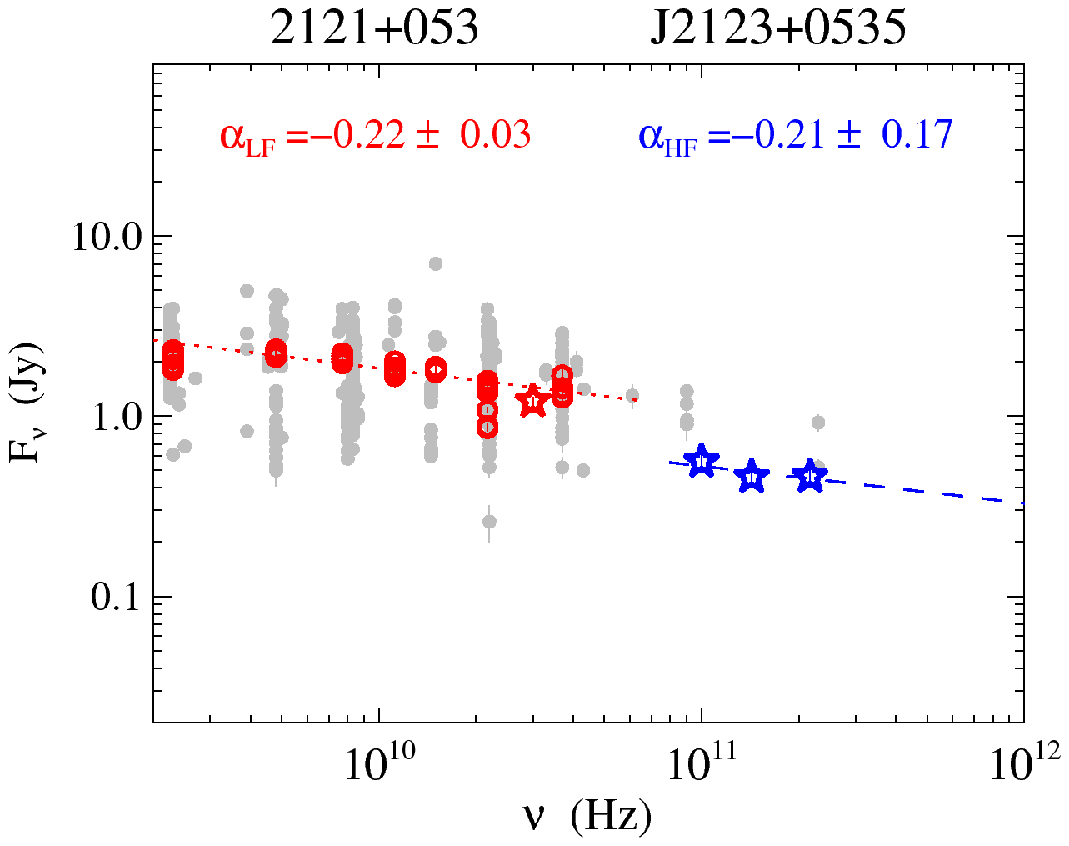}
 \caption{2121+053}
\end{figure*}

\begin{figure*}
\includegraphics[scale=0.8]{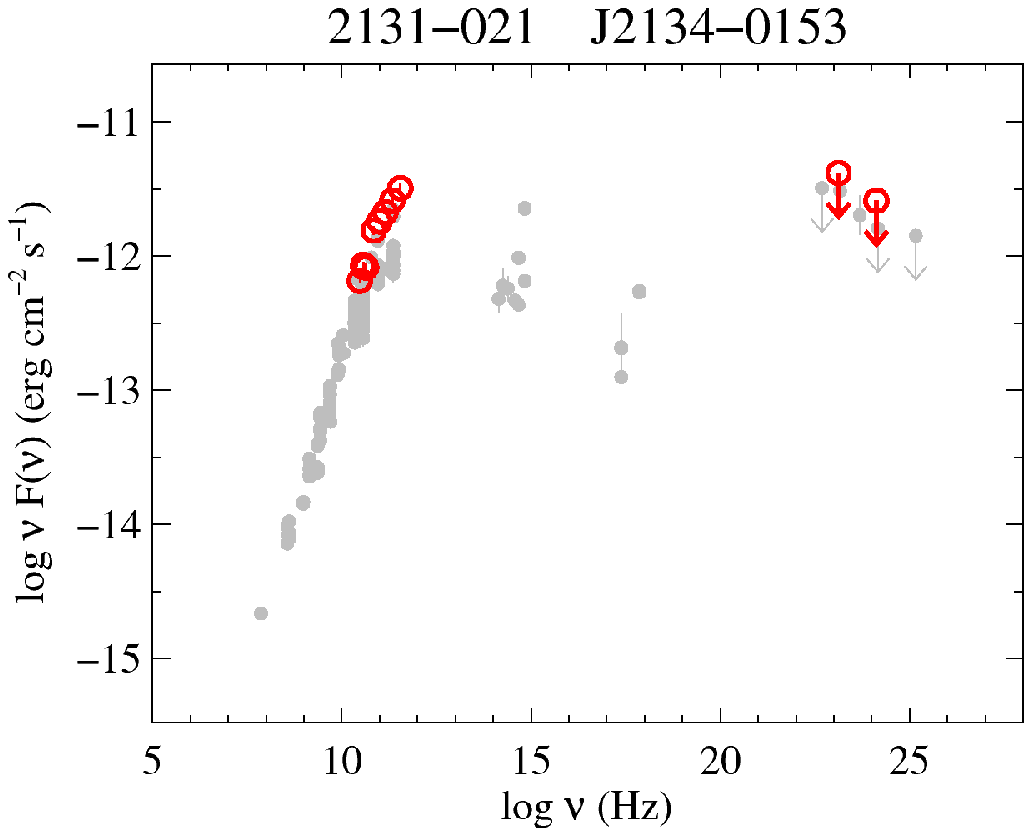}
\includegraphics[scale=0.8]{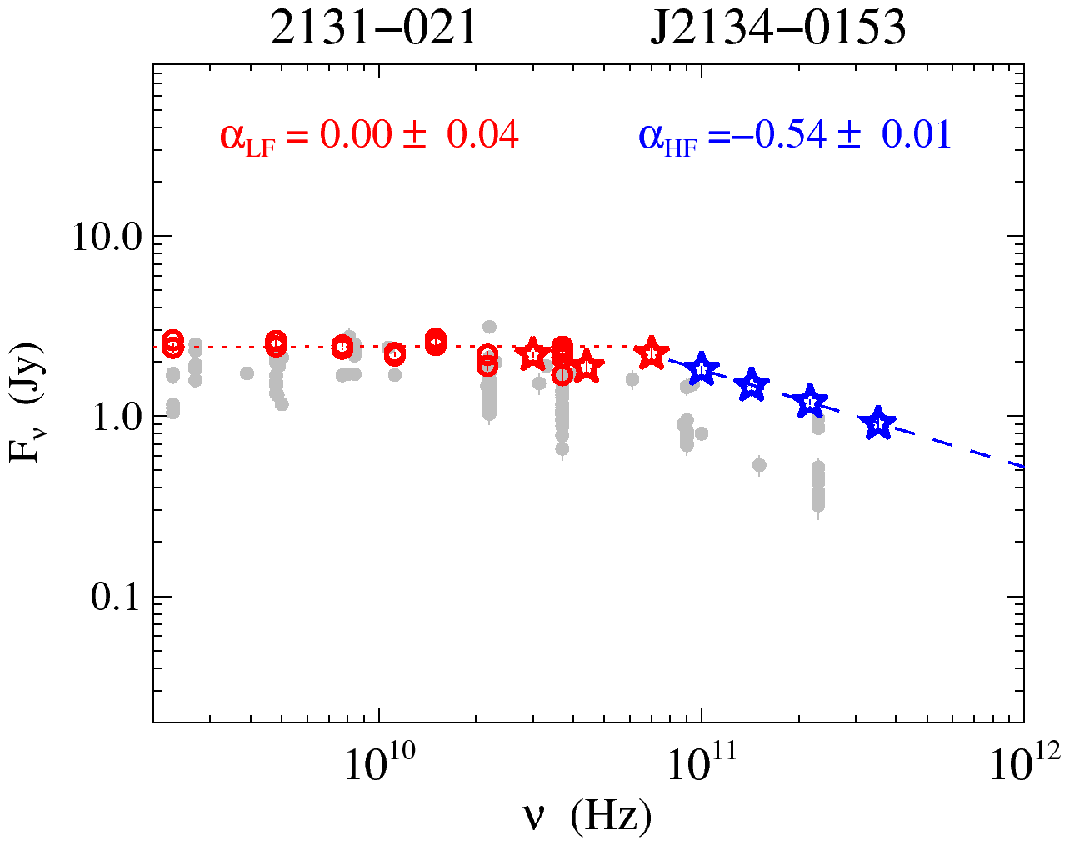}
 \caption{2131$-$021}
\end{figure*}
 
 \clearpage
 
\begin{figure*}
\includegraphics[scale=0.8]{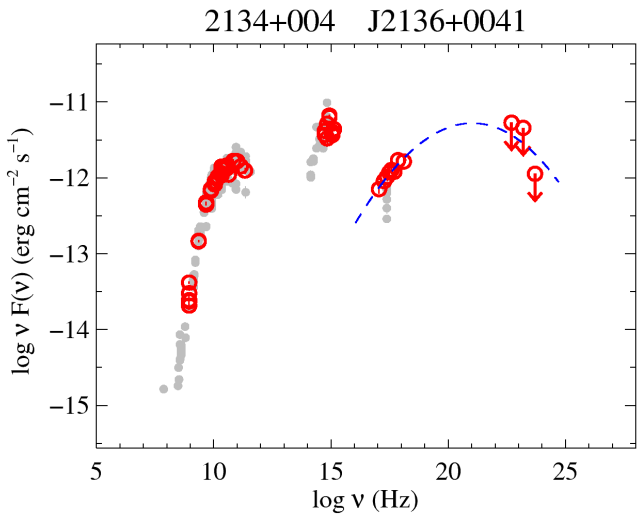}
\includegraphics[scale=0.8]{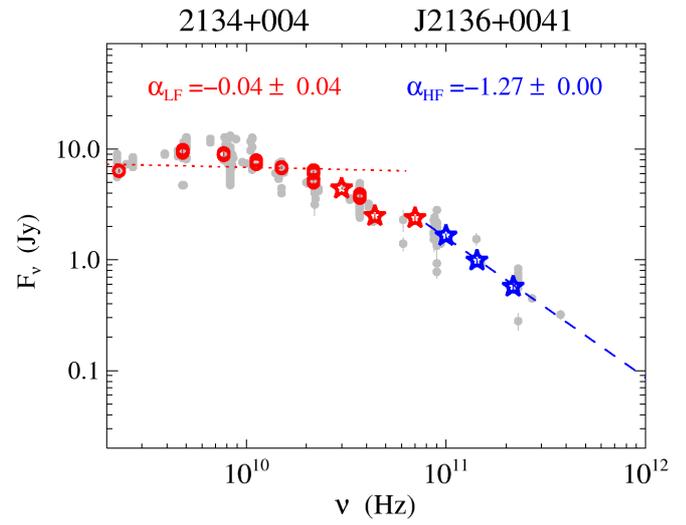}
 \caption{2134+004}
\end{figure*}

\begin{figure*}
\includegraphics[scale=0.8]{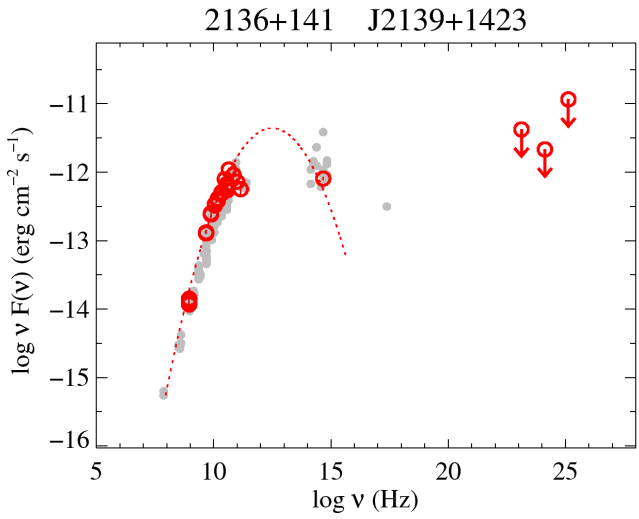}
\includegraphics[scale=0.8]{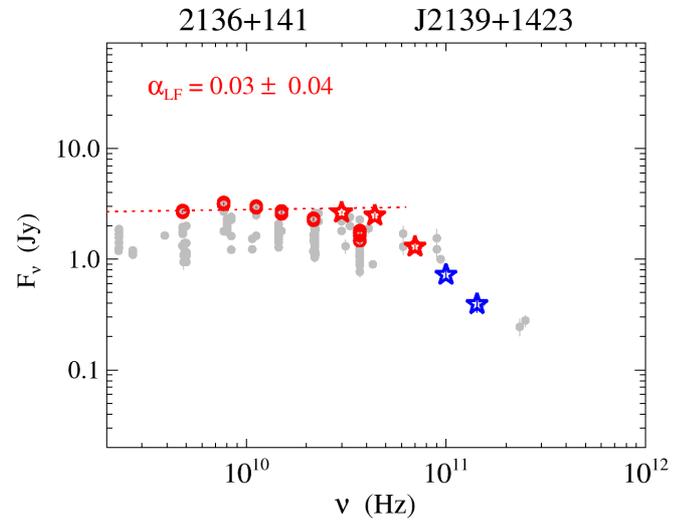}
 \caption{2136+141}
\end{figure*}

\begin{figure*}
\includegraphics[scale=0.8]{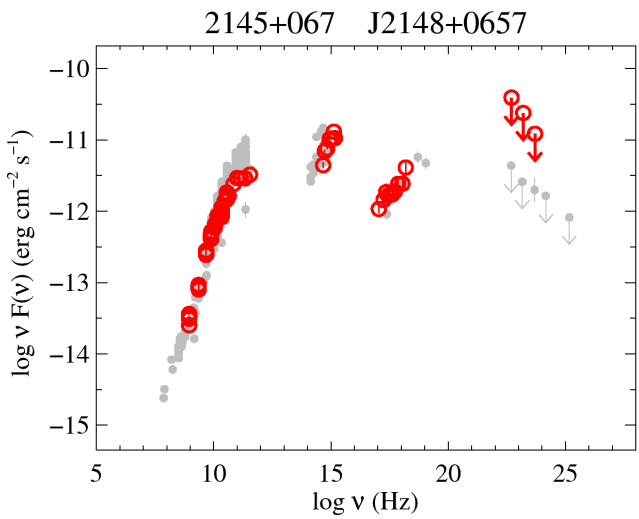}
\includegraphics[scale=0.8]{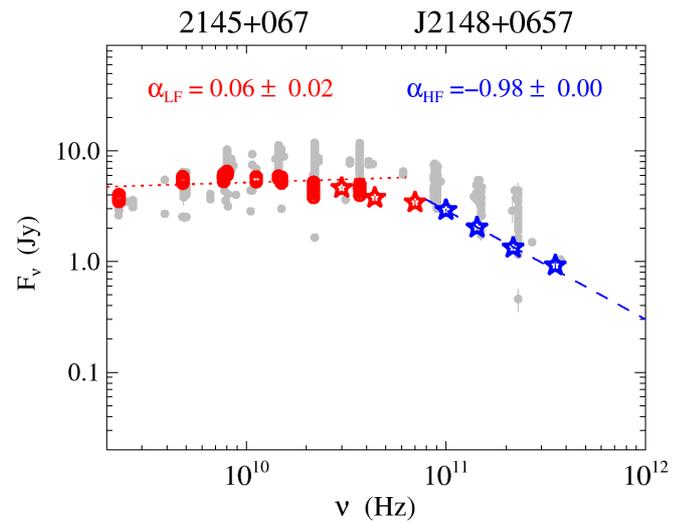}
 \caption{2145+067}
\end{figure*}
 
 \clearpage
 
\begin{figure*}
\includegraphics[scale=0.8]{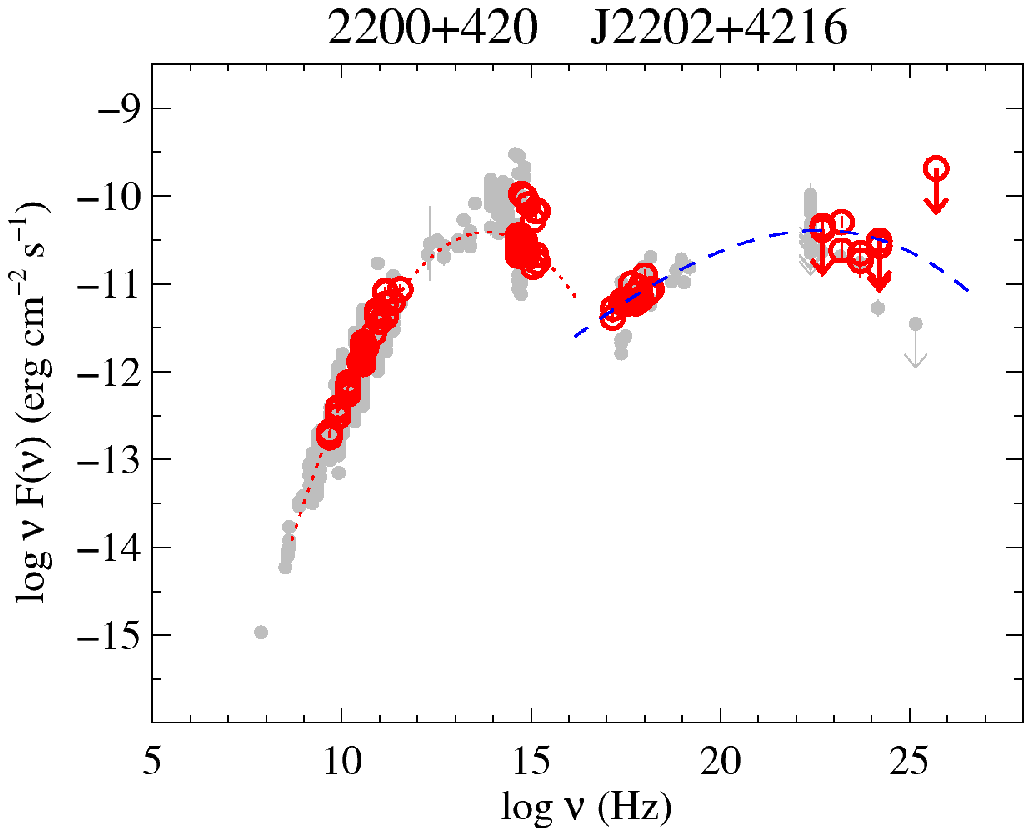}
\includegraphics[scale=0.8]{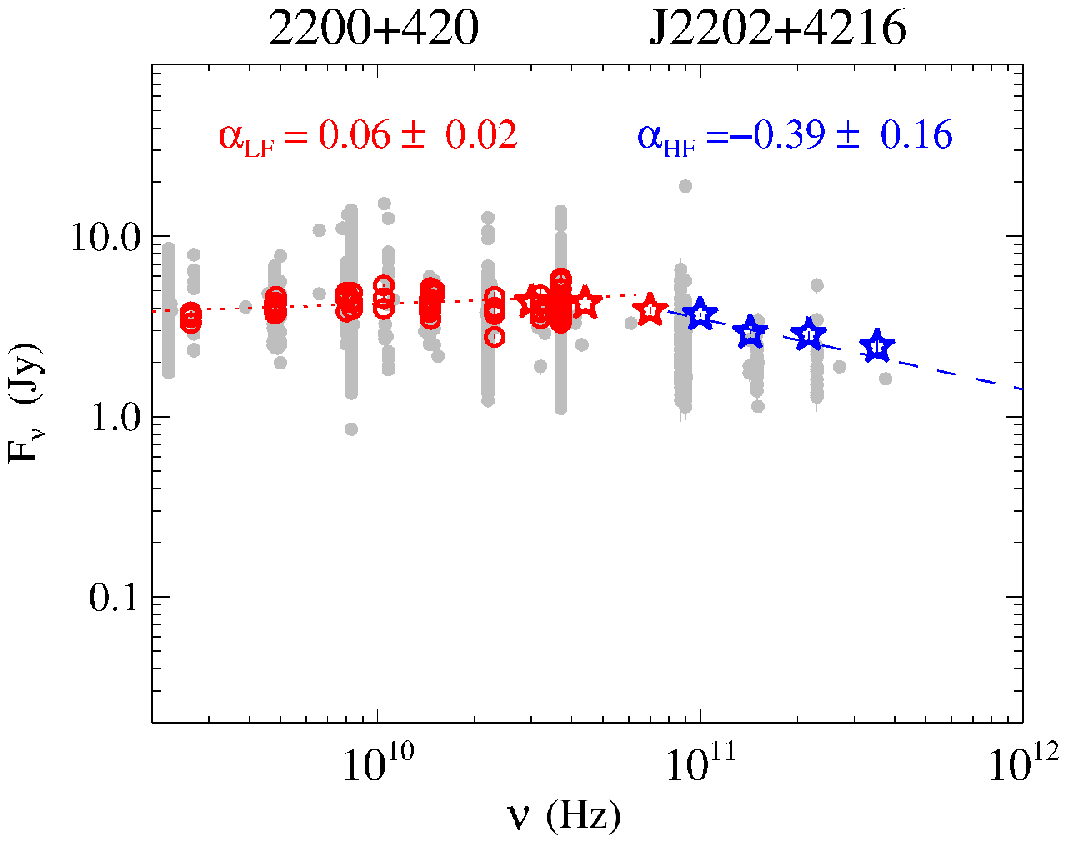}
 \caption{2200+420}
\end{figure*}

\begin{figure*}
\includegraphics[scale=0.8]{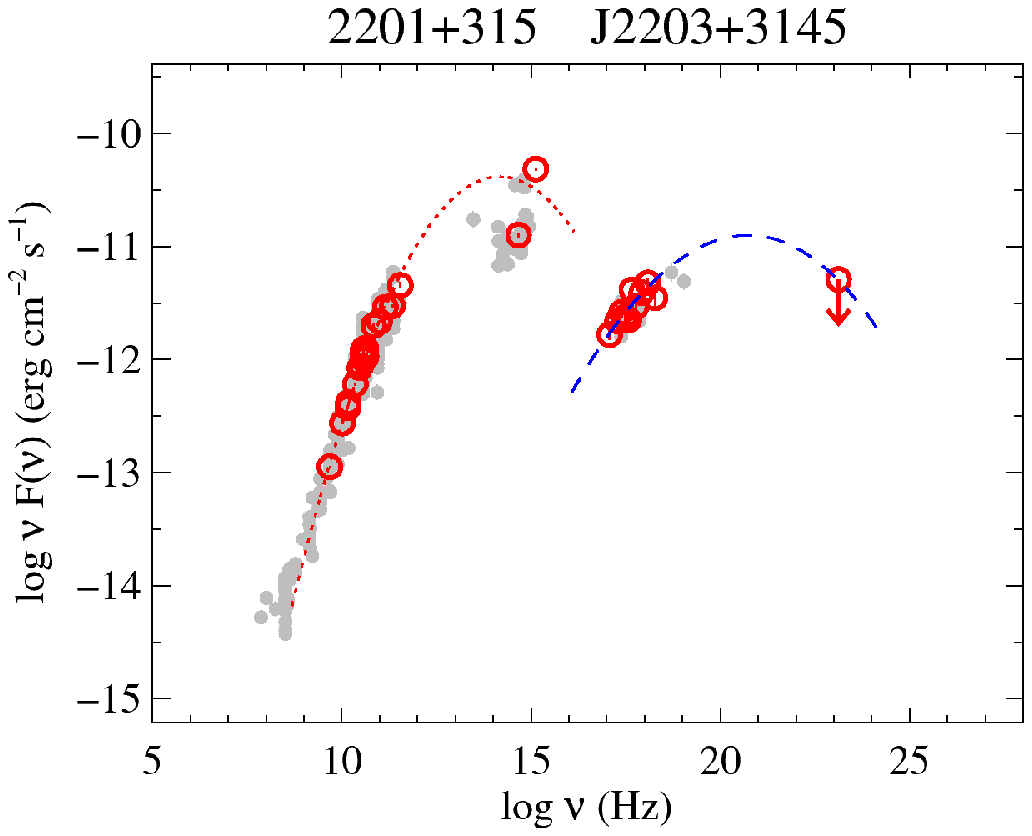}
\includegraphics[scale=0.8]{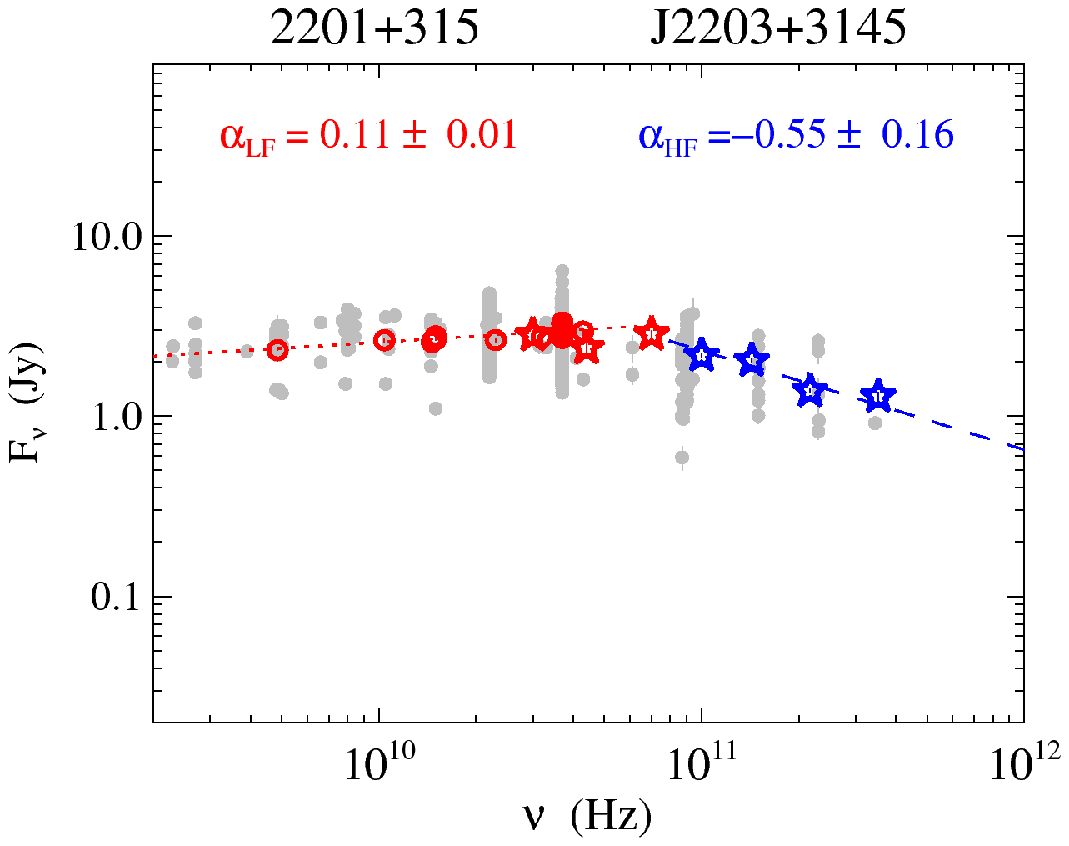}
 \caption{2201+315}
\end{figure*}

\begin{figure*}
\includegraphics[scale=0.8]{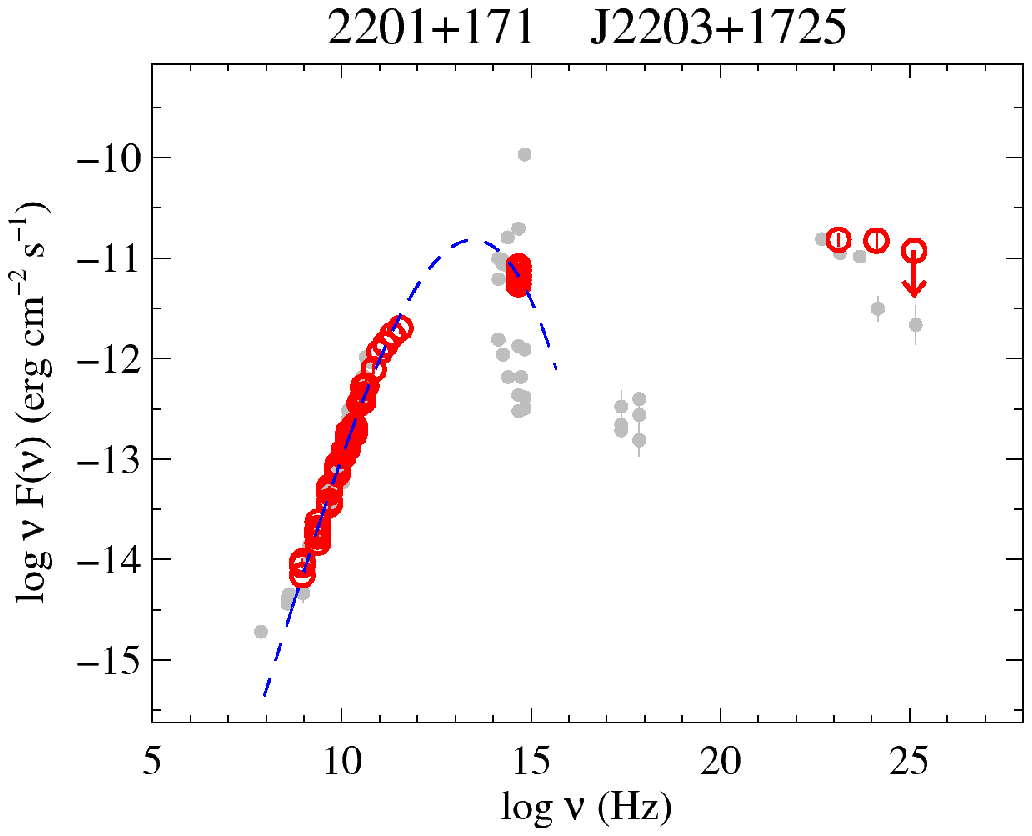}
\includegraphics[scale=0.8]{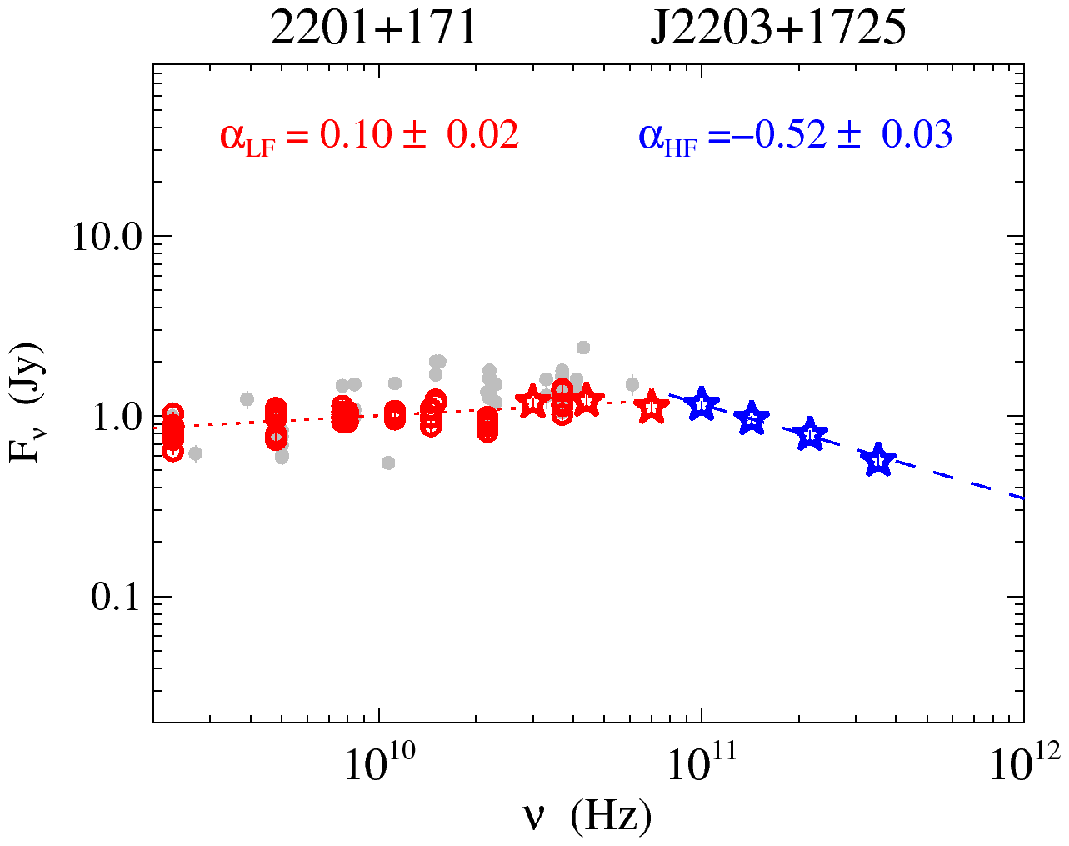}
 \caption{2201+171}
\end{figure*}
 
 \clearpage
 
\begin{figure*}
\includegraphics[scale=0.8]{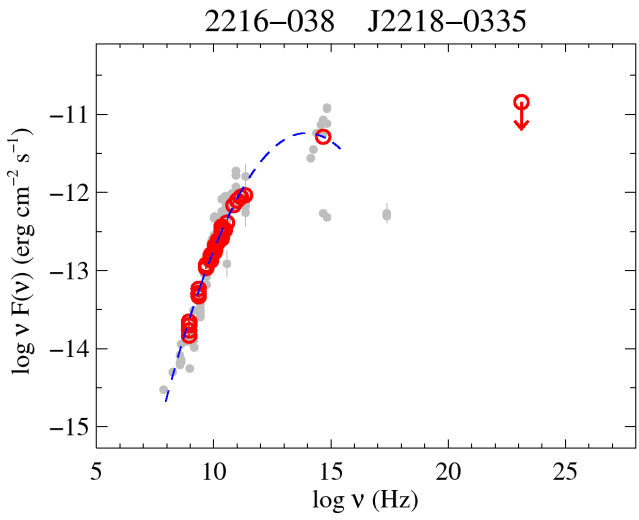}
\includegraphics[scale=0.8]{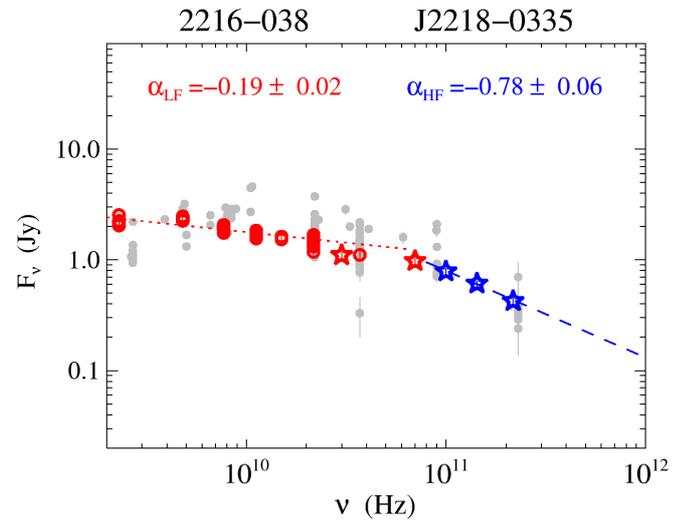}
 \caption{2216$-$038}
\end{figure*}

\begin{figure*}
\includegraphics[scale=0.8]{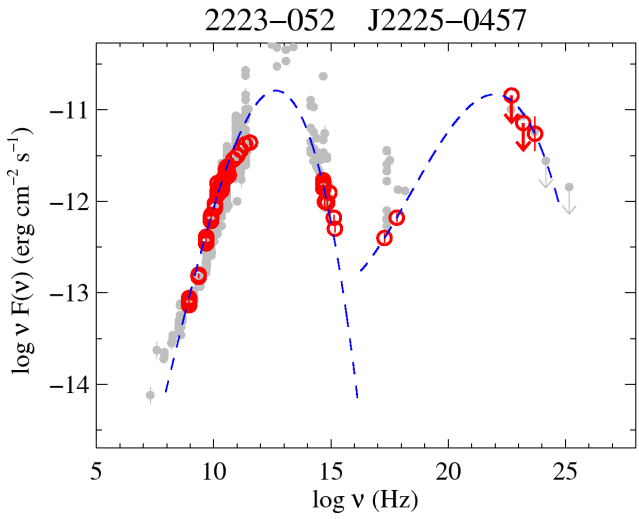}
\includegraphics[scale=0.8]{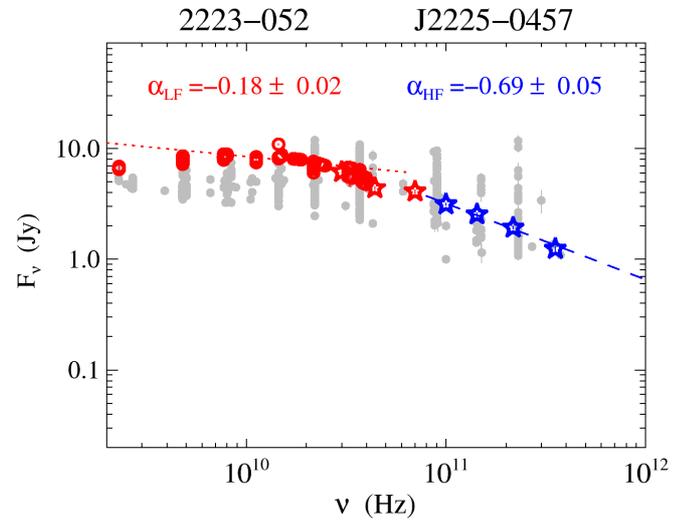}
 \caption{2223$-$052}
\end{figure*}

\begin{figure*}
\includegraphics[scale=0.8]{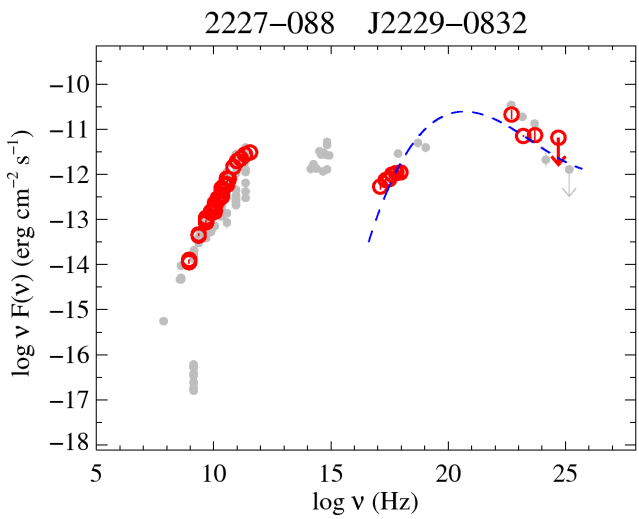}
\includegraphics[scale=0.8]{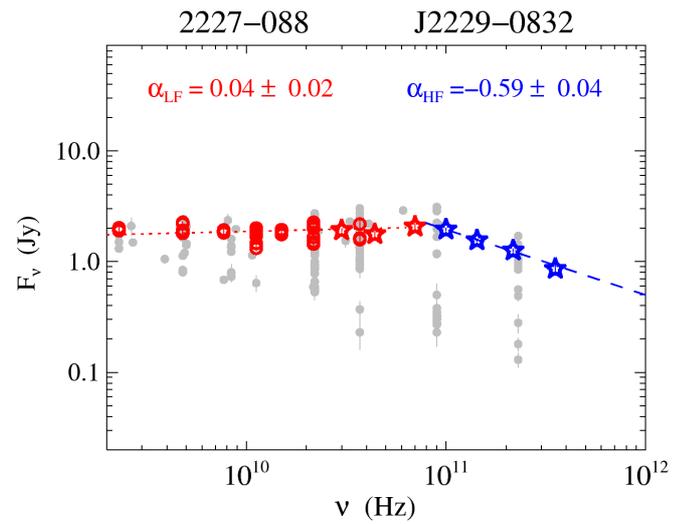}
 \caption{2227$-$088}
\end{figure*}
 
 \clearpage
 
\begin{figure*}
\includegraphics[scale=0.8]{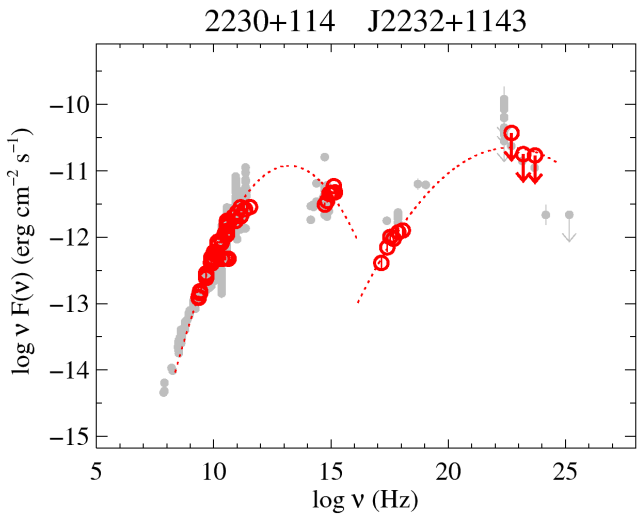}
\includegraphics[scale=0.8]{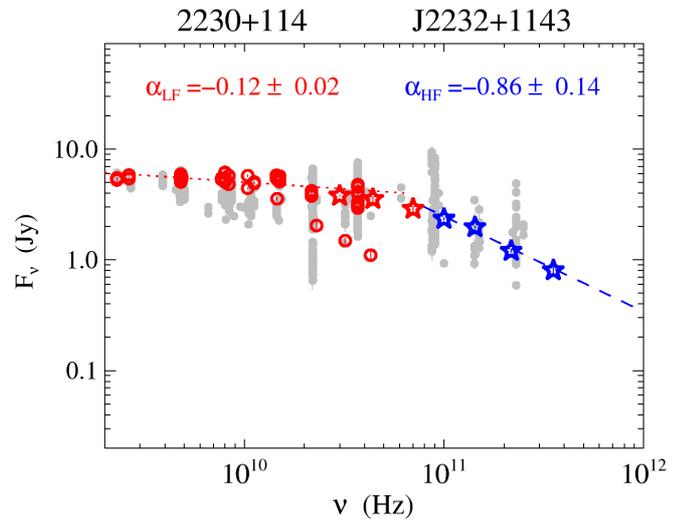}
 \caption{2230+114}
\end{figure*}

\begin{figure*}
\includegraphics[scale=0.8]{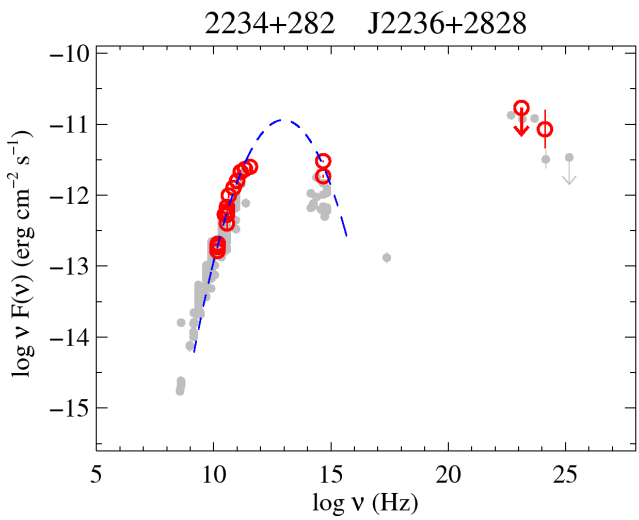}
\includegraphics[scale=0.8]{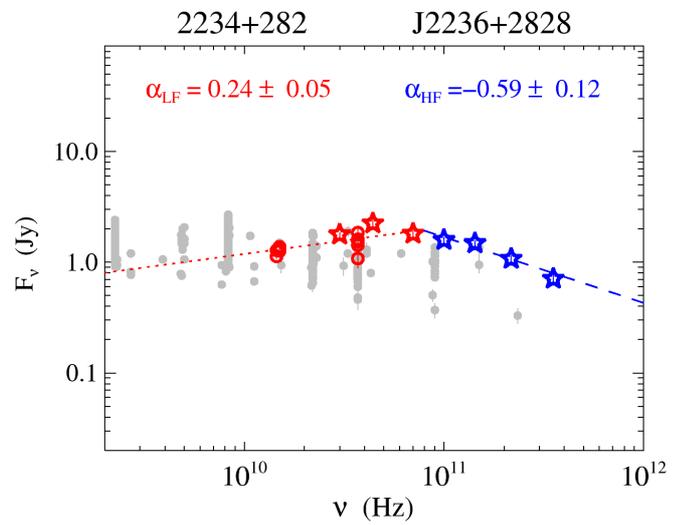}
 \caption{2234+282}
\end{figure*}

\begin{figure*}
\includegraphics[scale=0.8]{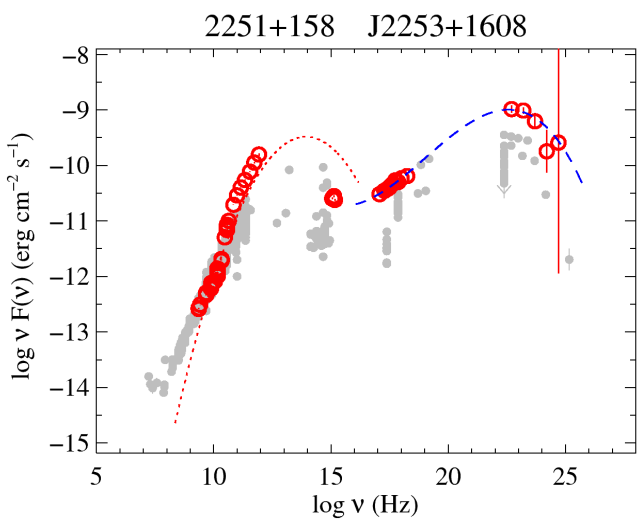}
\includegraphics[scale=0.8]{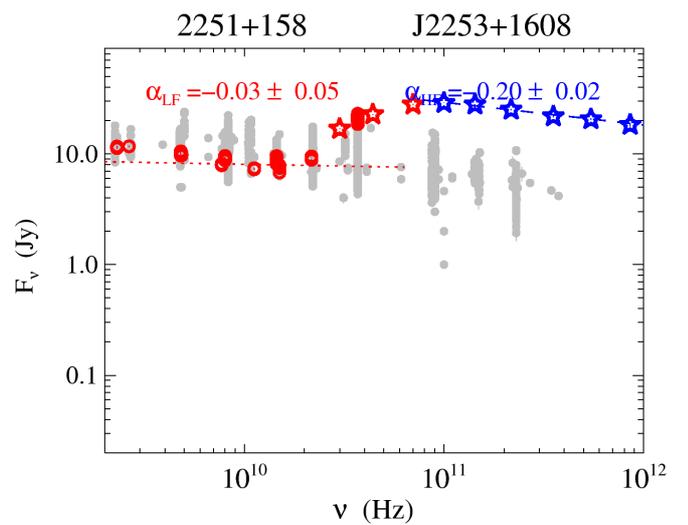}
 \caption{2251+158}
\label{2251_sed}
\end{figure*}
 
 \clearpage
 
\begin{figure*}
\includegraphics[scale=0.8]{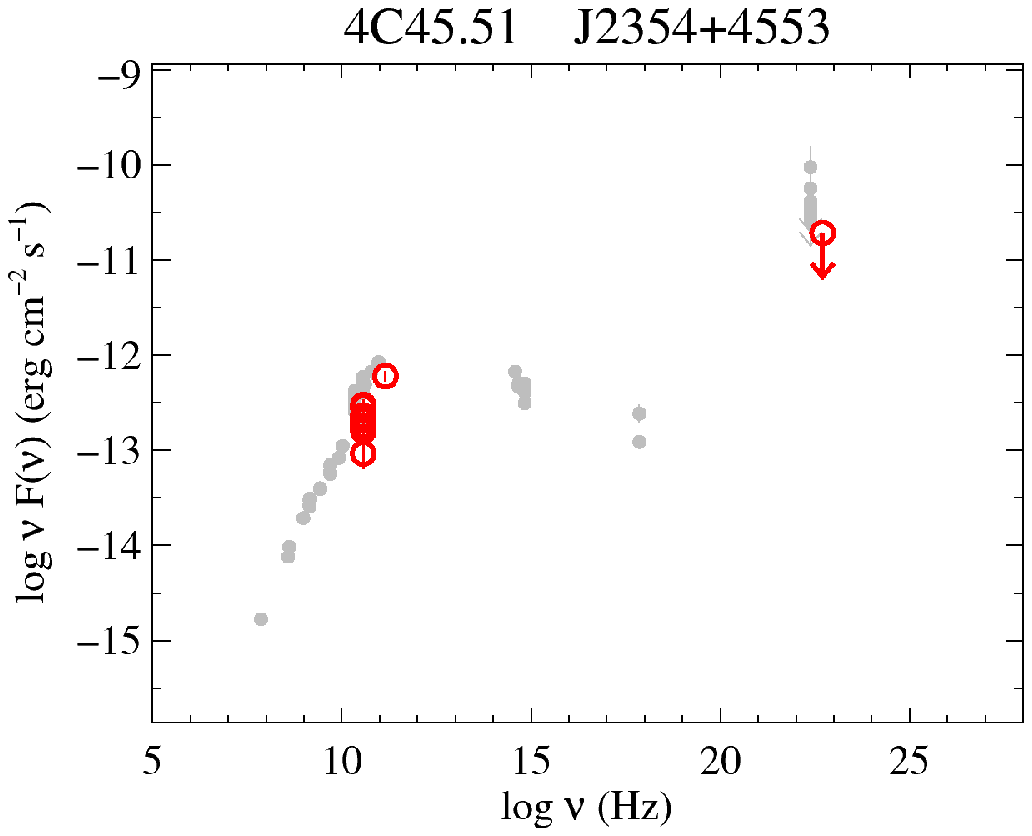}
\includegraphics[scale=0.8]{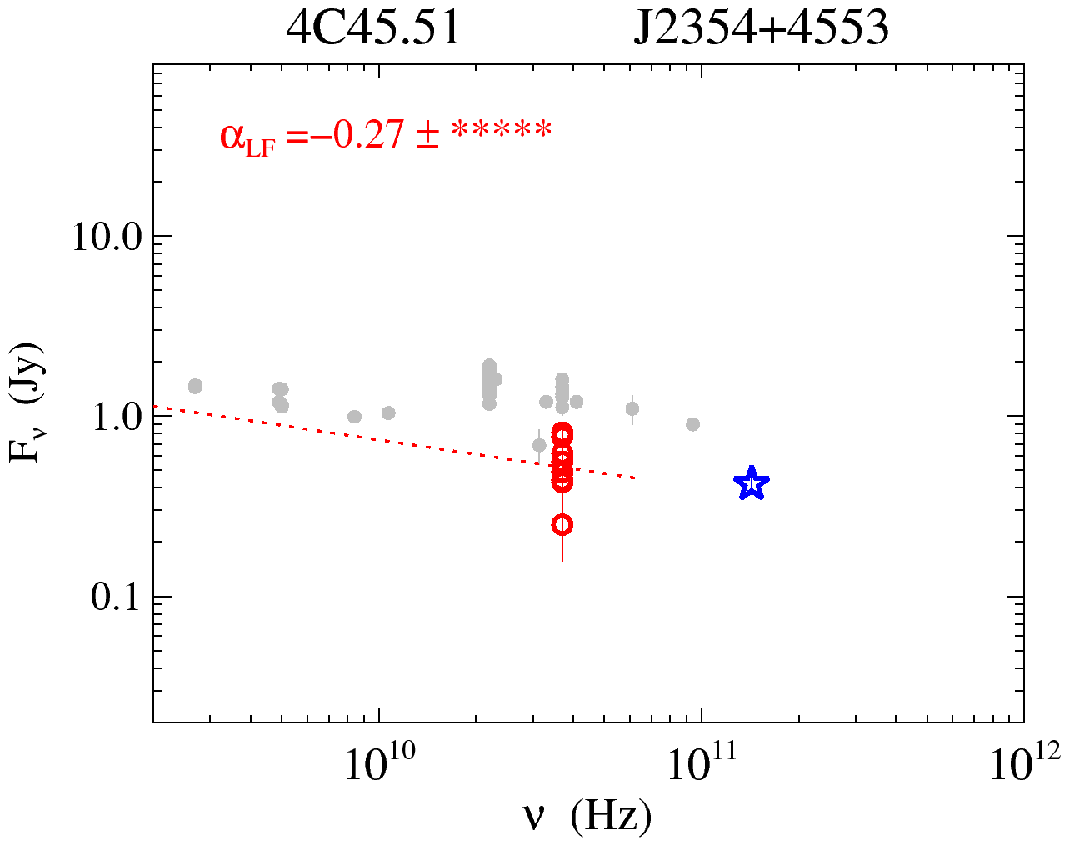}
 \caption{4C45.51}
\end{figure*}

\begin{figure*}
\includegraphics[scale=0.8]{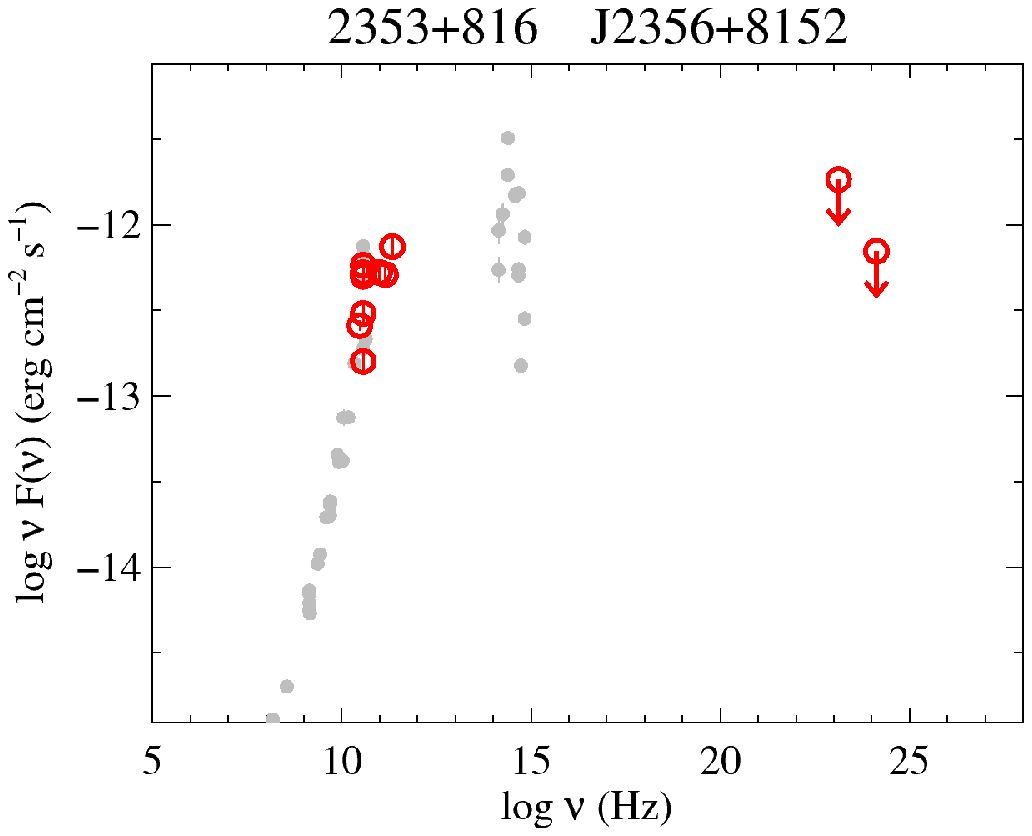}
\includegraphics[scale=0.8]{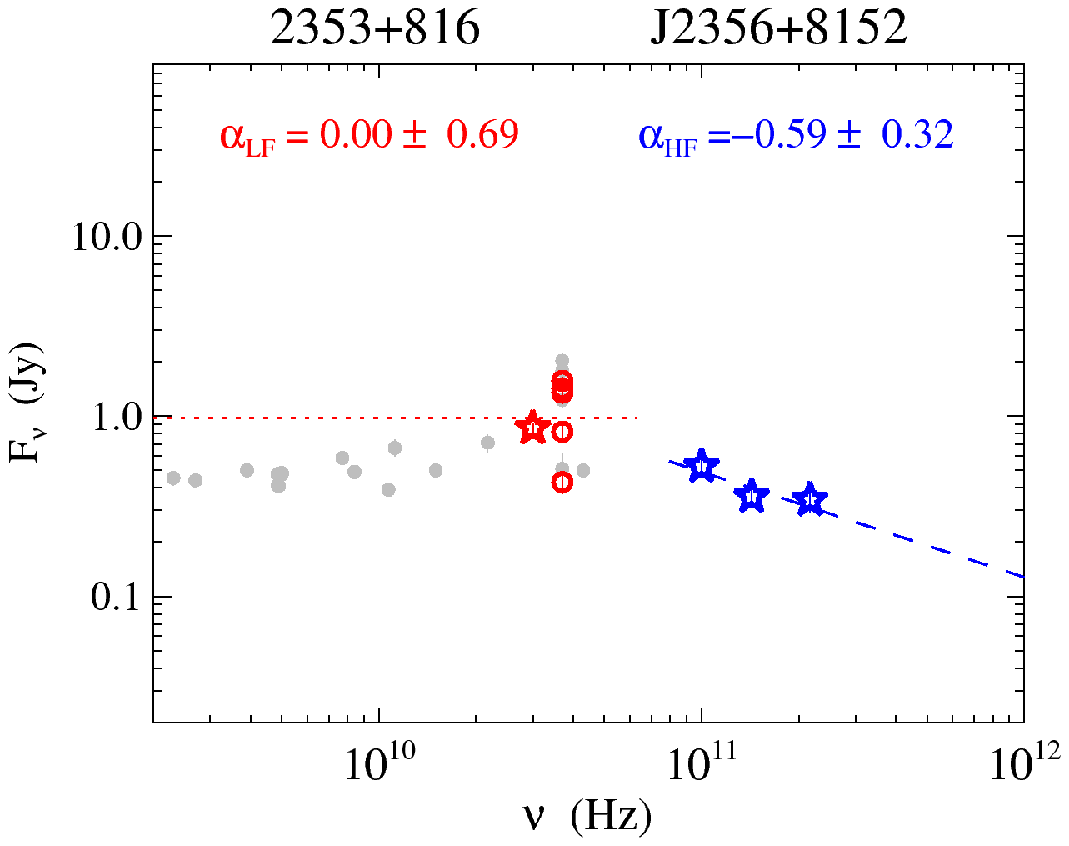}
 \caption{2353+816}
\label{last}
\end{figure*}

\end{document}